\documentclass[aps,preprint,nofootinbib,preprintnumbers,eqsecnum,superscriptaddress]{revtex4}
\usepackage{color}

\usepackage[normalem]{ulem}
\usepackage{enumerate}
\usepackage{amsfonts}
\usepackage{yfonts}
\usepackage{tcolorbox}
\usepackage{subfigure}
\usepackage{psfrag}

\usepackage[flushleft]{threeparttable}
\usepackage{lscape}
\definecolor{Mygrey}{gray}{0.8}
\definecolor{Mywhite}{gray}{1.0}
\usepackage{epsfig}
\usepackage[latin1]{inputenc}
\usepackage{float}
\usepackage{graphicx}
\usepackage{cancel}
\usepackage{mathrsfs}
\usepackage{amssymb}
\usepackage{amsfonts}
\usepackage{amsmath}
\usepackage{slashed}

\usepackage{braket}

\usepackage{graphicx}
\usepackage{bm}

\usepackage{bbold}

\usepackage{svg}

\usepackage[colorlinks=true,
      linkcolor=black,
      urlcolor=blue,
      filecolor=black,
      citecolor=red]{hyperref}

\def\({\left(}
\def\){\right)}
\def\[{\left[}
\def\]{\right]}
\def\<{\langle}
\def\>{\rangle}
\newcommand{\bmat}{\begin{bmatrix}}
	\newcommand{\emat}{\end{bmatrix}}

\newcommand{\be}{\begin{equation}}
	\newcommand{\ee}{\end{equation}}
\newcommand{\bea}{\begin{eqnarray}}
	\newcommand{\eea}{\end{eqnarray}}
\newcommand{\bwt}{\begin{widetext}}
	\newcommand{\ewt}{\end{widetext}}

\newcommand{\bi}{\begin{itemize}}
	\newcommand{\ei}{\end{itemize}}
\newcommand{\ben}{\begin{enumerate}}
	\newcommand{\een}{\end{enumerate}}
\newcommand{\bca}{\begin{cases}}
	\newcommand{\eca}{\end{cases}}
\newcommand{\bln}{\begin{align}}
	\newcommand{\eln}{\end{align}}
\newcommand{\bst}{\begin{split}}
	\newcommand{\est}{\end{split}}

\newcommand{\bchi}{\bar{\chi}}

\def\oversortoftilde#1{\mathop{\vbox{\m@th\ialign{##\crcr\noalign{\kern3\p@}%
      \sortoftildefill\crcr\noalign{\kern3\p@\nointerlineskip}%
      $\hfil\displaystyle{#1}\hfil$\crcr}}}\limits}

\def\sortoftildefill{$\m@th \setbox\z@\hbox{$\braceld$}%
  \braceld\leaders\vrule \@height\ht\z@ \@depth\z@\hfill\braceru$}

\usepackage{bbold}

\usepackage{multirow}
\usepackage[normalem]{ulem}
\useunder{\uline}{\ul}{}

\usepackage{scalerel,stackengine}
\stackMath
\newcommand\reallywidehat[1]{%
\savestack{\tmpbox}{\stretchto{%
  \scaleto{%
    \scalerel*[\widthof{\ensuremath{#1}}]{\kern-.6pt\bigwedge\kern-.6pt}%
    {\rule[-\textheight/2]{1    ex}{\textheight}}%WIDTH-LIMITED BIG WEDGE
  }{\textheight}% 
}{0.5ex}}%
\stackon[1pt]{#1}{\tmpbox}%
}
\usepackage{tcolorbox}

\allowdisplaybreaks

\newmuskip\pFqmuskip
\newcommand*\pFq[6][8]{%
  \begingroup % only local assignments
  \pFqmuskip=#1mu\relax
  \mathcode`\,=\string"8000
  \begingroup\lccode`\~=`\,
  \lowercase{\endgroup\let~}\pFqcomma
  {}_{#2}F_{#3}{\left[\genfrac..{0pt}{}{#4}{#5};#6\right]}%
  \endgroup
}
\newcommand{\pFqcomma}{\mskip\pFqmuskip}

\usepackage{tikz}
\tikzset{Witten diagram/.style={execute at begin picture={%
\draw[blue,fill=blue!20] circle[radius=\pgfkeysvalueof{/tikz/Witten/radius}];
\path node (X){\phantom{X}};
},baseline={(X.base)}},vertex/.style={circle,fill,inner sep=1.5pt,node
contents={}},
Witten/.cd,radius/.initial=1.5cm}

\usepackage{tikz-cd}
\usetikzlibrary{cd}
\newcommand{\ndiv}{\hspace{-4pt}\not|\hspace{2pt}}

\begin{document} 
\title{Two- \& Three-character solutions to MLDEs and Ramanujan-Eisenstein Identities for Fricke Groups}
	\author{Arpit Das}
	\email{arpit.das@durham.ac.uk}
	\affiliation{Centre for Particle Theory, Department of Mathematical Sciences, Durham University,
		South Road, Durham DH1 3LE, UK\\}
	\author{Naveen Balaji Umasankar}
	\email{naveen.umasankar@yale.edu}
	\affiliation{Department of Physics, Yale University, 217 Prospect St, New Haven, CT 06511}
        %\affiliation{University of Amsterdam, Institute for Theoretical Physics (ITFA), Science Park 904, 1098 XH Amsterdam.}
	
	\begin{abstract}
		\noindent In this work we extend the study of \cite{Umasankar:2022kzs} by investigating two- and three-character MLDEs for Fricke groups at prime levels. We have constructed these higher-character MLDEs by using a \textit{novel} Serre-Ramanujan type derivative operator which maps $k$-forms to $(k+2)$-forms in $\Gamma^{+}_0(p)$, for prime $p\in\mathbb{N}$. We found that this \textit{novel} derivative construction enabled us to write down a general prescription for obtaining \textit{Ramanujan-Eisenstein} identities for these groups. We discovered several novel single-, two-, and three-character admissible solutions for Fricke groups at levels $2$ and $3$ after solving the MLDEs among which we have realized some in terms of Mckay-Thompson series and others in terms of modular forms of the corresponding Hecke groups. Among these solutions, we have identified interesting non-trivial bilinear identities. Furthermore, we could construct \textit{putative} partition functions for these theories based on these bilinear pairings, which could have a range of lattice interpretations. We also present and discuss modular re-parameterization of MLDE and their solutions for Fricke groups of prime levels. 		
	\end{abstract}
	\vfill
	\today
	
	\maketitle
	\tableofcontents
 
\section{Introduction}
\noindent A 2D rational conformal field theory, RCFT, has a finite set of holomorphic characters, denoted by $\chi_i(\tau)$ and a partition function of the form \cite{Anderson:1987ge, Moore:1988qv, Mathur:1988na, Schellekens:1996tg, Hampapura:2015cea, Das:2022uoe}:
\be\label{eq1}
Z(\tau,\Bar{\tau})=\sum_{i,j=0}^{n-1}M_{ij}\,\bchi_i(\Bar{\tau})\chi_j(\tau) = |\chi_0|^2 + \sum_{i=1}^{n-1}Y_i|\chi_i|^2
\ee
where $n\in\mathbb{Z}$ labels the number of linearly independent characters. These characters are vector-valued modular forms (vvmfs) (see \cite{Bantay:2005vk, Mason:2007, Gannon:2013jua}), and we refer to $n$ as the \textit{dimension} of the vvmf. When $n=1$, there is only the identity character. In this case, we will refer to the resulting theory as a meromorphic CFT \footnote{In a meromorphic CFT, the partition function can be modular invariant up to a phase.}.
\\
\\
An RCFT classification program, called the Mathur-Mukhi-Sen (MMS) program, initiated in \cite{Mathur:1988rx, Mathur:1988na, Mathur:1988gt} is pursued by both mathematicians and physicists in recent times \cite{Naculich:1988xv, Kiritsis:1988kq, Bantay:2005vk, Mason:2007, Mason:2008, Bantay:2007zz, Tuite:2008pt, Bantay:2010uy, Marks:2011, Gannon:2013jua, Kawasetsu:2014, Hampapura:2015cea, Franc:2016, Gaberdiel:2016zke, Hampapura:2016mmz, Arike:2016ana, Tener:2016lcn, Mason:2018, Harvey:2018rdc, Chandra:2018pjq, Chandra:2018ezv,Bae:2018qfh,Bae:2018qym,franc2020classification,Bae:2020xzl, Mukhi:2020gnj, Kaidi:2020ecu, Das:2020wsi, Das:2021uvd, Mukhi:2022bte, Das:2022slz, Das:2022uoe, Kaidi:2021ent,Bae:2021mej,Duan:2022ltz, Rayhaun:2023pgc, Gowdigere:2023xnm, Pan:2023jjw}. This classification is based on the fact that characters are vvmfs of weight 0 and that they solve modular linear differential equations \cite{Mathur:1988na},
\begin{align}\label{mlde00}
    \left(\mathcal{D}^n + \sum\limits_{r=0}^{n-1}\phi_r\mathcal{D}^r\right)\chi(\tau) = 0
\end{align}
where, $\mathcal{D}_k\equiv \frac{\partial_\tau}{2\pi i} - \frac{k}{12}E_2(\tau) = q\partial_q - \frac{k}{12}E_2(q)$ is the {\it Serre-Ramanujan} derivative for $\text{SL}(2,\mathbb{Z})$ which maps weight $k$-forms to weight $(k+2)$-forms. The characters $\chi(\tau)$s above are weight $0$ modular functions\footnote{The difference between modular forms and modular functions is that, modular forms are holomorphic in the upper half plane while modular functions are allowed to be meromorphic.}. $\phi_r$ above are weight $2(n-r)$ modular functions. Note that, $E_2(\tau)$ is the weight 2 Eisenstein series of $\text{SL}(2,\mathbb{Z})$ and $\tau$ lives in the upper half plane. Furthermore, $q\equiv e^{2\pi i\tau}$ maps the upper half plane to the unit disk. A review of modular forms can be found in Sec. \ref{pri}.

\noindent Eq. \eqref{mlde00} is called a Modular Linear Differential Equation (MLDE). Like ordinary differential equations (ODEs), an MLDE is characterized by its order and the structure of poles of its coefficient functions. The order of an MLDE also equals the number of linearly independent solutions of it or in the language of RCFTs -- the number of linearly independent characters. The pole structure of the MLDE is captured by the poles of the coefficient functions, $\phi_r$, in the upper half-plane or by the zeros of the Wronskian\footnote{The Wronskian of the MLDE is defined in the same way as that of an ODE except here the ordinary derivative operator is replaced by the {\it Serre-Ramanujan} derivative operator.}. It turns out that the zeros of the Wronskian in the upper half plane are captured by an integer, $\ell$, called the Wronskian index \cite{Mathur:1988na}.

\noindent Equations of the form given in \eqref{mlde00} have finitely many parameters and these can be varied to scan for solutions that satisfy the basic criteria to be those of an RCFT. These criteria correspond to the fact that each character is holomorphic in $q=e^{2\pi i\tau}$ except as $q\to 0$, and have an expansion of the form:
\be
\chi_i(\tau)=q^{\alpha_i}\left(a_{0,i}+a_{1,i} \,q+a_{2,i}\,q^2+\cdots\right)
\label{charexp}
\ee
If the vvmf corresponds to a genuine RCFT then these exponents can be identified with the central charge and (chiral) conformal dimensions via:
\be
\alpha_i=-\frac{c}{24}+h_i
\ee
with $h_0=0$ corresponding to the identity character of the RCFT. Using, what is called, the valence formula for modular forms of $\text{SL}(2,\mathbb{Z})$, we can relate the Wronskian index $\ell$ to the exponents above as \cite{Mathur:1988na},
\begin{align}\label{RRthm}
    \sum\limits_{i=0}^{n-1} = \frac{n(n-1)}{12} - \frac{\ell}{6}
\end{align}
The coefficients $a_{r,i},r\ge 1$, in \eqref{charexp}, should be non-negative integers for some choice of a positive integer $a_{0,i}$ that provides the overall normalization of each character. We must choose $a_{0,0}=1$ (non-degeneracy of the vacuum or uniqueness of the ground state), while for $i\ne 0$ we take $a_{0,i}$ to be the minimum integer such that the $q$-series for the corresponding character has integral coefficients. In \cite{Chandra:2018pjq} character sets with the above properties were called \textit{admissible}. It is not, in general, the case that admissible characters correspond to an RCFT\footnote{If one looks at $(1,6)$ MLDE then as solutions to this equation, there exists an infinite set of admissible solutions labeled by an integral parameter $\mathcal{N}$ where $\mathcal{N}\geq -744$. The admissible solutions are of the form $j+\mathcal{N}$ where $j$ is the Klein-j invariant. However, from this infinite set only $71$ values of $\mathcal{N}$ correspond to genuine RCFTs (see \cite{Schellekens:1992db}).}. We also define $m_i\equiv a_{i,0}$ and note that for a CFT, $m_1$, corresponds to the number of spin-1 generators (Kac-Moody currents) in the chiral algebra. Also for $i\ne 0$, we write $D_i=a_{0,i}$.  In a CFT these are the ground-state degeneracies of the various modules other than the identity. 
\\
In \cite{Umasankar:2022kzs}, the study of MLDEs was extended to some congruence subgroups of $\text{SL}(2,\mathbb{Z})$ and also some other similar groups where in particular, Fricke and Hecke groups were studied since their theory of modular forms is well understood. Riemann-Roch relations were studied for Fricke groups (of prime divisor levels of the Monster group) following which admissible single-character solutions were reported. This paper extends the work by presenting a more exhaustive analysis of the single-character solutions, and reports on admissible two- and three-character solutions to MLDEs in $\Gamma_{0}^{+}(2)$ and $\Gamma_{0}^{+}(3)$.\\
\\
The main results of the paper are as follows:
\begin{itemize}
    \item Detailed set up of Serre-Ramanujan covariant derivatives and MLDEs for Fricke and Hecke groups.
    \item Ramanujan-Eisenstein identities for Fricke groups $\Gamma_{0}^{+}(p)$, for $p = 2,3,5,7,11$.
    \item Modular re-parameterization schemes for Fricke groups for two- and three-characters. 
    \item Admissible two-character solutions for modular forms in the Fricke groups $\Gamma_{0}^{+}(p)$ for $p = 2,3$ up to the {\it bulk point}\footnote{Here, {\it bulk point} refers to the value of the Wronskian index $\ell$, at which the Wronskian of the corresponding MLDE develops a full zero in the upper-half plane.}.
    \item Admissible three-character solutions to the $(3,0)$ MLDE in the Fricke groups $\Gamma_{0}^{+}(p)$ for $p = 2,3$, in the sense of Sec. 3.4 of \cite{Franc:2016}.
    \item Identification of numerous bilinear identities among the two-character solutions.
    \item Construction of some \textit{putative} partition functions for two-character solutions.
    \item Identification of certain Fricke two-character solutions with sequences of the conjugacy classes of the Monster group and as closed-form expressions in terms of modular forms of the corresponding Hecke group.
    \item Generalization to the relation between $\Theta$-series of the odd Leech lattice and $\Theta_{2^{+}}$ that corresponds to the series obtained via a generalized single-character solution.
\end{itemize}

\noindent
\textbf{Outline}\\

\noindent In the next section, we provide a quick introduction to modular forms for physicists that comprises basic definitions and motivations for certain foundational concepts. This section also includes an overview of the necessary definitions of the Fricke groups of levels $p = 2,3$ required for reading the paper. In section \ref{sec:Two-character_Hecke_Fricke}, we provide a general prescription to construct Serre-Ramanujan derivatives for Fricke and Hecke groups following which we present the re-parameterized MLDEs at the two-character levels for Fricke groups of levels $p = 2,3, 5, 7$ and at the three-character level for level $p =2,3$. In sections \ref{sec:Gamma_0_2+} and \ref{sec:Gamma_0_3+}, we present an exhaustive single-character analysis and report results on admissible two- and three-character solutions for $\Gamma_{0}^{+}(2)$ and $\Gamma_{0}^{+}(3)$ respectively. In section \ref{sec:Lattice}, we generalize the lattice relation found in $\Gamma_{0}^{+}(2)$ and present some interesting lattice theta function realizations of modular forms of Fricke and Hecke groups. Section \ref{sec:Discussion} expands on some interesting results found in the two- and three-character analysis. In section \ref{sec:Future_work}, we finish by providing a detailed list of interesting questions worth tackling in future work. All fundamental domains were obtained using the Mathematica program in \cite{kainberger}. Several appendices complement the theory and results of the main sections.

\section{Preliminaries}\label{pri}
\subsection{Lighting review of the theory of modular forms}
\subsubsection{The modular group and its cousins}
\noindent We define the special linear group over the integers $\text{SL}(2,\mathbb{Z})$ as the set of all $2\times 2$ matrices with real entries and unit determinant, i.e.
\begin{equation}
    \text{SL}(2,\mathbb{Z}) = \left\{\left.\begin{pmatrix}
    a & b\\ c& d
    \end{pmatrix}\right\vert a,b,c,d\in\mathbb{Z},ad-bc=1\right\}.
\end{equation}
The projective special linear group  over the integers (or the modular group $\text{PSL}(2,\mathbb{Z})$) on the other hand, is nothing but the special linear group quotiented by $\mathbb{Z}_{2}$, i.e.
\begin{equation}
    \text{PSL}(2,\mathbb{Z}) = \text{SL}(2,\mathbb{Z})/\mathbb{Z}_{2} = \left\{\left.\begin{pmatrix}
    a & b\\ c& d
    \end{pmatrix}\right\vert a,b,c,d\in\mathbb{Z},ad-bc=1\right\}/\{\pm 1\}.
\end{equation}
We loosely refer to both $\text{PSL}(2,\mathbb{Z})$ and $\text{SL}(2,\mathbb{Z})$ as the modular group since they are equal up to a quotient $\mathbb{Z}_{2}$. Now let us see why we need to consider the projective special linear group $\text{PSL}(2,\mathbb{Z})$). For this consider the action of the an element $\gamma = \left(\begin{smallmatrix}a & b\\ c & d \end{smallmatrix}\right)\in\text{SL}(2,\mathbb{Z})$ on the Riemann sphere $\widehat{\mathbb{C}}\equiv \mathbb{C}\cup \{i\infty\}$. This action is defined by a M\"obius transform (or a fractional linear transformation) given by, 
\begin{align}\label{action_SL2Z}
    \begin{split}
    \gamma(\tau) \equiv& \frac{a\tau + b}{c\tau + d},\\
    \gamma(\infty) =& \lim\limits_{y\to \infty}\frac{a(x+iy) + b}{c(x+iy) + d} = \begin{cases}
    \frac{a}{c},\ c\neq 0,\\
    0,\ c=0.
    \end{cases}
    \end{split}
\end{align}
Now note that the fractional linear transformation given by some matrix $\gamma\in\text{SL}(2,\mathbb{Z})$ is equivalent to that given by $-\gamma$. Indeed, we have,
\begin{align}
    \gamma(\tau) = \frac{a\tau + b}{c\tau + d} = \frac{-a\tau-b}{-c\tau-d} = (-\gamma)(\tau), \label{psl2z_rel}
\end{align}
Thus, we do not have a direct correspondence between fractional linear transformations and elements of $\text{SL}(2,\mathbb{Z})$. This, therefore, suggests defining a group where we quotient out the sign of the matrix from $\text{SL}(2,\mathbb{Z})$. This way, we have a natural one-to-one correspondence between elements of $\text{PSL}(2,\mathbb{Z})$ and fractional linear transformations. The modular group is comprised of the following three matrices
\begin{align}
    I = \mathbb{1}_2,\ S = \begin{pmatrix}
    0 & -1\\ 1 & 0
    \end{pmatrix},\ T = \begin{pmatrix}
    1 & 1\\ 0 & 1
    \end{pmatrix},
\end{align}
where $\mathbb{1}_{2}$ is the $2\times 2$ identity matrix. We denote the discrete subgroups of $\text{SL}(2,\mathbb{Z})$ as $\Gamma$ with the convention that $\Gamma(1) = \text{SL}(2,\mathbb{Z})$. The special set of discrete subgroups of our interest are the principal congruence subgroups of level $N$ denoted by $\Gamma(N)$, where $N\in\mathbb{N}$, defined as follows
\begin{align}
    \begin{split}
        \Gamma(N) =& \{\gamma\in\Gamma(1)\vert\ \gamma = \mathbb{1}_{2}\ (\text{mod}\ N)\}\\
        =& \left.\left\{\begin{pmatrix}
    a & b\\ c & d
    \end{pmatrix}\in\text{SL}(2,\mathbb{Z})\right\vert a, d\equiv 1\  (\text{mod}\ N),\ b, c\equiv 0\  (\text{mod}\ N)\right\},
    \end{split} 
\end{align}
The index of principal congruence subgroup $\Gamma(N)$ in $\text{SL}(2,\mathbb{Z})$ refers to the number of (left or right cosets) of $\Gamma(N)$ in $\text{SL}(2,\mathbb{Z})$ and is defined as follows
\begin{align}
    \mu \equiv \left[\text{SL}(2,\mathbb{Z}):\Gamma(N)\right]= N^{3}\prod\limits_{p\vert N}(1-p^{-2}),
\end{align}
where $p\vert N$ denotes the prime divisors of $N$. Every congruence subgroup $\Gamma$ has a finite index in $\text{SL}(2,\mathbb{Z})$. There exist several other subgroups of $\Gamma$ that are defined below
\begin{align}\label{congruence_subgroup_definitions}
    \begin{split}
    \Gamma_{0}(N) =& \left\{\left.\begin{pmatrix}
    a & b\\ c & d
    \end{pmatrix}\in\text{SL}(2,\mathbb{Z})\right\vert c\equiv 0\ (\text{mod}\ N)\right\} = \left\{\begin{pmatrix}
    * & *\\ 0 & *
    \end{pmatrix}\ \text{mod}\ N\right\},\\
    \Gamma^{0}(N) =& \left\{\begin{pmatrix}
    * & 0\\ * & *
    \end{pmatrix}\ \text{mod}\ N\right\},\ \ \  \Gamma_{0}^{0}(N) = \left\{\begin{pmatrix}
    * & 0\\ 0 & *
    \end{pmatrix}\ \text{mod}\ N\right\},\\
    \Gamma_{1}(N) =& \left\{\begin{pmatrix}
    1 & *\\ 0 & 1
    \end{pmatrix}\ \text{mod}\ N\right\},\ \ \ 
    \Gamma^{1}(N) = \left\{\begin{pmatrix}
    1 & 0\\ * & 1
    \end{pmatrix}\ \text{mod}\ N\right\},
    \end{split}
\end{align}
where ``$*$'' stands for ``unspecified''. It follows from the definitions that $\Gamma(N)\subset\Gamma_{1}(N)\subset\Gamma_{0}(N)\subset\text{SL}(2,\mathbb{Z})$, and analogously, $\Gamma(N)\subset\Gamma^{1}(N)\subset\Gamma^{0}(N)\subset\text{SL}(2,\mathbb{Z})$. We shall refer to groups $\Gamma_{0}(N)$ as Hecke groups throughout this paper\footnote{There is a small subtlety in the naming we are ignoring here. A Hecke group, $G(\lambda)$, with $\lambda>0$, is a subgroup of $\text{SL}(2,\mathbb{R})$, where $\lambda = 4\cos^{2}\left(\tfrac{\pi}{p}\right)$. These groups come in two flavors- the arithmetic and the non-arithmetic with the former being the ones where $\lambda\in\mathbb{Z}$ that correspond to the Hecke congruence groups $\Gamma_{0}(\lambda)$ and the latter being the ones with $\lambda\not\in\mathbb{Z}$.}. Since $S^{-1}\left(\begin{smallmatrix} a & b\\ c & d\end{smallmatrix}\right)S = \left(\begin{smallmatrix} d&-c\\-b&a\end{smallmatrix}\right)$, we can write $\Gamma^{0}(N) = S^{-1}\Gamma_{0}(N)S$, i.e. $\Gamma_{0}(N)$ and $\Gamma^{0}(N)$ are conjugate groups of $\text{SL}(2,\mathbb{Z})$. From the definition \ref{congruence_subgroup_definitions}, we also see that we can write $\Gamma^{0}_{0}(N) = \Gamma_{0}(N)\cap\Gamma^{0}(N)$. The index of the Hecke group in the modular group is given by
\begin{align}\label{index_Hecke}
    \mu_{0} = \left[\text{SL}(2,\mathbb{Z}):\Gamma_{0}(N)\right]= N\prod\limits_{p\vert N}(1+p^{-1}).
\end{align}
For $p\in\mathbb{P}$, the right cosets of the Hecke group in $\text{SL}(2,\mathbb{Z})$ are given by
\begin{align}\label{right_cosets}
    \left\{\Gamma_{0}(p)\right\}\cup \left\{\Gamma_{0}(p)ST^{k}:\ 0\leq k<p\right\}.
\end{align}
For a positive integer $N$, the Fricke group denoted by $\Gamma^{+}_{0}(N)$ is a supergroup of the Hekce group defined as follows
\begin{equation}\label{Fricke}
    \Gamma_{0}^{+}(N) \equiv \Gamma_{0}(N)\cup \Gamma_{0}(N)W_{N},
\end{equation}
where $W_{N}$, am element of  is called the Fricke involution with the following action
\begin{align}\label{Fricke_involution}
   \begin{split}
    W_{N} = \frac{1}{\sqrt{N}}\begin{pmatrix}
    0 & -1\\ N & 0
   \end{pmatrix},\\
   W_{N}:\tau\mapsto -\frac{1}{N\tau},
   \end{split}
\end{align}
with $W_{N}^{4} = \mathbb{1}_{2}$ and $W_{N}^{2} = \left(TW_{N}\right)^{2N} = -\mathbb{1}_{2}$ for $N\in\mathbb{Z}_{+}$. From the definition above, we see that special points on the fundamental domain of the Fricke groups are mapped to themselves. We refer to these points as Fricke involution points defined as $\tau_{F} = \tfrac{i}{\sqrt{N}}$, where $N$ corresponds to the level of the group. The Fricke involution is a special case of the Atkin-Lehner involution defined as follows
\begin{align}
    A_{l} = \frac{1}{\sqrt{l}}\begin{pmatrix}
    l\cdot a & b\\ N c& l\cdot d
  \end{pmatrix},
\end{align}
for $a,b,c,d\in\mathbb{Z}$ such that $\text{det}\ A_{l} =1$, and where the divisor $l$ and $\tfrac{N}{l}$ are coprime. The group generated by a Hecke group of level $N$ and the $A_{l}$ involution is called the Atkin-Lehner group, denoted by $\Gamma_{0}^{*}(N)$.  Unlike the Hecke groups, the Fricke and Atkin-Lehner groups are not subgroups of $\text{SL}(2,\mathbb{Z})$. When $N\in\mathbb{P}$ and when $N\leq 5$, the Atkin-Lehner groups are equal to the Fricke groups, i.e. $\Gamma_{0}^{*}(N) = \Gamma_{0}^{+}(N)$. This distinction becomes important when we consider non-prime groups of higher levels in subsequent sections. Finding $\mu_{0}^{+}$, the Fricke group index is a bit involved and hence, we only present the results in this work when relevant and direct the reader to \cite{Umasankar:2022kzs} for the specifics. %Lastly, it is interesting to note that since $\Gamma_{0}^{+}(N)$ contains the elements $T$ and $W_{N}$ for any $N$, we find that $-\mathbb{1}_{2}$, $T$ and $W_{N}TW_{N}^{-1}$ generate the Hecke group at level $N = 4$. 

\subsubsection{An invitation to modular forms}
\noindent The word ``modular" refers to the moduli space of complex curves (i.e. Riemann surfaces) of genus $1$. Such a curve can be represented as $\mathbb{C}/\Lambda$ where $\Lambda\subset\mathbb{C}$ is a lattice. Let us first consider the following definition of a modular form $f$ of weight $k$
\begin{equation}\label{modulardef}
    f\left(\frac{az+b}{cz+d}\right) = (cz+d)^{k}f(z),\ \ \text{Im}(z)>0,\ \begin{pmatrix}
    a & b\\ c & d
    \end{pmatrix}\in\text{SL}(2,\mathbb{Z}).
\end{equation}
We will soon return to this definition and make it more rigorous but for the moment let us use this as a working definition to get further along. Consider now the case when the weight is null, i.e. $k=0$. We have
\begin{equation}\label{modular_def_k=0}
    f\left(\frac{az+b}{cz+d}\right) = f(z).
\end{equation}
This tells us that the modular form $f$ is invariant under the action of $\text{SL}(2,\mathbb{Z})$. The implication of us considering $\text{Im}(z)>0$ is that the modular form $f$ is a function that is invariant under the upper-half plane $\mathbb{H}^{2}$. Formally, the upper-half plane is defined as follows
\begin{equation}
    \mathbb{H}^{2} = \{z\in\mathbb{C}\vert \text{Im}(z)>0\}  = \{x+iy\in\mathbb{C}\vert x,y\in\mathbb{R},y>0\}.
\end{equation}
But we are to ask ourselves why we should even be interested in such functions. The reason is that these are in fact functions of elliptic curves. Here, we have a function that goes from the set of elliptic curves to the complex numbers, i.e.
\begin{equation}
    f:\{E\}\to \mathbb{C}.
\end{equation}
Note that we have used $\{E\}$ indicating a set of elliptic curves as opposed to just one since in the latter case $f$ would be referred to as an elliptic function. Let $\omega_{1}, \omega_{2}$ be two periods. We now want to use these to probe for the set of all integer linear combinations, i.e. $n\omega_{1} + m\omega_{2}$ with $n,m\in\mathbb{Z}$. These linear combinations form a lattice $\Lambda$. Formally, a lattice in the complex plane $\mathbb{C}$ is a subgroup $\Lambda\subset\mathbb{C}$ of the following form
\begin{equation}
      \Lambda\equiv \langle\omega_{1},\omega_{2}\rangle = n\omega_{1} + m\omega_{2},
\end{equation}
where $m,n\in\mathbb{Z}$ and $\omega_{1},\omega_{2}\in\mathbb{C}$ are $\mathbb{R}$-linearly independent. An elliptic function is one that is periodic with respect to the lattice $\langle\omega_{1},\omega_{2}\rangle$, i.e. for a function $g(z)$, being elliptic implies 
\begin{equation}
    g(z) = g(\omega_{1} + z) = g(\omega_{2} + z).   
\end{equation}
Consider the function $g(\omega_{1},\omega_{2})$. Now, rescaling the lattices as $\omega_{1}\sim \lambda\omega_{1}$ and $\omega_{2}\sim \lambda\omega_{2}$ should not change the elliptic curve. Thus, we have
\begin{equation}
    g(\omega_{1},\omega_{2}) = g(\lambda\omega_{1},\lambda\omega_{2}),\ \lambda\in\mathbb{C}^{\times},
\end{equation}
where $\mathbb{C}^{\times} = \mathbb{C}\backslash\{0\}$ denotes the group units of the complex numbers $\mathbb{C}$. Two lattices $\Lambda$ and $\Tilde{\Lambda}$ are called homothetic if the $\Tilde{\Lambda} = \lambda\Lambda$ in which case, we write $\Tilde{\Lambda} \simeq \Lambda$. We can also consider the change 
\begin{align}\label{shift_omega}
    \begin{split}
        \omega_{1} \to \Tilde{\omega}_{1} =&  a\omega_{1} + b\omega_{2},\\
    \omega_{2}\to\Tilde{\omega}_{2} =&  c\omega_{1} + d\omega_{2},
    \end{split}
\end{align}
where $\left(\begin{smallmatrix}a & b\\ c & d\end{smallmatrix}\right)\in\text{GL}(2,\mathbb{Z})$ with $\text{det} = \pm 1$. We can now ask ourselves if the following claim holds
\begin{equation}\label{elliptic_claim}
    g(\Tilde{\omega}_{1},\Tilde{\omega}_{2})\overset{!}{=}g(\omega_{1},\omega_{2}).
\end{equation}
To simplify notation, we rid ourselves of two variables and just consider 
\begin{equation}
    f(z) = g(1,z),
\end{equation}
where we have chosen a basis with $\omega_{1} =1$ and $\omega_{2} = z$. We can reintepret $z$ to be $\tfrac{\omega_{2}}{\omega_{1}}$ and write
\begin{equation}
    g(\omega_{1},\omega_{2}) = f\left(\frac{\omega_{2}}{\omega_{1}}\right).
\end{equation}
With this map between $f$ and $g$, notice that the invariance \ref{modular_def_k=0} translates to the claim \ref{elliptic_claim} holding good. With the shift \ref{shift_omega}, we have
\begin{align}
    \begin{split}
    f\left(\frac{az+b}{cz+d}\right) =& f\left(\frac{a\omega_{1}+b\omega_{2}}{c\omega_{1} + d\omega_{2}}\right) = g(a\omega_{1} + b\omega_{2},c\omega_{2} + d\omega_{2},\omega_{1}),\\
    f(z) =& f\left(\frac{\omega_{2}}{\omega_{1}}\right) = g(\omega_{1},\omega_{2}),\\
    f\left(\frac{az+b}{cz+d}\right) =& f(z)\Longleftrightarrow g(\Tilde{\omega}_{1},\Tilde{\omega}_{2}) = g(\omega_{1},\omega_{2}).
    \end{split}
\end{align}
We can now search for functions $f$ that satisfy \ref{modular_def_k=0} such that these are holomorphic and not constant. As a first step, let us search for $1$-forms instead that satisfy
\begin{equation}
    f(z)dz = f\left(\frac{az+b}{cz+d}\right)d\left(\frac{az+b}{cz+d}\right).
\end{equation}
Working this out gives us the definition of a modular function of weight $2$,
\begin{equation}
    f\left(\frac{az+b}{cz+d}\right) = (cz+d)^{2}f(z).
\end{equation}
It is now natural to see how modular functions of weight $k$ are motivated. We look at forms of the type $f(z)(dz)^{\tfrac{k}{2}}$ and ask when this is left invariant. This leads us to the definition \ref{modulardef}. Consider now the forms $f_{1}(z)(dz)^{\tfrac{k}{2}}$ and $f_{2}(z)(dz)^{\tfrac{k}{2}}$. Since these are invariant by themselves, so should their ratio be and hence, we have
\begin{equation}
    \frac{f_{1}(z)(dz)^{\tfrac{k}{2}}}{f_{2}(z)(dz)^{\tfrac{k}{2}}} = \frac{f_{1}(z)}{f_{2}(z)},
\end{equation}
which is a ratio of two modular forms of the same weight. This is called a modular function which inhabits all the properties of the modular forms in the ratio. This tells us that in order to find invariant modular functions we are to probe for modular forms of similar weight and then take their ratio. We will call a function $h$ of weight $k$ to be a homogeneous function of a lattice $\Lambda$ if it satisfies
\begin{equation}\label{homo_function}
      h(\lambda\Lambda) = \lambda^{-k}h(\Lambda),\ \ \lambda\in\mathbb{C}^{\times},\ \Lambda\in\mathbb{C}.
\end{equation}
In terms of basis $\{\omega_{1},\omega_{2}\}$, the definition \ref{homo_function} reads
\begin{equation}
    h(\lambda\omega_{1},\lambda\omega_{2}) = \lambda^{-k}h(\omega_{1},\omega_{2}).
\end{equation}
Invariance under the shift \ref{shift_omega} demands
\begin{equation}
    h(a\omega_{1}+b\omega_{2}, c\omega_{1}+d\omega_{2}) = h(\omega_{1},\omega_{2}).
\end{equation}
Yet again, defining $f(z) = h(1,z)$ so that $h(\omega_{1},\omega_{2}) = f\left(\tfrac{\omega_{2}}{\omega_{1}}\right) = f(z)$, we obtain the functional equation \ref{modulardef} for $f(z)$. Thus, we have just seen the following three different ways to understand and define modular forms of weight $k$:
\begin{enumerate}
    \item They are functions that correspond to the functional equation \ref{modulardef}.
    \item They are functions that correspond to invariant modular forms $f(z)(dz)^{\tfrac{k}{2}}$, and
    \item They can be thought of as being homogeneous functions of lattices spanned by $\langle\omega_{1},\omega_{2}\rangle$ of weight $k$.
\end{enumerate}

\subsubsection{Formal definitions}
\begin{figure}[htb!]
    \centering
    \includegraphics[width = 13.5cm]{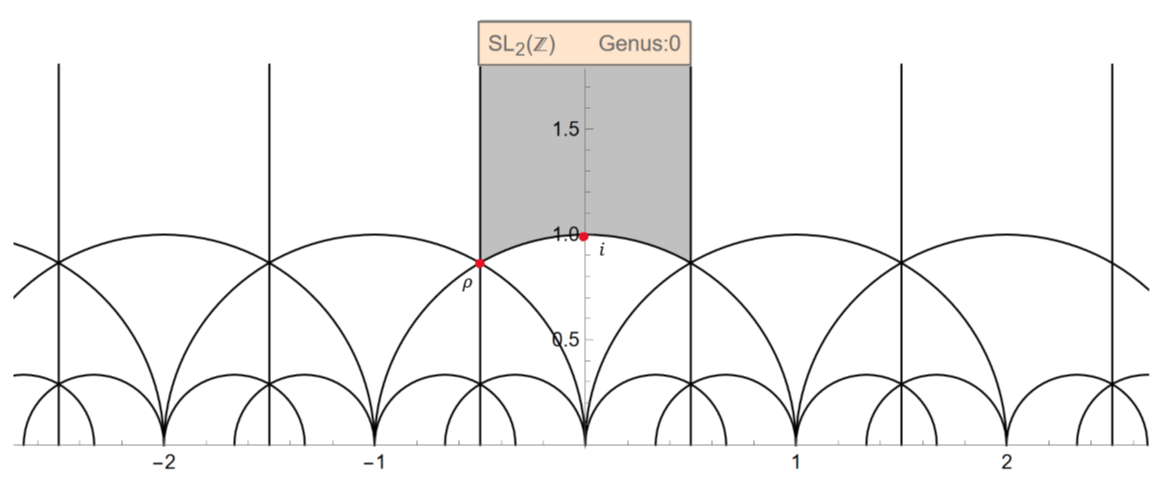}
    \caption{The fundamental domain of $\text{SL}(2,\mathbb{Z}) = \langle S,T\rangle$, as defined in \ref{fundamental_domain_SL2Z}, with elliptic points at $i,\rho$, and a cusp at $\tau = i\infty$. The points on vertical strips in the shaded region are identified by $\tau\mapsto \tau + 1$ and those on the circular segment about the point $\tau = i$ are identified by $\tau\mapsto -\tfrac{1}{\tau}$.}
    \label{fig:SL2Z_fundamental_domain}
\end{figure}

\begin{figure}[htb!]
    \centering
    \includegraphics[width = 13.5cm]{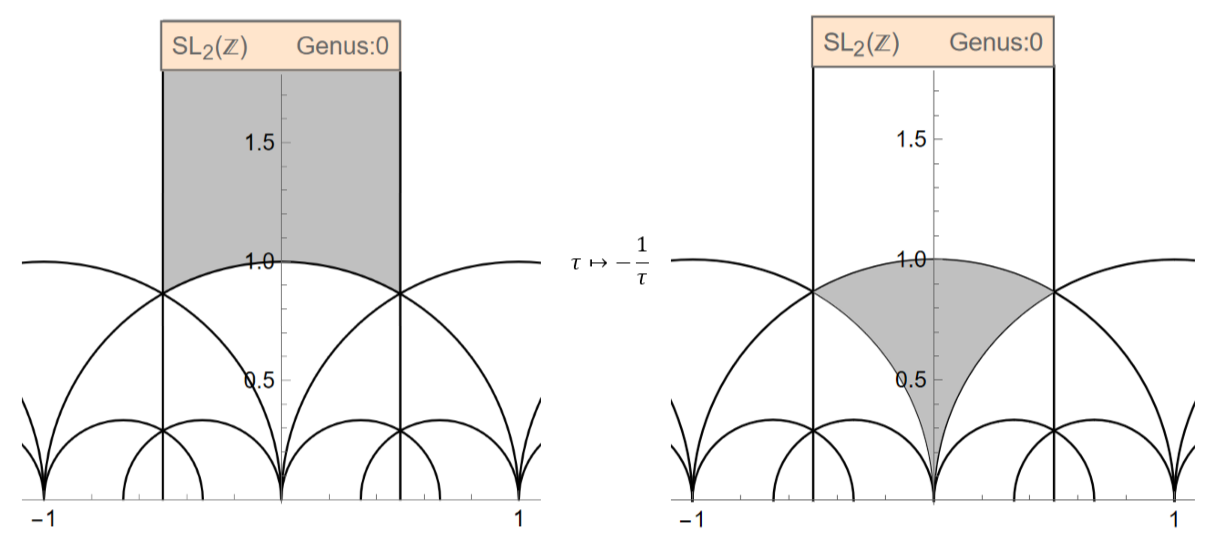}
    \caption{The equivalent fundamental domain (right) after the application of the $S$-transform to all the points in the fundamental domain (left).}
    \label{fig:SL2Z_cusp}
\end{figure}

\noindent\begin{tcolorbox}[colback=gray!5!white,colframe=teal!75!black,title=Fundamental domain]
    The fundamental domain for the action of the group $\text{SL}(2,\mathbb{Z})$ on $\mathbb{H}^{2}$ is denoted by $\mathcal{F}$ and is defined by
    \begin{align}\label{fundamental_domain_SL2Z}
        \mathcal{F} \equiv \left\{-\frac{1}{2}<\text{Re}(z)<\frac{1}{2},\ \vert\tau\vert>1\right\}\cup\left\{-\frac{1}{2}\leq \text{Re}(z)\leq 0,\ \vert\tau\vert=1\right\}.
    \end{align}
\end{tcolorbox}
\noindent The fundamental domain $\mathcal{F}$ is visualized in \ref{fig:SL2Z_fundamental_domain} where the elliptic points are at $i$ and $\rho\equiv \tfrac{-1+\sqrt{3}}{2}$.
\\
Identifying points on the boundary of $\mathcal{F}$ by $\tau\mapsto \tau + 1$ and $\tau\mapsto-\tfrac{1}{\tau}$ yields a cylinder-like shape that is topologically homeomorphic to an open disk. Adding the cusp at $i\infty$ helps us to turn the open disk into a sphere. To better visualize the cusp, we can simply remap the fundamental domain using the transform $\tau\mapsto-\tfrac{1}{\tau}$ to map the cusp at $i\infty$ to $0$. This is shown in figure \ref{fig:SL2Z_cusp}.
\\
\begin{tcolorbox}[colback=gray!5!white,colframe=teal!75!black,title=Modular functions and modular forms]
  Let $\gamma= \left(\begin{smallmatrix}
  a & b\\ c & d
  \end{smallmatrix}\right)$, acting by $\gamma: \tau\mapsto \frac{a\tau + b}{c\tau + d}$, vary throughout $\Gamma$. If $f$ is meromorphic on $\mathbb{H}^{2}$ and there exists $k\in\mathbb{Z}$ such that
  \\
    \begin{equation}\label{modular_form_def}
      f(\gamma(\tau)) = (c\tau + d)^{k}f(\tau),\ \forall \gamma\in\Gamma
    \end{equation}
  \\
  and in addition, $f$ is meromorphic at $\infty$, we say that $f$ is a modular function of weight $k$ for $\Gamma$. If $f$ is holomorphic on $\mathbb{H}^{2}$ and at $\infty$, we say that $f$ is a modular form of weight $k$ for $\Gamma$, and if it is in addition vanishes at $\infty$, $f$ is a cusp form of weight $k$ for $\Gamma$.
\end{tcolorbox}
\noindent A cusp for a discrete subgroup $\Gamma$ is defined to be $\infty$ or any other rational number $\tfrac{a}{c}$. From \ref{action_SL2Z}, we see that $\tfrac{a}{c}$ is clearly equivalent to $\infty$ under the action of $\text{SL}(2,\mathbb{Z})$. A weakly holomorphic modular form of weight $k$ for $\Gamma\in\text{SL}(2,\mathbb{Z})$ is a holomorphic function in $\mathbb{H}^{2}$ that satisfies \ref{modular_form_def} and that has a $q$-expansion of the form
\\
\begin{equation}\label{f(tau)}
    f(\tau) = \sum\limits_{n\geq n_{0}}a(n)q^{n},
\end{equation}
\\
where $q = e^{2\pi i\tau}$ and $n_{0} = \text{ord}_{\infty}(f)$, i.e. of infinite multiplicative order of $f$. 
The multiplicative order of an element $a$ is the smallest positive integer $m$ (if at all it exists) such that $a^{m} = e$, where $e$ is the identity element of the given set. If no such $m$ exists, then the order is said to be infinite. Here, for $f(\tau)$ given in \ref{f(tau)}, we ask ourselves if there exists a positive integer $m$ such that $f^{m}(\tau) = 1$. By definition, the multiplicative order of $f$ is infinite which implies that there exists no positive integer $m$. Based on the values of $n_{0}$, we can classify $f(\tau)$ as follows
\begin{align}
    \begin{split}
    f(\tau) = \begin{cases}
    \text{holomorphic},\ &\text{if}\ n_{0}\geq 0\\
    \text{cusp form},\ &\text{if}\ n_{0}\geq 1.
    \end{cases}
    \end{split}
\end{align}
The space of all holomorphic modular forms of weight $k$ that belong to a group $\Gamma$ is denoted by $\mathcal{M}_{k}(\Gamma)$, the space of all cusp forms is denoted by $\mathcal{S}_{k}(\Gamma)$, and the space of all weakly holomorphic modular forms is denoted by $\mathcal{M}_{k}^{!}(\Gamma)$. Any non-zero $f\in\mathcal{M}^{!}_{k}(\text{SL}(2,\mathbb{Z})$ satisfies the following valence formula
\begin{equation}\label{valence_formula_SL2Z}
  \text{ord}_{\infty}(f) + \frac{1}{2}\text{ord}_{i}(f) + \frac{1}{3}\text{ord}_{\rho}(f) + \sum\limits_{\substack{p\in\text{SL}(2,\mathbb{Z})\backslash\mathbb{H}^{2}\\ p\neq i,\rho}}\text{ord}_{p}(f) =   \frac{k}{12},
\end{equation}
where $\mathcal{F}$ is the fundamental domain defined in \ref{fundamental_domain_SL2Z}. The right-hand side of the equation tells us the number of zeros inside $\mathcal{F}$ which, for a general group $\Gamma$, is defined as follows
\begin{align}\label{no_zeros}
    \# = \text{weight}\cdot\frac{\text{index}}{12}.
\end{align}
Since the index of $\text{SL}(2,\mathbb{Z})$ in itself is $1$, for a weight $k$ modular form the number of zeros is $\tfrac{k}{12}$. 
\begin{tcolorbox}[colback=gray!5!white,colframe=teal!75!black,title=Dimension of the space of modular and cusp forms]
 If $k\geq 4$ is a non-negative integer, then the dimension of the space of cusp forms and modular forms of $\text{SL}(2,\mathbb{Z})$ is
 \begin{align}\label{dim_SL2Z}
     \begin{split}
         \text{dim}\ \mathcal{S}_{k}(\text{SL}(2,\mathbb{Z})) =& \begin{cases}
         \left\lfloor\left.\frac{k}{12}\right\rfloor\right. - 1,\ \text{if}\ k\equiv 2\ (\text{mod}\ 12),\\
         \left\lfloor\left.\frac{k}{12}\right\rfloor\right.,\ \ \ \ \ \ \text{if}\ k\not\equiv 2\ (\text{mod}\ 12).
         \end{cases}\\
         \text{dim}\ \mathcal{M}_{k}(\text{SL}(2,\mathbb{Z})) =& 1 + \text{dim}\ \mathcal{S}_{k}(\text{SL}(2,\mathbb{Z})).
     \end{split}
 \end{align}
 It is also convenient to club the two cases of $k$ and write the dimension of the space of cusp forms as follows
 \begin{align}
     \text{dim}\ \mathcal{S}_{k}(\text{SL}(2,\mathbb{Z})) = \left\lfloor\left.\frac{k}{3}\right\rfloor\right. + \left\lfloor\left.\frac{k}{4}\right\rfloor\right. - \frac{k}{2}.
 \end{align}
\end{tcolorbox}

\subsubsection{The Eisenstein series}
\begin{figure}
    \centering
    \includegraphics[width=12cm]{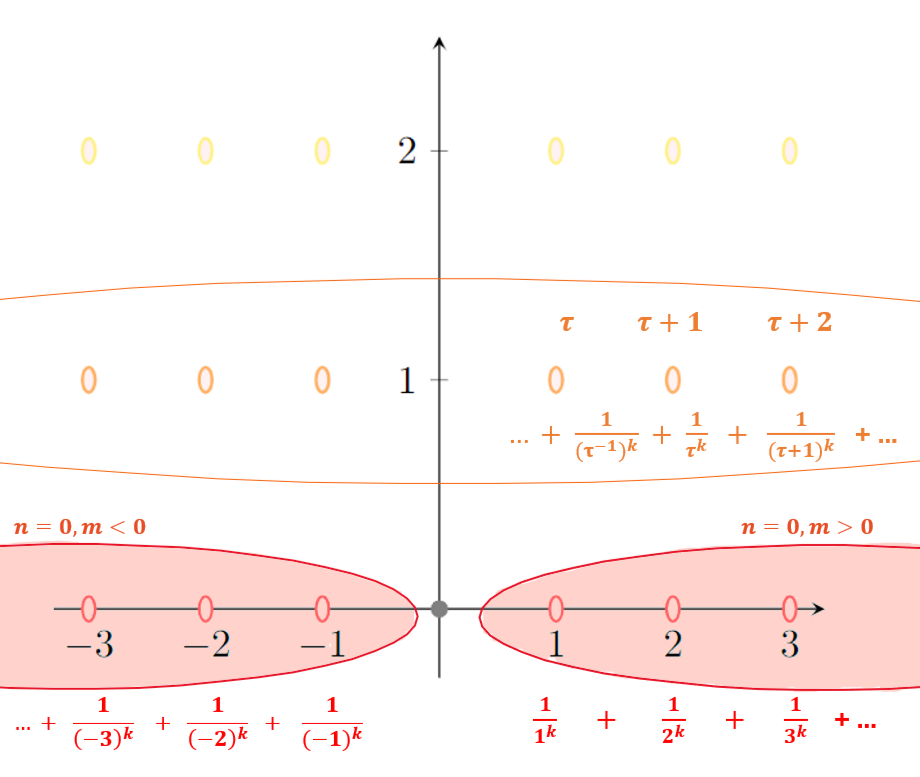}
    \caption{Visualizing the Eisenstein series.}
    \label{fig:Eisenstein_series}
\end{figure}

\noindent Consider the Weierstrass elliptic function, $\wp$ defined as follows
\begin{align}
    \wp(z,\omega_{1},\omega_{2})\equiv \frac{1}{z^{2}} + \sum\limits_{\substack{m,n\in\mathbb{Z}\\ (m,n)\neq (0,0)}}\left(\frac{1}{(z-m\omega_{1} - n\omega_{2})^{2}} - \frac{1}{(m\omega_{1} - n\omega_{2})^{2}}\right),
\end{align}
where the first piece inside the sum is invariant under transformations by $\omega_{1}$ and $\omega_{2}$ when summed over and the second piece is one that is added in by hand since it behaves as a fudge factor that ensures convergence of the sum. Note that the term $(m\omega_{1} - n\omega_{2})^{-2}\to \infty$ when $(m,n) = (0,0)$ and it is for this reason that we have excluded this case in the sum and added the $z^{-2}$ at the beginning. The $\wp$-function is homogeneous, i.e.
\begin{align}\label{wp_def}
    \wp(\lambda z,\lambda\omega_{1},\lambda\omega_{2}) = \lambda^{-2}\wp(z,\omega_{1},\omega_{2}).
\end{align}
We can now use the above equation to construct modular forms and to do this we first require the Laurent series expansion in $z$ of the $\wp$-function,
\begin{align}
    \wp(z) = \frac{1}{z^{2}} + a_{2}z^{2} + z_{4}z^{4} + \ldots
\end{align}
Here, coefficients $a_{2} = a_{2}(\omega_{1},\omega_{2})$ and $a_{4} = a_{4}(\omega_{1},\omega_{2})$. Performing a rescaling by $\lambda$, we get
\begin{align}
    \begin{split}
        \wp(z)\mapsto \wp(\lambda z) =& \frac{1}{(\lambda z)^{2}} + a_{2}(\lambda_{1}\omega_{1},\lambda\omega_{2})(\lambda z)^{2} + a_{4}(\lambda_{1}\omega_{1},\lambda\omega_{2})(\lambda z)^{4} + \ldots\\
        =& \lambda^{-2}\left(\frac{1}{z^{2}} + a_{2}(\omega_{1},\omega_{2})z^{2} + a_{4}(\omega_{2},\omega_{4})z^{4} + \ldots\right).
    \end{split} 
\end{align}
Homogeneity of $\wp(z)$ tells us that the Fourier coefficients $a_{2}(\omega_{1},\omega_{2}),a_{4}(\omega_{1},\omega_{2}),\ldots$ are also homogeneous but at a different degree, i.e.
\begin{align}
    \begin{split}
        a_{2}:\ a_{2}(\lambda\omega_{1},\lambda\omega_{2})=& \lambda^{-4}a_{2}(\omega_{1},\omega_{2}),\\
        a_{4}:\ a_{4}(\lambda\omega_{1},\lambda\omega_{2})=& \lambda^{-6}a_{4}(\omega_{1},\omega_{2}),\\
        &\vdots
    \end{split}
\end{align}
Thus, in general, the coefficients $a_{2}(1,\tau),a_{4}(1,\tau),a_{6}(1,\tau),\ldots$ are modular forms of weight $4,6,8,\ldots$ respectively. Now, expanding the first piece inside the sum \ref{wp_def}, we get
\\
\begin{align}
    \frac{1}{(z-m\omega_{1}-n\omega_{2})^{2}} = \frac{1}{(m\omega_{1} + n\omega_{2})^{2}} + \frac{2z}{(m\omega_{1} + n\omega_{2})^{3}} + \frac{3z}{(m\omega_{1} + n\omega_{2})^{4}} + \ldots
\end{align}
From this we see that the coefficients of $z^{k-2}$ in $\wp(z,\omega_{1},\omega_{2})$ are of the form, $\text{constant}\times\sum_{(m,n)\neq (0,0)}(m\omega_{1} + n\omega_{2})^{-k}$. Fixing $(\omega_{1},\omega_{2}) = (1,\tau)$, we obtain the definition of the holomorphic Eisenstein series (of weight $k$),
\begin{align}
    G_{k}(\tau) \equiv \sum\limits_{(m,n)\neq(0,0)}\frac{1}{(m + n\tau)^{k}}.
\end{align}
This function is valid for $k\geq 4$ and $k\in2\mathbb{Z}$, and is invariant under the T-transform, $T:\tau\to \tau + 1$. We note that the Eisenstein series is defined only for even $k$ because if $k$ were odd then the series would be null-valued since all the terms cancel out in pairs. Thanks to modular invariance, we can expand $G_{k}(z)$ as a function of the form $c_{0} + c_{1}q + c_{2}q^{2} + \ldots$, where $q = e^{2\pi i\tau}$ and $\{c_{i}\}$ are the Fourier coefficients which are to be found. To fix these coefficients we visualize the sum $G_{k}(\tau)$ on the upper-half plane as shown in figure \ref{fig:Eisenstein_series}. The real line can be partitioned into $(-\infty,0)\cup(0,\infty)$ which corresponds to $n=0,m<0$ and $n=0,m>0$ respectively. The point $\tau = 0$ is not included since $(m,n)\neq (0,0)$. The contribution of the strip $(0,\infty)$ to the sum is $\tfrac{1}{1^{k}} + \tfrac{1}{2^{k}} + \ldots = \zeta(k)$ and similarly, the contribution of the strip $(-\infty,0)$ to the sum is also $\zeta(k)$. Next, for the strip above the real line with points $\ldots, \tau-2,\tau-1,\tau, \tau+1,\tau+2,\ldots$ contributes $\tfrac{1}{(\tau^{-1})^{k}} + \tfrac{1}{\tau^{k}} + \tfrac{1}{(\tau + 1)^{k}} + \ldots$. The other contributions of strips above can be found by simply replacing $\tau\to \mathbb{Z}\tau$. All we are to do now is to evaluate the following two sums:
\begin{align*}
    (1)\ \sum\limits_{m\in\mathbb{Z}}\frac{1}{(m + \tau)^{k}},\ \ \ \ \ \ \ \ \ (2)\ \zeta(k) = \sum\limits_{m>0}\frac{1}{m^{k}}. 
\end{align*}
\\
\\
Starting with the first sum, we turn to the function $\tfrac{\pi}{\tan(\pi \tau)}$ which has poles at $\ldots, -2,-1,0,1,2,\ldots$, all of which have unit residue. This function can be expressed as a sum of partial fractions as shown below
\begin{align}
    \frac{\pi}{\tan(\pi \tau)} = \sum\limits_{m\in\mathbb{Z}}\left(\frac{1}{\tau - m} + \frac{1}{m}\right).
\end{align}
The term $m^{-1}$ plays the role of the fudge factor here that ensures convergence. Note that we are to omit the case $m=0$ since the sum would blow up otherwise. Evaluating the sum, we get
\begin{align}
         \frac{\pi}{\tan(\pi \tau)} = -i \pi \left(\frac{1 + e^{2\pi i\tau}}{1 - e^{2\pi i\tau}}\right) = -i\pi \left(1 + 2q + 2q^{2} + \ldots\right).
\end{align}
Taking the derivative, $\tfrac{d}{d\tau}$, $(k-1)$-times with $\tfrac{1}{2\pi i}\tfrac{d}{d\tau} = q\tfrac{d}{dq}$, we arrive at
\begin{align}
    (k-1)!\sum\limits_{m\in\mathbb{Z}}\frac{1}{(\tau - m)^{k}} = (-2\pi i)^{k}\left(1^{k-1}q + 2^{k-1}q^{2} + \ldots\right).
\end{align}
This fixes the first sum. We now turn to the second sum which is nothing but the Riemann $\zeta$-function where $k$ is even. It can be shown that the $\zeta$-function can be expressed in terms of the Bernoulli numbers $B_{k}$ as follows
\begin{align}
    \zeta(k) = (-1)^{\frac{k}{2} +1}\frac{B_{k}}{2}\frac{(2\pi)^{k}}{k!}.
\end{align}
The Bernoulli numbers $B_{k}$ are defined as the coefficients in the following power series
\begin{equation}
    \frac{x}{e^{x}-1} = \sum\limits_{k=0}^{\infty}B_{k}\frac{x^{k}}{k!},\ \ \vert x\vert<2\pi.
\end{equation}
They have the property that $B_{2k+1}=0$ for $k\geq 1$. The first few Bernoulli numbers is tabulated in table \ref{tab:Bernoulli}.
\\
\begin{table}[htb!]
    \centering
    \begin{tabular}{||c|c|c||}
    \hline
    Weight $k$  & Bernoulli number $B_{k}$ & $A_{k} = -\tfrac{2k}{B_{k}}$\\[0.5ex]
    \hline\hline
    $0$ & $+1$ & $0$\\[0.5ex]
    $1$ & $-\frac{1}{2}$ & $+4$\\[0.5ex]
    $2$ & $+\frac{1}{6}$ & $-24$\\[0.5ex]
    $4$ & $-\frac{1}{30}$ & $-240$\\[0.5ex]
    $6$ & $+\frac{1}{42}$ & $-504$\\[0.5ex]
    $8$ & $-\frac{1}{30}$ & $+480$\\[0.5ex]
    $10$ & $+\frac{5}{66}$ & $-264$\\[0.5ex]
    $12$ & $-\frac{691}{2730}$ & $+\frac{65520}{691}$\\[0.5ex]
    $14$ & $+\frac{7}{6}$ & $-24$\\[1ex]
    \hline
    \end{tabular}
    \caption{The first few Bernoulli numbers and the value of $A_{k}$ calculated using them.}
    \label{tab:Bernoulli}
\end{table}
\noindent 
Putting the results of the sums together, we finally have
\begin{align}
    \begin{split}
        G_{k}(\tau) =& 2\zeta(k) + 2\frac{(2\pi i)^{k}}{(k-1)!}\sum\limits_{n\geq 1}\sum\limits_{d\geq 1}d^{k-1}q^{nd}\\
        =& (-1)^{\frac{k}{2} +1}\frac{(2\pi)^{k}B_{k}}{k!} + 2\frac{(2\pi i)^{k}}{(k-1)!}\sum\limits_{n\geq 1}\sigma_{k-1}(n)q^{n},
    \end{split}
\end{align}
where the factor $2$ accounts for the fact that we are summing over positive and negative $n$, the sum over $d$ is a result of the sum over $m$, and $\sigma_{k-1}(n) = \sum\limits_{d\vert n}d^{k-1}$. This can be written in the following neater form
\begin{align}
    G_{k}(\tau) = 2\zeta(k)E_{k}(\tau),
\end{align}
where $E_{k}(\tau)$ is the Eisenstein series of weight $k$ given by
\begin{align}
    E_{k}(\tau) = 1 + A_{k}\sum\limits_{n\geq 1}\frac{n^{k-1}q^{n}}{1 - q^{n}} = 
    1 + A_{k}\sum\limits_{n\geq 1}\sigma_{k-1}(n)q^{n},\ \ A_{k} = -\frac{2k}{B_{k}}.
\end{align}
Using the values of $A_{k}$ in table \ref{tab:Bernoulli}, the first few $E_{k}(z)$ read
\begin{align}
    \begin{split}
    E_{2}(\tau) =& 1 - 24\sum\limits_{n=1}^{\infty}\frac{nq^{n}}{1-q^{n}} = 1 \ - \ \ 24\sum\limits_{n=1}^{\infty}\sigma_{1}(n)q^{n} = 1 - 24q -  72q^{2} -96q^{3} - 168q^{4} + \ldots,\\
    E_{4}(\tau) =& 1 + 240\sum\limits_{n=1}^{\infty}\frac{n^{3}q^{n}}{1-q^{n}} = 1 +  240\sum\limits_{n=1}^{\infty}\sigma_{3}(n)q^{n} = 1 + 240q + 2160q^{2} + 6720 q^{3} + 17520q^{4} + \ldots,\\
    E_{6}(\tau) =& 1 - 504\sum\limits_{n=1}^{\infty}\frac{n^{5}q^{n}}{1-q^{n}} = 1 - 504\sum\limits_{n=1}^{\infty}\sigma_{5}(n)q^{n} =  1 -  504q - 16632q^{2}  - 122976q^{3} - 532728q^{4} + \ldots.
    \end{split}
\end{align}
We now make an important note. The Eisenstein series of weight $2$, $E_{2}(\tau)$, is not a modular form since this series does not converge for $k=2$ (recall that $G_{k}(\tau)$ was defined for $k\geq 4$). The function $E_{2}(\tau)$ is rather called a quasi-modular form since it transforms as follows under a modular transformation $\gamma(\tau) = \tfrac{a\tau + b}{c\tau + d}$,
\begin{align}\label{E_k_transformation}
    \begin{split}
        E_{k}(p\gamma(\tau)) =& (c\tau + d)^{2}E_{k}(p\tau) + \frac{12c}{2\pi ip}(c\tau + d),\\
        E_{2}(\gamma(\tau)) =& (c\tau + d)^{2}E_{2}(\tau) + \frac{12c}{2\pi i}(c\tau + d).
    \end{split}
\end{align}
We also note that the Eisenstein series discussed here are those  defined on the full modular group $\text{SL}(2,\mathbb{Z})$. There also exist various other kinds of Eisenstein series for congruence subgroups and Hecke groups. Some of these relevant to this work will be defined in subsequent sections. Under the S-transformation, $E_{2}(\tau)$ behaves as follows
\begin{align}
    E_{2}\left(-\frac{1}{\tau}\right) = \tau^{2}E_{2}(\tau) + \frac{6\tau}{\pi i},
\end{align}
where the factor $\tfrac{6\tau}{\pi i}$ prevents $E_{2}(\tau)$ from being a modular form. We can remedy this by removing this factor and thus, obtaining a modular form out of the quasimodular $E_{2}(\tau)$. Consider the non-holomorphic function $\tfrac{1}{\text{Im}(\tau)}$, its behaviour under a modular transformation reads
\begin{align}
    \frac{1}{\text{Im}(\tau)} = \frac{(c\tau + d)^{2}}{\text{Im}(\tau)} + \frac{2c}{i}(c\tau + d).
\end{align}
It follows from this that under the S-transformation we have, $\tfrac{1}{\text{Im}\left(-\tfrac{1}{\tau}\right)} = \tfrac{\tau^{2}}{\text{Im}(\tau)} - 2i\tau$. With this, we can cook up the following almost-holomorphic modular form
\begin{align}
    E^{*}_{2}(\tau,\overline{\tau}) = E_{2}(\tau) - \frac{3}{\pi\text{Im}(\tau)} = E_{2}(\tau) + \frac{12}{2\pi i(\tau - \overline{\tau})},
\end{align}
where we used $\text{Im}(\tau - \overline{\tau}) = \tfrac{\tau - \overline{\tau}}{2i}$. Although this is not holomorphic, it can easily be shown that it transforms like a modular form as shown below
\begin{align}
    \begin{split}
        E^{*}_{2}\left(\gamma(\tau),\gamma(\overline{\tau})\right) =& E_{2}(\gamma(\tau)) - \frac{3}{\pi\text{Im}(\gamma(\tau))}\\
        =& (c\tau + d)^{2}E_{2}(\tau) + \frac{12}{2\pi i}(c\tau + d) - (c\tau + d)^{2}\frac{3}{\pi\text{Im}(\tau)} - \frac{12}{2\pi i}(c\tau + d)\\
        =& (c\tau + d)^{2}E_{2}^{*}(\tau,\overline{\tau}).
    \end{split}
\end{align}
For any given almost-holomorphic modular form $g(\tau, \overline{\tau})$, we can obtain the associated quasimodular form $g(\tau)$ by the limit, $g(\tau) = \lim\limits_{\overline{\tau} \to \infty}g(\tau,\overline{\tau})$.
\begin{tcolorbox}[colback=gray!5!white,colframe=teal!75!black,title=Quasimodular forms]
A quasimodular form of a group $\Gamma$ that is of weight $k$ and depth $r$ is defined as a holomorphic function such that for all $\gamma\in\Gamma$, $(c\tau + d)^{-k}f(\gamma(\tau))$ can be expressed as a polynomial of degree $r$ in $\tfrac{c}{c\tau + d}$ with holomorphic coefficients, i.e.
\begin{align}\label{quasi_def}
    f(\gamma(\tau)) = (c\tau + d)^{k}\sum\limits_{n = 0}^{r}f_{n}(\tau)\left(\frac{c}{c\tau + d}\right)^{n},
\end{align}
where $f_{n}(\tau)$ are holomorphic functions. 
\end{tcolorbox}
\noindent 
From this definition, we find that all quasimodular forms of depth $r = 0$ are modular forms. At depth $r = 1$, we find 
\begin{align}
    f(\gamma(\tau)) = (c\tau + d)^{k}f_{0}(\tau) + f_{1}(\tau)c(c\tau + d)^{k-1}.    
\end{align}
With $k = 2$, $f_{0}(\tau) = E_{2}(\tau)$, $f_{1}(\tau) = \tfrac{12}{2\pi i}$, we find that $E_{2}(\tau)$ to be a quasimodular form of weight $2$ and depth $r = 1$. The space of quasimodular forms in $\Gamma$ of weight $k$ and depth $r$ is denoted by $\mathcal{QM}_{k}(\Gamma)$.

\subsubsection{Modular functions}
\noindent Using these modular forms as a basis, we can define the different modular functions of weight $0$ by taking the ratio of two weight $k$ forms. First, consider the ratio of two weight $8$ forms, a possible candidate is the following
\begin{align}
    \frac{E_{4}^{2}(\tau)}{E_{8}(\tau)} = \frac{1 + 240\sum\limits_{n=1}^{\infty}\sigma_{3}(n)q^{n}}{1 + 480\sum\limits_{n=1}^{\infty}\sigma_{7}(n)q^{n}} = \frac{\left(1 + 240q + 2160q^{2} + \ldots\right)^{2}}{1 + 480q + 61920q^{2} + \ldots} = \frac{1 + 480q + 61920q^{2} + \ldots}{1 + 480q + 61920q^{2} + \ldots} = 1.
\end{align}
From this we learn that $E_{4}^{2}(\tau) = E_{8}(\tau)$. Now, consider the ratio of weight $10$ forms for which a possible candidate is the following
\begin{align}
    \begin{split}
        \frac{E_{4}(\tau)E_{6}(z)}{E_{10}(\tau)} =& \frac{\left(1 + 240\sum\limits_{n=1}^{\infty}\sigma_{3}(n)q^{n}\right)\left(1 - 504\sum\limits_{n=1}^{\infty}\sigma_{5}(n)q^{n}\right)}{1 - 264\sum\limits_{n=1}^{\infty}\sigma_{9}(n)q^{n}}\\
        =& \frac{\left(1 + 240q + \ldots\right)\left(1 - 504q + \ldots\right)}{1 - 264q + \ldots} = \frac{1 - 264q + \ldots}{1 - 264q + \ldots} = 1.
    \end{split}
\end{align}
Next, for weight $12$, we find that there are three possible candidates of weight $4$: $E_{4}(\tau)E_{6}(\tau) = E_{10}(\tau)$, but we find that $E_{4}^{3}(\tau)\neq E_{12}(\tau)\neq E_{6}^{2}(\tau)$. The reason for this lies in the fact that the dimensions of the spaces of modular forms of weight $12$ are different from weight $4$ and weight $10$. From \ref{dim_SL2Z}, we find
\\
\begin{align}
    \begin{split}
        \text{dim}\ \mathcal{M}_{4}\left(\text{SL}(2,\mathbb{Z})\right) =& 1,\ i = 4,6,8,10\\ \text{dim}\ \mathcal{M}_{12}\left(\text{SL}(2,\mathbb{Z})\right) =& 2.
    \end{split}
\end{align}
From this, we conclude that any two modular forms of weights $8$ must be equal. The same also applies to modular forms of weight $10$. Since the space $\mathcal{M}_{12}(\text{SL}(2,\mathbb{Z})$ is two-dimensional, we can separate them out and this is why $E_{4}^{3}\neq E_{6}^{2} \neq E_{12}$. Hence, for the case of weight $12$, there must exist a linear relation among forms $E_{4}^{3}$, $E_{6}^{2}$, and $E_{12}$ although all their individual $q$-series expansions are different. Taking the ratio of the Eisenstein series of different weights, we can build new modular forms. We define the discriminant $\Delta(\tau)$ as a cusp form of weight $12$ as follows
\begin{align}\label{cusp_form}
    \begin{split}
        \Delta(\tau) =& \frac{1}{1728}\left(E_{4}(\tau)^{3} - E_{6}(\tau)^{2}\right) = q\prod\limits_{n\geq 1}\left(1 -q^{n}\right)^{24} = \sum\limits_{n\geq 1}z(n)q^{n}\\
        =& q - 24q^{2} + 252 q^{3} - 1472 q^{4} + \ldots,
    \end{split}
\end{align}
where the definition involving the $q$-product is called Jacobi's product formula. The Dedekind $\eta$-function is defined in the usual way,
\begin{align}
    \eta(\tau) \equiv q^{\frac{1}{24}}\prod\limits_{n=1}^{\infty}(1-q)^{n},
\end{align}
and possesses the following behaviour
under the $T$- and $S$-transformations
\begin{align}
    \begin{split}
        \eta(\tau)\overset{T:\tau\mapsto \tau+1}{\longrightarrow} \eta(\tau+1) =& e^{\frac{2\pi i}{24}}\eta(\tau),\\
        \eta(\tau)\overset{S:\tau\mapsto -\tfrac{1}{\tau}}{\longrightarrow} \eta\left(-\frac{1}{\tau}\right) =& \sqrt{\frac{\tau}{i}}\eta(\tau).
    \end{split}
\end{align}
Furthermore, under a modular transformation $\gamma = \tfrac{a\tau + b}{c\tau + d}$, the Dedekind $\eta$-function reads
\\
\begin{align}
    \eta(\gamma(\tau)) = \epsilon(a,b,c,d)\sqrt{\frac{c\tau+d}{i}}\eta(\tau),
\end{align}
where $\epsilon(a,b,c,d)$ is one of the $24^{\text{th}}$ roots of unity. The cusp form \ref{cusp_form}, in terms of $\eta(\tau)$ is just $\Delta(\tau) = \eta^{24}(\tau)$.  The discriminant $\Delta$ is closely related to the quasi-modular form $E_{2}$ which can be seen by taking the logarithmic derivative of the former as shown below
\\
\begin{align}
  \begin{split}
      \frac{1}{2\pi i}\frac{d}{d\tau}\ln\Delta(\tau) =& 24\frac{1}{2\pi i}\frac{d}{d\tau}\ln\eta(\tau) =  \frac{1}{2\pi i}\left(\frac{\pi i}{12} + 2\pi i\sum\limits_{n=1}^{\infty}\frac{nq^{n}}{q-q^{n}}\right)\\
      =& \frac{1}{2\pi i}\left(\frac{\pi i}{12} + 2\pi i\sum\limits_{n=1}^{\infty}n\sum\limits_{m=1}^{\infty}q^{nm}\right) = \frac{1}{2\pi i}\left(\frac{\pi i}{12} + 2\pi i\sum\limits_{n=1}^{\infty}\sum\limits_{m=1}^{\infty}nq^{nm}\right)\\
      =& \frac{1}{2\pi i}\left(\frac{\pi i}{12} + 2\pi i\sum\limits_{n=1}^{\infty}\left(\sum\limits_{0<d\vert n}d\right)q^{n}\right) = E_{2}(\tau).
  \end{split}  
\end{align}
The fundamental relation established here is
\begin{align}
    \frac{\Delta'(\tau)}{\Delta(\tau)} =  E_{2}(\tau).
\end{align}
It turns out that this follows for all congruence subgroups but with a small twist being that the derivative of the cusp form of $\Gamma_{0}(p)$, where $p\in\mathbb{P}$, is equal to the product of itself with the weight $2$ (quasi-modular) Eisenstein series of the corresponding Fricke group $E_{2}^{(p^{+})}(\tau)$ with $p\in\mathbb{P}$, i.e. 
\begin{align}\label{general_cusp_derivative}
    \frac{\Delta'(\tau)}{\Delta(\tau)} = E_{2}^{(p^{+})}(\tau),\ \Delta(\tau)\in\mathcal{S}_{k}(\Gamma_{0}(p)).
\end{align}
We also have the following useful formula for the weight $2$ modular form of a Hecke group of level $N$, 
\begin{align}
    \begin{split}
    E_{2,N}(\tau) =& -\frac{24}{N-1}q\frac{d}{dq}\ln\left(\frac{\eta(N)}{\eta(N\tau)}\right)\\
    =& \frac{1}{N-1}\left(NE_{2}(N\tau) - E_{2}(\tau)\right),
    \end{split}
\end{align}
where we have used $\eta'(\tau) = \tfrac{\pi i}{12}\eta(\tau)E_{2}(\tau)$. From \ref{E_k_transformation}, we see that the factor $\tfrac{12c}{2\pi i}(c\tau + d)$ cancel under a modular transformation thus making $E_{2,N}(\tau)$ a modular form. A function $f(\tau)$ of the following form
\begin{align}
    f(\tau) = \prod\limits_{\alpha\vert N}\left(\eta(\alpha\tau)\right)^{r_{\alpha}},
\end{align}
where $N\geq 1$ and each $r_{\alpha}$ is an integer, is called an $\eta$-quotient. If each $r_{\alpha}\geq 0$, then the function $f(\tau)$ is called an $\eta$-product. 
\begin{tcolorbox}[colback=gray!5!white,colframe=teal!75!black,title=The $j$-function]
 The $j$-function (also called Klein's invariant) is a modular function of weight $0$ given by
    \begin{align}\label{j-function}
        \begin{split}
            j(\tau) =& \frac{E_{4}(\tau)^{3}}{\Delta(\tau)} = q^{-1} + 744 + \sum\limits_{n\geq 1}c(n)q^{n}\\
            =& q^{-1} + 744 + 196884 q + 21493760 q^{2} + 864299970 q^{3} + 20245856256 q^{4} + \ldots
        \end{split}
    \end{align}
\end{tcolorbox}
\noindent 
The factor $1728$ that appears in the denominator of the cusp form in \ref{cusp_form} and hence, the numerator of the $j$-function in \ref{j-function} is the least positive integer that ensures that the $q$-expansion of $j(z)$ has integral coefficients, i.e. coefficients of the algebraic expansion being integers. The $j$-function is also the ``Hauptmodul'', which is German for the principal modulus, of $\text{SL}(2,\mathbb{Z})$. Notice that $j(\tau)$ has a simple zero at the cusp $\tau = i\infty$ ($q=0$) and a simple pole at the cusp $\tau = 0$ ($q = 1$). This is in fact a characteristic of the Hauptmoduls of genus-zero Hecke subgroups $\Gamma_{0}(N)$. The Hauptmoduls of Hecke and Fricke groups come as $\eta$-products. For example, when $N=7$, we have $f(\tau) = \eta^{r_{1}}(\tau)\eta^{r_{7}}(7\tau)$. For $r_{1} = r_{7} = 4$, $f(\tau) = j_{7}(\tau)$, the Hauptmodul of $\Gamma_{0}(7)$. 
From \ref{dim_SL2Z}, we notice that
\begin{align}
    \text{dim}\left(\frac{\mathcal{M}_{k}(\text{SL}(2,\mathbb{Z}))}{\mathcal{S}_{k}(\text{SL}(2,\mathbb{Z}))}\right) \leq 1,
\end{align}
and hence, we can decompose the space of modular forms as follows
\begin{align}\label{M_basis_decomp}
    \mathcal{M}_{k}(\text{SL}(2,\mathbb{Z})) = \mathbb{C}E_{k}\oplus \mathcal{S}_{k}(\text{SL}(2,\mathbb{Z})).
\end{align}
\begin{tcolorbox}[colback=gray!5!white,colframe=teal!75!black,title=Basis decomposition of the spaces of forms]
 For every even integer $k\geq 4$, the space of cusp forms on $\text{SL}(2,\mathbb{Z})$ can be written as follows
 \begin{align}
     \mathcal{S}_{k}(\text{SL}(2,\mathbb{Z})) = \Delta \mathcal{M}_{k-12}(\text{SL}(2,\mathbb{Z})).
 \end{align}
 Using this in \ref{M_basis_decomp}, we have the following basis decomposition
 \begin{align}
    \mathcal{M}_{k}(\text{SL}(2,\mathbb{Z})) = E_{k-12n}\left(\mathbb{C}\left(E_{4})^{3}\right)^{n}\oplus\mathbb{C}\left(E_{4})^{3}\right)^{n-1}\Delta\oplus\mathbb{C}\left(E_{4})^{3}\right)^{n-2}\Delta^{2}\oplus\ldots\oplus\mathbb{C}\left(\Delta\right)^{n} \right),
\end{align}
where $n = \text{dim}\ \mathcal{M}_{k}(\text{SL}(2,\mathbb{Z})) - 1$.
\end{tcolorbox}
\noindent
Going back to the case of $k=12$, we now want to establish a relation between the Eisenstein series of weight $12$. To do this, we first notice that we have the following basis decomposition 
\begin{align}
    \mathcal{M}_{12}(\text{SL}(2,\mathbb{Z})) = \mathbb{C}E_{4}^{3}\oplus \mathbb{C}\Delta
\end{align}
Consider the candidate $\mathcal{C}(\tau) = aE_{4}^{3}(\tau) + b\Delta(\tau)$, where $a,b\in\mathbb{C}$. The $q$-series expansions of this candidate and the Eisenstein series $E_{12}(\tau)$ read
\begin{align}
    \begin{split}
    \mathcal{C}(\tau) =&  aE_{4}^{3}(\tau) + b\Delta(\tau)\\
    =& a + (720 a + b) q + (179280 a - 24 b) q^2 + (16954560 a + 252 b) q^3 + \ldots,\\
    E_{12}(\tau) =& 1 + \frac{65520}{691}\sum\limits_{n=1}^{\infty}\sigma_{11}q^{n} = 1 + \frac{65520}{691}\sum\limits_{n=1}^{\infty}\frac{n^{11}q^{n}}{1-q^{n}}\\
    =& 1 + \frac{1}{691}\left(65520q + 134250480q^{2} + 11606736960q^{3} + \ldots\right).
    \end{split}
\end{align}
Comparing the two series, we fix coefficients $a = 1$, $b = -\tfrac{432000}{691}$. Thus, we can establish the following linear relation between modular forms of weight $12$
\begin{align}
    441 E_{4}^{3}(\tau) + 250E_{6}^{2}(\tau) = 691E_{12}(\tau).
\end{align}
We can find similar relations for modular forms of higher weight. 

\subsubsection{Hecke opertators}
\noindent Let $f(\tau)$ be a modular function. We can now realize the function $f(2\tau)$ as the action of $\alpha = \left(\begin{smallmatrix}2 & 0\\ 0 & 1 \end{smallmatrix}\right)$. This, however, is not invariant under the action of $\text{SL}(2,\mathbb{Z})$ but is rather invariant under $\text{SL}(2,\mathbb{Z})$ conjugated by $\alpha$, i.e. $\Tilde{\gamma} = \alpha^{-1}\gamma\alpha =  \left(\begin{smallmatrix} a & \tfrac{b}{2}\\ 2c & d\end{smallmatrix}\right)$. The intersection of $\Tilde{\gamma}$ with $\gamma\in\text{SL}(2, \mathbb{Z})$ yields $\left(\begin{smallmatrix}a & b\\ c & d\end{smallmatrix}\right)$ with $c\in2\mathbb{Z}$ or equivalently, $c \equiv 0\ (\text{mod}\ 2)$. Now, from \ref{congruence_subgroup_definitions}, we see that matrices of this form define the Hecke group $\Gamma_{0}(2) = \langle T,S^{-1}T^{-2}S\rangle$ and from \ref{index_Hecke}, we find $\mu_{0} = 3$. Hence, we see that the function $f(2\tau)$ is invariant under an index-$3$ subgroup of the modular group rather than the full group itself. Now, in order to make the modular function invariant under $\text{SL}(2,\mathbb{Z})$, we can simply sum over the cosets $\Gamma_{0}(2)\backslash\text{SL}(2,\mathbb{Z})$. The coset representatives of $\Gamma_{0}(N)\backslash\text{SL}(2,\mathbb{Z})$ are given by the following set upper triangular matrices,
\begin{align}\label{X_n}
    X_{n} = \left\{\left.\begin{pmatrix}
    a & b\\ 0 & d\end{pmatrix}\in\mathbb{M}_{n}(2,\mathbb{Z})\right\vert 0\leq b<d\right\},
\end{align}
where $\mathbb{M}_{n}(2,\mathbb{Z})$ denotes the set of $2\times 2$ matrices with integer entries and determinant $n$. For $n = 2$, we have $X_{n} = \left\{\left(\begin{smallmatrix}2 & 0\\ 0 & 1\end{smallmatrix}\right),\left(\begin{smallmatrix}1 & 1\\ 0 & 2\end{smallmatrix}\right),\left(\begin{smallmatrix}1 & 0\\ 0 & 2\end{smallmatrix}\right)\right\}$ and these correspond to the following modular function invariant under $\text{SL}(2,\mathbb{Z})$,
\begin{align}\label{invariant_combination}
    \mathcal{F}(\tau) = f(2\tau) + f\left(\frac{\tau + 1}{2}\right) + f\left(\frac{\tau}{2}\right).
\end{align}
This provides a natural motivation for the Hecke operator $T_{n}$ that can be thought of as operators that act on a modular function $f(\tau)$ to yield an invariant combination of modular functions. Here $\mathcal{F}(\tau)  = T_{2}f(\tau)$. For the next couple of $n$ values in \ref{X_n}, the action of the Hecke operator $T_{n}$ on $f(\tau)$ reads
\begin{align}
    \begin{split}
        X_{3} =& \left\{\left(\begin{smallmatrix}3 & 0\\ 0 & 1\end{smallmatrix}\right),\left(\begin{smallmatrix}1 & 1\\ 0 & 3\end{smallmatrix}\right), \left(\begin{smallmatrix}1 & 2\\ 0 & 3\end{smallmatrix}\right), \left(\begin{smallmatrix}1 & 0\\ 0 & 3\end{smallmatrix}\right)\right\},\\  T_{3}f(\tau) =& f(3\tau) + f\left(\frac{\tau + 1}{3}\right) + f\left(\frac{\tau + 2}{3}\right) + f\left(\frac{\tau}{3}\right),\\
        X_{4} =& \left\{\left(\begin{smallmatrix}4 & 0\\ 0 & 1\end{smallmatrix}\right),\left(\begin{smallmatrix}1 & 1\\ 0 & 4\end{smallmatrix}\right), \left(\begin{smallmatrix}1 & 2\\ 0 & 4\end{smallmatrix}\right), \left(\begin{smallmatrix}1 & 3\\ 0 & 4\end{smallmatrix}\right), \left(\begin{smallmatrix}2 & 0\\ 0 & 2\end{smallmatrix}\right), \left(\begin{smallmatrix}2 & 1\\ 0 & 2\end{smallmatrix}\right), \left(\begin{smallmatrix}1 & 0\\ 0 & 4\end{smallmatrix}\right)\right\},\\ T_{4}f(\tau) =& f(4\tau) + f\left(\frac{\tau + 1}{4}\right) + f\left(\frac{\tau + 2}{4}\right) + f\left(\frac{\tau + 3}{4}\right)\\
        &{} \ \ \ \ \ \ \ + f(\tau) + f\left(\frac{2\tau + 1}{2}\right) + f\left(\frac{\tau}{4}\right),\\
        &\vdots
    \end{split}
\end{align}
Furthermore, it is easy to see that in the case of $p\in\mathbb{P}$, then the Hecke operator acts as follows
\begin{align}
    T_{p}:f(\tau)  \mapsto f(p\tau) + f\left(\frac{\tau + 1}{p}\right) + \ldots + f\left(\frac{\tau + p-1}{p}\right) + f\left(\frac{\tau}{p}\right). 
\end{align}
With $f(\tau) = \sum_{n}a_{n}q^{n}$, we find 
\begin{align}
    \begin{split}
    &f\left(\frac{\tau + 1}{p}\right) + \ldots + f\left(\frac{\tau + p-1}{p}\right) + f\left(\frac{\tau}{p}\right) = \begin{cases}
    p\sum\limits_{n}a_{n}q^{\tfrac{n}{p}},\ p\vert n,\\
    0,\ \ \ \ \ \ \ \ \ \ \ \ p\not\vert n,
    \end{cases}\\
    &f(p\tau) = \sum_{n}a_{n}q^{np},
    \end{split}   
\end{align}
For the case of $p = 3$, we have the following expansions
\begin{align}
    \begin{split}
        f(\tau) =& \ldots + a_{-3}q^{-3} + a_{-2}q^{-2} + a_{-1}q^{-1} + a_{0} + a_{1}q\  + a_{2}q^{2} + a_{3}q^{3} + \ldots,\\
        f(3\tau) =& \ldots + a_{-3}q^{-9} + a_{-2}q^{-6} + a_{-1}q^{-3} + a_{0} + a_{1}q^{3} + a_{2}q^{6} + a_{3}q^{9} + \ldots,\\
        \sum\limits_{i=0}^{3-1}f\left(\frac{\tau + i}{3}\right) =& \ldots + 3a_{-3}q^{-1}\ \ \ \ \ \ \ \ \ \ \ +\ \ \ \ \ \ \ \ \ \ \ \  a_{0}\ \ \ \ \ \ \ \ \ \ +\ \ \ \ \ \ \ \ \ 3a_{3}q + \ldots,
    \end{split}
\end{align}
from which we notice that the action of the last set of functions on any prime pulls the $q$-series expansion inwards by $\tfrac{n}{p}$ places while also multiplying all the coefficients by $p$, and the action of the second function pushes $q$-series expansion outwards by $np$ places. 
\begin{tcolorbox}[colback=gray!5!white,colframe=teal!75!black,title=Hecke operator]
 We can succinctly define the Hecke operator $T'_{\ell} = \ell T_{\ell}$ as follows
\begin{align}
    \begin{split}
        T_{\ell} =& \begin{cases}
            f(\ell \tau) + \sum\limits_{m=0}^{\ell - 1}f\left(\frac{\tau + i}{\ell}\right),\ &\ell\in\mathbb{P},\\
            \sum\limits_{d\vert \ell}\sum\limits_{m=0}^{d- 1}f\left(\frac{\ell\tau + md}{d^{2}}\right),\ &\ell\in\mathbb{Z}_{+},
        \end{cases}
    \end{split}
\end{align}
and the coefficients $a_{-\ell}$, for $0\leq \ell\leq k$, come from 
\begin{align}
    \begin{split}
    q^{-k}\prod\limits_{n=2}^{\infty}\left(1-q^{n}\right)^{-1} =& \sum\limits_{\ell = -k}^{\infty}A_{l}q^{\ell},\\
    =& q^{-k}\left(1 + q^{2} + q^{3} + 2q^{4} + 2q^{5} + 4q^{6} + \ldots\right).
    \end{split}
\end{align}
\end{tcolorbox}
\noindent
Suppose now we want to use non-zero weight modular forms instead of modular functions. Then, we simply consider the invariant form $f(\tau)\left(d\tau\right)^{\tfrac{k}{2}}$ for the Hecke operator to act on. For the case of $p=2$, instead of the invariant linear combination shown in \ref{invariant_combination}, we now have
\begin{align}
    \begin{split}
        \mathscr{F}(\tau) =& f(2\tau)\left(d\ 2\tau\right)^{\tfrac{k}{2}} + f\left(\frac{\tau + 1}{2}\right)\left(d\ \left(\frac{\tau + 1}{2}\right)\right)^{\tfrac{k}{2}} + f\left(\frac{\tau}{2}\right)\left(d\ \frac{\tau}{2}\right)^{\tfrac{k}{2}},\\
        =& \left[2^{\tfrac{k}{2}}f(2\tau) + 2^{-\tfrac{k}{2}}f\left(\frac{\tau + 1}{2}\right) + 2^{-\tfrac{k}{2}}f\left(\frac{\tau}{2}\right)\right]\left(d\tau\right)^{\tfrac{k}{2}},
    \end{split}
\end{align}
from which we see that the inclusion of $\left(d\tau\right)^{\tfrac{k}{2}}$ results in all the powers of $p$ multiplying the coset representatives acting on the modular form. Formally, we can define the Hecke operator and its Fourier expansion as follows
\begin{align}
    \begin{split}
        T_{n} \equiv& \sum\limits_{\substack{ad = n\\ 0\leq b< d}}f\left(\frac{a\tau + b}{d}\right),\\
        T_{n}f(\tau)=& \sum\limits_{n}\left(\sum\limits_{d\vert (m,n)}d^{k-1}c\left(\frac{mn}{d^{2}}\right)\right)q^{n},
    \end{split}
\end{align}
where $(m,n)\equiv \text{gcd}(m,n)$.

\subsubsection{Eigenspace decomposition}
\noindent The action of the Atkin-Lehner operator $A_{l}$ on $f\in\Gamma_{0}(N)$ reads $\left(f\vert_{k}A_{l}\right)(\tau) = \pm f(\tau)$, where $\vert_{k}$ is called the slash operator defined as follows
\begin{align}
    \left(f\vert_{k}\gamma\right)(\tau) = (c\tau + d)^{-k}f(\gamma(\tau)),\ \forall\gamma\in\Gamma.
\end{align}
To understand this, consider the special case of the Fricke operator, We can express the action of $W_{N}$ as follows
\begin{align}\label{eigenvalue_equation}
    \left(f\vert_{k}W_{N}\right)(\tau) =  N^{-\tfrac{k}{2}}\tau^{-k}f\left(-\frac{1}{N\tau}\right).
\end{align}
Let $\mathcal{M}_{k}(\Gamma_{0}^{+}(N))$ denote the space of all modular forms of weight $k$ in $\Gamma_{0}(N)$ (note, not the Fricke group) such that \ref{eigenvalue_equation} holds. More formally, for $\varepsilon\in G_{N}$, where $G_{N}$ is the group of characters of $\Gamma_{0}^{+}(N)/\Gamma_{0}(N)\cong\mathbb{Z}/2\mathbb{Z}$, we have
\begin{align}
    \mathcal{M}_{k}(\Gamma_{0}^{+}(N),\varepsilon) = \{f\in\mathcal{M}_{k}(\Gamma_{0}(N)\vert\  f\vert_{k}a_{l} = \varepsilon(A_{l})f\ \text{for all}\ l\},
\end{align}
where $l$ is the Hall divisor and $a_{l}$ is any element in $A_{l}$ such that $A_{l} = A_{l}\Gamma_{0}(N)$. The quotient group $\Gamma^{*}_{0}(N)/\Gamma_{0}(N)$, of which $G_{N}$ is a special case, is a group of order $2^{\omega(N)}$ where $\omega(N)$ denotes the number of prime factors dividing $N$. Now, the Atkin-Lehner eigenspace decomposition can be written as follows
\begin{align}
    \mathcal{M}_{k}(\Gamma_{0}(N)) = \bigoplus\limits_{\varepsilon\in G_{N}}\mathcal{M}_{k}(\Gamma_{0}^{+}(N),\varepsilon)
\end{align}
The dimension of these eigenspaces for $N = 2,3$ take the following form
\begin{align}
    \begin{split}
        \text{dim}\ \mathcal{M}_{k}(\Gamma_{0}^{+}(N),+) =& \text{dim}\ \mathcal{M}_{k}(\Gamma_{0}^{+}(N))\\
        \text{dim}\ \mathcal{M}_{k}(\Gamma_{0}^{+}(N),-) =& \text{dim}\ \mathcal{M}_{k}(\Gamma_{0}(N)) - \text{dim}\ \mathcal{M}_{k}(\Gamma_{0}^{+}(N)).
    \end{split}
\end{align}
We refer the reader to \cite{Lin2021QuasimodularFA} for more details.

\subsection{\texorpdfstring{$\mathbf{\Gamma_{0}^{+}(2)}$}{ Γ0(2)+}}
\noindent The Fricke group of level $2$ is generated by $\Gamma_{0}^{+}(2) = \langle\left(\begin{smallmatrix}1 & 1\\ 0 & 1\end{smallmatrix}\right), W_{2}\rangle$. The following relation exists between the generators, $W_{2}^{4} = \left(W_{2}^{3}T\right)^{8} = \mathbb{1}_{2}$. A non-zero $f\in\mathcal{M}_{k}^{!}(\Gamma_{0}^{+}(2))$ satisfies the following valence formula
\begin{align}\label{valence_Fricke_2}
    \nu_{\infty}(f) + \frac{1}{2}\nu_{\tfrac{i}{\sqrt{2}}}(f) + \frac{1}{4}\nu_{\rho_{2}}(f) + \sum\limits_{\substack{p\in\Gamma_{0}^{+}(2)\backslash\mathbb{H}^{2}\\ p\neq \tfrac{i}{\sqrt{2}},\rho_{2}}}\nu_{p}(f) =   \frac{k}{8},
\end{align}
where $\tfrac{i}{\sqrt{2}}$ and $\rho_{2} = \tfrac{-1+i}{2}$ are elliptic points and this group has a cusp at $\tau = \infty$. The dimension of the space of modular forms reads
\begin{align}\label{dimension_Fricke_2}
        \text{dim}\ \mathcal{M}_{k}(\Gamma_{0}^{+}(2)) = \begin{cases}\left\lfloor\left.\frac{k}{8}\right\rfloor\right.,\ &k \equiv 2\ (\text{mod}\ 8)\\
        \left\lfloor\left.\frac{k}{8}\right\rfloor\right. + 1,\ &k\not\equiv2\ (\text{mod}\ 8),
        \end{cases}
\end{align}
where $k>2$ and $k\in2\mathbb{Z}$. From \cite{Umasankar:2022kzs, Junichi}, we know that the Riemann-Roch theorem takes the following form
\begin{align}\label{Riemann_Roch_Gamma_0_2+}
    \begin{split}
        \sum\limits_{i=0}^{n-1}\alpha_{i} =& -\frac{nc}{24} + \sum\limits_{i}\Delta_{i}\\
        =&\frac{1}{8}n(n-1) - \frac{1}{4}\ell.
    \end{split}
\end{align}
The Eisenstein series of $\Gamma_{0}^{+}(2)$,  which for weight $k = 2$ is a quasi-modular form, is defined as follows
\begin{align}
    E_{k}^{(2^{+})}(\tau) \equiv \frac{2^{\tfrac{k}{2}}E_{k}(2\tau) + E_{k}(\tau)}{2^{\tfrac{k}{2}} + 1}, \ k\geq 4.
\end{align}
We note that only at this level do we observe the relation $E_{10}^{(2^{+})} = E_{4}^{(2^{+})}E_{6}^{(2^{+})}$. Using the transformation formula $E_{k}(p\gamma(\tau)) = (c\tau + d)^{2}E_{k}(p\tau) + \tfrac{12c}{2\pi ip}(c\tau + d),\ \gamma\in\Gamma_{0}^{+}(2)$, we find
\begin{align}
    \begin{split}
        E_{2}^{(2^{+})}(\gamma(\tau)) =& (c\tau + d)^{2}E_{2}^{(2^{+})}(\tau) + \frac{8c}{2\pi i}(c\tau + d),
    \end{split}
\end{align}
with the action $\gamma(\tau) = \tfrac{a\tau + b}{c\tau + d}$. The Hauptmodul and the cusp form of this group are defined as follows
\begin{align}\label{Hauptmodul_Gamma_0_2+}
    \begin{split}
        j_{2^{+}}(\tau) =& \left(\frac{\left(E_{4}^{(2^{+})}\right)^{2}}{\Delta_{2}}\right)(\tau) = \left(\left(\frac{\eta(\tau)}{\eta(2\tau)}\right)^{12} + 2^{6}\left(\frac{\eta(2\tau)}{\eta(\tau)}\right)^{12}\right)^{2},\\
        \Delta_{2}(\tau) =& \left(\eta(\tau)\eta(2\tau)\right)^{8}\in\mathcal{S}_{8}(\Gamma_{0}^{+}(2)).
    \end{split}
\end{align}

\subsection{\texorpdfstring{$\mathbf{\Gamma_{0}^{+}(3)}$}{ Γ0(3)+}}
\noindent The Fricke group of level $3$ is generated by $\Gamma_{0}^{+}(3) = \langle\left(\begin{smallmatrix}1 & 1\\ 0 & 1\end{smallmatrix}\right), W_{3}\rangle$. The following relation exists between the generators, $W_{3}^{4} = \left(W_{3}^{3}T\right)^{12} = \mathbb{1}_{2}$. A non-zero $f\in\mathcal{M}_{k}^{!}(\Gamma_{0}^{+}(3))$ satisfies the following valence formula
\begin{align}\label{valence_Fricke_3}
    \nu_{\infty}(f) + \frac{1}{2}\nu_{\tfrac{i}{\sqrt{3}}}(f) + \frac{1}{4}\nu_{\rho_{3}}(f) + \sum\limits_{\substack{p\in\Gamma_{0}^{+}(3)\backslash\mathbb{H}^{2}\\ p\neq \tfrac{i}{\sqrt{3}},\rho_{3}}}\nu_{p}(f) =   \frac{k}{6},
\end{align}
where $\tfrac{i}{\sqrt{3}}$ and $\rho_{3} = -\tfrac{1}{2}+\tfrac{i}{2\sqrt{3}}$ are elliptic points and this group has a cusp at $\tau = \infty$. The dimension of the space of modular forms reads
\begin{align}\label{dimension_Fricke_3}
        \text{dim}\ \mathcal{S}_{k}(\Gamma_{0}^{+}(3)) = \begin{cases}\left\lfloor\left.\frac{k}{6}\right\rfloor\right.,\ &k \equiv 2,6\ (\text{mod}\ 12)\\
        \left\lfloor\left.\frac{k}{6}\right\rfloor\right. + 1,\ &k\not\equiv2,6\ (\text{mod}\ 12),
        \end{cases}
\end{align}
where $k>2$ and $k\in2\mathbb{Z}$. From  \cite{Umasankar:2022kzs, Junichi}, we know that the Riemann-Roch theorem takes the following form
\begin{align}
    \begin{split}
        \sum\limits_{i=0}^{n-1}\alpha_{i} =& -\frac{nc}{24} + \sum\limits_{i}\Delta_{i}\\
        =& \frac{1}{6}n(n-1) - \frac{1}{3}\ell.
    \end{split}
\end{align}
The Eisenstein series of $\Gamma_{0}^{+}(3)$, which for weight $k = 2$ is a quasi-modular form, is defined as follows
\begin{align}
       E_{k}^{(3^{+})}(\tau)\equiv\frac{3^{\tfrac{k}{2}}E_{k}(3\tau) + E_{k}(\tau)}{3^{\tfrac{k}{2}} + 1},\ k\geq 4,\\
\end{align}
When $k = 2$, this transforms as follows under $\gamma\in\Gamma_{0}^{+}(3)$
\begin{align}
    \begin{split}
        E_{2}^{(3^{+})}(\gamma(\tau)) =& (c\tau + d)^{2}E_{2}^{(3^{+})}(\tau) + \frac{6c}{2\pi i}(c\tau + d),
    \end{split}
\end{align}
The Hauptmodul and other modular forms in this group are defined as follows
\begin{align}
        j_{3^{+}}(\tau) =& \left(\frac{\left(E_{2,3^{'}}\right)^{3}}{\Delta_{3}}\right)(\tau) = \left(\left(\frac{\eta(\tau)}{\eta(3\tau)}\right)^{6} + 3^{3}\left(\frac{\eta(3\tau)}{\eta(\tau)}\right)^{6}\right)^{2},\nonumber\\ 
        \Delta_{3}(\tau) =& \left(\eta(\tau)\eta(3\tau)\right)^{6}\in\mathcal{S}_{6}(\Gamma_{0}(3)),\nonumber\\
        \Delta_{3^{+}, 8}(\tau) =& \frac{41}{1728}\left(\left(E_{4}^{(3^{+})}\right)^{2} - E_{8}^{(3^{+})}\right)(\tau)\in\mathcal{S}_{8}(\Gamma_{0}^{+}(3)),\nonumber\\
        \Delta_{3^{+}, 10}(\tau) =& \frac{61}{432}\left(E_{4}^{(3^{+})}E_{6}^{(3^{+})} - E_{10}^{(3^{+})}\right)(\tau)\in\mathcal{S}_{10}(\Gamma_{0}^{+}(3)),\nonumber\\
        \Delta_{3^{+},12}(\tau) =& \left( \Delta_{3^{+},8}E_{4}^{(3^{+})}\right)(\tau)\in\mathcal{S}_{12}(\Gamma_{0}^{+}(3)),\nonumber\\ \Delta_{3^{+}, 14}(\tau) =&  \left( \Delta_{3^{+},10}E_{4}^{(3^{+})}\right)(\tau)\in\mathcal{S}_{14}(\Gamma_{0}^{+}(3)).
\end{align}
\section{MLDEs for Hecke and Fricke groups}\label{sec:Two-character_Hecke_Fricke}
\noindent In this section, we outline a general procedure to obtain \textit{novel} Serre-Ramanujan type derivative operators for Hecke and Fricke groups. Our main focus will be to construct such derivative operators for Fricke groups but this procedure can be readily generalized to other related groups such as congruence subgroups. As we shall see, these \textit{novel} derivative operators not only help us to construct the MLDEs for Fricke groups but also will enable us to write down Ramanujan-Eisenstein identities for these groups.
\\
%As for the case of higher-character theories, we ought to be careful in the formulation of the MLDE for higher-character theories since the form of the Ramanujan-Serre covariant derivative would change. We will now expand on this. 
\subsection{Introduction}
\noindent The most general ODE that is invariant under $\Gamma_{0}^{+}(p)$ reads
\begin{align}\label{general_MLDE}
    \left[\omega_{2\ell}(\tau)\mathcal{D}^{n} + \omega_{2\ell + 2}(\tau)\mathcal{D}^{n-1} + \ldots + \omega_{2(n-1+\ell)}(\tau)\mathcal{D} + \omega_{2(n+\ell)}(\tau)\right]f(\tau) = 0,
\end{align}
where the functions $\omega_{k}$ are holomorphic modular forms of weight $k$ and the derivatives here are the \textit{modified} Serre-Ramanujan covariant derivatives given by
\begin{align}\label{Ramanujan-Serre}
    \begin{split}
        \mathcal{D} \equiv& \frac{1}{2\pi i}\frac{d}{d\tau} - \kappa(r) E^{(p^{+})}_{2}(\tau),\\
        \mathcal{D}^{n} =& \mathcal{D}_{r+2n-2}\circ\ldots\circ\mathcal{D}_{r+2}\circ\mathcal{D}_{r},
    \end{split}
\end{align}
where $r$ is the weight of the modular form on which the derivative acts and $\kappa(r)$ (the number of zeros, $\#$, from \ref{no_zeros}), which is obtained from the valence formula, is tabulated in \ref{tab:Fricke_valence}.
\\
\begin{table}[htb!]
    \centering
    \begin{tabular}{||c|c|c|c|c|c|c|c|c|c|c|c|c|c|c||}
    \hline
    $p$ & 2 & 3 & 5 & 7 & 11 & 13 & 17 & 19 & 23 & 29 & 31 & 41 & 59 & 71\\ [0.5ex]
    \hline\hline
    $\kappa(r)$ & $\tfrac{r}{8}$ & $\tfrac{r}{6}$ & $\tfrac{r}{4}$ & $\tfrac{r}{3}$ & $\tfrac{r}{2}$ & $\tfrac{7r}{12}$ & $\tfrac{3r}{4}$ & $\tfrac{5r}{6}$ & $r$ & $\tfrac{5r}{4}$ & $\tfrac{4r}{3}$ & $\tfrac{7r}{4}$ & $\tfrac{5r}{2}$ & $3r$\\ [1ex]
    \hline
    \end{tabular}
    \caption{Values of $\kappa(r) = \tfrac{r\overline{\mu}_{0}^{+}}{12}$ computed using index data in table \ref{tab:Fricke_index_data}.}
    \label{tab:Fricke_valence}
\end{table}
\\
The solution to the ODE \ref{general_MLDE} is given by the Frobenius ansatz which reads
\begin{align}
    \chi_{i} = q^{\alpha_{i}}\sum\limits_{n=0}^{\infty}a_{n}^{(i)}q^{n},
\end{align}
where the exponents $\alpha_{i}$, the Wronskian index $\ell$, and the order $n$ of the differential equation are related by an expression that follows from the Riemann-Roch theorem.
Note that the Fricke groups considered here do not possess any modular form of weight $2$ (see corresponding basis of these groups in \cite{Junichi}), this can be seen from the dimension of the space of modular forms. The covariant derivative constructed here transforms a weight $r$ modular form to a weight $r+2$ modular form. Since each Fricke group has a distinct valence formula, each Fricke level would have a different formulation of the Ramanujan-Serre covariant derivatives. Let us now show the difference between the $\text{SL}(2,\mathbb{Z})$ Serre-Ramanujan covariant derivative and those in Fricke groups. The $\text{SL}(2,\mathbb{Z})$ covariant derivative is defined as follows
\begin{align}\label{Ramanujan-Serre_SL2Z}
    \mathcal{D}_{\Gamma(1)} \equiv& \frac{1}{2\pi i}\frac{d}{d\tau} - \frac{r}{12}E_{2}(\tau).
\end{align}
Consider now the Eisenstein series of a Fricke group of level $p\in\mathbb{P}$ and weight $k$,
\begin{align}
    E^{(p^{+})}_{k}(\tau)\equiv \frac{p^{\tfrac{k}{2}}E_{k}(p\tau) + E_{k}(\tau)}{p^{\tfrac{k}{2}} + 1}.
\end{align}
Using this in \ref{Ramanujan-Serre} with $k = 2$, we find
\begin{align}
    \mathcal{D}_{\Gamma^{+}_{0}(p)} \equiv& \left(\frac{1}{2\pi i}\frac{d}{d\tau} - \frac{\kappa(r)}{p + 1}E_{2}(\tau)\right) - \frac{\kappa(r)}{p + 1}E_{2}(p\tau).
\end{align}
Since $p\neq 1$ and no $\kappa(r)$ value corresponding to a prime level $p$ conspires with $\tfrac{1}{p+1}$ to yield the desired factor of $\tfrac{r}{12}$, it is simply impossible to express the derivatives $\mathcal{D}_{\Gamma^{+}_{0}(p)}$ in terms of $\mathcal{D}_{\Gamma(1)}$. It turns out that we obtain the factor $\tfrac{r}{12}$ in \ref{Ramanujan-Serre_SL2Z} only at the Fricke level $N = 12$. The Eisenstein series of weight $2$ at this level is defined as follows \cite{Junichi}
\begin{align}
    E_{2,12^{+}}(\tau)\equiv \frac{12E_{2}(12\tau) - 6E_{2}(6\tau) + 4E_{2}(4\tau) + 3E_{2}(3\tau) - 2E_{2}(2\tau) + E_{2}(\tau)}{4}.
\end{align}
Using this in \ref{Ramanujan-Serre}, with $\kappa(r) = \tfrac{r}{2}$, we find
\begin{align}\label{Gamma_12+_derivative}
    \begin{split}
        \mathcal{D}_{\Gamma^{+}_{0}(12)} \equiv& \left(\frac{1}{2\pi i}\frac{d}{d\tau} - \frac{r}{12}E_{2}(\tau)\right) - \frac{r}{12}\left(12E_{2}(12\tau) - 6E_{2}(6\tau) + 4E_{2}(4\tau) + 3E_{2}(3\tau) - 2E_{2}(2\tau)\right)\\
        =& \mathcal{D}_{\Gamma(1)} - \frac{r}{12}\mathcal{E}_{2}(\tau),
    \end{split}
\end{align}
where we have used $\mathcal{E}_{2}(\tau)$ to denote the other terms containing weight $2$ Eisenstein series outside the parenthesis in the first line of \ref{Gamma_12+_derivative}. From this brief discussion, we see that when we deal with the MLDEs of higher-character theories, it is important to consider the Eisenstein series associated with the Fricke group in building the covariant derivatives although they are built out of
$\text{SL}(2,\mathbb{Z})$ Eisenstein series. It is easy to see from \ref{Gamma_12+_derivative} that the other terms in $\mathcal{E}_{2}(\tau)$ can have considerable influence on the solutions to the MLDE. Solving the $(n,\ell) = (2,0)$ MLDE for $\Gamma_{0}^{+}(2)$ with $\mathcal{D}_{\Gamma(1)}$ instead of $\mathcal{D}_{\Gamma_{0}^{+}(2)}$ yields spurious solutions (i.e. solutions with negative Fourier coefficients). The algorithm we follow to classify CFTs using MLDEs is the following:
\begin{enumerate}
    \item Postulate MLDE for values of $n$ and fix the number of free parameters $\mu_{i}$.
    \item Find the solutions to the MLDE as a power series in $q = e^{2\pi i\tau}$ of the form
    \begin{align}\label{q-series}
        \chi_{j}(\tau) = q^{\alpha_{j}}\left(m_{0}^{(j)} + m_{1}^{(j)}q + m_{2}^{(j)}q^{2} + \ldots\right),
    \end{align}
    where the exponent $\alpha_{0} = -\tfrac{c}{24}$ is negative in a unitary theory.
    \item Write down the q-series expansions for the coefficient functions present in the MLDE. Plug the q-series in \ref{q-series} along with the q-series of the coefficient functions into the MLDE to obtain a recursion relation. Putting ``$i=0$" in the recursion relation gives us the \textit{indicial} equation.   
    \item Set up recursion relations to obtain coefficients $m_{n}^{(j)}$ (for $i\geq 1$).
    \item Verify that coefficients $m_{n}^{(j)}$ remain non-negative integers to higher orders in $q$. This test passes the character as an admissible one and the corresponding candidate theory to be viable.
    \item Check if the candidate CFT is actually a well-defined consistent theory\footnote{In the MLDE approach to RCFT, this procedure involves first computing the fusion coefficients using the {\it Verlinde} formula (see \cite{DiFrancesco:1997nk}) and checking if all these coefficients are non-negative. Even if all the fusion coefficients turn out to be non-negative, the admissible solution in question may not still describe a genuine RCFT. So, one moves to the next step which is to check if the characters of this admissible solution can be bilinearly paired with the characters of another known RCFT to result in a known meromorphic single-character RCFT. If this happens then the coset procedure of \cite{Gaberdiel:2016zke} can be used to identify the chiral algebra of the admissible solution using Lie algebra embeddings. Furthermore, the fusion algebra and modular data -- $S$ and $T$ matrices, etc. -- of this admissible solution must belong to some {\it Modular Tensor Category} (see \cite{rowell2009classification}).}.
\end{enumerate}
The general form of the MLDE for $n=2$ and $\ell\neq0$ reads
\begin{align}
    \left[\omega_{2\ell}(\tau)\mathcal{D}^{2} + \omega_{2\ell + 2}(\tau)\mathcal{D} + \omega_{2\ell + 4}(\tau)\right]f(\tau) = 0.
\end{align}
The free parameters $\mu_{i}$ in this MLDE come as coefficients to the modular forms and the number of these that show up in the MLDE can be determined as follows
\begin{align}
        \#(\mu) =\text{dim}\ \mathcal{M}_{2\ell}(\Gamma_{0}^{+}(p)) + \text{dim}\ \mathcal{M}_{2\ell + 2}(\Gamma_{0}^{+}(p)) + \text{dim}\ \mathcal{M}_{2\ell + 4}(\Gamma_{0}^{+}(p)) -1.
\end{align}
Knowledge of the dimensions of spaces of modular forms for each group would help in determining specific formulae for $\#(\mu)$.

\subsection{Building the covariant derivative}
\noindent There are certain subtleties we need to be careful of when formulating the covariant derivative that acts on modular forms belonging to Hecke groups. The key difference is that, unlike Fricke groups, Hecke groups possess non-zero dimensional space of weight $2$ modular forms. Thanks to this we don't need to include a quasi-modular form in the covariant derivative to preserve mapping from weight $r$ to weight $r+2$ forms. This being said, we could also define the covariant derivatives with the Eisenstein series of weight $2$ associated with the cusps to act as the quasi-modular forms. A distinct difference between Fricke and Hecke groups is that the latter possess a cusp at $\tau = 0$ in addition to others listed in table \ref{tab: Hecke_cusp_list}.
\\
\begin{table}[htb!]
    \centering
    \begin{tabular}{||c|c|c|c|c|c|c|c||}
     \hline
      $N$ & 4 & 6 & 8 & 9 & 10 & 12 \\[0.5ex]
      \hline\hline
      $\tau$ & $-\tfrac{1}{2}$ & $-\tfrac{1}{2}$, $-\tfrac{1}{3}$ & $-\tfrac{1}{2}$, $-\tfrac{1}{4}$ & $-\tfrac{1}{3}$, $-\tfrac{1}{3}$ & $-\tfrac{1}{2}$, $-\tfrac{1}{5}$ & $-\tfrac{1}{2}$, $-\tfrac{1}{3}$, $-\tfrac{1}{4}$, $-\tfrac{1}{6}$\\[1ex]
    \hline
    \end{tabular}
    \caption{The non-trivial cusps of Hecke groups $\Gamma_{0}(N)$ apart from the ones at $\tau = 0,i\infty$ that is common to all levels. The levels $N=2,3,7,11$ do not possess any non-trivial cusps.}
    \label{tab:  Hecke_cusp_list}
\end{table}
\\
Hence, the first choice for a Ramanujan-Serre covariant derivative for $\Gamma_{0}(N)$ is one that includes the Eisenstein series associated with the cusp at $\tau = 0$. Let us make the following definition
\begin{align}
    \begin{split}
        \mathcal{D} \equiv& \frac{1}{2\pi i}\frac{d}{d\tau} - \upsilon(r) E^{0}_{2,N}(\tau),\\
        \mathcal{D}^{n} =& \mathcal{D}_{r+2n-2}\circ\ldots\circ\mathcal{D}_{r+2}\circ\mathcal{D}_{r},
    \end{split}
\end{align}
where the Eisenstein series $E_{k,N}^{0}(\tau)$ is defined as follows \cite{Junichi}
\begin{align}
    E^{0}_{k,N}(\tau) = \begin{cases}
    -2^{\tfrac{k}{2} - \left\lfloor\left.\tfrac{N}{4}\right\rfloor\right.\tfrac{k}{2}}\frac{\left(E_{k}(2\tau) - E_{k}(\tau)\right)}{2^{k} - 1},\ &N = 2,4,8,\\
    \frac{-N^{\tfrac{k}{2}}\left(E_{k}(N\tau) - E_{k}(\tau)\right)}{N^{k} - 1},\ &N\in\mathbb{P},\\
    \frac{6^{\tfrac{k}{2}}\left(E_{k}(6\tau) - E_{k}(3\tau) - E_{k}(2\tau) + E_{k}(\tau)\right)}{(3^{k} - 1)(2^{k} - 1)},\ &N = 6,\\
    \frac{10^{\tfrac{k}{2}}\left(E_{k}(10\tau) - E_{k}(5\tau) - E_{k}(2\tau) + E_{k}(\tau)\right)}{(5^{k} - 1)(2^{k} - 1)},\ &N = 10,\\
    2^{-\tfrac{k}{2}}E^{0}_{k,6}(\tau),\ &N = 12,\\
    \end{cases}
\end{align}
Another definition of the Ramanujan-Serre covariant derivatives in the $\Gamma_{0}(N)$ is one including the Eisenstein series associated with the cusp at $\tau = i\infty$ as follows
\begin{align}
    \begin{split}
        \mathcal{D} \equiv& \frac{1}{2\pi i}\frac{d}{d\tau} - \upsilon(r) E^{\infty}_{2,N}(\tau),\\
        \mathcal{D}^{n} =& \mathcal{D}_{r+2n-2}\circ\ldots\circ\mathcal{D}_{r+2}\circ\mathcal{D}_{r},
    \end{split}
\end{align}
where $E^{\infty}_{k,N}(\tau)$ is the Eisenstein series associated with the cusp $\infty$ is defined as follows \cite{Junichi}
\begin{align}
    E^{\infty}_{k,N}(\tau) = \begin{cases}
    \frac{2^{k}E_{k}(N\tau) - E_{k}\left(\tfrac{N\tau}{2}\right)}{2^{k} - 1},\ &N = 2,4,8,\\
    \frac{N^{k}E_{k}(N\tau) - E_{k}(\tau)}{N^{k} - 1},\ &N\in\mathbb{P},\\
    \frac{6^{k}E_{k}(6\tau) - 3^{k}(3\tau) - 2^{k}E_{k}(2\tau) + E_{k}(\tau)}{(3^{k} - 1)(2^{k} - 1)},\ &N = 6,\\
    \frac{10^{k}E_{k}(10\tau) - 5^{k}E_{k}(5\tau) - 2^{k}E_{k}(2\tau) + E_{k}(\tau)}{(5^{k} - 1)(2^{k} - 1)},\ &N = 10,\\
    E^{\infty}_{k,6}(2\tau),\ &N = 12,\\
    \end{cases}
\end{align}
for $k\geq 4$. The values of $\upsilon(r)$ for levels $N\leq 12$ and prime divisor levels of $\mathbb{M}$ $N>12$ are tabulated in \ref{tab:Hecke_valence}.
\begin{table}[htb!]
    \centering
    \begin{tabular}{||c|c|c|c|c|c|c|c|c|c|c|c|c|c|c|c|c|c|c|c|c|c||}
    \hline
    $N$ & 2 & 3 & 4 & 5 & 6 & 7 & 8 & 9 & 10 & 11 & 12 & 13 & 17 & 19 & 23 & 29 & 31 & 41 & 59 & 71\\ [0.5ex]
    \hline\hline
    $\upsilon(r)$ & $\tfrac{r}{4}$ & $\tfrac{r}{3}$ & $\tfrac{r}{2}$ & $\tfrac{r}{2}$ & $r$ & $\tfrac{2r}{3}$ & $r$ & $r$ & $\tfrac{3r}{2}$ & $r$ & $2r$ & $\tfrac{7r}{6}$ & $\tfrac{3r}{2}$ & $\tfrac{5r}{2}$ & $2r$ & $\tfrac{5r}{2}$ & $\tfrac{8r}{3}$ & $\tfrac{7r}{2}$ & $5r$ & $6r$\\ [1ex]
    \hline
    \end{tabular}
    \caption{Values of $\upsilon(r) = \tfrac{r\overline{\mu}_{0}}{12}$ computed index data in table \ref{tab:Hecke_index_data}.}
    \label{tab:Hecke_valence}
\end{table}
\\
The choices, $E_{k,N}^{0}(\tau)$ and $E_{k,N}^{\infty}(\tau)$ however, do not yield correct covariant derivatives. For example, at level $N = 4$, we see from table \ref{tab:  Hecke_cusp_list} that the fundamental domain has a cusp at $\tau = -\tfrac{1}{2}$ and there exists an Eisenstein series associated with this cusp using which we can define yet another Ramanujan-Serre covariant derivative. Thus, we would have four unique MLDEs at level $N = 4$ and similarly, five unique MLDEs at levels $N = 6,8,9,10$, and seven unique MLDEs at level $N = 12$ (The definitions of the Eisenstein series associated with the non-trivial cusps can be found in \cite{Junichi}). 

\subsection{Quasimodular forms of Hecke groups}
\noindent To build sensible quasimodular forms for other Hecke levels, we use the following isomorphism \cite{Rose2014IntroductionTM}
\begin{align}
    \mathcal{QM}_{k}(\Gamma) \cong \mathcal{M}_{k}(\Gamma) \otimes_{\mathbb{C}}\mathbb{C}\left[G_{2}(\tau)\right],
\end{align}
It follows from this that a quasimodular form in $\Gamma\subset\text{SL}(2,\mathbb{Z})$ of weight $k$ and depth lesser than $r$ can be constructed as follows
\begin{align}
    Q(\tau) = f(\tau) - f_{r}(\tau)\left(\frac{E_{2}(\tau)}{12}\right)^{r},
\end{align}
where $f(\tau)$ is a quasimodular form of weight $k$ and depth $r$ and $f_{r}(\tau)$ is the coefficient in \ref{quasi_def}. As an example, let us cook up a quasimodular form for $\Gamma_{0}(6)$. At depth $r = 1$ and weight $k = 2$, we have 
\begin{align}
   Q(\tau) = f(\tau) - f_{1}(\tau)\frac{E_{2}(\tau)}{12}. 
\end{align}
If we were to make the choice, $Q(\tau) = \left(a_{1}\left(\Delta_{6}^{\infty}\right)^{2} + a_{2}\left(\Delta_{6}^{\infty}\Delta_{6}^{0}\right)  + a_{3}\left(\Delta_{6}^{0}\right)^{2}\right)(\tau)$ (see \cite{Junichi} for definitions), we find it to be a quasimodular form of weight $2$ and depth $0$, or simply a modular form. Now, setting $a_{1} = 22, a_{2} = 11, a_{3} = \tfrac{11}{12}$, we obtain the following expression for the quasimodular form in $\Gamma_{0}(6)$,
\begin{align}
    \begin{split}
        f(\tau) =& \frac{11}{12}\left(24\left(\Delta_{6}^{\infty}\right)^{2} + 12\left(\Delta_{6}^{\infty}\Delta_{6}^{0}\right)  + \left(\Delta_{6}^{0}\right)^{2}\right)(\tau) + \frac{1}{12}E_{2}(\tau)\\
        =& 1 + 24 q^3 + 24 q^6 + 96 q^9 + 24 q^{12} + 144 q^{15} + \ldots
    \end{split}
\end{align}
It is interesting to note that
\begin{align}\label{to_generalize}
    Q_{\Gamma_{0}(6)}(\tau) = \left(24\left(\Delta_{6}^{\infty}\right)^{2} + 12\left(\Delta_{6}^{\infty}\Delta_{6}^{0}\right)  + \left(\Delta_{6}^{0}\right)^{2}\right)(\tau) = \sum\limits_{n=0}^{\infty}c_{\text{D}_{4}}q^{3n},
\end{align}
where $c_{\textbf{D}_{4}}$ are the coefficients of the theta series for the $D_{4}$ lattice (OEIS sequence A$004011$ \cite{A004011}). Redoing the calculations for the group $\Gamma_{0}(4)$, we find that the quasimodular form is given by
\begin{align}
    f(\tau) = \frac{11}{12}\Theta_{\textbf{D}_{4}}(\tau) + \frac{1}{12}E_{2}(\tau).
\end{align}
Repeating the calculation at level $N=8$, we find the following relation between the part of the quasimodular form excluding $E_{2}(\tau)$ and the $D_{4}$ lattice theta series
\begin{align}
    Q_{\Gamma_{0}(8)}(\tau) 
 = \left(8\left(\Delta_{8}^{\infty}\right)^{2} + 8\left(\Delta_{8}^{\infty}\Delta_{8}^{0}\right) + \left(\Delta_{8}^{0}\right)^{2}\right)(\tau) = \sum\limits_{n=0}^{\infty}c_{\textbf{D}_{4}}q^{4n}.
\end{align}
For the three levels $N = 4, 6,8$, we have
\begin{align}\label{quasi_mod_ansatz}
    f_{\Gamma_{0}(N)}(\tau) =  + \frac{1}{12}E_{2}(\tau) + \frac{11}{12}\begin{cases}\sum\limits_{n=0}^{\infty}c_{\textbf{D}_{4}}q^{n} ,\ &N = 4\\
    \sum\limits_{n=0}^{\infty}c_{\textbf{D}_{4}}q^{\frac{N}{2}n},\ &N =6,8
    \end{cases}. 
\end{align}
This type of behavior in coefficients does not show up for the next levels $N = 10, 12$.
\section{Modular re-parameterization in \texorpdfstring{$\mathbf{\Gamma^{+}_{0}(p)}$}{Γ(p)+}}
\noindent  Taking a hint from the trend of the definitions and the method followed for modular re-parameterization as in \cite{Franc:2016}, one might be tempted to state that this should work for all prime-level Fricke groups. This, unfortunately, does not turn out to be true. We note that the procedure we followed to obtain MLDEs for level $p = 2$ and $ p = 3$ in \ref{reparameterized_MLDE_2+} and \ref{reparameterized_MLDE_3+_new} respectively mimics that done in the case of $\text{SL}(2,\mathbb{Z})$ (see, for example, \cite{Franc:2016}). A drastic difference at the group level we immediately notice is that levels $p\geq 5$, unlike $p = 2,3$, possess more than one elliptic point, excluding the Fricke involution point $\rho_{F_{p}} = \tfrac{i}{\sqrt{p}}$. At levels $p = 5,13$, we notice that the value of the Hauptmodules at the Fricke involution point becomes irrational (see table \ref{tab: Fricke_involution_point_limits}) and this affects the definition of $K_{p^{+}}(\tau)\equiv \tfrac{j_{p^{+}}(\rho_{F_{p}})}{j_{p^{+}}(\tau)}$, which in turn affects the way in which we define the Eisenstein series. The following expressions hold only when $p = 2,3$
\begin{align}
    E_{4}^{(p^{+})} = \frac{A_{p^{+}}^{2}}{1 - K_{p^{+}}},\ \ \ E_{6}^{(p^{+})} = \frac{A_{p^{+}}^{3}}{1 - K_{p^{+}}}.
\end{align}
\begin{table}[htb!]
    \centering
    \begin{tabular}{||c|c|c|c|c|c||}
    \hline
     $p$ & 2 & 3 & 5 & 7 & 13\\[1ex]
     \hline\hline
     $j_{p^{+}}(\rho_{F_{p}})$ & 256 & 108 & 22 + 10$\sqrt{5}$ & 27 & $2\sqrt{13}$\\ [1ex]
     \hline
    \end{tabular}
    \caption{The values of the Hauptmodules at the Fricke involution points for select levels of Fricke groups that possess integral coefficients in the $q$-series expansion of the Hauptmodules.}
    \label{tab: Fricke_involution_point_limits}
\end{table}
\\
Let us consider the following general ansatz for the re-parameterized $(n,\ell) = (2,0)$ MLDE in $\Gamma_{0}^{+}(p)$
\begin{align}\label{general_ansatz}
    \left[\theta_{K_{p^{+}}}^{2} + \left(\mathcal{D}A_{p^{+}}\right)\theta_{K_{p^{+}}} + \sum\limits_{i = 1}^{\nu_{p^{+}}}\mu_{i}\mathcal{T}_{p^{+}, i}\right]f(\tau) = 0,
\end{align}
where $\nu_{p^{+}} = \text{dim}\ \mathcal{M}_{4}(\Gamma_{0}^{+}(p))$ and $\mathcal{T}_{p^{+}, i}$ are weight $4$ modular forms which are the Eisenstein series $E_{4}^{(p^{+})}$ for levels $p= 2,3,5,7$, but are more complicated for level $p = 13$. See \cite{Umasankar:2022kzs} for an idea of constructing such modular forms for $\Gamma_{0}^{+}(13)$. Due to the irrational nature of $K_{5^{+}}(\tau)$ and the complicated relation between the Eisenstein series and the Hauptmodul, it turns out that re-parameterization following the ansatz \ref{general_ansatz} does not perform well at level $p = 5$. This will be discussed in further detail in subsequent sections. To set up the re-parameterized MLDE, we require Ramanujan identities at each Fricke level which we will present here. We note that although the first three identities for levels $p = 2,3$, shown in the sections below match the expressions originally derived by Zudilin in \cite{Zudilin2003TheHE}, the fact that the Ramanujan-Serre covariant derivative can be used as a building block to create these identities, is the emphasis we place here.

\subsection{Recipe for Ramanujan-Eisenstein identities}
\noindent We can use the Ramanujan-Serre covariant derivatives we have built previously for the Fricke groups as a guide to help us find Ramanujan-Eisenstein identities or simply the Ramanujan identities. The algorithm we shall use is the following:
\begin{enumerate}
    \item Choose Fricke group and identify the correct value of $\kappa(r)$ from table \ref{tab:Fricke_valence}.
    \item Choose an Eisenstein series whose identity we want to find and by looking at the space of modular forms of weight corresponding to the weight of the product of the Eisenstein series under consideration and the weight $2$ quasi-modular form $E_{2}^{(p^{+})}(\tau)$. Next, build a linear combination, all of which have the multiple $\kappa(r)$. Now, set $r = 1$ if the Eisenstein series being considered is $E_{2}^{(p^{+})}(\tau)$.
    \item Lastly, fix the coefficients of the linear combination by comparison to the $q$-series expansion of the $q$-derivative of the Eisenstein series considered.
\end{enumerate}
The Ramanujan identities in the $\text{SL}(2,\mathbb{Z})$ case read
\begin{align}\label{Ramanujan_identities_SL2Z}
    \begin{split}
        \mathcal{D}_{2}E_{2}(\tau) =& -\frac{1}{12}E_{4}(\tau),\\
        \mathcal{D}_{4}E_{4}(\tau) =& -\frac{1}{3}E_{6}(\tau),\\
        \mathcal{D}_{6}E_{6}(\tau) =& -\frac{1}{2}E_{4}^{2}(\tau).
    \end{split}
\end{align}
Since the space of modular forms for $\Gamma(1) = \text{SL}(2,\mathbb{Z})$ is given by the polynomial ring $\mathcal{M}_{k}(\Gamma(1)) = \mathbb{C}\left[E_{4}, E_{6}\right]$, we can build Eisenstein series of higher weights using powers of those of weights $4$ and $6$, i.e. $E_{2k}(\tau) = \left(E_{4}^{a}E_{6}^{b}\right)(\tau)$, where $a, b\in\mathbb{Z}$. This tells us that Ramanujan identities for higher weight Eisenstein series can be obtained by simply using the identities \ref{Ramanujan_identities_SL2Z}. Consider for example the Eisenstein series of weights $8$, $10$, and $12$ that can be built as follows
\begin{align}
    \begin{split}
        E_{8}(\tau) =& E_{4}^{2}(\tau),\ \ \ \
        E_{10}(\tau) = \left(E_{4}E_{6}\right)(\tau),\\
        E_{12}(\tau) =& \frac{1}{691}\left(441E_{4}^{3} + 250E_{6}^{2}\right)(\tau).
    \end{split}
\end{align}
Using \ref{Ramanujan_identities_SL2Z}, we find the following identities
\begin{align}
    \begin{split}
        \mathcal{D}_{8}E_{8}(\tau) =& -\frac{2}{3}E_{10}(\tau),\\
        \mathcal{D}_{10}E_{10}(\tau) =& -\frac{1}{2}E_{4}^{3}(\tau) - \frac{1}{3}E_{6}^{2}(\tau),\\
        \mathcal{D}_{12}E_{12}(\tau) =& -\left(E_{4}^{2}E_{6}\right)(\tau).
    \end{split}
\end{align}
This method, however, is not applicable to the congruence groups of interest since the ring of modular forms is more complicated. We are to study the space of modular forms of a particular weight for each group in order to derive identities for the Eisenstein series of higher weights.

\subsection{Level 2}
\noindent Next, consider the group $\Gamma_{0}^{+}(2)$ with $\kappa(r) = \tfrac{r}{8}$. Say we are interested in finding the Ramanujan identity associated with the Eisenstein series $E_{6}^{(2^{+})}(\tau)$. Firstly, we note that $r = 6$ and the common factor to the linear combination is $\kappa(r) = \tfrac{3}{4}$. Now, since the weight of the product $E_{2}^{(2^{+})}E_{6}^{(2^{+})}$ is $8$, we pick the linear combination $a\left(E_{4}^{(2^{+})}\right)^{2} + b\Delta_{2}$. Comparing the $q$-series expansions, we find the expression shown in \ref{Ramanujan_p=2}. The general Ramanujan identities for $\Gamma_{0}^{+}(2)$ are found to be 
\begin{align}\label{Ramanujan_p=2}
        \mathcal{D}_{2}E_{2}^{(2^{+})}(\tau) =& -\frac{1}{8}E_{4}^{(2^{+})}(\tau),\nonumber\\
        q\frac{d}{dq}E_{2}^{(2^{+})}(\tau) =& \frac{1}{8}\left(\left(E_{2}^{(2^{+})}\right)^{2} - E_{4}^{(2^{+})}\right)(\tau),\nonumber\\\\
        \mathcal{D}_{4}E_{4}^{(2^{+})}(\tau) =& -\frac{1}{2}E_{6^{(2^{+)}}}(\tau),\nonumber\\ 
        q\frac{d}{dq}E_{4}^{(2^{+})}(\tau) =& \frac{1}{2}\left(E_{2}^{(2^{+})}E_{4}^{(2^{+})} - E_{6}^{(2^{+})}\right)(\tau),\nonumber\\\\ 
        \mathcal{D}_{6}E_{6}^{(2^{+})}(\tau) =& -\frac{3}{4}\left(E_{4}^{(2^{+})}\right)^{2} + 64\Delta_{2}(\tau),\nonumber\\ 
        q\frac{d}{dq}E_{6}^{(2^{+})}(\tau) =& \frac{3}{4}\left(E_{2}^{(2^{+})}E_{6}^{(2^{+})} - \left(E_{4}^{(2^{+})}\right)^{2} + \frac{256}{3}\Delta_{2}\right)(\tau),\nonumber\\\\ 
        \mathcal{D}_{8}E_{8}^{(2^{+})}(\tau) =& -E_{10}^{(2^{+})}(\tau),\nonumber\\ 
        q\frac{d}{dq}E_{8}^{(2^{+})}(\tau) =& \left(E_{2}^{(2^{+})}E_{8}^{(2^{+})} - E_{10}^{(2^{+})}\right)(\tau),\nonumber\\\\ 
        \mathcal{D}_{10}E_{10}^{(2^{+})}(\tau) =& -\frac{5}{4}\left(E_{4}^{(2^{+})}\right)^{2}(\tau) - 192\left(E_{4}^{(2^{+})}\Delta_{2}\right)(\tau),\nonumber\\ 
        q\frac{d}{dq}E_{10}^{(2^{+})}(\tau) =& \frac{5}{4}\left(E_{2}^{(2^{+})}E_{10}^{(2^{+})} - \left(E_{4}^{(2^{+})}\right)^{3} - \frac{768}{5}E_{4}^{(2^{+})}\Delta_{2}\right)(\tau),\nonumber\\\\ 
        \mathcal{D}_{12}E_{12}^{(2^{+})}(\tau) =& -\frac{3}{2}\left(E_{4}^{(2^{+})}\right)^{2}(\tau)E_{6}^{(2^{+})}(\tau) + \frac{49248}{691}\left(E_{6}^{(2^{+})}\Delta_{2}\right)(\tau),\nonumber\\ 
        q\frac{d}{dq}E_{12}^{(2^{+})}(\tau) =& \frac{3}{2}\left(E_{2}^{(2^{+})}E_{12}^{(2^{+})} - \left(E_{4}^{(2^{+})}\right)^{2}E_{6}^{(2^{+})} + \frac{32832}{691}E_{6}^{(2^{+})}\Delta_{2}\right)(\tau).
\end{align}
The behaviour of the Hauptmodul $j_{2^{+}}(\tau)$ near points $\tau = \rho_{2}, \tfrac{i}{\sqrt{2}}, i\infty$ reads
\begin{align}\label{Hauptmodul_limits_2+}
    \begin{split}
        j_{2^{+}}(\rho_{2}) \to& \ 0,\\ 
        j_{2^{+}}\left(\frac{i}{\sqrt{2}}\right) \to&\  256,\\
        j_{2^{+}}\left(i\infty\right) \to&\  \infty.
    \end{split}
\end{align}
Now, the $(n,\ell) = (2,0)$ MLDE for $\Gamma_{0}^{+}(2)$ takes the following form
\begin{align}
    \left[\mathcal{D}^{2} + \mu E_{4}^{(2^{+})}(\tau)\right]f(\tau) = 0,
\end{align}
where we have used the fact that  $\text{dim}\ \mathcal{M}_{2}(\Gamma_{0}^{+}(2)) = 0$ and $\omega_{4}(\tau) = \mu E_{4}^{(2^{+})}(\tau)$. Substituting for the covariant derivative, the ODE now reads
\begin{align}\label{reparameterized_MLDE}
    \left[\theta_{q}^{2} - \frac{1}{4}E_{2}^{(2^{+})}\theta_{q} + \mu E_{4}^{(2^{+})}\right]f(\tau) = 0.
\end{align}
where we have defined $\theta_{a}\equiv a\tfrac{d}{da}$ with $q\tfrac{d}{dq} = \tfrac{1}{2\pi i}\frac{d}{d\tau}$. The $q$-derivative of $j_{2^{+}}(\tau)$ reads
\begin{align}\label{j_2+_derivatives}
    \begin{split}
    q\frac{d}{dq}j_{2^{+}}(\tau) =& \frac{1}{2\pi i}\frac{d}{d\tau}j_{2^{+}}(\tau)\\
    =& -\left(\frac{E_{6}^{(2^{+})}}{E_{4}^{(2^{+})}}j_{2^{+}}\right)(\tau),\\
    q^{2}\frac{d^{2}}{dq^{2}}j_{2^{+}}(\tau) =& \frac{1}{(2\pi i)^{2}}\frac{d^{2}}{d\tau^{2}}j_{2^{+}}(\tau)\\
    =& \left[\left(\frac{3}{4}E_{4}^{(2^{+})} + \frac{1}{2}\left(\frac{E_{6}^{(2^{+})}}{E_{4}^{(2^{+})}}\right)^{2}- \frac{1}{4}\frac{E_{2}^{(2^{+})}E_{6}^{(2^{+})}}{E_{4}^{(2^{+})}}\right)j_{2^{+}} - 64\frac{\Delta_{2}}{E_{4}^{(2^{+})}}\right](\tau).
    \end{split}
\end{align}
We now make the following definitions
\begin{align}
        \theta_{q}\equiv A_{2^{+}}\theta_{K_{2^{+}}},\ \ A_{2^{+}}\equiv \frac{E_{6}^{(2^{+})}}{E_{4}^{(2^{+})}},\ \ K_{2^{+}}\equiv \frac{256}{j_{2^{+}}}.
\end{align}
The $q$-derivatives of $A_{2^{+}}$ and $K_{2^{+}}$ are found to be
\begin{align}
    \begin{split}
        \theta_{q}A_{2^{+}} =& -\frac{3}{4}E_{4}^{(2^{+})} + 64\frac{\Delta_{2}}{E_{4}^{(2^{+})}} + \frac{1}{2}\left(\frac{E_{6}^{(2^{+})}}{E_{4}^{(2^{+})}}\right)^{2} + \frac{E_{2}^{(2^{+})}E_{6}^{(2^{+})}}{E_{4}^{(2^{+})}},\\
    \theta_{q}K_{2^{+}} =& A_{2^{+}}K_{2^{+}}.
    \end{split}
\end{align}
We find the following relations between the weight $4$ and weight $6$ Eisenstein series and the newly defined variables,
\begin{align}\label{repara_definitions_2+}
        E_{4}^{(2^{+})} = \frac{A_{2^{+}}^{2}}{1 - K_{2^{+}}},\ \ \ 
        E_{6}^{(2^{+})} =\frac{A_{2^{+}}^{3}}{1 - K_{2^{+}}},\ \ \ \Delta_{2} = \frac{1}{256}\frac{A_{2^{+}}^{4}K_{2^{+}}}{(1-K_{2^{+}})^{2}}.
\end{align}
Now, the covariant derivative of $A_{2^{+}}$ is found to be 
\begin{align}
    \begin{split}
        \mathcal{D}A_{2^{+}} =& -\frac{3}{4}E_{4}^{(2^{+})} + 64\frac{\Delta_{2}}{E_{4}^{(2^{+})}} + \frac{1}{2}\left(\frac{E_{6}^{(2^{+})}}{E_{4}^{(2^{+})}}\right)^{2}\\
        =& -A_{2^{+}}^{2}\frac{1 + K_{2^{+}}}{4(1 - K_{2^{+}})}.
    \end{split} 
\end{align}
Using these results, we find
\begin{align}\label{derivatives_repara_2+}
    \begin{split}
        \mathcal{D} =& \theta_{q} = A_{2^{+}}\theta_{K_{2^{+}}},\\
        \mathcal{D}^{2} =& \left(\theta_{q} - \frac{1}{4}E_{2}^{(2^{+})}\right)\theta_{q} = A_{2^{+}}^{2}\theta_{K_{2^{+}}}^{2} + \left(\mathcal{D}A_{2^{+}}\right)\theta_{K_{2^{+}}}.
    \end{split}
\end{align}
We can now recast the MLDE \ref{reparameterized_MLDE} as follows
\begin{align}\label{reparameterized_MLDE_2+}
    \left[\theta_{K_{2^{+}}}^{2} - \frac{1 + K_{2^{+}}}{4(1 - K_{2^{+}})}\theta_{K_{2^{+}}} + \frac{\mu}{1 - K_{2^{+}}}\right]f(K_{2^{+}}) = 0.
\end{align}
This ODE is now of the hypergeometric type and can be solved to obtain 
\begin{align}
    f(\tau) = c_{1}\left(-K_{2^{+}}\right)^{\alpha}\pFq{2}{1}{\alpha - \frac{1}{4},\  \alpha}{2\alpha + \frac{1}{4}}{K_{2^{+}}} + c_{2}\left(-K_{2^{+}}\right)^{\beta}\pFq{2}{1}{\beta - \frac{1}{4},\  \beta}{2\beta + \frac{1}{4}}{K_{2^{+}}},
\end{align}
where we have defined $\alpha \equiv \tfrac{1}{8}\left(3 - \sqrt{25 - 64\mu}\right)$ and $\beta\equiv \tfrac{1}{8}\left(5 + \sqrt{25 - 64\mu}\right)$.
\newline

\subsection{Level 3}
The Ramanujan identities for $\Gamma_{0}^{+}(3)$ are found to be
\begin{align}\label{Ramanujan_p=3}
        \mathcal{D}_{2}E_{2}^{(3^{+})}(\tau) =& -\frac{1}{6}E_{4}^{(3^{+})}(\tau),\nonumber\\ 
        q\frac{d}{dq}E_{2}^{(3^{+})}(\tau) =& \frac{1}{6}\left(\left(E_{2}^{(3^{+})}\right)^{2} - E_{4}^{(3^{+})}\right)(\tau),\nonumber\\\\ 
        \mathcal{D}_{4}E_{4}^{(3^{+})}(\tau) =& -\frac{2}{3}E_{6}^{(3^{+})}(\tau),\nonumber\\ 
        q\frac{d}{dq}E_{4}^{(3^{+})}(\tau) =& \frac{2}{3}\left(E_{2}^{(3^{+})}E_{4}^{(3^{+})} - E_{6}^{(3^{+})}\right)(\tau),\nonumber\\\\ 
        \mathcal{D}_{6}E_{6}^{(3^{+})}(\tau) =& -\frac{1}{2}\left(E_{4}^{(3^{+})}\right)^{2}(\tau) - \frac{1}{2}\frac{\left(E_{6}^{(3^{+})}\right)^{2}(\tau)}{E_{4}^{(3^{+})}(\tau)},\nonumber\\ 
        q\frac{d}{dq}E_{6}^{(3^{+})}(\tau) =& \left(E_{2}^{(3^{+})}E_{6}^{(3^{+})} - \frac{1}{2}\left(E_{4}^{(2^{+})}\right)^{2} - \frac{1}{2}\frac{\left(E_{6}^{(3^{+})}\right)^{2}}{E_{4}^{(3^{+})}}\right)(\tau),\nonumber\\\\ 
        \mathcal{D}_{8}E_{8}^{(3^{+})}(\tau) =& -\frac{4}{3}\left(E_{4}^{(3^{+})}E_{6}^{(3^{+})}\right)(\tau) + \frac{8064}{205}\frac{E_{6}^{(3^{+})}(\tau)}{\left(E_{4}^{(3^{+})}\right)^{\tfrac{1}{2}}(\tau)}\Delta_{3}(\tau),\nonumber\\ 
        q\frac{d}{dq}E_{8}^{(3^{+})}(\tau) =& \frac{4}{3}\left(E_{2}^{(3^{+})}E_{8}^{(3^{+})} - E_{4}^{(3^{+})}E_{6}^{(3^{+})} + \frac{6048}{205}\frac{E_{6}^{(3^{+})}}{\left(E_{4}^{(3^{+})}\right)^{\tfrac{1}{2}}}\Delta_{3}\right)(\tau),\nonumber\\\\ 
        \mathcal{D}_{10}E_{10}^{(3^{+})}(\tau) =& -\frac{5}{3}\left(E_{4}^{(3^{+})}\right)^{3}(\tau) + \frac{7974}{61}\left(E_{4}^{(3^{+})}\right)^{\tfrac{3}{2}}(\tau)\Delta_{3}(\tau) - \frac{7776}{61}\Delta_{3}^{2}(\tau),\nonumber\\ 
        q\frac{d}{dq}E_{10}^{(3^{+})}(\tau) =& \frac{5}{3}\left(E_{2}^{(3^{+})}E_{10}^{(3^{+})} - \left(E_{4}^{(3^{+})}\right)^{3} + \frac{23922}{305}\left(E_{4}^{(3^{+})}\right)^{\tfrac{3}{2}}\Delta_{3} - \frac{23328}{305}\Delta_{3}^{2}\right)(\tau),\nonumber\\ 
        =& \frac{5}{3}\left(E_{2}^{(3^{+})}E_{10}^{(3^{+})} - E_{12}^{(3^{+})} + \frac{76660529}{492323680}\left(\left(E_{4}^{(3^{+})}\right)^{3} - E_{4}^{(3^{+})}E_{8}^{(3^{+})}\right) - \frac{465230304}{15385115}\Delta_{3}^{2}\right)(\tau),\nonumber\\\\ 
        \mathcal{D}_{12}E_{12}^{(3^{+})}(\tau) =& -2\left(E_{4}^{(3^{+})}\right)^{2}(\tau)E_{6}^{(3^{+})}(\tau) + \frac{3625344}{50443}\left(E_{4}^{(3^{+})}\right)^{\tfrac{1}{2}}(\tau)E_{6}^{(3^{+})}(\tau)\Delta_{3}(\tau),\nonumber\\ 
        q\frac{d}{dq}E_{12}^{(3^{+})}(\tau) =& 2\left(E_{2}^{(3^{+})}E_{12}^{(3^{+})} - \left(E_{4}^{(3^{+})}\right)^{2}E_{6}^{(3^{+})} + \frac{1812672}{50443}\left(E_{4}^{(3^{+})}\right)^{\tfrac{1}{2}}E_{6}^{(3^{+})}\Delta_{3}\right)(\tau).
\end{align}
\noindent The behaviour of the Hauptmodul $j_{3^{+}}(\tau)$ near points $\tau = \rho_{3}, \tfrac{i}{\sqrt{3}}, i\infty$ reads
\begin{align}\label{Hauptmodul_limits_3+}
    \begin{split}
        j_{3^{+}}(\rho_{3}) \to& \ 0,\\ 
        j_{3^{+}}\left(\frac{i}{\sqrt{3}}\right) \to&\  108,\\
        j_{3^{+}}\left(i\infty\right) \to&\  \infty.
    \end{split}
\end{align}
Now, the $(n,\ell) = (2,0)$ MLDE for $\Gamma_{0}^{+}(3)$ takes the following form
\begin{align}
    \left[\mathcal{D}^{2} + \mu E_{4}^{(3^{+})}(\tau)\right]f(\tau) = 0.
\end{align}
where we have used $\text{dim}\ \mathcal{M}_{2}(\Gamma_{0}^{+}(3)) = 0$ and $\omega_{4}(\tau) = \mu E_{4}^{(3^{+})}(\tau)$. Substituting for the covariant derivative, the ODE now reads
\begin{align}\label{reparameterized_MLDE_3+}
    \left[\theta_{q}^{2} - \frac{1}{3}E_{2}^{(3^{+})}\theta_{q} + \mu E_{4}^{(3^{+})}\right]f(\tau) = 0.
\end{align}
We now make the following definitions
\begin{align}
        \theta_{q}\equiv A_{3^{+}}\theta_{K_{3^{+}}},\ \ A_{3^{+}}\equiv \frac{E_{6}^{(3^{+})}}{E_{4}^{(3^{+})}},\ \ K_{3^{+}}\equiv \frac{108}{j_{3^{+}}}.
\end{align}
We find the following relations between the weight $4$ and weight $6$ Eisenstein series and the newly defined variables,
\begin{align}
        E_{4}^{(3^{+})} = \frac{A_{3^{+}}^{2}}{1 - K_{3^{+}}},\ \ \ 
        E_{6}^{(3^{+})} = \frac{A_{3^{+}}^{3}}{1 - K_{3^{+}}}.
\end{align}
Now, the covariant derivative of $A_{3^{+}}$ is found to be 
\begin{align}
    \begin{split}
        DA_{3^{+}} =& -\frac{1}{2}E_{4}^{(3^{+})} + \frac{1}{6}\left(\frac{E_{6}^{(3^{+})}}{E_{4}^{(3^{+})}}\right)^{2}\\
        =& -A_{3^{+}}^{2}\frac{2 + K_{3^{+}}}{6(1 - K_{3^{+}})}.
    \end{split} 
\end{align}
Using this result, we can recast the MLDE \ref{reparameterized_MLDE_3+} as follows
\begin{align}\label{reparameterized_MLDE_3+_new}
    \left[\theta_{K_{3^{+}}}^{2} - \frac{2 + K_{3^{+}}}{6(1 - K_{3^{+}})}\theta_{K_{3^{+}}} + \frac{\mu}{1 - K_{3^{+}}}\right]f(K_{3^{+}}) = 0.
\end{align}
This ODE is now of the hypergeometric type and can be solved to obtain 
\begin{align}
    f(\tau) = c_{1}\left(-K_{2^{+}}\right)^{\alpha}\pFq{2}{1}{\alpha - \frac{5}{6},\  \alpha}{2\alpha - \frac{1}{3}}{K_{3^{+}}} + c_{2}\left(-K_{2^{+}}\right)^{\beta}\pFq{2}{1}{\beta - \frac{5}{6},\  \beta}{2\beta - \frac{1}{3}}{K_{3^{+}}},
\end{align}
where we have defined $\alpha \equiv \tfrac{1}{3}\left(2 - \sqrt{4 - 9\mu}\right)$ and $\beta\equiv \tfrac{1}{3}\left(2 + \sqrt{4 - 9\mu}\right)$.

\subsection{Level 5}
\noindent The Ramanujan identities for $\Gamma_{0}^{+}(5)$ are found to be (see appendix \ref{appendix: Mod_5_and_7} for definitions of modular forms belonging to this group)
\begin{align}
        \mathcal{D}_{2}E_{2}^{(5^{+})}(\tau) =& -\frac{1}{4}E_{4}^{(5^{+})}(\tau) + \frac{4}{13}\Delta_{5}(\tau),\nonumber\\ 
        q\frac{d}{dq} E_{2}^{(5^{+})}(\tau) =&\frac{1}{4}\left(\left(E_{2}^{(5^{+})}\right)^{2} - E_{4}^{(5^{+})} + \frac{16}{13}\Delta_{5}\right)(\tau),\nonumber\\\\ 
        \mathcal{D}_{4}E_{4}^{(5^{+})}(\tau) =& -E_{6}^{(5^{+})}(\tau),\nonumber\\ 
        q\frac{d}{dq} E_{4}^{(5^{+})}(\tau) =& \left(E_{2}^{(5^{+})}E_{4}^{(5^{+})} - E_{6}^{(5^{+})}\right)(\tau),\nonumber\\\\ 
        \mathcal{D}_{6}E_{6}^{(5^{+})}(\tau) =& -\frac{3}{2}\left(E_{4}^{(5^{+})}\right)^{2}(\tau) + \frac{464}{13}\left(E_{4}^{(5^{+})}\Delta_{5}\right)(\tau) + \frac{20000}{169}\Delta_{5}^{2}(\tau),\nonumber\\ 
        q\frac{d}{dq} E_{6}^{(5^{+})}(\tau) =& \frac{3}{2}\left(E_{2}^{(5^{+})}E_{6}^{(5^{+})} - \left(E_{4}^{(5^{+})}\right)^{2} + \frac{928}{39}E_{4}^{5^{+}}\Delta_{5} + \frac{40000}{507}\Delta_{5}^{2} \right)(\tau),\nonumber\\\\ 
        \mathcal{D}_{8}E_{8}^{(5^{+})}(\tau) =& -2\left(E_{4}^{(5^{+})}(\tau)E_{6}^{(5^{+})}(\tau)\right)(\tau) + \frac{72000}{4069}\left(E_{6}^{(5^{+})}\Delta_{5}\right)(\tau),\nonumber\\ 
        q\frac{d}{dq} E_{8}^{(5^{+})}(\tau) =& 2\left(E_{2}^{(5^{+})}E_{8}^{(5^{+})} - E_{4}^{(5^{+})}E_{6}^{(5^{+})} + \frac{36000}{4069}E_{6}^{(5^{+})}\Delta_{5}\right)(\tau),\nonumber\\\\ 
        \mathcal{D}_{10}E_{10}^{(5^{+})}(\tau) =& -\frac{5}{2}\left(E_{4}^{(5^{+})}\right)^{3}(\tau) + \frac{537488}{6773}\left(E_{4}^{(5^{+})}\right)^{2}(\tau)\Delta_{5}(\tau) + \frac{14556000}{88049}E_{4}^{(5^{+})}(\tau)\Delta_{5}^{2}(\tau)\nonumber\\ 
        &{}- \frac{307368000}{1144637}\Delta_{5}^{3}(\tau),\nonumber\\ 
        q\frac{d}{dq} E_{10}^{(5^{+})}(\tau) =& \frac{5}{2}\left(E_{2}^{(5^{+})}E_{10}^{(5^{+})} - \left(E_{4}^{(5^{+})}\right)^{3} + \frac{1074976}{33865}\left(E_{4}^{(5^{+})}\right)^{2}\Delta_{5}\right.\nonumber\\
        &{}\ \ \ \ \ \ \ \ \ \ \ \ \ \ \ \ \ \ \ \left.+ \frac{5822400}{88049}E_{4}^{(5^{+})}\Delta_{5}^{2} - \frac{122947200}{1144637}\Delta_{5}^{3}\right)(\tau),\nonumber\\\\ 
        \mathcal{D}_{12}E_{12}^{(5^{+})}(\tau) =& -3\left(E_{4}^{(5^{+})}\right)^{2}E_{6}^{(5^{+})} + \frac{298944000}{5398783}\left(E_{4}^{(5^{+})}E_{6}^{(5^{+})}\Delta_{5}\right)(\tau)\nonumber\\ 
        &+ \frac{646164000}{70184179}\left(E_{6}^{(5^{+})}\Delta_{5}^{2}\right)(\tau),\\ 
        q\frac{d}{dq} E_{12}^{(5^{+})}(\tau) =& 3\left(E_{2}^{(5^{+})} E_{12}^{(5^{+})} - \left( E_{4}^{(5^{+})}\right)^{2} E_{6}^{(5^{+})} + \frac{99648000}{5398783}E_{4}^{(5^{+})}E_{6}^{(5^{+})}\Delta_{5}\right.\nonumber\\ 
        &{}\ \ \ \ \ \ \ \ \ \ \ \ \ \ \ \ \ \ \ \ \ \ \ \ \ \ \ \ \ \ \ \ \ \ \ \ \ \ \ \ \ \left.+ \frac{2153088000}{70184179}E_{6}^{(5^{+})}\Delta_{5}^{2}\right)(\tau).
\end{align}
The behaviour of the Hauptmodul $j_{5^{+}}(\tau)$ near points $\tau = \rho_{5,1}, \rho_{5,2}, \tfrac{i}{\sqrt{5}}, i\infty$ reads
\begin{align}
    \begin{split}
        j_{5^{+}}(\rho_{5,1})\to&\ 22 + 10\sqrt{5},\\
        j_{5^{+}}(\rho_{5,2})\to&\ 44,\\
        j_{5^{+}}\left(\frac{i}{\sqrt{5}}\right)\to&\ 22 + 10\sqrt{5},\\
        j_{5^{+}}(i\infty)\to&\ \infty.
    \end{split}
\end{align}
We find the derivative of the Hauptmodul to be
\begin{align}\label{derivative_1}
    \begin{split}
    q\frac{d}{dq}j_{5^{+}}(\tau) =& -\frac{E_{6}^{(5^{+})}}{E_{4}^{(5^{+})}}\left(j_{5^{+}} - \frac{36}{13}\right)(\tau)\\
    =& -j_{5^{+}}\left(\frac{E_{6}^{(5^{+})}}{E_{2,5^{'}}^{2}}\right)(\tau).
    \end{split}
\end{align}
It is easy to show that the re-parameterized derivatives take the following form
\begin{align}
    \begin{split}
        \theta_{q} =& -\left(\frac{E_{6}^{(5^{+})}}{E_{2,5^{'}}^{2}}\right)\theta_{j_{5^{+}}},\\
        \mathcal{D}^{2} =& \left(\theta_{q} - \frac{1}{2}E_{2}^{(5^{+})}\right)\theta_{q}\\
        =& \left(\frac{E_{6}^{(5^{+})}}{E_{2,5^{'}}^{2}}\right)^{2}\theta_{j_{5^{+}}}^{2} + \left(\frac{3}{2}E_{2,5^{'}}^{2} -8\frac{\Delta_{2}^{2}}{E_{2,5^{'}}^{2}} - 44\Delta_{5} - \frac{\left(E_{6}^{(5^{+})}\right)^{2}}{E_{2,5^{'}}^{4}}\right)\theta_{j_{5^{+}}},
    \end{split}
\end{align}
where we used $q\tfrac{d}{dq}\Delta_{5}(\tau) = \left(\Delta_{5}E_{2}^{(5^{+})}\right)(\tau)$. We find that at this level, it is easier to re-parameterize the MLDE in terms of the Hauptmodul rather than in terms of $K_{5^{+}}(\tau)$ since the Eisenstein series is related to this inverse Haupotmodul via Heun's functions as opposed to a straightforward relation we could derive earlier by mimicking the $\text{SL}(2, \mathbb{Z})$ construction. This is explored further in appendix \ref{appendix:B}. From \cite{Sakai2014TheAO}, we have the following expressions for the Eisenstein series and the cusp form in terms of Heun's function,
\begin{align}
    \begin{split}
        \Delta_{5}(\tau) =& \frac{1}{j_{5^{+}}}H\ell_{5}^{4}(\tau),\\    
        E_{2,5^{'}}^{2}(\tau) =& H\ell_{5}^{4}(\tau),\\
        E_{6}^{(5^{+})}(\tau) =& \sqrt{1 - \frac{44}{j_{5^{+}}} - \frac{16}{j_{5^{+}}^{2}}}\ H\ell_{5}^{6}(\tau),\\
        H\ell_{5}(\tau) \equiv& H\ell\left(\frac{11 + 5\sqrt{5}}{11 - 5\sqrt{5}}, -\frac{3(11 + 5\sqrt{5})}{8};\frac{1}{4},\frac{3}{4},1,\frac{1}{2};K_{5^{+}}(\tau)\right).
    \end{split}
\end{align}
From this, we find the following useful relations
\begin{align}
    \begin{split}
        \left(\frac{\Delta_{5}}{E_{2,5^{'}}^{2}}\right)(\tau) =& \mathfrak{a} = \frac{1}{j_{5^{+}}(\tau)},\\
        \left(\frac{\left(E_{6}^{(5^{+})}\right)^{2}}{\Delta_{5}^{3}}\right)(\tau) =& \mathfrak{b} = \left(j_{5^{+}}^{3} - 44j_{5^{+}}^{2} - 16j_{5^{+}}\right)(\tau),\\
        \left(\frac{\left(E_{6}^{(5^{+})}\right)^{2}}{E^{6}_{2,5^{'}}}\right)(\tau) =& \mathfrak{c}  = \left(1 - \frac{44}{j_{5^{+}}} - \frac{16}{j_{5^{+}}^{2}}\right)(\tau).
    \end{split}
\end{align}
Now, since $\text{dim}\ \mathcal{M}_{4}(\Gamma_{0}^{+}(5)) = 2$, we choose $\omega_{4}(\tau) = \mu_{1}E_{2,5^{'}}^{2}(\tau) + \mu_{2}\Delta_{5}(\tau)$ with which the re-parameterized MLDE reads
\begin{align}
    \begin{split}
        \left[\theta_{j_{5^{+}}}^{2} + \left(\frac{3}{2}\frac{1}{\mathfrak{c}} - 8\frac{\mathfrak{a}^{2}}{\mathfrak{c}} - 44\frac{\mathfrak{a}}{\mathfrak{c}} - 1\right)\theta_{j_{5^{+}}} + \frac{\mu_{1}}{\mathfrak{c}} + \frac{\mu_{2}\mathfrak{a}}{\mathfrak{c}}\right]f(\tau) = 0.
    \end{split}
\end{align}
This can be simplified to read
\begin{align}
    \left[\theta_{j_{5^{+}}}^{2} + \left(\frac{j_{5^{+}}^{2} + 16}{2(j_{5^{+}}^{2} - 44j_{5^{+}} - 16)}\right)\theta_{j_{5^{+}}} + \frac{j_{5^{+}}(j_{5^{+}}\mu_{1} + \mu_{2})}{(j_{5^{+}}^{2} - 44j_{5^{+}} - 16)}\right]f(\tau) = 0.
\end{align}
The solution to this ODE is expressed in terms of Heun's function as shown below
\begin{align}
    \begin{split}
        f(\tau) =& c_{1}\ H\ell\left[-\frac{\left(2j_{5^{+}}(\rho_{5,1}) + 4\right)}{4},\frac{\mu_{2}j_{5^{+}}(\rho_{5,1})}{16};\frac{\alpha_{-}}{4}, \frac{\alpha_{+}}{4},\frac{1}{2},\frac{1}{2};\frac{1}{2};-\frac{j_{5^{+}}(\rho_{5,1})j_{5^{+}}(\tau)}{16}\right]\\
        \\+& c_{2}\ H\ell\left[-\frac{\left(2j_{5^{+}}(\rho_{5,1}) + 4\right)}{4},\frac{(\mu_{2} - 11)j_{5^{+}}(\rho_{5,1})}{16};\frac{\beta_{+}}{4}, \frac{\beta_{-}}{4},\frac{3}{2},\frac{1}{2};-\frac{j_{5^{+}}(\rho_{5,1})j_{5^{+}}(\tau)}{16}\right]k(j_{5^{+}}),
    \end{split} 
\end{align}
where $\alpha_{\pm}\equiv \tfrac{1 \pm \sqrt{1-16\mu_{1}}}{4}$ and $\beta_{\pm}\equiv \tfrac{3 \pm \sqrt{1-16\mu_{1}}}{4}$ and  $k(j_{5^{+}}) \equiv \left(\tfrac{j_{5^{+}}(\tau)}{j_{5^{+}}(\rho_{5,2}) - j_{5^{+}}(\rho_{5,1})}\right)^{\frac{1}{2}}$.

\subsection{Level 7}
\noindent The Ramanujan identities for $\Gamma_{0}^{+}(7)$ are found to be (see appendix \ref{appendix: Mod_5_and_7} for definitions of modular forms belonging to this group)
\begin{align}
        \mathcal{D}_{2}E_{2}^{(7^{+})}(\tau)  =& -\frac{1}{3}E_{4}^{(7^{+})}(\tau) + \frac{3}{5}\Delta_{7^{+},4}(\tau),\nonumber\\
        q\frac{d}{dq}E_{2}^{(7^{+})}(\tau) =& \frac{1}{3}\left(\left(E_{2}^{(7^{+})}\right)^{2} - E_{4}^{(7^{+})} + \frac{9}{5}\Delta_{7^{+},4}\right)(\tau),\nonumber\\\\
        \mathcal{D}_{4}E_{4}^{(7^{+})}(\tau) =& -\frac{4}{3}E_{6}^{(7^{+})}(\tau) + \frac{96}{215}\Delta_{7^{+},6}(\tau),\nonumber\\
        q\frac{d}{dq}E_{4}^{(7^{+})}(\tau) =& \frac{4}{3}\left(E_{2}^{(7^{+})}E_{4}^{(7^{+})} - E_{6}^{(7^{+})} + \frac{72}{215}\Delta_{7^{+},6}\right)(\tau),\nonumber\\\\
        \mathcal{D}_{6}E_{6}^{(7^{+})}(\tau) =& -2\left(E_{4}^{(7^{+})}\right)^{2}(\tau) + \frac{5733}{215}\left(E_{4}^{(7^{+})}\Delta_{7^{+},6}\right)(\tau) + \frac{136269}{1075}\Delta_{7^{+},4}^{2}(\tau),\nonumber\\
        q\frac{d}{dq}E_{6}^{(7^{+})}(\tau) =& 2\left(E_{2}^{(7^{+})}E_{6}^{(7^{+})} - \left(E_{4}^{(7^{+})}\right)^{2} + \frac{5733}{430}E_{4}^{(7^{+})}\Delta_{7^{+},4} + \frac{136269}{2150}\Delta_{7^{+},4}^{2}\right)(\tau),\nonumber\\\\
        \mathcal{D}_{8}E_{8}^{(7^{+})}(\tau) =& -\frac{8}{3}\left(E_{4}^{(7^{+})}E_{6}^{(7^{+})}\right)(\tau) + \frac{4276032}{258215}\left(E_{4}^{(7^{+})}\Delta_{7^{+},6}\right)(\tau) + \frac{1862784}{30025}\Delta_{7^{+},10}(\tau),\nonumber\\
        q\frac{d}{dq}E_{8}^{(7^{+})}(\tau) =& \frac{8}{3}\left(E_{2}^{(7^{+})}E_{8}^{(7^{+})} - E_{4}^{(7^{+})}E_{6}^{(7^{+})} + \frac{1603512}{258215}E_{4}^{(7^{+})}\Delta_{7^{+},6} + \frac{698544}{30025}\Delta_{7^{+},10}\right)(\tau),\nonumber\\\\
        \mathcal{D}_{10}E_{10}^{(7^{+})}(\tau) =& -\frac{10}{3}\left(\frac{E_{4}^{(7^{+})}}{\Delta_{7^{+},4}}\right)^{4}(\tau)\Delta_{7}^{4}(\tau) + \frac{27143}{573}\left(\frac{E_{4}^{(7^{+})}}{\Delta_{7^{+},4}}\right)^{3}(\tau)\Delta_{7}^{4}(\tau)\nonumber\\
        & + \frac{372057}{955}\left(\frac{E_{4}^{(7^{+})}}{\Delta_{7^{+},4}}\right)^{2}(\tau)\Delta_{7}^{4}(\tau) +\frac{310464}{955}\left(\frac{E_{4}^{(7^{+})}}{\Delta_{7^{+},4}}\right)^{2}(\tau)\Delta_{7}^{4}(\tau) + \frac{25239312}{23875}\Delta_{7}^{4}(\tau),\nonumber\\
        q\frac{d}{dq}E_{10}^{(7^{+})}(\tau) =& \frac{10}{3}\left(E_{2}^{(7^{+})}E_{10}^{(7^{+})} - \left(\frac{E_{4}^{(7^{+})}}{\Delta_{7^{+},4}}\right)^{4}\Delta_{7}^{4} + \frac{27143}{1910}\left(\frac{E_{4}^{(7^{+})}}{\Delta_{7^{+},4}}\right)^{3}\Delta_{7}^{4}\right.\nonumber\\
        &{}\ \ \ \ \ \ \ \ \ \left. + \frac{1116171}{9550}\left(\frac{E_{4}^{(7^{+})}}{\Delta_{7^{+},4}}\right)^{2}\Delta_{7}^{4} + \frac{465696}{4775}\left(\frac{E_{4}^{(7^{+})}}{\Delta_{7^{+},4}}\right)\Delta_{7}^{4} 
        + \frac{37858968}{119375}\Delta_{7}^{4}
        \right)(\tau),\nonumber\\\\
        \mathcal{D}_{12}E_{12}^{(7^{+})}(\tau) =& 4\left(\frac{E_{4}^{(7^{+})}}{\Delta_{7^{+},4}}\right)^{\tfrac{2}{3}}(\tau)\Delta_{7^{+},6}(\tau)\Delta_{7}^{\tfrac{8}{3}}(\tau)\left(-\left(\frac{E_{4}^{(7^{+})}}{\Delta_{7^{+},4}}\right)^{3}(\tau) + a_{1}\left(\frac{E_{4}^{(7^{+})}}{\Delta_{7^{+},4}}\right)^{2}\right.\nonumber\\
        &\ \ \ \ \ \ \ \ \ \ \ \ \ \ \ \ \ \ \ \ \ \ \ \ \ \ \ \ \ \ \ \ \ \ \ \ \ \ \ \ \ \ \ \ \ \ \ \ \ \ \ \ \ \ \ \ \ \ \ \ \ \ \ \ \left.+ a_{2}\left(\frac{E_{4}^{(7^{+})}}{\Delta_{7^{+},4}}\right)(\tau) + a_{3}\right),
\end{align}
where the coefficients $a_{i}$ are rational number that can be fixed by comparing the $q$-series expansions. The behaviour of the Hauptmodul $j_{7^{+}}(\tau)$ near points $\tau = \rho_{7,1}, \rho_{7,2}, \tfrac{i}{\sqrt{7}}, i\infty$ reads
\begin{align}
    \begin{split}
        j_{7^{+}}(\rho_{7,1})\to&\ 0,\\
        j_{7^{+}}(\rho_{7,2})\to&\ 0,\\
        j_{7^{+}}\left(\frac{i}{\sqrt{7}}\right)\to&\ 27,\\
        j_{7^{+}}(i\infty)\to&\ \infty.
    \end{split}
\end{align}
We make the following definitions
\begin{align}
    \begin{split}
        \widetilde{E_{4,7}}(\tau) \equiv& E_{2,7^{'}}^{2}(\tau),\ \ \ \widetilde{E_{6,7}} (\tau)\equiv \left(\sqrt{E_{2,7^{'}}^{3}}\frac{\Delta_{7^{+},6}}{\Delta_{7}}\right)(\tau),\\
        K_{7^{+}}(\tau) \equiv& \frac{27}{j_{7^{+}}(\tau)},\ \ \ A_{7^{+}}(\tau) \equiv \left(\frac{\widetilde{E_{6,7}}}{\widetilde{E_{4,7}}}\right)(\tau) = \left(\frac{\Delta_{7^{+},10}}{E_{2,7^{'}}\Delta_{7}^{2}}\right)(\tau).
    \end{split}
\end{align}
Using identities found in appendix \ref{appendix:modular_forms}, the derivative of the Hauptmodul, $K_{7^{+}}(\tau)\equiv \tfrac{27}{j_{7^{+}}(\tau)}$, and $A_{7^{+}}(\tau)$ read
\begin{align}
    \begin{split}
        q\frac{d}{dq}j_{7^{+}}(\tau) =& -j_{7^{+}}\left(\frac{\Delta_{7^{+},10}}{\Delta^{2}_{7}E_{2,7^{'}}}\right)(\tau)\\
        q\frac{d}{dq}K_{7^{+}}(\tau) =& A_{7^{+}}(\tau)K_{7^{+}}(\tau),\\
        q\frac{d}{dq}A_{7^{+}}(\tau) =& A_{7^{+}}\left(q\frac{d}{dq}\log E_{4,7^{'}}-\frac{2}{3}E_{2}^{(7^{+})} -\frac{4}{3} \frac{E_{4,7^{'}}}{E_{2,7^{'}}} \right)(\tau).
    \end{split}
\end{align}
Using this it is straightforward to show that the re-parameterized derivatives read
\begin{align}
    \begin{split}
        \theta_{q} =& A_{7^{+}}\theta_{K_{7^{+}}},\\
        \mathcal{D}^{2} =& \left(\theta_{q} - \frac{2}{3}E_{2}^{(7^{+})}\right)\theta_{q}\\
        =& A_{7^{+}}^{2}\theta_{K_{7^{+}}}^{2} + A_{7^{+}}\left(q\frac{d}{dq}\log E_{4,7^{'}} - \frac{4}{3}\frac{E_{4,7^{'}}}{E_{2,7^{'}}}\right)\theta_{K_{7^{+}}},
    \end{split}
\end{align}
We do not present the simplification of the re-parameterized MLDE here but discuss a possible method to find the same in appendix \ref{appendix:level_7+_simplifications}.

\subsection{Levels 11 and 13}
\noindent We reserve the derivation of the re-parameterized MLDE for groups $\Gamma_{0}^{+}(11)$ and $\Gamma_{0}^{+}(13)$ for future work and report the Ramanujan identities at this level that haven't been explored in the literature previously. We note here that the dimension of the space of weight $4$ modular forms at level $13$ being three-dimensional complicates the derivation of the re-parameterized MLDE since we had previously found a basis decomposition in \cite{Umasankar:2022kzs} which was two-dimensional. This suggests that there is something more complicated going on at level $13$ and a detailed analysis of the space of modular forms is required prior to us setting up the MLDE. This issue persists at level $11$ since the space of weight $4$ modular forms again turns out to be three-dimensional, but the basis decomposition of this space is well-understood \cite{Junichi}, and hence, setting up the MLDE wouldn't be a cumbersome task. The first few Ramanujan identities for $\Gamma_{0}^{+}(11)$ were found to be (see \cite{Umasankar:2022kzs} for basis decomposition and definitions)
\begin{align}
     \begin{split}
        q\frac{d}{dq}E_{2}^{(11^{+})}(\tau) =& \frac{1}{2}\left(\left(E_{2}^{(11^{+})}\right)^{2} - E_{2,11^{'}}^{2} + \frac{24}{5}E_{2,11^{'}}\Delta_{11} + \frac{56}{25}\Delta_{11}^{2}\right)(\tau),\\
        q\frac{d}{dq}E_{4}^{(11^{+})}(\tau) =& 2\left(E_{2}^{(11^{+})}E_{4}^{(11^{+})} - E_{4,11^{'}}E_{2,11^{'}} + \frac{432}{305}E_{4,11^{'}}\Delta_{11}\right)(\tau),\\
        q\frac{d}{dq}E_{6}^{(11^{+})}(\tau) =& 3\left(E_{2}^{(11^{+})}E_{6}^{(11^{+})} - E_{2,11^{'}}^{4} + \frac{6578}{555}E_{2,11^{'}}^{3}\Delta_{11} + \frac{30896}{2775}E_{2,11^{'}}^{2}\Delta_{11}^{2}\right.\\
        &{}\ \ \ \ \ \ \ \ \ \ \ \ \ \ \ \ \ \ \ \ \ \ \ \ \ \ \ \ \ \ \ \ \left.- \frac{24906}{4625}E_{2,11^{'}}\Delta_{11}^{3} + \frac{37504}{69375}\Delta_{11}^{4}\right)(\tau).
    \end{split}
\end{align}
The modular forms $E_{k,11^{'}}(\tau)$ can be related to the Eisenstein series in $\Gamma_{0}^{+}(11)$ to obtain a simplified expression. Although it is difficult to obtain identities for the Eisenstein series of weights $4$ and $6$ in level $13$, the identity corresponding to $E_{2}^{(13^{+})}$ is easy to construct. This reads (see \cite{Umasankar:2022kzs} for basis decomposition and definitions)
\begin{align}
        q\frac{d}{dq}E_{2}^{(13^{+})}(\tau) =\frac{7}{12}\left(\left(E_{2}^{(13^{+})}\right)^{2} - E_{4}^{(13^{+})} + \frac{36}{49}\left(E_{2,13^{'}}^{2} - E_{4}^{(13^{+})}\right)\right)(\tau).
\end{align}

\subsection{Three-character re-parameterization: levels 2 \& 3}
\noindent The $(n,\ell) = (3,0)$ MLDE for $\Gamma_{0}^{+}(2)$ takes the following form
\begin{align}
    \left[\mathcal{D}^{3}  + \mu_{1}E_{4}^{(2^{+})}(\tau)\mathcal{D} + \mu_{2}E_{6}^{(2^{+})}(\tau)\right]f(\tau) = 0.
\end{align}
The third-order covariant derivative is found to be
\begin{align}
    \begin{split}
        \mathcal{D}^{3} =& \mathcal{D}_{(4)}\circ\mathcal{D}_{(2)}\circ\mathcal{D}_{(0)} = \mathcal{D}_{4}\circ\mathcal{D}^{2}\\
        =& \left(\theta_{q} - \frac{1}{2}E_{2}^{(2^{+})}\right)\left(\theta_{q} - \frac{1}{4}E_{2}^{(2^{+)}}\right)\theta_{q}\\
        =& A_{2^{+}}^{3}\theta_{K_{2^{+}}}^{3} - A_{2^{+}}^{3}\frac{3(1 + K_{2^{+}})}{4(1-K_{2^{+}})}\theta_{K_{2}^{+}}^{2} + \frac{1}{8}A_{2^{+}}^{3}\theta_{K_{2^{+}}}.
    \end{split}
\end{align}
With this and \ref{derivatives_repara_2+}, we find the third-order re-parameterized MLDE for $\Gamma_{0}^{+}(2)$ to be
\begin{align}
    \left[\theta_{K_{2^{+}}}^{3} - \frac{3(1 + K_{2^{+}})}{4(1-K_{2^{+}})}\theta_{K_{2}^{+}}^{2} + \frac{8\mu_{1} + 1 - K_{2^{+}}}{8(1-K_{2^{+}})}\theta_{K_{2^{+}}} + \frac{\mu_{2}}{1-K_{2^{+}}}\right]f(K_{2^{+}}) = 0.
\end{align}
The solution to this ODE, which we don't mention here, is found to be given by ${}_{3}F_{2}$ hypergeometric functions.\\

\noindent The $(n,\ell) = (3,0)$ $\Gamma_{0}^{+}(3)$ MLDE takes the following form
\begin{align}
    \left[\mathcal{D}^{3}  + \mu_{1}E_{4}^{(3^{+})}(\tau)\mathcal{D} + \mu_{2}E_{6}^{(3^{+})}(\tau)\right]f(\tau) = 0.
\end{align}
The third-order covariant derivative is found to be
\begin{align}
    \begin{split}
        \mathcal{D}^{3} =& \mathcal{D}_{(4)}\circ\mathcal{D}_{(2)}\circ\mathcal{D}_{(0)} = \mathcal{D}_{4}\circ\mathcal{D}^{2}\\
        =& \left(\theta_{q} - \frac{2}{3}E_{2}^{(3^{+})}\right)\left(\theta_{q} - \frac{1}{3}E_{2}^{(3^{+)}}\right)\theta_{q}\\
        =& A_{3^{+}}^{3}\theta_{K_{3^{+}}}^{3} - A_{3^{+}}^{3}\frac{2 + K_{3^{+}}}{2(1-K_{3^{+}})}\theta_{K_{3}^{+}}^{2} -A_{3^{+}}^{3}\frac{K_{3^{+}} - 4}{18(1-K_{3^{+}})}\theta_{K_{3^{+}}}.
    \end{split}
\end{align}
With this, the third-order re-parameterized MLDE for $\Gamma_{0}^{+}(3)$ reads
\begin{align}
    \left[\theta_{K_{3^{+}}}^{3} - \frac{2 + K_{3^{+}}}{2(1-K_{2^{+}})}\theta_{K_{3}^{+}}^{2} + \frac{18\mu_{1} + 4 - K_{3^{+}}}{18(1-K_{3^{+}})}\theta_{K_{3^{+}}} + \frac{\mu_{2}}{1-K_{3^{+}}}\right]f(K_{3^{+}}) = 0,
\end{align}
whose solution is again given by ${}_{3}F_{2}$ hypergeometric functions.
\section{\texorpdfstring{$\mathbf{\Gamma_{0}^{+}(2)}$}{Γ0(2)+}}\label{sec:Gamma_0_2+}
\subsection{Single-character solutions}
\noindent From \cite{Umasankar:2022kzs}, we find the following form for the character in a $n = 1$ theory
\begin{align}\label{single-character_normie}
    \begin{split}
    &\chi(\tau) = j_{2^{+}}^{w_{\rho}}(\tau),\\
    &c = 24w_{\rho} = 6\ell,
    \end{split}
\end{align}
where $w_{\rho}\in\left\{0,1,\tfrac{3}{2}\right\}$ corresponding to the three characters at $\ell = 0$, $\ell = 4$, and $\ell = 6$ respectively. Here, we will present the admissible solutions at the other values of $\ell$ and a more detailed analysis of the $\ell = 4$ \textit{movable pole}\footnote{Whenever the RHS of the valence formula is $\geq 1$, we shall refer to the MLDE corresponding to that $\ell$ value as a \textit{movable pole} MLDE (for more details see \cite{Das:2023qns}).}.

\subsubsection{\texorpdfstring{$\ell = 1$}{l=1}}
\noindent For $\ell=1$, the RHS of valence formula \ref{valence_Fricke_2} reads $\frac{1}{4}$ (since $k=2\ell$). The MLDE takes the following form,
\begin{align}\label{F2_l1_p1_l=1}
    \left[\mathcal{D} + \frac{1}{4}\left(\mu_1\frac{\left(E_{4}^{(2+)}\right)^2}{E_{6}^{(2+)}}+\mu_2\frac{\Delta_2}{E_{6}^{(2+)}}\right)\right]\chi(\tau) = 0.
\end{align}
Now using the indicial equation we can set $\mu_1=1$. Consider now the ansatz, $\chi=j_{2^{+}}^{\tfrac{1}{4}}$. One can readily check that this is a solution with $\mu_2=-256$. We note here that this solution re-appears as a solution to a $(2,0)$ MLDE paired up with an unstable character.

\subsubsection{\texorpdfstring{$\ell = 2$}{l=2}}
\noindent The MLDE takes the following form,
\begin{align}
    &\left[\mathcal{D}+\mu_1\frac{E_6^{(2+)}}{E_4^{(2+)}}\right]\chi(\tau) = 0, \label{1char_F2_l2}\\
    &\left[E_4^{(2+)}\mathcal{D}+\mu_1E_6^{(2+)}\right]\chi(\tau) = 0. \label{1char_F2_l2_aliter}
\end{align}
The indicial equation ($\alpha_0+\mu_1=0$) dictates $\mu_1=\frac{1}{2}$. Consider the ansatz, $\chi=j_{2^{+}}^{\tfrac{1}{2}}$. Then, $\Tilde{\partial}_\tau j_{2^{+}}^{\tfrac{1}{2}} = -\frac{1}{2}j_{2^{+}}^{\tfrac{1}{2}}\frac{E_6^{(2+)}}{E_4^{(2+)}}$. Note, that this $\chi$ indeed satisfies \ref{1char_F2_l2} with $\mu_1=\frac{1}{2}$.

\subsubsection{\texorpdfstring{$\ell = 3$}{l=3}}
\noindent When $\ell=3$, the RHS of valence formula \ref{valence_Fricke_2} reads $\frac{3}{4}$. The MLDE takes the following form,
\begin{align}
    \left[\mathcal{D} + \frac{1}{4}\left(\mu_1\frac{\left(E_{4}^{(2+)}\right)^2}{E_{6}^{(2+)}}+\mu_2\frac{\Delta_2}{E_{6}^{(2+)}}\right)\right]\chi(\tau) = 0. \label{F2_l1_p1_l=3}
\end{align}
Now using the indicial equation we can set $\mu_1=3$. Consider now the ansatz, $\chi=j_{2^{+}}^{\tfrac{3}{4}}$. One can readily check that this is a solution with $\mu_2=-768$.

\subsubsection{\texorpdfstring{$\ell = 4$}{l=4}}
\noindent From Riemann-Roch, we notice that, when $\ell=4$ we get the \textit{movable pole} MLDE. The MLDE takes the following form,
\begin{align}
    \left[\theta_{q}+\mu_1\frac{E_{10}^{(2+)}}{(E_{4}^{(2+)})^2+\mu_2\Delta_2}\right]\chi(\tau) &= 0, \label{1char_F2_l4} \\
    \left[\left((E_{4}^{(2+)})^2+\mu_2\Delta_2\right)\theta_{q} + \mu_1 E_{10}^{(2+)}\right]\chi(\tau) &= 0. \label{1char_F2_l4_aliter}
\end{align}
The indicial equation dictates $\mu_1=1$. The recursion relation reads,
\begin{align}
    \sum\limits_{k=0}^i\left[\left(E_{4}^{(2+)}\right)^2_{,k}m_{i-k}(i-k-1) + \mu_2\Delta_{2,k}m_{i-k}(i-k-1)+E^{(2^{+})}_{10,k}m_{i-k}\right] = 0. \label{recurF2l42char}
\end{align}
Putting $i=1$ in \ref{recurF2l42char} gives,
\begin{align}
    m_1 = 104 + \mu_2, \label{m1eqn_F2_2char_l4}
\end{align}
which implies that $\mu_2\in\mathbb{Z}$. Now putting $i=2$ in \ref{recurF2l42char} yields,
\begin{align}
    m_2 = 4372. \label{m2eqn_F2_2char_l4}
\end{align}
One can check up to higher orders and check that,
\begin{align}\label{j_104_c=24}
    \chi(\tau) &= q^{-1}\left(1 + (104+\mathcal{N})q + 4372q^2 + 96256q^3 + 1240002q^4 + \mathcal{O}(q^5)\right) \nonumber\\
    &= j_{2^{+}}(\tau) + \mathcal{N} \, \, \, \, (\text{with}\ \mathcal{N}\geq -104).
\end{align}
A more direct way to see the above is the following. Say, we consider $j_{2^{+}} + \mathcal{N}$ to be an ansatz to \ref{1char_F2_l4_aliter}, substitution of which yields
\begin{align}
    -j_{2^{+}}E_6^{(2+)}E_4^{(2+)} - \mu_2\Delta_2 j_{2^{+}}\frac{E_6^{(2+)}}{E_4^{(2+)}} + j_{2^{+}}E_6^{(2+)}E_4^{(2+)} + \mathcal{N}E_6^{(2+)}E_4^{(2+)} =& 0, \nonumber\\
    \Rightarrow\, \, (\mu_2-\mathcal{N})E_6^{(2+)}E_4^{(2+)} =& 0. \label{const}
\end{align}
This implies that $j_{2^{+}} + \mathcal{N}$ is a solution to the MLDE in \ref{1char_F2_l4} if and only if $\mathcal{N}=\mu_2$ and by integrality of $m_1$ in \ref{m1eqn_F2_2char_l4}, with $m_1\geq 0$, we have that $\mathcal{N}\in\mathbb{Z}$ and $\mathcal{N}\geq -104$.

\subsubsection{Tensor-Product formula}
\noindent Consider two different admissible $p$-character and $q$-character solutions. Let us say the Wronskian indices are $m$ and $n$ respectively. The tensor product solution will have $pq$ number of characters and say its Wronskian index is $\Tilde{\ell}$, then,
\begin{align}
    \Tilde{\ell} = \frac{pq}{2}(p-1)(q-1) + pn + qm, \label{hampa_mukhi_tensor}
\end{align}
which is exactly similar to the \textit{Hampapura-Mukhi} formula for the $\text{SL}(2,\mathbb{Z})$ case (see \cite{Hampapura:2015cea}). Using \ref{hampa_mukhi_tensor} we immediately note that $\left(j_{2^{+}}^{1/4}\right)^{\otimes n}$ is a $(1,n)$ admissible solution. Similarly, one can readily see that ${}^{(2)}\mathcal{W}\otimes\left(j_{2^{+}}^{1/4}\right)^{\otimes n}$ is a $(2,2n)$ admissible solution (where ${}^{(2)}\mathcal{W}$ is a two-character admissible solution). Also, ${}^{(3)}\mathcal{W}\otimes\left(j_{2^{+}}^{1/4}\right)^{\otimes n}$ is a $(3,3n)$ admissible solution (where ${}^{(3)}\mathcal{W}$ is a three-character admissible solution). Motivated by the $\text{SL}(2,\mathbb{Z})$ case, we predict that the most general single-character solution can be written as, (with $c=6N$ and $N\in\mathbb{Z}$), (see \cite{Das:2022slz}),
\begin{align}
    \begin{split}
        \chi^\mathcal{H}(\tau) =& j_{2^{+}}^{\tfrac{N}{4}-s}\left[j_{2^{+}}^s + a_1\,j_{2^{+}}^{s-1}+ a_2 \, j_{2^{+}}^{s-2} + \ldots + a_s\right],\\
        =&j_{2^{+}}^{w_{\rho}}j_{2^{+}}^{-s} \, \mathfrak{P}_{s}(j_{2^{+}}),
    \end{split}
\label{gen_j}
\end{align}
where we have defined $s \equiv \left \lfloor \frac{N}{4} \right \rfloor$, $c = 24w_{\rho} = 6N$, and $\mathfrak{P}_{s}(j_{2^{+}})$ is a monic polynomial of degree $s$ in $j_{2^{+}}$. 
\subsection{Two-character solutions}
\noindent From the dimension of the space of modular forms listed in \ref{dimension_Fricke_2}, we obtain the following expression for the number of free parameters
\begin{align}\label{number_of_parameters_Gamma_0_2+}
    \#(\mu) = \begin{cases}
        \left\lfloor\left.\frac{\ell}{4}\right\rfloor\right. + \left\lfloor\left.\frac{\ell + 1}{4}\right\rfloor\right. + 
    \left\lfloor\left.\frac{\ell + 2}{4} \right\rfloor\right. + 2,\ &2\ell\not\equiv 2\ (\text{mod}\ 8),\ 2\ell>2,\\
    \left\lfloor\left.\frac{\ell}{4}\right\rfloor\right. + \left\lfloor\left.\frac{\ell + 1}{4}\right\rfloor\right. + \left\lfloor\left.\frac{\ell + 2}{4}\right\rfloor\right. -1,\  &2\ell\equiv 2\ (\text{mod}\ 8),\ 2\ell>2.
    \end{cases}
\end{align}

\subsubsection{\texorpdfstring{$\ell = 0$}{l = 0}}
\noindent The $n = 2$ MLDE reads
\begin{align}\label{MLDE_n=2_Fricke_2}
    \left[\mathcal{D}^{2} + \phi_{1}(\tau) \mathcal{D} + \phi_{0}(\tau)\right]f(\tau) = 0,
\end{align}
where $\phi_{0}(\tau)$ and $\phi_{1}(\tau)$ are modular forms of weights $4$ and $2$ respectively. The space $\mathcal{M}_{2}(\Gamma_{0}^{+}(2))$ is zero-dimensional and there are no modular forms of weight $2$ in $\Gamma_{0}^{+}(2)$. Hence, we set $\phi_{1}(\tau) = 0$. The space $\mathcal{M}_{4}(\Gamma_{0}^{+}(2))$ is one-dimensional and hence, we set $\phi_{0}(\tau) = \mu_{1} E_{4}^{(2^{+})}(\tau)$. For these choices, the second-order MLDE takes the following form
\begin{align}\label{MLDE_n=2_Gamma_0_2+}
    \left[\mathcal{D}^{2} + \mu_{1} E_{4}^{(2^{+})}(\tau)\right]f(\tau) = 0,
\end{align}
where $\mu_{1}$ is an independent parameter. Now, since the covariant derivative transforms a weight $r$ modular form into one of weight $r+2$, the double covariant derivative of a weight $0$ form is
\begin{align}
    \begin{split}
        \mathcal{D}^{2} = \mathcal{D}_{(2)}\mathcal{D}_{(0)} =& \left(\frac{1}{2\pi i}\frac{d}{d\tau} - \frac{1}{4}E^{(2^{+})}_{2}(\tau)\right)\frac{1}{2\pi i}\frac{d}{d\tau}\\
        =& \Tilde{\partial}^{2} - \frac{1}{4}E^{(2^{+})}_{2}(\tau)\Tilde{\partial}.
    \end{split}
\end{align}
where $\Tilde{\partial}=\frac{1}{2\pi i}\frac{\partial}{\partial\tau}$. \\
\\
The MLDE \ref{MLDE_n=2_Gamma_0_2+} now reads
\begin{align}\label{MLDE_n=2_Gamma_0_2+_expanded}
    \left[\Tilde{\partial}^{2} - \frac{1}{4}E^{(2^{+})}_{2}(\tau)\Tilde{\partial} + \mu_{1} E_{4}^{(2^{+})}(\tau)\right]f(\tau) = 0.
\end{align}
This equation can be solved by making the following mode expansion substitution for the character $f(\tau)$ and the other modular forms,
\begin{align}\label{series_defs_Gamma_0_2+}
    \begin{split}
        f(\tau) =& q^{\alpha}\sum\limits_{n= 0}^{\infty}f_{n}q^{n},\\
        E_{2}^{(2^{+})}(\tau) =& \sum\limits_{n=0}^{\infty}E_{2,n}^{(2^{+})}q^{n}\\
        =& 1 - 8 q - 40 q^2 - 32 q^3 - 104 q^4 + \ldots,\\
        E_{4}^{(2^{+})}(\tau) =& \sum\limits_{n=0}^{\infty}E_{4,n}^{(2^{+})}q^{n}\\
        =& 1 + 48 q + 624 q^2 + 1344 q^3 + 5232 q^4 + \ldots
    \end{split}
\end{align}
Substituting these Fourier expansions we get the following recursion relation,
\begin{align}
    &\left[(\alpha+i)^2 - \frac{1}{4}(\alpha+i) + \mu_1\right]a_i = \sum\limits_{k=1}^i\left[\frac{1}{4}(\alpha+i-k)E^{(2^{+})}_{2,k}-\mu_1E_{4,k}^{(2^{+})}\right]a_{i-k}, \label{recursion0}
\end{align}
where $a_i$s are Fourier coefficients of the character $q$-series expansion. The indicial equation is obtained by setting $i=0$ and $k=0$ in \ref{recursion0},
\begin{align}
    \alpha^2 - \frac{1}{4}\alpha + \mu_1 = 0. \label{indicial0}
\end{align}
Setting $\alpha=\alpha_0$ in \ref{indicial0} we get,
\begin{align}
    \mu_1 = \frac{1}{4}\left(\alpha_0-4\alpha_0^2\right) = -\frac{c(c+6)}{576}, \label{param0}
\end{align}
Now setting $i=1$ in \ref{recursion0} we get the following quadratic equation in $\alpha_0$,
\begin{align}
    9a_1 + 168\alpha_0 + 24a_1\alpha_0 - 576\alpha_0^2 = 0, \label{m1_eqn_0}
\end{align}
which can be recast as below if we identify $N=-24\alpha_0=c$,
\begin{align}
    -9a_1 + 7N + a_1N + N^2 = 0, \label{N_eqn_0}
\end{align}
$a_1$ can now be expressed in terms of $N$ as,
\begin{align}
    a_1 = \frac{N(N+7)}{9-N}, \label{a1_eqn_0}
\end{align}
which sets an upper bound on the central charge $c<9$. Furthermore, applying the integer root theorem to \ref{N_eqn_0}, we see that $c=N\in\mathbb{Z}$. Now, checking for higher orders (up to $\mathcal{O}(q^{5000})$) for both characters, for $0<c<9$, we find the following admissible character solutions,
\begin{itemize}
    \item $(c,h)=\left(1,\frac{1}{3}\right)$: 
    \begin{align}\label{c=1_Fricke_2_2,0}
        \begin{split}
        \chi_0 =& q^{-\tfrac{1}{24}}(1 + q + 2q^2 + q^3 + 3q^4 + 3q^5 + 5q^6 + 5q^7 + 8q^8 + 8q^9 + 12q^{10} + \mathcal{O}(q^{11})),\\
        \chi_{\frac{1}{3}} =& q^{\tfrac{7}{24}}(1 + q^2 + q^3 + 2q^4 + q^5 + 3q^6 + 2q^7 + 5q^8 + 4q^9 + 7q^{10} + \mathcal{O}(q^{11})).
        \end{split}
    \end{align}
    \item $(c,h)=\left(3,\frac{1}{2}\right)$: 
    \begin{align}\label{c=3_Fricke_2_2,0}
        \begin{split}
        \chi_0 =& q^{-\tfrac{1}{8}}(1 + 5q + 11q^2 + 20q^3 + 41q^4 + 76q^5 + 127q^6 + 211q^7 + 342q^8 + 541q^9\\
        &{}\ \ \ \ \ \ \ \ + 838q^{10} + \mathcal{O}(q^{11})),\\
        \chi_{\frac{1}{2}} =& q^{\tfrac{3}{8}}(1 + q + 5q^2 + 6q^3 + 16q^4 + 21q^5 + 46q^6 + 61q^7 + 117q^8 + 157q^{9}\\
        &{}\ \ \ \ \ \ + 273 q^{10} +  \mathcal{O}(q^{11}))
        \end{split}
    \end{align}
    \item $(c,h)=\left(5,\frac{2}{3}\right)$: 
    \begin{align}\label{c=5_Fricke_2_2,0}
        \begin{split}
        \chi_0 =& q^{-\tfrac{5}{24}}(1 + 15q + 40q^2 + 135q^3 + 290q^4 + 726q^5 + 1415q^6 + 3000q^7 + 5485q^8\\
        &{}\ \ \ \ \ \ \ \ \ + 10565q^9 + 18407q^{10} + \mathcal{O}(q^{11})),\\
        \chi_{\frac{2}{3}} =& q^{\tfrac{11}{24}}(5 + 11q + 55q^2 + 100q^3 + 290q^4 + 535q^5 + 1235q^6 + 2160q^7 + 4400q^8\\
        &{}\ \ \ \ \ \ \ + 7465q^{9} + 13945q^{10} + \mathcal{O}(q^{11})).
        \end{split}
    \end{align}
\end{itemize}
We also find an \textit{identity only} admissible solution, which only has a nice identity character but an unstable non-trivial character\footnote{By unstable character, we mean a character which has fractional or negative coefficients and is not a quasi-character.}. This is found at $(c,h)=\left(6,\frac{3}{4}\right)$.
\begin{align}\label{8C_character_1}
    \begin{split}
        \chi_0 =& q^{-\tfrac{1}{4}}(1 + 26q + 79q^2 + 326q^3 + 755q^4 + 2106q^5 + 4460q^6 + 10284q^7\\
        &{}\ \ \ \ \ \ \ \ + 20165q^8 + 41640q^9 + 77352q^{10} + \mathcal{O}(q^{11}))\\
        =& j_{2^{+}}^{\tfrac{1}{4}},
    \end{split}
\end{align}
which is the $\ell = 1$ single-character solution reappearing. This surprisingly turns out to be the McKay-Thompson series of the $8C$ conjugacy class of the Monster $\mathbb{M}$ (OEIS sequence A052241 \cite{A052241}). A detailed derivation of the admissible solutions is provided in 
\ref{appendix:detailed_derivation}. For $c=7,8$, both the characters are unstable and for $c=2,4$ there exists no $a_1\in\mathbb{Z}$ solutions to \ref{N_eqn_0}. Consider now the $c=6$, \textit{identity only} solution. It turns out $\chi_{0}^{(c=6)} = j_{2^{+}}^{\tfrac{1}{4}}$. So, a $(1,1)$ solution has appeared as a $(2,0)$ solution, a phenomenon which keeps happening in the $\text{SL}(2,\mathbb{Z})$ case. So, we can use this phenomena: \textit{$j_{2^{+}}^x$ where $x\in\{\frac{1}{4},\frac{1}{2},\frac{3}{4}\}$ will always appear as a particular solution to a $(n, \ell)$ MLDE}, to reduce the no. of parameters of the $(n, \ell)$ MLDE by one. For $c=6$, $\mu_1 = -\frac{1}{8}$ (using \ref{param0}). With this value of $\mu_1$, \ref{reparameterized_MLDE_2+} should yield a solution $\chi=j_{2^{+}}^{\tfrac{1}{4}}$. It is interesting to note that the identity character in \ref{c=1_Fricke_2_2,0} can be realized in terms of modular forms of the Hecke group $\Gamma_{0}(2)$ as follows
\begin{align}\label{Hecke_1_c=1}
    \begin{split}
    \chi_{0}^{(c = 1)} =& \left(\frac{\left(j_{2}\Delta_{0}^{\infty}\right)(3\tau)}{\left(j_{2}\Delta_{2}^{\infty}\right)^{\tfrac{1}{3}}(\tau)}\right)^\frac{1}{8}\left(\frac{1}{\Delta_{2}(\tau)}\right)^{\tfrac{1}{24}},\\
    \Delta_{2}^{\infty}(\tau) =& \frac{\eta^{16}(2\tau)}{\eta^{8}(\tau)}\in\mathcal{M}_{4}(\Gamma_{0}(2)),\ \ \Delta_{2}^{0}(\tau) =  \frac{\eta^{16}(\tau)}{\eta^{8}(2\tau)}\in\mathcal{M}_{4}(\Gamma_{0}(2)),\\
    \Delta_{2}(\tau) =& \Delta_{2}^{\infty}\Delta_{2}^{0}\in\mathcal{S}_{8}(\Gamma_{0}(2)),\ \ \ \ \ \ \ j_{2}(\tau) = \left(\frac{\eta(\tau)}{\eta(2\tau)}\right)^{24}.
    \end{split}
\end{align}
Also, the $q$-series expansion of this character is the sequence of the number of partitions of $n$ in which no part appears more than twice and no two parts differ by unity (OEIS sequence \cite{A070047}).

\subsubsection{Bilinear pairs}
\noindent Let ${}^{(2)}\mathcal{W}_1$, ${}^{(2)}\mathcal{W}_3$ and ${}^{(2)}\mathcal{W}_5$ denote the $c\in\{1,2,5\}$ admissible character-like (2,0) solutions. Note that we have the following bilinear identities,
\begin{align}
    j_{2^{+}}^{\tfrac{1}{4}} =& \chi_0^{{}^{(2)}\mathcal{W}_1}\chi_0^{{}^{(2)}\mathcal{W}_5} + 2 \, \chi_{\frac{1}{3}}^{{}^{(2)}\mathcal{W}_1}\chi_{\frac{2}{3}}^{{}^{(2)}\mathcal{W}_5}, \label{bilin0} \\
    j_{2^{+}}^{\tfrac{1}{4}} =& \chi_0^{{}^{(2)}\mathcal{W}_3}\chi_0^{{}^{(2)}\mathcal{W}_3} + 16 \, \chi_{\frac{1}{2}}^{{}^{(2)}\mathcal{W}_3}\chi_{\frac{1}{2}}^{{}^{(2)}\mathcal{W}_3}, \label{bilin1}
\end{align}
where \ref{bilin1} is a self-dual relation and where the above bilinears are for the following relation: $(2,0)\overset{c^\mathcal{H}=6}{\underset{n_1=1}{\longleftrightarrow}}(2,0)$ (where $n_1=\text{sum of conformal dimensions of the bilinear pair}$). These bilinear relations can probably help us to write down the modular invariant partition function for the $(2,0)$ admissible solutions. For instance, the below could be true,
\begin{align}
    \mathcal{Z}_1 =&\left|\chi_0^{{}^{(2)}\mathcal{W}_1}\right|^2 + 2 \, \left|\chi_{\frac{1}{3}}^{{}^{(2)}\mathcal{W}_1}\right|^2, \label{partintn1} \\
    \mathcal{Z}_3 =& \left|\chi_0^{{}^{(2)}\mathcal{W}_3}\right|^2 + 16 \, \left|\chi_{\frac{1}{2}}^{{}^{(2)}\mathcal{W}_3}\right|^2, \label{partintn3} \\
    \mathcal{Z}_5 =& \left|\chi_0^{{}^{(2)}\mathcal{W}_5}\right|^2 + 2 \, \left|\chi_{\frac{2}{3}}^{{}^{(2)}\mathcal{W}_5}\right|^2. \label{partintn5}
\end{align}

\subsubsection{\texorpdfstring{$\ell = 1$}{l = 1}}
\noindent The MLDE in this case takes the following form
\begin{align}
    \left[\mathcal{D}^2+\left(\mu_1\frac{\left(E_4^{(2+)}\right)^2}{E_6^{(2+)}}+\mu_2\frac{\Delta_2}{E_6^{(2+)}}\right)\mathcal{D}+\mu_3\frac{E_{10}^{(2+)}}{E_{6}^{(2+)}}\right]\chi(\tau) =& 0, \label{mldeF2_2char_l1} \\
    \left[E_6^{(2+)}\theta_{q}^{2} + \left(E_6^{(2+)} - \frac{E_6^{(2+)}E_2^{2+}}{4} + \mu_1\left(E_4^{(2+)}\right)^2 + \mu_2\Delta_2\right)\theta_{q} + \mu_3 E_{10}^{(2+)} \right]\chi(\tau) =& 0. \label{qmldeF2_2char_l1}
\end{align}
We obtain the following recursion relation from the above MLDE
\begin{align}
    &\sum\limits_{k=0}^i\left[E_{6,k}^{(2+)}(\alpha^j+i-k)(\alpha^j+i-k-1) + E_{6,k}^{(2+)}(\alpha^j+i-k)-\frac{E_{2,k}^{(2^{+})}E_{2,k}^{(2^{+})}}{4}(\alpha^j+i-k)\right. \nonumber\\ 
    &\left.\mu_1 (E_{4}^{(2^{+})})^2_{,k} (\alpha^j+i-k) + \mu_2\Delta_{2,k}(\alpha^j+i-k) + \mu_3 E^{(2^{+})}_{10,k}\right]a^j_{i-k} = 0, \label{recursion_l1F2_2char}
\end{align}
which gives the following indicial equation,
\begin{align}
    \alpha^2 + \left(\mu_1-\frac{1}{4}\right)\alpha + \mu_3 = 0, \label{indicial_l1F2_2char}
\end{align}
From Riemann-Roch (sum of exponents $= 0$), we see that $\mu_1=\frac{1}{4}$. Now, let us determine the value of $\mu_2$ by transforming the $\tau$-MLDE in \ref{mldeF2_2char_l1} into the corresponding $j_{2^{+}}$-MLDE (about $\tau=\rho_2$),
\begin{align}
    \left[j_{2^{+}}^2\partial_{j_{2^{+}}}^2 + \frac{j_{2^{+}}}{2}\partial_{j_{2^{+}}} +  \frac{j_{2^{+}}}{j_{2^{+}}-256}\left(\frac{j_{2^{+}}}{2}-(64+\mu_2)\right)\partial_{j_{2^{+}}} + \frac{\mu_3 j_{2^{+}}}{j_{2^{+}}-256}\right]\chi = 0, \label{j2+mlde_p2_l1}
\end{align}
where we have set $\mu_1=\frac{1}{4}$ from the $\ell=1$ indicial equation (see \ref{indicial_l1F2_2char}). Now, note that about $\tau=\rho_2$, we have, $\chi_0\sim j_{2^{+}}^x$ and $j_{2^{+}}\to 0$, with $x=\frac{1}{4}$ for $\ell=1$ (see single-character solutions). Upon substituting $\chi=j_{2^{+}}^x$ about $j_{2^{+}}=0$, the $j_{2^{+}}$-indicial equation (which is basically, coefficient of $j_{2^{+}}^x$ $=\left.0\right|_{j_{2^{+}}=0}$) reads
\begin{align}
    x\left(x+\frac{\mu_2}{256}-\frac{1}{4}\right) = 0, \label{jindip2l1F2}
\end{align}
which yields $\mu_2 = 0$ for $x=\frac{1}{4}$. \\

\noindent Putting $i=1$ in \ref{recursion_l1F2_2char} yields\footnote{We shall at times identify $m_i\equiv a^0_i$.},
\begin{align}
    &m_1 + 40 \alpha_0 + 2 m_1 \alpha_0 - 48 \alpha_0^2 = 0, \label{m1al1}\\
    &-12 m_1 + 20 N + m_1 N + N^2 = 0 \, \, \, \, (\text{where}, \ N\equiv-24\alpha_0 = c). \label{diophNl1p2}
\end{align}
Applying integer root theorem to \ref{diophNl1p2} we note that $N=c\in\mathbb{Z}$.\\

With the above definition of $\alpha_0$ we can now find $m_1$ in terms of $N$,
\begin{align}
    m_1 = \frac{N(N+20)}{12-N}, \label{m1N_p2F2l1}
\end{align}
which implies that $0<c< 12$. Scanning this space of $c$, we observe that there exist the following admissible solutions:
\begin{itemize}
    \item $(c,h)=\left(4,\frac{1}{3}\right)$: 
    \begin{align}
        \chi_0 &= q^{-\tfrac{1}{6}}(1 + 12q + 70q^2 + 1304q^3 + 45969q^4 + 2110036q^5 + 109853542q^6 \nonumber\\
        &{}\ \ \ \ \ \ \ \ \ \ \ \ + 6204772264q^7 + 371363466708q^8 + \mathcal{O}(q^{9})), \\
        \chi_{\frac{1}{3}} &= q^{\tfrac{1}{6}}(1 - 4q - 86q^2 - 2712q^3 - 110221q^4 - 5321572q^5 - 285692130q^6 \nonumber\\
        &{}\ \ \ \ \ \ \ \ \ \ - 16484830680q^7 - 1002347627239q^8 - \mathcal{O}(q^{9})),
    \end{align}
    where $\chi_0$ can be interpreted as a $c=4$ single-character solution. Also, note that, $\chi_{\frac{1}{3}}$ is a quasi-character of type II (see \cite{Chandra:2018pjq}). One can build this out of the general form of a single-character solution in $\Gamma_{0}^{+}(2)$ given in \ref{gen_j}.
    \item $(c,h)=\left(6,\frac{1}{2}\right)$: 
    \begin{align}\label{8C_character_2}
        \begin{split}
        \chi_0 =& j_{2^{+}}^{\tfrac{1}{4}},\\
        \chi_{\frac{1}{2}} =& \text{unstable}.
        \end{split}
    \end{align}
\end{itemize}
Akin to the $(c,h) = \left(6, \tfrac{3}{4}\right)$ identity character in \ref{8C_character_1}, this identity character's $q$-series expansion turns out to be the McKay-Thompson series of the $8C$ conjugacy class of $\mathbb{M}$. Note that, we can make use of the fact that $j_{2^{+}}^x$ with $x\in\{\frac{1}{2},\frac{3}{4}\}$ can also appear as a solution to the above $(2,1)$ MLDE and hence, we get for these values of $x$, $\mu_2=-64$ and $\mu_2=-128$ respectively.
\begin{itemize}
    \item When $\mu_2=-64$, we get from the $i=1$ equation: $(m_1+c)\left(1-\frac{c}{12}\right)=0$ which implies $c=12$. So, this choice of $\mu_2$ only leads to one unitary single-character solution: $j_{2^{+}}^{\tfrac{1}{2}}$.
    \item When $\mu_2=-128$, we get from the $i=1$ equation: $m_1=\frac{N(N-44)}{12-N}$ which means the allowed range of central charge is, $12<c<45$ (where $N=-24\alpha_0$). Scanning this space of $c$, we only obtain $j_{2^{+}}^{\tfrac{3}{4}}$ and three type II quasi-character solutions for $c=20,28,44$. Thus, we conclude that this value of $\mu_2$ has no admissible two-character solutions.
\end{itemize}

\subsubsection{\texorpdfstring{$\ell = 2$}{l = 2}}
\noindent Let us consider the case with $\ell = 2$ and two characters $n = 2$. The MLDE in this case reads
\begin{align}\label{MLDE_n=2,l=2_Fricke_2}
    \left[\omega_{4}(\tau)\mathcal{D}^{2} + \omega_{6}(\tau) \mathcal{D} + \omega_{8}(\tau)\right]\chi(\tau) = 0.
\end{align}
Since $2\ell = 4 \not\equiv 2\ (\text{mod}\ 8)$, using \ref{number_of_parameters_Gamma_0_2+}, we see that we have three free parameters to deal with here. The basis decomposition of the space of modular forms in $\Gamma_{0}^{+}(2)$ of weights $6$ and $8$ read
\begin{align}
    \begin{split}
        \mathcal{M}_{6}(\Gamma_{0}^{+}(2)) =& E_{6}^{(2^{+})},\\
        \mathcal{M}_{8}(\Gamma_{0}^{+}(2)) =& \mathbb{C}\left(E_{4}^{(2^{+})}\right)^{2}\oplus\mathbb{C}\Delta_{2}.
    \end{split}
\end{align}
We make the following choices for $\omega_{4}(\tau)$, $\omega_{6}(\tau)$, and $\omega_{8}(\tau)$
\begin{align}
    \begin{split}
        \omega_{4}(\tau) =& \nu_{1}E_{4}^{(2^{+})}(\tau),\\
        \omega_{6}(\tau) =& \nu_{2}E_{6}^{(2^{+})}(\tau),\\
        \omega_{8}(\tau) =& \nu_{3}\left(E_{4}^{(2^{+})}\right)^{2}(\tau) + \nu_{4}\Delta_{2}(\tau).
    \end{split}
\end{align}
Making these substitutions, we obtain the following MLDE 
\begin{align}\label{n=2,l=2_MLDE}
    \left[\Tilde{\partial}^{2} -\frac{1}{4}\left(\frac{E_{2}^{(2^{+})}}{E_{4}^{(2^{+})}}\right)\Tilde{\partial} + \mu_{1}\left(\frac{E_{6}^{(2^{+})}}{E_{4}^{(2^{+})}}\right)\Tilde{\partial} + \mu_{2}E_{4}^{(2^{+})} + \mu_{3}\left(\frac{\Delta_{2}}{E_{4}^{(2^{+})}}\right)\right]\chi(\tau) = 0,
\end{align}
where the three free parameters are $\mu_{1} \equiv \tfrac{\nu_{2}}{\nu_{1}}$, $\mu_{2} \equiv \tfrac{\nu_{3}}{\nu_{1}}$, and $\mu_{3}\equiv \tfrac{\nu_{4}}{\nu_{1}}$. 
Substituting mode expansions, we find
\begin{align}
    &\sum\limits_{k=0}^i\left[E_{4,k}^{(2+)}(\alpha^j+i-k)(\alpha^j+i-k-1) + E_{4,k}^{(2+)}(\alpha^j+i-k)-\frac{E_{4,k}^{(2+)}E_{2,k}^{2+}}{4}(\alpha^j+i-k)\right. \nonumber\\ 
    &\left.\mu_1E_{6,k}^{(2+)}(\alpha^j+i-k) + \mu_2(E_4^{2+})_{,k}^2 + \mu_3\Delta_{2,k}\right]a^j_{i-k} = 0, \label{recursion_l2F2_2char}
\end{align}
From this, we obtain the following indicial equation,
\begin{align}\label{indicial_l2F2_2char}
    \begin{split}
        &\alpha^{2} + \alpha\left(\mu_{1} - \frac{1}{4}\right) + \mu_{2} = 0,\\
        \alpha_{0} =& \frac{1}{2}\left(-\Tilde{\mu}_{1} - \sqrt{\Tilde{\mu}_{1}^{2} - 4\mu_{2}}\right),\\
        \alpha_{1} =& \frac{1}{2}\left(-\Tilde{\mu}_{1} + \sqrt{\Tilde{\mu}_{1}^{2} - 4\mu_{2}}\right),
    \end{split}
\end{align}
where $\Tilde{\mu}_{1} \equiv \mu_{1} - \tfrac{1}{4}$. Using the Riemann-Roch identity \ref{Riemann_Roch_Gamma_0_2+} with $(n,\ell) = (2,0)$, we find
\begin{align}\label{RR_alphas_l=2}
    \alpha_{0} + \alpha_{1} = -\frac{1}{4}.
\end{align}
From \ref{indicial_l2F2_2char}, we have $\alpha_{0} + \alpha_{1} = -\Tilde{\mu}_{1}$ and \ref{RR_alphas_l=2} implies $\mu_{1} = \tfrac{1}{2}$. The roots in \ref{indicial_l2F2_2char} now read
\begin{align}
    \begin{split}
        \alpha_{0} = \frac{1}{8}\left(-1 -\sqrt{1-64\mu}\right) \equiv \frac{1}{8}(-1 - x),\\
        \alpha_{1} = \frac{1}{8}\left(-1 +\sqrt{1-64\mu}\right) \equiv \frac{1}{8}(-1 + x),
    \end{split}
\end{align}
where we have set $x = \sqrt{1 - 64\mu}$. From this, we can find the central charge and the conformal dimension to be
\begin{align}
    c = 3(x+1),\ \ h = \frac{x}{4}.
\end{align}
We have now reduced the number of free parameters to two. We now perform modular re-parameterization to fix the parameter $\mu_{3}$. The MLDE \ref{MLDE_n=2_Fricke_2} with $\ell = 2$ has modular forms
\begin{align}
    \begin{split}
        \phi_{1} =&\mu_{1}\left(\frac{E_{6}^{(2^{+})}}{E_{4}^{(2^{+})}}\right) = \mu_{1}A_{2^{+}}^{2},\\
        \phi_{0} =& \mu_{2}E_{4}^{(2^{+})} + \mu_{3}\left(\frac{\Delta_{2}}{E_{4}^{(2^{+})}}\right) = \mu_{2}\frac{A_{2^{+}}^{2}}{1 - K_{2^{+}}} + \frac{\mu_{3}}{256}\frac{A_{2^{+}}^{2}K_{2^{+}}}{1 - K_{2^{+}}},
    \end{split}
\end{align}
where we used the definitions in \ref{repara_definitions_2+}.
We can recast the MLDE as follows
\begin{align}\label{reparameterized_l=2}
    \left[\theta_{K_{2^{+}}}^{2} + \left(\mu_{1} -\frac{1 + K_{2^{+}}}{4(1- K_{2^{+}})}\right)\theta_{K_{2^{+}}} +  \frac{\mu_{2}}{1-K_{2^{+}}} + \frac{\mu_{3}}{256}\frac{K_{2^{+}}}{1-K_{2^{+}}}\right]\chi(\tau) = 0.
\end{align}
Now, at $\ell = 2$, the single-character solution reads $f(\tau) = j_{2^{+}}^{\tfrac{1}{2}}(\tau) = 16K_{2^{+}}^{-\tfrac{1}{2}}(\tau)$. From \ref{Hauptmodul_limits_2+}, we have $j_{2^{+}}(\tau)\to 0$ at the elliptic point  $\tau = \rho_{2} = \tfrac{-1+i}{2}$, or equivalently, $K(\rho_{2})\to \infty$. With $\mu_{1} = \tfrac{1}{2}$, substituting $f(\tau) = (256K_{2^{+}}^{-1})^{\gamma}(\tau)$ in \ref{reparameterized_l=2}, we obtain the following indicial equation
\begin{align}
    \begin{split}
        \lim\limits_{K_{2^{+}}\to \infty}\left[\gamma^{2} - \left(\frac{1}{2} - \frac{1-K_{2^{+}}}{4(1-K_{2^{+}})}\right)\gamma + \frac{\mu_{2}}{(1-K_{2^{+}})} + \frac{\mu_{3}}{256}\frac{K_{2^{+}}}{(1-K_{2^{+}})}\right] =& 0,\\
        \gamma^{2} - \frac{3}{4}\gamma - \frac{1}{256}\mu_{3} =& 0.
    \end{split}
\end{align}
With $\gamma = \tfrac{1}{2}$, we find $\mu_{3} = -32$. Putting $i=1$ in \ref{recursion_l2F2_2char} yields\footnote{We shall at times identify $m_i\equiv a^0_i$.},
\begin{align}
    &-128 + 5 m_1 - 248 \alpha_0 + 8 m_1 \alpha_0 - 192 \alpha_0^2 = 0, \label{m1al}\\
    &384 - 15 m_1 - 31 N + m_1 N + N^2 = 0 \, \, \, \, (\text{where}, \ N\equiv -24\alpha_0 = c). \label{diophNl2p2}
\end{align}
Applying the integer root theorem to \ref{diophNl2p2} we note that $N=c\in\mathbb{Z}$. With the above definition of $\alpha_0$ we can now find $m_1$ in terms of $N$,
\begin{align}
    m_1 = (16-N)+\frac{144}{15-N}, \label{m1N_p2F2l2}
\end{align}
which implies that $0<c< 15$. Scanning this space of $c$, we observe that there exist the following admissible solutions.
\begin{itemize}\label{c=3,h=0_no_MT}
    \item $(c,h)=\left(3,0\right)$: 
    \begin{align}\label{c=3,single_char_n=2}
        \begin{split}
            \chi_0 = q^{-\tfrac{1}{8}}(1 +& 25q + 51q^2 + 196q^3 + 297q^4 + 780q^5 + 1223q^6\\
        +& 2551q^7 + 3798q^8 + 7201q^9 + 10582q^{10} + \mathcal{O}(q^{11})),
        \end{split}
    \end{align}
    where since $h=0$ both the characters are identical. This is a single-character solution which surprisingly is not ``yet" any $(1,\ell)$ solution and also does not possess a McKay-Thompson interpretation.
    \item $(c,h)=\left(6,\frac{1}{4}\right)$: 
    \begin{align}\label{8C_character_3}
        &\chi_0 = j_{2^{+}}^{\tfrac{1}{4}}, \\
        &\chi_{\frac{1}{4}} = \text{unstable} \nonumber
    \end{align}
    \item $(c,h)=\left(7,\frac{1}{3}\right)$: 
    \begin{align}\label{c=7_Fricke_2_2,2}
        &\chi_0 = j_{2^{+}}^{\tfrac{1}{4}}\otimes \chi_0^{{}^{(2)}\mathcal{W}_1} \nonumber\\
        &\chi_{\frac{1}{3}} = j_{2^{+}}^{\tfrac{1}{4}}\otimes \chi_{\frac{1}{3}}^{{}^{(2)}\mathcal{W}_1}
    \end{align}
    \item $(c,h)=\left(9,\frac{1}{2}\right)$: 
    \begin{align}
        &\chi_0 = j_{2^{+}}^{\tfrac{1}{2}}\otimes \chi_0^{{}^{(2)}\mathcal{W}_3} \nonumber\\
        &\chi_{\frac{1}{2}} = j_{2^{+}}^{\tfrac{1}{4}}\otimes \chi_{\frac{1}{2}}^{{}^{(2)}\mathcal{W}_3}
    \end{align}
    \item $(c,h)=\left(11,\frac{2}{3}\right)$: 
    \begin{align}
        &\chi_0 = j_{2^{+}}^{\tfrac{2}{3}}\otimes \chi_0^{{}^{(2)}\mathcal{W}_5} \nonumber\\
        &\chi_{\frac{2}{3}} = j_{2^{+}}^{\tfrac{1}{4}}\otimes \chi_{\frac{2}{3}}^{{}^{(2)}\mathcal{W}_5}
    \end{align}
    \item $(c,h)=\left(12,\frac{3}{4}\right)$: 
    \begin{align}
        \begin{split}\label{4B_character_1}
        \chi_0 =& j_{2^{+}}^{\tfrac{1}{2}}\\
        =& q^{-\tfrac{1}{2}}(1 + 52 q + 834 q^2 + 4760 q^3 + 24703 q^4 + 94980 q^5 + 343998 q^6 + 1077496 q^7\\ 
        &{}\ \ \ \ \ \ \ \ + 3222915 q^8 + \mathcal{O}(q^{9})),\\
        \chi_{\frac{3}{4}} =& \text{unstable}.
        \end{split}
    \end{align}
\end{itemize}
The identity character in \ref{8C_character_3}, along with \ref{8C_character_1} and \ref{8C_character_2} forms a set of central charge $c = 6$ admissible solutions that correspond to the $8C$ conjugacy class of $\mathbb{M}$. Also, the identity character in \ref{4B_character_1} surprisingly turns out to be the McKay-Thompson series of the $4B$ conjugacy class of $\mathbb{M}$ (OEIS sequence A007247 \cite{A007247}).
Let ${}^{(2)}\mathcal{W}_7$, ${}^{(2)}\mathcal{W}_9$ and ${}^{(2)}\mathcal{W}_{11}$ denote the $c\in\{7,9,11\}$ admissible character-like (2,2) solutions. Note that, here also, we have some bilinear identities,
\begin{align}
    &j_{2^{+}}^{\tfrac{3}{4}} = \chi_0^{{}^{(2)}\mathcal{W}_7}\chi_0^{{}^{(2)}\mathcal{W}_{11}} + 2 \, \chi_{\frac{1}{3}}^{{}^{(2)}\mathcal{W}_7}\chi_{\frac{2}{3}}^{{}^{(2)}\mathcal{W}_{11}}, \label{bilin2} \\
    &j_{2^{+}}^{\tfrac{1}{4}} = \chi_0^{{}^{(2)}\mathcal{W}_9}\chi_0^{{}^{(2)}\mathcal{W}_9} + 16 \, \chi_{\frac{1}{2}}^{{}^{(2)}\mathcal{W}_9}\chi_{\frac{1}{2}}^{{}^{(2)}\mathcal{W}_9}, \label{bilin3}
\end{align}
where \ref{bilin3} is a self-dual relation. Note, \ref{bilin2} and \ref{bilin3} are nothing but $j_{2^{+}}^{\tfrac{1}{2}}$ multiplied with \ref{bilin0} and \ref{bilin1} respectively. More bilinear pairs can be constructed by multiplying $j_{2^{+}}^{\tfrac{1}{4}}$ with \ref{bilin0} and \ref{bilin1}.\\
\noindent Note that, we can make use of the fact that $j_{2^{+}}^x$ with $x\in\{\frac{3}{4}\}$ can also appear as a solution to the above $(2,2)$ MLDE and hence we get for this value of $x$, $\mu_3=0$. With $\mu_3=0$, we get from the $i=1$ equation: $m_1=\frac{N(N-31)}{15-N}$ where $N=-24\alpha_0$. This gives the allowed range of central charge as, $15<c<32$. Scanning this space of $c$ we get the following admissible solutions.
\begin{itemize}
    \item $(c,h)=\left(19,\frac{4}{3}\right)$: 
    \begin{align}
        \begin{split}
        \chi_0 =& q^{-\tfrac{19}{24}}(1 + 57q + 2147q^2 + 31540q^3 + 260243q^4 + 1691798q^5 +  8887877q^6\\
        &{}\ \ \ \ \ \ \ \ \ +  41091167q^7 + 168938614q^8 + \mathcal{O}(q^{9})),\\
        \chi_{\frac{4}{3}} =& q^{\tfrac{13}{24}}(133 + 2717q + 33839q^2 + 246468q^3 + 1506510q^4 + 7478629q^5 + 33354310q^6\\
        &{}\ \ \ \ \ \ \ \ \ \ +  132591975q^7 + 489341675q^8 + \mathcal{O}(q^{9})).
        \end{split}
    \end{align}
    \item $(c,h)=\left(21,\frac{3}{2}\right)$: 
    \begin{align}
        \begin{split}
        \chi_0 =& q^{-\tfrac{7}{8}}(1 + 35q + 2394q^2 + 40873q^3 + 405426q^4 + 2946132q^5 + 17381133q^6\\
        &{}\ \ \ \ \ \ \ \ +88016962q^7 + 395953299q^8 + \mathcal{O}(q^{9})),\\
        \chi_{\frac{3}{2}} =& q^{\tfrac{5}{8}}(7 + 161q + 2160q^2 + 17409q^3 + 115003q^4 + 619101q^5 + 2962183q^6\\
        &{}\ \ \ \ \ \ +  12628973q^7 + 49681296q^8 + \mathcal{O}(q^{9})).
        \end{split}
    \end{align}
    \item $(c,h)=\left(23,\frac{5}{3}\right)$: 
    \begin{align}
        \begin{split}
        \chi_0 =& q^{-\tfrac{23}{24}}(1 + 23q + 3335q^2 + 67068q^3 + 793776q^4 + 6461689q^5\\
        &{}\ \ \ \ \ \ \ \ \ + 42601060q^6 + 236116965q^7 + 1157910689q^8 + \mathcal{O}(q^{9})),\\
        \chi_{\frac{5}{3}} =& q^{\tfrac{17}{24}}(506 + 12903q + 185725q^2 + 1639463q^3 + 11650696q^4 + 67590238q^5\\
        &{}\ \ \ \ \ \ \ \ \ \ + 345549517q^6 + 1572505630q^7 + 6570405101q^8 + \mathcal{O}(q^{9})).
        \end{split}
    \end{align}
\end{itemize}
For $\mu_3=-112$, we get from the $i=1$ equation: $m_1=\frac{N(N-10)}{18-N}$ where $N=-24\alpha_0$. This gives the allowed range of central charge as, $9<c<18$. Scanning this space of $c$ we find no admissible solutions. Also, we note that we did not get any quasi-character solutions.

\subsubsection{Non-trivial bilinear pairs}
\noindent We present some more bilinear pairs that are non-trivial in the sense that they cannot be derived from \ref{bilin0} and \ref{bilin1}. Let ${}^{(2)}\mathcal{W}_{19}$, ${}^{(2)}\mathcal{W}_{21}$ and ${}^{(2)}\mathcal{W}_{23}$ denote the $c\in\{19,21,23\}$ admissible character-like (2,2) solutions (obtained with $\mu_3=0$).
\begin{align}
    &j_{2^{+}} - 32 = \chi_0^{{}^{(2)}\mathcal{W}_5}\chi_0^{{}^{(2)}\mathcal{W}_{19}} + 2 \, \chi_{\frac{2}{3}}^{{}^{(2)}\mathcal{W}_5}\chi_{\frac{4}{3}}^{{}^{(2)}\mathcal{W}_{19}}, \label{bilin_non_0} \\
    &j_{2^{+}} - 64 = \chi_0^{{}^{(2)}\mathcal{W}_3}\chi_0^{{}^{(2)}\mathcal{W}_{21}} + 256 \, \chi_{\frac{1}{2}}^{{}^{(2)}\mathcal{W}_3}\chi_{\frac{3}{2}}^{{}^{(2)}\mathcal{W}_{21}}, \label{bilin_non_1} \\
    &j_{2^{+}} - 80 = \chi_0^{{}^{(2)}\mathcal{W}_1}\chi_0^{{}^{(2)}\mathcal{W}_{23}} + 2 \, \chi_{\frac{1}{3}}^{{}^{(2)}\mathcal{W}_1}\chi_{\frac{5}{3}}^{{}^{(2)}\mathcal{W}_{23}}, \label{bilin_non_2}
\end{align}
where the above bilinears are for the following relation: $(2,0)\overset{c^\mathcal{H}=24}{\underset{n_1=2}{\longleftrightarrow}}(2,2)$. Furthermore, we also have,
\begin{align}
    j_{2^{+}}^{\tfrac{3}{4}}(j_{2^{+}}-102) =& \chi_0^{{}^{(2)}\mathcal{W}_{19}}\chi_0^{{}^{(2)}\mathcal{W}_{23}} + 2 \, \chi_{\frac{4}{3}}^{{}^{(2)}\mathcal{W}_{19}}\chi_{\frac{5}{3}}^{{}^{(2)}\mathcal{W}_{23}} \nonumber\\
    =& q^{-\tfrac{7}{4}}\left(1 + 80 q + 6793 q^2 + 472680 q^3 + 18944362 q^4 + \mathcal{O}(q^{5})\right),\\\label{bilin_non_3}
    j_{2^{+}}^{\tfrac{3}{4}}(j_{2^{+}}-112) =& \chi_0^{{}^{(2)}\mathcal{W}_{21}}\chi_0^{{}^{(2)}\mathcal{W}_{21}} + 256 \, \chi_{\frac{3}{2}}^{{}^{(2)}\mathcal{W}_{21}}\chi_{\frac{3}{2}}^{{}^{(2)}\mathcal{W}_{21}}\\
    =& q^{-\tfrac{7}{4}}\left(1 +70 q + 6013 q^2 + 450030 q^3 + 18635582 q^4 + \mathcal{O}(q^{5})\right),\label{bilin_non_4}
\end{align}
where the above bilinears are for the following relation: $(2,2)\overset{c^\mathcal{H}=42}{\underset{n_1=3}{\longleftrightarrow}}(2,2)$.\footnote{A similar analysis should be performed with the $\frac{k}{12}$ derivative operator (the traditional one used in the case of $\text{SL}(2,\mathbb{Z})$). In this case, we can tune the $\mu_1$ parameter to even make this MLDE consistent with the Riemann-Roch. This is because in this case there exists a weight $2$ meromorphic form contribution to the MLDE due to the allowance of poles in the coefficient functions present in the MLDE. We reserve this for future work.}.

\subsubsection{\texorpdfstring{$\ell = 3$}{l = 3}}
\noindent The MLDE, in this case, reads\footnote{The generic form of the $(2,3)$ and $(2,1)$ MLDEs are identical with only the values of the parameters being different},
\begin{align}\label{mldeF2_2char_l3}
    \left[\mathcal{D}^2+\left(\mu_1\frac{\left(E_4^{(2+)}\right)^2}{E_6^{2+}}+\mu_2\frac{\Delta_2}{E_6^{2+}}\right)\mathcal{D}+\mu_3\frac{E_{10}^{(2+)}}{E_{6}^{(2+)}}\right]\chi = 0.
\end{align}
From above we obtain the following recursion relation,
\begin{align}
    &\sum\limits_{k=0}^i\left[E_{6,k}^{2+}(\alpha^j+i-k)(\alpha^j+i-k-1) + E_{6,k}^{2+}(\alpha^j+i-k)-\frac{E_{2,k}^{2+}E_{2,k}^{2+}}{4}(\alpha^j+i-k)\right. \nonumber\\ 
    &\left.\mu_1 (E_{4}^{(2+)})^2_{,k} (\alpha^j+i-k) + \mu_2\Delta_{2,k}(\alpha^j+i-k) + \mu_3 E_{10,k}\right]a^j_{i-k} = 0,
\end{align}
which gives the following indicial equation,
\begin{align}
    \alpha^2 + \left(\mu_1-\frac{1}{4}\right)\alpha + \mu_3 = 0, \label{indicial_l3F2_2char}
\end{align}
From Riemann-Roch (sum of exponents $= -\frac{1}{2}$), we see that $\mu_1=\frac{3}{4}$. Now, let us determine the value of $\mu_2$ by transforming the $\tau$-MLDE in \ref{mldeF2_2char_l3} into the corresponding $j_{2^{+}}$-MLDE (about $\tau=\rho_2$),
\begin{align}
    \left[(j_{2^{+}})^2\partial_{j_{2^{+}}}^2 + \frac{j_{2^{+}}}{2}\partial_{j_{2^{+}}} -  \frac{j_{2^{+}}}{j_{2^{+}}-256}\left((64+\mu_2)\right)\partial_{j_{2^{+}}} + \frac{\mu_3 j_{2^{+}}}{j_{2^{+}}-256}\right]\chi = 0, \label{j2+mlde_p2_l3}
\end{align}
where we have set $\mu_1=\frac{3}{4}$ from the $\ell=3$ indicial equation (see \ref{indicial_l3F2_2char}). Now, note that about $\tau=\rho_2$, we have, $\chi_0\sim (j_{2^{+}})^x$ and $j_{2^{+}}\to 0$, with $x=\frac{3}{4}$ for $\ell=3$ (see single-character solutions). So, substituting $\chi=(j_{2^{+}})^x$ about $j_{2^{+}}=0$, the $j_{2^{+}}$-indicial equation (which is basically, coefficient of $(j_{2^{+}})^x$ $=\left.0\right|_{j_{2^{+}}=0}$) reads,
\begin{align}
    x^2-\frac{x}{2}+\frac{1}{4}+\frac{\mu_2}{256} = 0, \label{jindip2l3F2}
\end{align}
which yields $\mu_2 = -112$ for $x=\frac{3}{4}$. Similar analysis gives, $(x,\mu_2)=(\frac{1}{4},-48)$, $(x,\mu_2)=(\frac{1}{2},-64)$ as solution to \ref{jindip2l3F2}.\\

\noindent For $\mu_2=-48$, we get from the $i=1$ equation: $m_1=\frac{N(N+22)}{18-N}$ where $N=-24\alpha_0$. This gives the allowed range of central charge as, $0<c<18$. Scanning this space of $c$ we get the following admissible solutions:
\begin{itemize}
    \item $(c,h)=\left(6,0\right)$: 
    \begin{align}\label{no_luck_1}
        \begin{split}
        \chi_0 =& q^{-\tfrac{1}{4}}(1 + 14q + 331q^2 + 4506q^3 + 112795q^4 + 3965662q^5\\
        &{}\ \ \ \ \ \ \ \ + 175520124q^6 + 8718679572q^7 + 470448806341q^8 + \mathcal{O}(q^{9})),\\
        \chi_{\text{non-id}} =& \text{unstable},
        \end{split}
    \end{align}
    where the above is a single-character solution (not yet a $(1,\ell)$ solution).
\end{itemize}
\noindent
For $\mu_2=-64$, we get from the $i=1$ equation: $m_1=\frac{N(N+14)}{18-N}$ where $N=-24\alpha_0$. This gives the allowed range of central charge as, $0<c<18$. Scanning this space of $c$ we get the following admissible solutions:
\begin{itemize}
    \item $(c,h)=\left(6,0\right)$: 
    \begin{align}\label{no_luck_2}
        \begin{split}
        \chi_0 =& q^{-\tfrac{1}{4}}(1 + 10q + 367q^2 + 6422q^3 + 169555q^4 + 6322026q^5\\
        &{}\ \ \ \ \ \ \ \ \ \ \ \ \ \ \  + 286760396q^6 + 14591718924q^7 + 802225180229q^8 + \mathcal{O}(q^{9})),\\
        \chi_{\text{non-id}} =& \text{unstable},
        \end{split}
    \end{align}
    where the above is a single-character solution (not yet a $(1,\ell)$ solution).
    \item $(c,h)=\left(12,\frac{1}{2}\right)$: 
    \begin{align}\label{4B_character_2}
        \begin{split}
        \chi_0 =& j_{2^{+}}^{\tfrac{1}{2}},\\
        \chi_{\text{non-id}} =& 1.
        \end{split}
    \end{align}
\end{itemize}
We note that although the identity characters in \ref{no_luck_1} and \ref{no_luck_2} possess an unstable non-trivial primary, these do not have a nice McKay-Thompson interpretation. This is similar to what we observed for the case of the identity-only solution $(c,h) = (3,0)$ in \ref{c=3,h=0_no_MT}. The identity character with a constant non-trivial primary in \ref{4B_character_2} along with \ref{4B_character_1} forms a set of central charge $c = 12$ admissible solutions that correspond to the $4B$ conjugacy class of $\mathbb{M}$.

\subsection{Three-character solutions}
\noindent We restrict our analysis to the $(n,\ell) = (3,0)$ MLDEs and reserve the other cases for future work. This modular invariant ODE in this case reads
\begin{align}
    \left[\mathcal{D}^{3} + \omega_{2}(\tau)\mathcal{D}^{2} + \omega_{4}(\tau)\mathcal{D}  +\omega_{6}(\tau)\right]\chi(\tau) = 0.
\end{align}
With choices $\omega_{4}(\tau) =\mu_{1} E_{4}^{(2^{+})}(\tau)$ and $\omega_{6}(\tau) = \mu_{2} E_{6}^{(2^{+})}(\tau)$ and in terms of derivatives $\Tilde{\partial}$, the MLDE reads
\begin{align}
        \left[\Tilde{\partial}^{3} - \frac{3}{4}\left(\Tilde{\partial}E_{2}^{(2^{+})}(\tau)\right)\Tilde{\partial}^{2} + \frac{1}{8}\left(E_{2}^{(2^{+})}\right)^{2} - \frac{1}{4}\left(\Tilde{\partial}E_{2}^{(2^{+})}\right)\Tilde{\partial} + \mu_{1}E_{4}^{(2^{+})}(\tau)\Tilde{\partial} + \mu_{2}E_{6}^{(2^{+})}(\tau)\right]\chi(\tau) = 0.
\end{align}
We now make use of the Ramanujan identity in \ref{Ramanujan_p=2} for $E_{2}^{(2^{+})}(\tau)$ to rewrite the MLDE as follows
\begin{align}
    \left[\Tilde{\partial}^{3} - \frac{3}{4}\left(\Tilde{\partial}E_{2}^{(2^{+})}(\tau)\right)\Tilde{\partial}^{2} +
    \left(\frac{3}{4}\left(\Tilde{\partial}E_{2}^{(2^{+})}\right) + \frac{1}{8}E_{4}^{(2^{+})} + \mu_{1}E_{4}^{(2^{+})}(\tau)\right)\Tilde{\partial} + \mu_{2}E_{6}^{(2^{+})}(\tau)\right]\chi(\tau) = 0.
\end{align}
The recursion relation reads,
\begin{align}
    &\left[(\alpha^j+i)(\alpha^j+i-1)(\alpha^j+i-2) + 3(\alpha^j+i)(\alpha^j+i-1) + (\alpha^j+i)\right]a^j_i \nonumber\\ 
    &+ \sum\limits_{k=0}^i\left[-\frac{3}{4}E_{2,k}^{(2+)}a^j_{i-k}(\alpha^j+i-k)(\alpha^j+i-k-1) - \frac{3}{4}E_{2,k}^{(2+)}a^j_{i-k}(\alpha^j+i-k)\right. \nonumber\\
    &\left.+ \frac{1}{32}E_{4,k}^{2+}a^j_{i-k}(\alpha^j+i-k) + \frac{3}{32}\left(E_{2}^{(2+)}\right)^2_{,k}a^j_{i-k}(\alpha^j+i-k) + \mu_1E_{4,k}^{(2+)}a^j_{i-k}(\alpha^j+i-k)\right. \nonumber\\
    &\left.+ \mu_2E_{6,k}^{(2+)}a^j_{i-k}\right] = 0. \label{recursion3_2}
\end{align}
From which the indicial equation reads,
\begin{align}
    \alpha^3 - \frac{3}{4}\alpha^{2} + \left(\mu_1+\frac{1}{8}\right)\alpha + \mu_2 = 0, \label{indicial3_2}
\end{align}
where we have defined $\Tilde{\mu}_{1}\equiv \mu_{1} + \tfrac{1}{8}$. Using the equation we obtain at $k = 1$, we find the following relations among the roots of the indicial equation, $\alpha_{i}, i = 0,1,2$, and the free parameters
\begin{align}
    \begin{split}
    \Tilde{\mu}_{1} =& \alpha_{0}\alpha_{1} + \alpha_{1}\alpha_{2} + \alpha_{0}\alpha_{2},\\
    \mu_{2} =& -\alpha_{0}\alpha_{1}\alpha_{2}.
    \end{split}
\end{align}
We can now fix $\mu_{2}$ from the indicial equation and express $\mu_{1}$ in terms of the roots $\alpha_{i}$ and the ratio $m_{1}$ obtain ed via the equation at $k = 1$. At $k = 2$, we obtain the following equation
\begin{align}
&-24 m_1^2 + 9 m_1 m_2 - 1004 m_1 \alpha_0 - 184 m_1^2 \alpha_0 + 1064 m_2 \alpha_0 + 12 m_1 m_2 \alpha_0 + 22560 \alpha_0^2 - 6192 m_1 \alpha_0^2 \nonumber\\
&- 672 m_1^2 \alpha_0^2 + 2016 m_2 \alpha_0^2 - 141696 \alpha_0^3 - 18304 m_1 \alpha_0^3 - 512 m_1^2 \alpha_0^3 + 1024 m_2 \alpha_0^3 + 150528 \alpha_0^4 = 0. \label{al3}
\end{align}
We observe that if we define the variable
\begin{align}
    N = -1176\alpha_{0},
\end{align}
then the polynomial equation simplifies to (after multiplying $7^6\times 3^3\times 2^2$ with Eq.(\ref{al3}) and using the above identification)
\begin{align}
&-304946208 m_1^2 + 114354828 m_1 m_2 + 10847718 m_1 N + 1988028 m_1^2 N - 11495988 m_2 N \nonumber\\
&- 129654 m_1 m_2 N + 207270 N^2 - 56889 m_1 N^2 - 6174 m_1^2 N^2 + 18522 m_2 N^2 + 1107 N^3 \nonumber\\
&+ 143 m_1 N^3 + 4 m_1^2 N^3 - 8 m_2 N^3 + N^4 = 0. \label{Diop1_3_2}
\end{align} 
The integer root theorem, $N=-1176\alpha_0$ above should be an integer and hence, $N=49c\in\mathbb{Z}$. this tells us that the central charge has to be a rational number whose denominator is always $49$. Since $\alpha_{1}$ and $\alpha_{2}$ are to be rational, this implies that the discriminant (which is an integer, say $k^2$ with $k\in\mathbb{Z}$) has to be the perfect square to be able to result in rational roots Thus, we have,
\begin{align}
    &46694888100 m_1^2 - 5006200248 m_1 N - 190591380 m_1^2 N + 77533092 N^2 \nonumber\\
    &+ 17114328 m_1 N^2 + 194481 m_1^2 N^2 - 22932 N^3 - 15582 m_1 N^3 - 143 N^4 = k^2. \label{Diop2_3_2}
\end{align}
From this, we find the three roots of the indicial equation to read
\begin{align}
    &\alpha_0 = -\frac{N}{1176}, \label{root0_30_F2} \\
    &\alpha_1 = \frac{129654 m_1 - 11466 N + 147 m_1 N - 13 N^2 - k}{2352(147 m_1 - 13 N)}, \label{roo1_30_F2} \\
    &\alpha_2 = \frac{129654 m_1 - 11466 N + 147 m_1 N - 13 N^2 + k}{2352(147 m_1 - 13 N)}. \label{roo2_30_F2}
\end{align}
This equation where all the unknown variables $N$, $m_{1}$, and $k$ are positive integers is known as a Diophantine equation.\\
\\
Now, solving \ref{Diop1_3_2} and \ref{Diop2_3_2}, for the range $1\leq N\leq 2058$, yields,
\begin{itemize}
    \item $(c,h_1,h_2)=\left(\frac{6}{7},\frac{5}{7},\frac{1}{7}\right)$: 
    \begin{align}
        \chi_0 &= q^{-\tfrac{1}{28}}(1 + q^2 + q^3 + 2q^4 + q^5 + 3q^6 + 2q^7 + 4q^8 + 3q^9 + 6q^{10} + \mathcal{O}(q^{11})), \nonumber\\
        \chi_{\frac{5}{7}} &= q^{\tfrac{19}{28}}(1 + q^2 + q^4 + q^5 + 2q^6 + q^7 + 3q^8 + 2q^9 + 4q^{10} + \mathcal{O}(q^{11})), \nonumber\\
        \chi_{\frac{1}{7}} &= q^{\tfrac{3}{28}}(1 + q + q^2 + q^3 + 2q^4 + 2q^5 + 3q^6 + 3q^7 + 5q^8 + 5q^9 + 7q^{10} + \mathcal{O}(q^{11})).
    \end{align}
    where the identity character above is similar to that of a Virasoro minimal model CFT by which we mean that $m_1=0$ for such models since they do not possess any Kac-Moody currents. From here on we shall use the notation minimal model to denote any admissible solution whose identity character has $m_1=0$.
    \item $(c,h_1,h_2)=\left(1,\frac{13}{24},\frac{1}{3}\right)$: 
    \begin{align}
        \chi_0 &= q^{-\tfrac{1}{24}}(1 + q + 2q^2 + q^3 + 3q^4 + 3q^5 + 5q^6 + 5q^7 + 8q^8 + 8q^9 + 12q^{10} + \mathcal{O}(q^{11})), \nonumber\\
        \chi_{\frac{13}{24}} &= \text{unstable}, \nonumber\\
        \chi_{\frac{1}{3}} &= q^{\tfrac{7}{24}}(1 + q^2 + q^3 + 2q^4 + q^5 + 3q^6 + 2q^7 + 5q^8 + 4q^9 + 7q^{10} + \mathcal{O}(q^{11})),
    \end{align}
    where above is a two-character solution paired with an unstable character. This is the $(2,0)$ two-character solution found in \ref{c=1_Fricke_2_2,0} that makes a reappearance here.
    \item $(c,h_1,h_2)=\left(2,\frac{2}{3},\frac{1}{3}\right)$: 
    \begin{align}
        \chi_0 &= q^{-\tfrac{1}{12}}(1 + 2q + 5q^2 + 6q^3 + 12q^4 + 16q^5 + 29q^6 + 38q^7 + 61q^8 + 80q^9\\ \nonumber
        &{}\ \ \ \ \ \ \ \ \ \ \ \ \ + 121q^{10} + \mathcal{O}(q^{11})), \nonumber\\
        \chi_{\frac{2}{3}} &= q^{\tfrac{7}{12}}(1 + 2q^2 + 2q^3 + 5q^4 + 4q^5 + 11q^6 + 10q^7 + 22q^8 + 22q^9 + 41q^{10} + \mathcal{O}(q^{11})), \nonumber\\
        \chi_{\frac{1}{3}} &= q^{\tfrac{1}{4}}(1 + q + 3q^2 + 3q^3 + 8q^4 + 9q^5 + 17q^6 + 20q^7 + 36q^8 + 43q^9 + 70q^{10} + \mathcal{O}(q^{11})).
    \end{align}
    We note here that the identity character can be expressed in terms of modular forms of $\Gamma_{0}(7)$, akin to what we saw in the two-character case at low central charge $c = 1$ in \ref{Hecke_1_c=1}, as follows
    \begin{align}\label{Hecke_1_c=2}
        \chi_{0}^{(c = 2)} = \left(\frac{\left(j_{2}\Delta_{2}^{\infty}\right)(3\tau)}{\left(j_{2}\Delta_{2}^{\infty}\right)^{\tfrac{1}{3}}(\tau)}\right)^{\tfrac{1}{4}}\left(\frac{1}{\Delta_{2}(\tau)}\right)^{\tfrac{1}{12}}.
    \end{align}
    \item $(c,h_1,h_2)=\left(3,\frac{5}{8},\frac{1}{2}\right)$: 
    \begin{align}
        \chi_0 &= q^{-\tfrac{1}{8}}(1 + 5q + 11q^2 + 20q^3 + 41q^4 + 76q^5 + 127q^6 + 211q^7 + 342q^8 + 541q^9 \nonumber\\
        &{}\ \ \ \ \ \ \ \ \ \ \ \ + 838q^{10} + \mathcal{O}(q^{11})), \nonumber\\
        \chi_{\frac{5}{8}} &= \text{unstable}, \nonumber\\
        \chi_{\frac{1}{2}} &= q^{\tfrac{3}{8}}(1 + q + 5q^2 + 6q^3 + 16q^4 + 21q^5 + 46q^6 + 61q^7 + 117q^8 + 157q^9\\ \nonumber
        {}&\ \ \ \ \ \ \ \ \ \ + 273q^{10} + \mathcal{O}(q^{11})).
    \end{align}
    This is the $(2,0)$ two-character solution we found in \ref{c=3_Fricke_2_2,0} that makes a reappearance here.
    \item $(c,h_1,h_2)=\left(3,0,\frac{9}{8}\right)$: 
    \begin{align}
        \chi_0 &= q^{-\tfrac{1}{8}}(1 + 25q + 51q^2 + 196q^3 + 297q^4 + 780q^5 \nonumber\\
        &{}\ \ \ \ \ \ \ \ \ \ + 1223q^6 + 2551q^7 + 3798q^8 + 7201q^9 + 10582q^{10} + \mathcal{O}(q^{11})), \nonumber\\
        \chi_{0} &=  q^{-\tfrac{1}{8}}(1 + 25q + 51q^2 + 196q^3 + 297q^4 + 780q^5 \nonumber\\
        &{}\ \ \ \ \ \ \ \ \ \ + 1223q^6 + 2551q^7 + 3798q^8 + 7201q^9 + 10582q^{10} + \mathcal{O}(q^{11})), \nonumber\\
        \chi_{\frac{9}{8}} &= \text{unstable}.
    \end{align}
    This is the single character identified in \ref{c=3,single_char_n=2} that makes a reappearance here. 
    \item $(c,h_1,h_2)=\left(\frac{30}{7},\frac{5}{7},\frac{4}{7}\right)$: 
    \begin{align}
        \chi_0 &= q^{-\tfrac{5}{28}}(1 + 10q + 25q^2 + 70q^3 + 145q^4 + 330q^5 + 610q^6 + 1200q^7\nonumber\\ 
        &{}\ \ \ \ \ \ \ \ \ \ \ \ \ + 2095q^8 + 3800q^9 + 6336q^{10} + \mathcal{O}(q^{11})), \nonumber\\
        \chi_{\frac{5}{7}} &= q^{15/28}(4 + 5q + 30q^2 + 45q^3 + 130q^4 + 204q^5 + 480q^6 + 745q^7\nonumber\\
        &{}\ \ \ \ \ \ \ \ \ \ \ \ \ \ + 1510q^8 + 2330q^9 + 4294q^{10} + \mathcal{O}(q^{11})), \nonumber\\
        \chi_{\frac{4}{7}} &= q^{11/28}(5 + 10q + 45q^2 + 74q^3 + 205q^4 + 350q^5 + 770q^6 + 1260q^7\nonumber\\
        &{}\ \ \ \ \ \ \ \ \ \ \ \ \ \ + 2470q^8 + 3950q^9 + 7115q^{10} + \mathcal{O}(q^{11})).
    \end{align}
    \item $(c,h_1,h_2)=\left(5,\frac{17}{24},\frac{2}{3}\right)$: 
    \begin{align}
        \chi_0 &= q^{-\tfrac{5}{24}}(1 + 15q + 40q^2 + 135q^3 + 290q^4 + 726q^5 \nonumber\\ 
        &{}\ \ \ \ \ \ \ \ \ \ \ \ \ + 1415q^6 + 3000q^7 + 5485q^8 + 10565q^9 + 18407q^{10} + \mathcal{O}(q^{11})), \nonumber\\
        \chi_{\frac{17}{24}} &= \text{unstable}, \nonumber\\
        \chi_{\frac{2}{3}} &= q^{\tfrac{11}{24}}(5 + 11q + 55q^2 + 100q^3 + 290q^4 + 535q^5 \nonumber\\
        &{}\ \ \ \ \ \ \ \ \ \ \ + 1235q^6 + 2160q^7 + 4400q^8 + 7465q^9 + 13945q^{10} + \mathcal{O}(q^{11})).
    \end{align}
    This is the $(2,0)$ two-character solution we found in \ref{c=5_Fricke_2_2,0} that makes a reappearance here.
    \item $(c,h_1,h_2)=\left(6,1,\frac{1}{2}\right)$:
    \begin{align}
        \chi_0 &= q^{-\tfrac{1}{4}}(1 + m_1q + m_2q^2 + \mathcal{O}(q^{3})), \nonumber\\
        \chi_{1} &= q^{\tfrac{3}{4}}(1 + 2q + 11q^2 + 22q^3 + 69q^4 + 134q^5 \nonumber\\
        &{}\ \ \ \ \ \ \ \ \ \ + 330q^6 + 616q^7 + 1324q^8 + 2382q^9 + 4675q^{10} + \mathcal{O}(q^{11})), \nonumber\\
        \chi_{\frac{1}{2}} &= q^{\tfrac{1}{4}}(1 + 6q + 21q^2 + 62q^3 + 162q^4 + 384q^5 \nonumber\\
        &{}\ \ \ \ \ \ \ \ \ \ + 855q^6 + 1806q^7 + 3648q^8 + 7110q^9 + 13434q^{10} + \mathcal{O}(q^{11}))
    \end{align}
    where for the above identity character we have, $m_1\in\mathbb{N}\cup \{0\}$, $m_2=27+2m_1$ and $k=2247336-86436 m_1$ for $m_1< 26$ and $k=-2247336+86436 m_1$ for $m_1\geq 27$.\footnote{This kind of infinite one-parameter solutions (here the parameter being $m_1$) were first observed in \cite{Das:2021uvd} where it was explained using an argument using Inhomogenous MLDE. We have checked that the same argument applies here too with the following replacement $\eta(\tau)^{2k}\leftrightarrow(\eta(\tau)\eta(2\tau))^k$.} For $m_1=26$, the id. character is just $j_{2^{+}}^{\frac{1}{4}}$
    \item $(c,h_1,h_2)=\left(7,\frac{31}{24},\frac{1}{3}\right)$: 
    \begin{align}
        \chi_0 &= q^{-\tfrac{7}{24}}(1 + 27q + 107q^2 + 458q^3 + 1268q^4 + 3673q^5 \nonumber\\ 
        &{}\ \ \ \ \ \ \ \ \ \ \ \ \ + 8722q^6 + 21061q^7 + 45251q^8 + 97657q^9 + 195561q^{10} + \mathcal{O}(q^{11})), \nonumber\\
        \chi_{\frac{31}{24}} &= \text{unstable}, \nonumber\\
        \chi_{\frac{1}{3}} &= q^{\tfrac{1}{24}}(1 + 26q + 80q^2 + 353q^3 + 862q^4 + 2564q^5 \nonumber\\
        &{}\ \ \ \ \ \ \ \ \ \ \ + 5728q^6 + 13956q^7 + 28861q^8 + 62621q^9 + 122250q^{10} + \mathcal{O}(q^{11}))
    \end{align}
    This is the $(2,2)$ two-character solution we found in \ref{c=7_Fricke_2_2,2} that makes a reappearance here.
    \item $(c,h_1,h_2)=\left(\frac{54}{7},\frac{9}{7},\frac{3}{7}\right)$: 
    \begin{align}
        \chi_0 &= q^{-\tfrac{9}{28}}(1 + 27q + 135q^2 + 591q^3 + 1836q^4 + 5481q^5 \nonumber\\ 
        &{}\ \ \ \ \ \ \ \ \ \ \ \ \ + 14037q^6 + 35046q^7 + 79866q^8 + 178083q^9 + 374328q^{10} + \mathcal{O}(q^{11})), \nonumber\\
        \chi_{\frac{9}{7}} &= q^{\tfrac{27}{28}}(26 + 81q + 405q^2 + 1047q^3 + 3375q^4 + 7884q^5 \nonumber\\
        &{}\ \ \ \ \ \ \ \ \ \ \ \ \ + 20352q^6 + 43983q^7 + 99927q^8 + 202994q^9 + 422901q^{10} + \mathcal{O}(q^{11})), \nonumber\\
        \chi_{\frac{3}{7}} &= q^{\tfrac{3}{28}}(3 + 54q + 189q^2 + 840q^3 + 2268q^4 + 6858q^5 \nonumber\\
        &{}\ \ \ \ \ \ \ \ \ \ \ + 16397q^6 + 40986q^7 + 89505q^8 + 199410q^9 + 407457q^{10} + \mathcal{O}(q^{11})).
    \end{align}
    \item $(c,h_1,h_2)=\left(\frac{54}{7},\frac{10}{7},\frac{2}{7}\right)$: 
    \begin{align}
        \chi_0 &= q^{-\tfrac{9}{28}}(1 + 30q + 135q^2 + 594q^3 + 1836q^4 + 5484q^5 \nonumber\\ 
        &{}\ \ \ \ \ \ \ \ \ \ \ \ \ + 14040q^6 + 35052q^7 + 79869q^8 + 178092q^9 + 374334q^{10} + \mathcal{O}(q^{11})), \nonumber\\
        \chi_{\frac{10}{7}} &= q^{\tfrac{31}{28}}(51 + 186q + 837q^2 + 2262q^3 + 6852q^4 + 16388q^5 \nonumber\\
        &{}\ \ \ \ \ \ \ \ \ \ \ \ \ + 40977q^6 + 89490q^7 + 199395q^8 + 407436q^9 + 838032q^{10} + \mathcal{O}(q^{11})), \nonumber\\
        \chi_{\frac{2}{7}} &= q^{-\tfrac{1}{28}}(3 + 26q + 84q^2 + 408q^3 + 1053q^4 + 3378q^5 \nonumber\\
        &{}\ \ \ \ \ \ \ \ \ \ \ \ \ + 7893q^6 + 20358q^7 + 43995q^8 + 99936q^9 + 203012q^{10} + \mathcal{O}(q^{11})).
    \end{align}
\item $(c,h_1,h_2)=\left(9,\frac{11}{8},\frac{1}{2}\right)$: 
    \begin{align}
        \chi_0 &= q^{-\tfrac{3}{8}}(1 + 31q + 220q^2 + 1027q^3 + 3815q^4 + 12189q^5 \nonumber\\ 
        &{}\ \ \ \ \ \ \ \ \ \ \ \ + 35157q^6 + 93733q^7 + 234357q^8 + 556019q^9 + 1262556q^{10} + \mathcal{O}(q^{11})), \nonumber\\
        \chi_{\frac{11}{8}} &= \text{unstable}, \nonumber\\
        \chi_{\frac{1}{2}} &= q^{\tfrac{1}{8}}(1 + 27q + 110q^2 + 541q^3 + 1648q^4 + 5402q^5 \nonumber\\
        &{}\ \ \ \ \ \ \ \ \ \ + 14153q^6 + 37936q^7 + 89648q^8 + 212550q^9 + 465321q^{10} + \mathcal{O}(q^{11})).
    \end{align}
    \item $(c,h_1,h_2)=\left(10,\frac{4}{3},\frac{2}{3}\right)$: 
    \begin{align}
        \chi_0 &= q^{-\tfrac{5}{12}}(1 + 30q + 305q^2 + 1470q^3 + 6230q^4 + 20952q^5 \nonumber\\ 
        &{}\ \ \ \ \ \ \ \ \ \ \ \ \ + 66035q^6 + 184830q^7 + 494290q^8 + 1228810q^9 + 2950340q^{10} + \mathcal{O}(q^{11})), \nonumber\\
        \chi_{\frac{4}{3}} &= q^{\tfrac{11}{12}}(25 + 110q + 671q^2 + 2210q^3 + 8125q^4 + 22730q^5 \nonumber\\
        &{}\ \ \ \ \ \ \ \ \ \ \ \ \ + 66020q^6 + 165620q^7 + 418470q^8 + 966350q^9 + 2222205q^{10} + \mathcal{O}(q^{11})), \nonumber\\
        \chi_{\frac{2}{3}} &= q^{\tfrac{1}{4}}(5 + 86q + 420q^2 + 2040q^3 + 6925q^4 + 23130q^5 \nonumber\\
        &{}\ \ \ \ \ \ \ \ \ \ + 65371q^6 + 180730q^7 + 453375q^8 + 1111940q^9 + 2560355q^{10} + \mathcal{O}(q^{11})).
    \end{align}
\item $(c,h_1,h_2)=\left(10,\frac{5}{3},\frac{1}{3}\right)$: 
    \begin{align}
        \chi_0 &= q^{-\tfrac{5}{12}}(1 + 40q + 305q^2 + 1490q^3 + 6250q^4 + 21002q^5 + 66075q^6 + 184940q^7\nonumber\\ 
        &{}\ \ \ \ \ \ \ \ \ \ \ \ \ + 494390q^8 + 1229030q^9 + 2950560q^{10} + \mathcal{O}(q^{11})), \nonumber\\
        \chi_{\frac{5}{3}} &= q^{\tfrac{5}{4}}(1 + 5q + 25q^2 + 85q^3 + 285q^4 + 806q^5 + 2230q^6 + 5595q^7\nonumber\\
        &{}\ \ \ \ \ \ \ \ \ \ \  + 13725q^8 + 31605q^9 + 71258q^{10} + \mathcal{O}(q^{11})), \nonumber\\
        \chi_{\frac{1}{3}} &= q^{-\tfrac{1}{12}}(5 + 60q + 245q^2 + 1372q^3 + 4480q^4 + 16330q^5 + 45605q^6 + 132230q^7 \nonumber\\
        &{}\ \ \ \ \ \ \ \ \ \ \ \ \ + 331545q^8 + 837340q^9 + 1933305q^{10} + \mathcal{O}(q^{11})).
    \end{align}
\item $(c,h_1,h_2)=\left(11,\frac{35}{24},\frac{2}{3}\right)$: 
    \begin{align}
        \chi_0 &= q^{-\tfrac{11}{24}}(1 + 41q + 509q^2 + 2686q^3 + 12605q^4 + 45402q^5 + 153461q^6 + 455033q^7\nonumber\\ 
        &{}\ \ \ \ \ \ \ \ \ \ \  \ \ + 1288031q^8 + 3367910q^9 + 8491985q^{10} + \mathcal{O}(q^{11})), \nonumber\\
        \chi_{\frac{35}{24}} &= \text{unstable}, \nonumber\\
        \chi_{\frac{2}{3}} &= q^{\tfrac{5}{24}}(5 + 141q + 736q^2 + 4029q^3 + 14596q^4 + 52740q^5 + 157646q^6 + 462885q^7\nonumber\\
        &{}\ \ \ \ \ \ \ \ \ \ \ + 1221334q^8 + 3151415q^9 + 7594605q^{10} + \mathcal{O}(q^{11}))
    \end{align}
\item $(c,h_1,h_2)=\left(\frac{78}{7},\frac{9}{7},\frac{6}{7}\right)$: 
    \begin{align}
        \chi_0 &= q^{-\tfrac{39}{84}}(1 + 26q + 403q^2 + 2002q^3 + 9906q^4 + 35204q^5 + 122161q^6 + 361062q^7\nonumber\\ 
        &{}\ \ \  \ \ \ \ \ \ \ \ \ \ + 1038219q^8 + 2718820q^9 + 6930937q^{10} + \mathcal{O}(q^{11})), \nonumber\\
        \chi_{\frac{9}{7}} &= q^{\tfrac{23}{28}}(52 + 299q + 2002q^2 + 7501q^3 + 29406q^4 + 90506q^5 + 278408q^6 + 751907q^7\nonumber\\
        &{}\ \ \ \ \ \ \ \ \ \ \ \ \  + 2002066q^8 + 4919317q^9 + 11872484q^{10} + \mathcal{O}(q^{11})), \nonumber\\
        \chi_{\frac{6}{7}} &= q^{\tfrac{11}{28}}(26 + 352q + 2054q^2 + 9880q^3 + 37180q^4 + 127764q^5 + 388765q^6 + 1115322q^7\nonumber\\
        &{}\ \ \ \ \ \ \ \ \ \ \ \ \  + 2969551q^8 + 7572838q^9 + 18351476q^{10} + \mathcal{O}(q^{11})).
    \end{align}
\item $(c,h_1,h_2)=\left(12,\frac{5}{4},1\right)$:
    \begin{align}
        \chi_0 &= q^{-\tfrac{1}{2}}(1 + m_1q + m_2q^2 + \mathcal{O}(q^{3})), \nonumber\\
        \chi_{\frac{5}{4}} &= \text{unstable}, \nonumber\\
        \chi_{1} &= q^{\tfrac{1}{4}}(1 + 12q + 78q^2 + 376q^3 + 1509q^4 + 5316q^5 \nonumber\\
        &{}\ \ \ \ \ \ \ \ \ \ \ + 16966q^6 + 50088q^7 + 138738q^8 + 364284q^9 + 913824q^{10} + \mathcal{O}(q^{11})),
    \end{align}
    where for the above identity character we have, $m_1\in\mathbb{N}\cup \{0\}$, $m_2=210+12m_1$ and $k=2247336-43218 m_1$ for $m_1< 52$ and $k=-2247336+43218 m_1$ for $m_1\geq 53$. For $m_1=52$, the id. character is just $j_{2^{+}}^{\frac{1}{2}}$.
    \item $(c,h_1,h_2)=\left(14,\frac{1}{3},\frac{13}{6}\right)$: 
    \begin{align}\label{3_char_unstable}
        \chi_0 &= \text{unstable},\nonumber\\
        \chi_{\frac{1}{3}} &= q^{-\tfrac{1}{4}}(1 + 26q + 79q^2 + 326q^3 + 755q^4 + 2106q^5 +4460q^6 + 10284q^7\nonumber\\
        &{}\ \ \ \ \ \ \ \ \ \ \ \ + 20165q^8 + 41640q^9 + 77352q^{10} + \mathcal{O}(q^{11})), \nonumber\\
        \chi_{\frac{13}{6}} &= \text{unstable}.
    \end{align}
This is the $c = 6$ single-character theory we found in \ref{8C_character_1} that makes a reappearance as a non-trivial primary of a higher central charge three-character theory with unstable identity and another non-trivial primary. We can actually re-interpret the central charge (using \textit{unitary presentation} kind of arguments of \cite{Harvey:2018rdc}) as $c_{\text{eff}}=c-24h_1 = 14-24\times\frac{1}{3} = 6$. Hence, when the $\frac{1}{3}$ non-trivial primary is re-interpreted as a $c=6$ theory with central charge equal to $c_{\text{eff}}=6$, it is indeed $j_{2^{+}}^{\frac{1}{4}}$.
\item $(c,h_1,h_2)=\left(18,\frac{1}{2},\frac{5}{2}\right)$: 
    \begin{align}
        \chi_0 &= \text{Log unstable}, \nonumber\\
        \chi_{\frac{1}{2}} &= \text{Log unstable}, \nonumber\\
        \chi_{\frac{5}{2}} &= q^{\tfrac{7}{4}}(1 + 10q + 71q^2 + 390q^3 + 1831q^4 + 7602q^5 + 28712q^6 + 100292q^7 \nonumber\\
        &{}\ \ \ \ \ \ \ \ \ \ \ \ + 328247q^8 + 1016012q^9 + 2996354q^{10} + \mathcal{O}(q^{11})).
    \end{align}
    Above, the term \textit{Log unstable} refers to some pathologies in the q-series coefficients of these characters when we try to obtain their coefficients from the recursion relation. At times, these happen when the corresponding hypergeometric expressions for these characters, of ${}_3F_2$ types, when expanded about $q=0$ result in a $\log[q]$ behaviour. However, in certain cases, there is no such instability and it is just a shortcoming of the recursion relation approach to obtain q-series coefficients of characters. This happens when one tries to obtain the q-series coefficients of $D_{4,1}$ WZW model in the $\text{SL}(2,\mathbb{Z})$ case. To remedy this, one uses the theta series representations for the characters of $D_{4,1}$ to find the q-series of its characters. However, for the congruence subgroups, we are not aware whether such theta series representations exist for the characters and hence, we have not been able to conclude whether these are really Log singularities in the q-series or just an artefact of the recursion relation. In any case, we call them Log unstable here and shall return in the future to understand this feature better in the realm of congruence subgroups. \footnote{Note that, we have only reported the Log unstable solutions for the three-character case and we postpone the analysis of such instabilities in the two-character case to future work. Indeed, these log instabilities do appear even in the two-character case whenever one of the conformal dimensions is a non-zero positive integer. For the $\text{SL}(2,\mathbb{Z})$ case, whenever a non-trivial primary had a conformal dimension $\in\mathbb{N}$, the theory is called a \textit{logarithmic} CFT. Here, we do not know how to interpret such things just as yet, and thus, we reserve this for future work too.}  
\item $(c,h_1,h_2)=\left(22,\frac{2}{3},\frac{17}{6}\right)$:
    \begin{align}
        \chi_0 &= \text{unstable}, \nonumber\\
        \chi_{\frac{2}{3}} &= q^{-\tfrac{1}{4}}(1 + 26q + 79q^2 + 326q^3 + 755q^4 + 2106q^5 + 4460q^6 + 10284q^7 \nonumber\\
        &{}\ \ \ \ \ \ \ \ \ \ \ \ + 20165q^8 + 41640q^9 + 77352q^{10} + \mathcal{O}(q^{11})), \nonumber\\
        \chi_{\frac{17}{6}} &= \text{unstable}.
    \end{align}
    The first non-trivial primary is precisely equal to the first non-trivial primary in \ref{3_char_unstable}, which is a reappearing single-character theory.
\item $(c,h_1,h_2)=\left(24,\frac{291}{292},\frac{201}{73}\right)$:
    \begin{align}\label{the_one_with_N_as15987}
        \chi_0 &= q^{-1}(1 + 16091q + 4372q^2 + 96256q^3 + 1240002q^4 + 10698752q^5 + 74428120q^6 \nonumber\\ 
        &{}\ \ \ \ \ \ \ \ \ \ \ + 431529984q^7 + 2206741887q^8 + 10117578752q^9 + 42616961892q^{10} + \mathcal{O}(q^{11})), \nonumber\\
        \chi_{\frac{291}{292}} &= \text{unstable}, \nonumber\\
        \chi_{\frac{201}{73}} &= \text{unstable}.
    \end{align}
The identity is an $\ell = 4$, $c = 24$ single-character theory $j_{2^{+}} + 15987$.
\item $(c,h_1,h_2)=\left(26,\frac{2}{3},\frac{10}{3}\right)$: 
    \begin{align}
        \chi_0 &= q^{-\tfrac{13}{12}}(1 + 286q + 6799q^2 + 102466q^3 + 1374100q^4 + 12916124q^5 + 98031453q^6\nonumber\\ 
        &{}\ \ \ \ \ \ \ \ \ + 619853026q^7 + 3426275008q^8 + 16919200870q^9 + 76289142098q^{10} + \mathcal{O}(q^{11})), \nonumber\\
        \chi_{\frac{2}{3}} &= q^{-\tfrac{5}{12}}(65 + 2782q + 26387q^2 + 366938q^3 + 3482752q^4 + 26387140q^5 + 174154539q^6\nonumber\\
        &{}\ \ \ \ \ \ \ \ \  + 989454102q^7 + 5051982756q^8 + 23348937742q^9 + 99790466750q^{10} + \mathcal{O}(q^{11})), \nonumber\\
        \chi_{\frac{10}{3}} &= q^{\tfrac{9}{4}}(13 + 198q + 1872q^2 + 13728q^3 + 83655q^4 + 447174q^5 + 2143011q^6 + 9418266q^7\nonumber\\
        &{}\ \ \ \ \ \ \  \ \ \ \ \ + 38395656q^8 + 146938090q^9 + 531665433q^{10} + \mathcal{O}(q^{11})).
    \end{align}
\item $(c,h_1,h_2)=\left(27,\frac{1}{8},4\right)$: 
    \begin{align}
        \chi_0 &= \text{Log unstable}, \nonumber\\
        \chi_{\frac{1}{8}} &= q^{-1}(69 + 9224q + 301668q^2 + 6641664q^3 + 85560138q^4 + 738213888q^5  \nonumber\\
        &{}\ \ \ \ \ \ \ \ \ \ \ \ \ + 5135540280q^6 + 29775568896q^7 + 152265190203q^8\\
        &{}\ \ \ \ \ \ \ \ \ \ \ \ \ + 698112933888q^9 + 2940570370548q^{10} + \mathcal{O}(q^{11})), \nonumber\\
        \chi_{4} &= \text{unstable}.
    \end{align}
\item $(c,h_1,h_2)=\left(30,\frac{1}{2},4\right)$: 
    \begin{align}
        \chi_0 &= \text{Log unstable}, \nonumber\\
        \chi_{\frac{1}{2}} &= q^{-\frac{3}{4}}(25 + 2110q + 51953q^2 + 1015950q^3 + 15480440q^4 + 163643540q^5  \nonumber\\
        &{}\ \ \ \ \ \ \ \ \ \ \ \ \ + 1372184025q^6 + 9575867298q^7 + 58118855135q^8\\
        &{}\ \ \ \ \ \ \ \ \ \ \ \ \ + 314673021160q^9 + 1550715669375q^{10} + \mathcal{O}(q^{11})), \nonumber\\
        \chi_{4} &= q^{\frac{11}{4}}(15 + 286q + 3085q^2 + 25370q^3 + 170775q^4 + 1000690q^5  \nonumber\\
        &{}\ \ \ \ \ \ \ \ \ \ \ \ + 5228922q^6 + 24959680q^7 + 110204890q^8\nonumber\\ 
        &{}\ \ \ \ \ \ \ \ \ \ \ \ + 455662830q^9 + 1777830395q^{10} + \mathcal{O}(q^{11})).
    \end{align}
    \item $(c,h_1,h_2)=\left(30,\frac{5}{6},\frac{11}{3}\right)$: 
    \begin{align}
        \chi_0 &= q^{-\tfrac{5}{4}}(1 + 670q + 21195q^2 + 261130q^3 + 4298745q^4 + 52456626q^5 \nonumber\\ 
        &{}\ \ \ \ \ \ \ \ \ \ \ \ +486596870q^6 + 3700866400q^7 + 23959194930q^8\\
        &{}\ \ \ \ \ \ \ \ \ \ \ \ + 137023596950q^9 + 706045556129q^{10} + \mathcal{O}(q^{11})), \nonumber\\
        \chi_{\frac{5}{6}} &= \text{unstable}, \nonumber\\
        \chi_{\frac{11}{3}} &= \text{unstable}.
    \end{align}
The above is a single-character solution (not yet a $(1,\ell)$ solution).
\item $(c,h_1,h_2)=\left(30,\frac{37}{38},\frac{67}{19}\right)$: 
    \begin{align}
        \chi_0 &= q^{-\tfrac{5}{4}}(1 + 3740q + 101015q^2 + 503660q^3 + 5299565q^4 + 54774476q^5 \nonumber\\ 
        &{}\ \ \ \ \ \ \ \ \ \ \ \ + 493062290q^6 + 3714558600q^7 + 23990766810q^8\\
        &{}\ \ \ \ \ \ \ \ \ \ \ \ + 137085503500q^9 + 706173390929q^{10} + \mathcal{O}(q^{11})), \nonumber\\
        \chi_{\frac{37}{38}} &= \text{unstable}, \nonumber\\
        \chi_{\frac{67}{19}} &= \text{unstable}.
    \end{align}
The above is a single-character solution (not yet a $(1,\ell)$ solution).
\item $(c,h_1,h_2)=\left(33,\frac{7}{8},4\right)$:
    \begin{align}
        \chi_0 &= \text{unstable}, \\
        \chi_{\frac{7}{8}} &= q^{-\tfrac{1}{2}}(1 + 52q + 834q^2 + 4760q^3 + 24703q^4 + 94980q^5 + 343998q^6 + 1077496q^7\nonumber\\ 
        &{}\ \ \ \ \ \ \ \ \ \ \ \ + 3222915q^8 + 8844712q^9 + 23381058q^{10} + \mathcal{O}(q^{11})), \nonumber\\
        \chi_{4} &= \text{unstable}.
    \end{align}
The single-character theory found in \ref{1char_F2_l2_aliter} makes a reappearance as a non-trivial primary of a higher central charge three-character theory with unstable identity and another non-trivial primary.
\item $(c,h_1,h_2)=\left(36,\frac{1}{4},5\right)$: 
    \begin{align}
        \chi_0 &= \text{Log unstable}, \\
        \chi_{\frac{1}{4}} &= q^{-\tfrac{5}{4}}(5 + 906q + 42431q^2 + 1112574q^3 + 20696981q^4 + 260437910q^5 \nonumber\\ 
        &{}\ \ \ \ \ \ \ \ \ \ \ \ + 2427837286q^6 + 18493431760q^7 + 119770840554q^8\nonumber\\
        &{}\ \ \ \ \ \ \ \ \ \ \ \ + 685068701490q^9 + 3530126012485q^{10} + \mathcal{O}(q^{11})), \nonumber\\
        \chi_{5} &= \text{unstable}.
    \end{align}
    \item $(c,h_1,h_2)=\left(36,\frac{1}{2},\frac{19}{4}\right)$:
    \begin{align}
        \chi_0 &= q^{-\tfrac{3}{2}}(1 + 412q + 23926q^2 + 628600q^3 + 11629865q^4 + 185255140q^5 \nonumber\\ 
        &{}\ \ \ \ \ \ \ \ \ \ \ \ + 2265641766q^6 + 22044942216q^7 + 179314657672q^8\nonumber\\
        &{}\ \ \ \ \ \ \ \ \ \ \ \ + 1260100453460q^9 + 7865573552644q^{10} + \mathcal{O}(q^{11})), \nonumber\\
        \chi_{\frac{1}{2}} &= q^{-1}(13 + 1608 \, q + 56836q^2 + 1251328q^3 + 16120026q^4 + 139083776q^5 \nonumber\\
        &{}\ \ \ \ \ \ \ \ \ \ \ \ \ + 967565560q^6 + 5609889792q^7 + 28687644531q^8\nonumber\\
        &{}\ \ \ \ \ \ \ \ \ \ \ \ \ + 131528523776q^9 + 554020504596q^{10} + \mathcal{O}(q^{11})), \nonumber\\
        \chi_{\frac{19}{4}} &= \text{unstable}.
    \end{align}
The non-trivial primary character, $\chi_{\frac{1}{2}}$, is $13j_{2^{+}}+256$ which can be thought of as twelve copies of $j_{2^{+}}$ single characters and one copy of the single character $j_{2^{+}} + 256$.
    \item $(c,h_1,h_2)=\left(36,\frac{23}{28},\frac{31}{7}\right)$: 
    \begin{align}
        \chi_0 &= q^{-\tfrac{3}{2}}(1 + 940q + 51382q^2 + 1068952q^3 + 14143145q^4 + 198298324q^5 \nonumber\\ 
        &{}\ \ \ \ \ \ \ \ \ \ \ \ + 2315791206q^6 + 22226573160q^7 + 179883575560q^8\nonumber\\
        &{}\ \ \ \ \  \ \ \ \ \ \ \  + 1261802152580q^9+ \mathcal{O}(q^{11})), \nonumber\\
        \chi_{\frac{23}{28}} &= \text{unstable}, \nonumber\\
        \chi_{\frac{31}{7}} &= \text{unstable}.
    \end{align}
The above is a single-character solution (not yet a $(1,\ell)$ solution).
\item $(c,h_1,h_2)=\left(36,\frac{11}{12},\frac{13}{3}\right)$: 
    \begin{align}
        \chi_0 &= q^{-3/2}(1 + 1884q + 100470q^2 + 1856248q^3 + 18636585q^4 + 221617956q^5 \nonumber\\ 
         &{}\ \ \ \ \ \ \ \ \ \ \ \ \ + 2405452326q^6 + 22551307272q^7 + 180900731784q^8\nonumber\\
        &{}\ \ \ \ \ \ \ \ \ \ \ \ \ + 1264844584340q^9 + 7878592968708q^{10} + \mathcal{O}(q^{11})), \nonumber\\
        \chi_{\frac{11}{12}} &= \text{unstable}, \nonumber\\
        \chi_{\frac{13}{3}} &= \text{unstable}.
    \end{align}
The above is a single-character solution (not yet a $(1,\ell)$ solution).
\item $(c,h_1,h_2)=\left(36,\frac{151}{156},\frac{167}{39}\right)$: 
    \begin{align}
        \chi_0 &= q^{-\tfrac{3}{2}}(1 + 4719q + 247890q^2 + 4220638q^3 + 32131185q^4 + 291650961q^5 \nonumber\\ 
        &{}\ \ \ \ \ \ \ \ \ \ \ \ + 2674720626q^6 + 23526541602q^7 + 183955432944q^8\nonumber\\
        &{}\ \ \ \ \ \ \ \ \ \ \ \  + 1273981548365q^9 + 7903667727228q^{10} + \mathcal{O}(q^{11})), \nonumber\\
        \chi_{\frac{151}{156}} &= \text{unstable}, \nonumber\\
        \chi_{\frac{167}{39}} &= \text{unstable}.
    \end{align}
The above is a single-character solution (not yet a $(1,\ell)$ solution).
\item $(c,h_1,h_2)=\left(36,\frac{193}{196},\frac{209}{49}\right)$: 
    \begin{align}
        \chi_0 &= q^{-\tfrac{3}{2}}(1 + 9760q + 510022q^2 + 8424832q^3 + 56126345q^4 + 416178784q^5 \nonumber\\ 
        &{}\ \ \ \ \ \ \ \ \ \ \ \ + 3153514806q^6 + 25260635520q^7 + 189387090280q^8\nonumber\\
        &{}\ \ \ \ \ \ \ \ \ \ \ \ + 1290228262880q^9 + 7948253920420q^{10}+ \mathcal{O}(q^{11})), \nonumber\\
        \chi_{\frac{193}{196}} &= \text{unstable}, \nonumber\\
        \chi_{\frac{209}{49}} &= \text{unstable}
    \end{align}
The above is a single-character solution (not yet a $(1,\ell)$ solution).
\item $(c,h_1,h_2)=\left(36,\frac{235}{236},\frac{251}{59}\right)$: 
    \begin{align}
        \chi_0 &= q^{-\tfrac{3}{2}}(1 + 34966q + 1820734q^2 + 29446636q^3 + 176106905q^4 + 1038842602q^5 \nonumber\\ 
        &{}\ \ \ \ \ \ \ \ \ \ \ \ + 5547580686q^6 + 33931449108q^7 + 216546454456q^8\nonumber\\
        &{}\ \ \ \ \ \ \ \ \ \ \ \ + 1371465058370q^9 + 8171193731092q^{10} + \mathcal{O}(q^{11})), \nonumber\\
        \chi_{\frac{235}{236}} &= \text{unstable}, \nonumber\\
        \chi_{\frac{251}{59}} &= \text{unstable}.
    \end{align}
The above is a single-character solution (not yet a $(1,\ell)$ solution).
\item $(c,h_1,h_2)=\left(39,\frac{5}{8},5\right)$: 
    \begin{align}
        \chi_0 &= \text{Log unstable}, \nonumber\\
        \chi_{\frac{5}{8}} &= q^{-1}(117 + 14216q + 511524q^2 + 11261952q^3 + 145080234q^4 + 1251753984q^5 \nonumber\\ 
        &{}\ \ \ \ \ \ \ \ \ \ \ \ + 8708090040q^6 + 50489008128q^7 + 258188800779q^8\nonumber\\
        &{}\ \ \ \ \ \ \ \ \ \ \ \ + 1183756713984q^9 + 4986184541364q^{10} + \mathcal{O}(q^{11})), \nonumber\\
        \chi_{5} &= \text{unstable}.
    \end{align}
    \item $(c,h_1,h_2)=\left(42,0,6\right)$: \begin{align}
        \chi_0 &= \text{Log unstable}, \nonumber\\
        \chi_{0} &= \text{Log unstable}, \nonumber\\
        \chi_{6} &= q^{\frac{17}{4}}(98 + 3196q + 53669q^2 + 642838q^3 + 6133855q^4 + 49500906q^5 \nonumber\\ 
        &{}\ \ \ \ \ \ \ \ \ \ \ \ + 350187964q^6 + 2224588380q^7 + 12909977905q^8\nonumber\\
        &{}\ \ \ \ \ \ \ \ \ \ \ \ + 69325524646q^9 + 347916765224q^{10} + \mathcal{O}(q^{11})).
    \end{align}
\end{itemize}
\section{\texorpdfstring{$\mathbf{\Gamma_{0}^{+}(3)}$}{Γ0(3)+}}\label{sec:Gamma_0_3+}
\subsection{Single character solutions}
\noindent From \cite{Umasankar:2022kzs}, we find the following form for the character in a $n = 1$ theory
\begin{align}
    \begin{split}
    &\chi(\tau) = j_{3^{+}}^{w_{\rho}}(\tau),\\
    &c = 24w_{\rho} = 8\ell,
    \end{split}
\end{align}
where $w_{\rho}\in\left\{0,\tfrac{4}{3},2\right\}$ corresponding to the three characters at $\ell = 0$, $\ell = 4$, and $\ell = 6$ respectively. Here, we will present the admissible solutions at $\ell$ values all the way to the \textit{movable pole}.

\subsubsection{\texorpdfstring{$\ell = 1$}{l=1}}
\noindent
For $\ell=1$, the RHS of Valence Formula reads $\frac{1}{3}$ (since $k=2\ell$). The MLDE takes the following form,
\begin{align}
    \left[\mathcal{D} + \mu_1\frac{E_4^{(3+)}E_6^{(3+)}}{\left(E_4^{(3+)}\right)^2+\mu\Delta_{3^{+},8}} + \mu_2\frac{\Delta_{3^{+},10}}{(E_4^{(3+)})^2+\mu\Delta_{3^{+},8}}\right]\chi(\tau) = 0. \label{F3_l1_p1}
\end{align}
Now using the indicial equation we can set $\mu_1=\frac{1}{3}$. Now let us take a guess, $\chi=j_{3^{+}}^{\tfrac{1}{3}}$. One can readily check that this is an admissible central charge $c = 8$ solution with $\mu_2=\frac{1}{3}$. The $q$-series expansion of this character reads
\begin{align}\label{9a_character_OG}
    \begin{split}
    \chi = q^{-\tfrac{1}{3}}(1 +&14 q + 65 q^2 + 156 q^3 + 456 q^4 + 1066 q^5 + 2250 q^6 + 4720 q^7\\
    +& 9426 q^8 + 17590 q^9 + \mathcal{O}(q^{10})),
    \end{split}
\end{align}
which turns out to be the McKay-Thompson series of the $9a$ conjugacy class of $\mathbb{M}$ (OEIS sequence A058092 \cite{A058092}).

\subsubsection{\texorpdfstring{$\ell = 2$}{l=2}}
\noindent The MLDE takes the following form,
\begin{align}
    &\left[\mathcal{D}+\mu_1\frac{E_6^{(3+)}}{E_4^{(3+)}}\right]\chi(\tau) = 0, \label{1char_F3_l2} \\
    &\left[E_4^{3+}\mathcal{D}+\mu_1E_6^{3+}\right]\chi(\tau) = 0. \label{1char_F3_l2_aliter}
\end{align}
The indicial equation ($\alpha_0+\mu_1=0$) dictates $\mu_1=\frac{2}{3}$. Consider the ansatz, $\chi=j_{3^{+}}^{\tfrac{2}{3}}$. Then, $\Tilde{\partial}_\tau j_{3^{+}}^{\tfrac{2}{3}} = -\frac{2}{3}j_{3^{+}}^{\tfrac{2}{3}}\frac{E_6^{(3+)}}{E_4^{(3+)}}$. Note, that this $\chi$ indeed satisfies \ref{1char_F3_l2} with $\mu_1=\frac{2}{3}$.

\subsubsection{\texorpdfstring{$\ell = 3$}{l=3}}
\noindent From Riemann-Roch, we notice that this is the \textit{movable pole} MLDE which takes the following form
\begin{align}
    &\left[\mathcal{D}+\mu_1\frac{\left(E_4^{(3+)}\right)^2}{E_6^{(3+)}} + \mu_2\frac{\Delta_{3^{+},8}}{E_6^{(3+)}}\right]\chi(\tau) = 0, \label{1char_F3_l3} \\
    &\left[E_6^{(3+)}\mathcal{D}+\mu_1\left(E_4^{(3+)}\right)^2 + \mu_2\Delta_{3^{+},8}\right]\chi(\tau) = 0. \label{1char_F3_l3_aliter}
\end{align}
Note that in above $\mathcal{D}=\theta_{q}$. Also, the indicial equation ($\alpha_0+\mu_1=0$) dictates $\mu_1=1$.\\
Consider now the ansatz, $\chi=j_{3^{+}}+\mathcal{N}$. Substituting this in ODE \ref{1char_F3_l3_aliter} yields
\begin{align}
    &\left(E_6^{(3+)}\mathcal{D}+\mu_{1}\left(E_4^{(3+)}\right)^2 + \mu_{2}\Delta_{3^{+},8}\right)(j_{3^{+}}+\mathcal{N}) = \Delta_{3^{+},8}\left[(108+b+\mathcal{N})j_{3^{+}}+\mathcal{N}b\right]
\end{align}
So, we see that $j_{3^{+}}+\mathcal{N}$ is a solution to the above differential equation if and only if,
\begin{align}
    &b=0 \, \, \, \text{and} \, \, \, \mathcal{N} = -108 \longrightarrow  \text{soln.:} \, \, j_{3^{+}}-108, \label{sol1} \\
    &b=-108 \, \, \, \text{and} \, \, \, \mathcal{N} = 0 \longrightarrow  \text{soln.:} \, \, j_{3^{+}}, \label{sol2}
\end{align}
Note that, $j_{3^{+}}-108$ has $m_1<0$ and hence is not an admissible solution. So, the only admissible solution is: $j_{3^{+}}$ which happens when $b=-108$ and $\mathcal{N}=0$. Thus, unlike the $\text{SL}(2,\mathbb{Z})$ or the Fricke level $2$ case, here we do not get $j_{3^{+}}+\mathcal{N}$ as a solution to the \textit{movable pole} MLDE. This probably happens because of the following reason. In the \textit{movable pole} MLDE of both $\text{SL}(2,\mathbb{Z})$ and $\Gamma^{+}_0(2)$ which appears at $\ell=6$ and $\ell=4$ respectively, the \textit{movable pole} shows up in the denominator ($D^r = E_4^3+\mu\Delta$ and $D^r=\left(E_4^{(2+)}\right)^2+\mu\Delta_2$ respectively) while in $\Gamma^{+}_0(3)$ the \textit{movable pole} appears in the numerator ($N^r=\left(E_4^{(3+)}\right)^2+\mu\Delta_{3^{+},8}$).

\subsection{Two-character solutions}
\noindent From the dimension of the space of modular forms in \ref{dimension_Fricke_3}, we obtain the following expression for the number of free parameters
\begin{align}\label{number_of_parameters_Gamma_0_3+}
    \begin{split}
    \#(\mu) =  \begin{cases}
    \left\lfloor\left.\frac{\ell}{3}\right\rfloor\right. + \left\lfloor\left.\frac{\ell + 1}{3}\right\rfloor\right. + \left\lfloor\left.\frac{\ell + 2}{3}\right\rfloor\right. + 2,\ &2\ell\not\equiv 2,6\ (\text{mod}\ 12),\ 2\ell>2\\
    \left\lfloor\left.\frac{\ell}{3}\right\rfloor\right. + \left\lfloor\left.\frac{\ell + 1}{3}\right\rfloor\right. + \left\lfloor\left.\frac{\ell + 2}{3}\right\rfloor\right. -1,\ &2\ell\equiv 2, 6\ (\text{mod}\ 12),\ 2\ell>2.
    \end{cases}
    \end{split}
\end{align}
On the other hand, for the case of $k \equiv 2,6\ (\text{mod}\ 12), k\in2\mathbb{Z}$, we are to use the following expression
\begin{align}\label{number_of_parameters_Gamma_0_3+_2,6mod12}
    \#(\mu) = \left\lfloor\left.\frac{\ell}{3}\right\rfloor\right. + \left\lfloor\left.\frac{\ell + 1}{3}\right\rfloor\right. + \left\lfloor\left.\frac{\ell + 2}{3}\right\rfloor\right. -1,\  2\ell\equiv 2, 6\ (\text{mod}\ 12).
\end{align}

\subsubsection{\texorpdfstring{$\ell = 0$}{l = 0}}
\noindent The space $\mathcal{M}_{2}(\Gamma_{0}^{+}(3))$ is zero-dimensional and there are no modular forms of weight $2$ in $\Gamma_{0}^{+}(3)$. Hence, we set $\phi_{1}(\tau) = 0$. The space $\mathcal{M}_{4}(\Gamma_{0}^{+}(2))$ is one-dimensional and hence, we set $\phi_{0}(\tau) = \mu E_{4}^{(3^{+})}(\tau)$. For these choices, the second-order MLDE takes the following form
\begin{align}\label{MLDE_n=2_Gamma_0_3+}
    \left[\mathcal{D}^{2} + \mu_{1} E_{4}^{(3^{+})}(\tau)\right]\chi(\tau) = 0,
\end{align}
where $\mu_{1}$ is an independent parameter. Now, since the covariant derivative transforms a weight $r$ modular form into one of weight $r+2$, the double covariant derivative of a weight $0$ form is
\begin{align}
    \begin{split}
        \mathcal{D}^{2} = \mathcal{D}_{(2)}\mathcal{D}_{(0)} =& \left(\frac{1}{2\pi i}\frac{d}{d\tau} - \frac{1}{3}E^{(3^{+})}_{2}(\tau)\right)\frac{1}{2\pi i}\frac{d}{d\tau}\\
        =& \Tilde{\partial}^{2} - \frac{1}{3}E^{(2^{+})}_{2}(\tau)\Tilde{\partial}.
    \end{split}
\end{align}
The MLDE \ref{MLDE_n=2_Gamma_0_3+} now reads
\begin{align}\label{MLDE_n=2_Gamma_0_3+_expanded}
    \left[\Tilde{\partial}^{2} - \frac{1}{3}E^{(3^{+})}_{2}(\tau)\Tilde{\partial} + \mu_{1} E_{4}^{(3^{+})}(\tau)\right]\chi(\tau) = 0.
\end{align}
This equation can be solved by making the following mode expansion substitution for the character $\chi(\tau)$ and the other modular forms,
\begin{align}\label{series_defs_Gamma_0_3+}
        \chi(\tau) =& q^{\alpha}\sum\limits_{n= 0}^{\infty}\chi_{n}q^{n},\\
        E_{2}^{(3^{+})}(\tau) =& \sum\limits_{n=0}^{\infty}E_{2,n}^{(3^{+})}q^{n}\nonumber\\
        =& 1 - 6 q - 18 q^2 - 42 q^3 - 42 q^4 + \ldots,\nonumber\\
        E_{4}^{(2^{+})}(\tau) =& \sum\limits_{n=0}^{\infty}E_{4,n}^{(3^{+})}q^{n}\\
        =& 1 + 24 q + 216 q^2 + 888 q^3 + 1752 q^4 + \ldots
\end{align}
Substituting these expansions in the MLDE, we obtain
Substituting these expansions in the MLDE, we obtain
\begin{align}\label{MLDE_mode_Gamma_0_3+} 
    q^{\alpha}\sum\limits_{n=0}^{\infty}\chi_{n}q^{n}\left[(n+\alpha)^{2} - (n + \alpha)\sum\limits_{m=0}^{\infty}\frac{1}{3}E^{(3^{+})}_{2,m}q^{m} + \mu\sum\limits_{m=0}^{\infty} E_{4,m}^{(3^{+})}q^{m}\right] = 0.
\end{align}
When $n=0,m=0$, with $E^{(3^{+})}_{2,0} = E^{(3^{+})}_{4,0} = 1$, we obtain the following indicial equation
\begin{align}
    \alpha^{2} -\frac{1}{3} \alpha +  \mu = 0.
\end{align}
Solving this indicial equation, we have
\begin{align}
    \begin{split}
        \alpha_{0} =& \frac{1}{6}\left(1 - \sqrt{1 - 36\mu}\right) \equiv \frac{1}{6}(1 - x),\\
        \alpha_{1} =& \frac{1}{6}\left(1 + \sqrt{1 - 36\mu}\right) \equiv \frac{1}{6}(1 + x),
    \end{split}
\end{align}
where we have set $x = \sqrt{1 - 36\mu}$. The smaller solution $\alpha_{0} = \tfrac{1}{5}(1-x)$ corresponds to the identity character which behaves as $\chi(\tau) \sim q^{\tfrac{1-x}{6}}\left(1 + \mathcal{O}(q)\right)$. We know that the identity character $\chi_{0}$, associated with a primary of weight $h_{0} = 0$ behaves as $\chi_{0}(\tau)\sim q^{-\tfrac{c}{24}}\left(1 + \mathcal{O}(q)\right)$. Comparing the two behaviours, we obtain the following expression for the central charge
\begin{align}\label{central_charge_x_3+}
    c = 4(x-1).
\end{align}
To find the conformal dimension $h$, we compare the behaviours with the larger solution for $\alpha$, i.e. $\chi(\tau)\sim q^{\tfrac{1 + x}{6}}\left(1  +\mathcal{O}(q)\right)$ and $\chi(\tau)\sim q^{-\tfrac{c}{24} + h}\left(1 + \mathcal{O}(q)\right)$. This gives us
\begin{align}
    h = \frac{x}{3}.
\end{align}
Using the Cauchy product formulae, we obtain the following recurrence relation
\begin{align}\label{recursion_l=0_Gamma_0_3+}
    \chi_{n} = \left((n + \alpha)^{2} - \frac{1}{3}(n + \alpha) + \mu\right)^{-1}\sum\limits_{m=1}^{n}\left(\frac{(n + \alpha - m)}{3}E^{(3^{+})}_{2,m} - \mu E^{(3^{+})}_{4,m}\right)\chi_{n-m}.
\end{align}
When $n = 1$, we solve for the ratio $\tfrac{\chi_{1}}{\chi_{0}}$ in terms of $\alpha$ with coefficients $E^{(3^{+})}_{2,1} = -6$ and $E_{4,1}^{(3^{+})} = 24$ to obtain 
\begin{align}
    m_{1}^{(i)} + 14\alpha_{0} + 3m_{1}\alpha_{0} - 36\alpha_{0}^{2} = 0.
\end{align}
for $ i = 0,1$ corresponding to ratios of $m^{(0)}_{1}$ and $m^{(1)}_{1}$ taken with respect to values $\alpha_{0}$ and $\alpha_{1}$ respectively. Restricting to $i=0$ and assuming that $\alpha_{0} = -\tfrac{c}{24}$, we can recast this equation by identifying $N=-12\alpha_0=\tfrac{c}{2}$,
\begin{align}
    -4m_1 + 5N + m_1N + N^2 = 0, \label{N_eqn_3}
\end{align}
$a_1$ can now be expressed in terms of $N$ as,
\begin{align}
    m_1 = \frac{N(N+5)}{4-N}, \label{a1_eqn_3}
\end{align}
which sets an upper bound on the central charge $c<8$. Furthermore, applying the integer root theorem to \ref{N_eqn_3}, we see that $c=2N$ and the central charge, in this case, is restricted to be an even integer, i.e.
\begin{align}
    m_{1} = \frac{c(10+c)}{8-c}.
\end{align}
Dropping the subscript associated with the index $i$, we see that for $m_{1}\geq 0$, we require $c<8$. Rewriting this, we have
\begin{align}
    c^{2} + c(m_{1} + 10) = 8m_{1}.
\end{align}
This tells us that for $(n,\ell) = (2,0)$ in $\Gamma_{0}^{+}(3)$, we have $c\in\mathbb{Z}$ and $c<8$. For $c$ to be rational, we demand the determinant, $\sqrt{25 + 26m_{1} + m_{1}^{2}}$ , to be rational. We further demand that the square root is an integer and this gives us
\begin{align}\label{m_1_3+}
    (m_{1} + 13)^{2} - 144 = k^{2},
\end{align}
where $k^{2}\in\mathbb{Z}$. We now set $p$ to be
\begin{align}
    p = 13 + m_{1} - k,
\end{align}
and we recast \ref{m_1_3+} as follows
\begin{align}
    m_{1} + 13 = \frac{144 + p^{2}}{2p} = \frac{72}{p} + \frac{p}{2}.
\end{align}
Restricting $k$ to be positive, we see that $p<13$. We conclude that all possible values of $m_{1}$ are found by those values of $p$ below $13$ that divide $72$ and are even. The list of these values reads $p = \{2,4,6,8,12\}$. We ignore the case corresponding to $p = 12$ since $m_{1} = -1$. Table \ref{tab:theory_Fricke_3+} contains CFT data.
\begin{table}[htb!]
    \centering
    \begin{tabular}{||c|c|c|c|c||}
    \hline
    $p$ & 2 & 4 & 6 & 8\\ [0.5ex]
    \hline\hline
    $\mu_{1}$  & $-\tfrac{7}{48}$ & $-\tfrac{1}{12}$ & $-\tfrac{5}{144}$ & $0$\\[0.5ex]
    \hline\hline
    $m_{1}$ & 24 & 7 & 2 & 0\\[0.5ex]
    \hline\hline
    $c$ & 6 & 4 & 2 & 0\\[0.5ex]
    \hline\hline
    $h$ & $\tfrac{5}{6}$ & $\tfrac{2}{3}$ & $\tfrac{1}{2}$ & $\tfrac{1}{3}$\\[1ex]
    \hline
    \end{tabular}
    \caption{$c$ and $\Delta$ data corresponding to the Fricke group $\Gamma_{0}^{+}(3)$ for the choice $\phi_{0} = \mu_{1} E^{(3^{+})}_{4}(\tau)$ with $\ell = 0$.}
    \label{tab:theory_Fricke_3+}
\end{table}
\noindent
Using recursion relation \ref{MLDE_mode_Gamma_0_3+}, when $n = 2$, we obtain
\begin{align}
     m_{2}^{(i)} \equiv \frac{\chi_{2}^{(i)}}{\chi_{0}^{(i)}} = \frac{9\alpha_{i}(36\alpha_{i} -13)}{6\alpha_{i} + 5} + \frac{3m_{1}^{(i)}(-1 + \alpha_{i}(-5 + 12\alpha_{i}))}{6\alpha_{i} + 5}.
\end{align}We discard the case $p = 8$ since further computations show that all coefficients $m_{i}$ are null valued thus making this a trivial solution. The values of $m_{2}$ fr $i = 0$ is tabulated in table \ref{tab:theory_Fricke_3+_m2}.
\begin{table}[htb!]
    \centering
    \begin{tabular}{||c|c|c|c||}
    \hline
    $p$ & 2 & 4 & 6\\[0.5ex]
    \hline\hline
    $c$ & 6 & 4 & 2\\[0.5ex]
    \hline\hline
    $m_{2}$ & $\tfrac{243}{7}$& 8 & 2\\[1ex]
    \hline
    \end{tabular}
    \caption{Values of $m_{2}$ for corresponding to the Fricke group $\Gamma_{0}^{+}(3)$ for the choice $\phi_{0} = \mu_{1} E^{(3^{+})}_{4}(\tau)$ with $\ell = 0$.}
    \label{tab:theory_Fricke_3+_m2}
\end{table}
\noindent
In a similar fashion, computing coefficients for $i = 0,1$, we obtain the coefficients of the two characters. Checking to higher-order (up to $\mathcal{O}(q^{5000})$) for both characters, for $0<c<8$ (even $c$), we find the admissible character shown below.
\begin{itemize}
    \item $(c,h)=\left(2,\frac{1}{2}\right)$: 
    \begin{align}\label{c=2_Fricke_3}
        \begin{split}
        \chi_0 =& q^{-\tfrac{1}{12}}(1 + 2q + 2q^2 + 4q^3 + 5q^4 + 6q^5 + 11q^6 + 14q^7 + 17q^8 + 24q^9 + 31q^{10} + \mathcal{O}(q^{11})),\\
        \chi_{\frac{1}{2}} =& q^{\tfrac{5}{12}}(1 + q^2 + 2q^3 + 2q^4 + 2q^5 + 5q^6 + 4q^7 + 7q^8 + 10q^9 + 11q^{10} + \mathcal{O}(q^{11})).
        \end{split}
    \end{align}
\end{itemize}
\noindent We also find an \textit{identity only} admissible solution, which only has a nice identity character but an unstable non-trivial character. This is found at $(c,h)=\left(4,\frac{2}{3}\right)$.
\begin{align}\label{18b_character_1}
   \begin{split}
    \chi_0 = q^{-\tfrac{1}{6}}(1 +& 7q + 8q^2 + 22q^3 + 42q^4 + 63q^5 + 106q^6 + 190q^7 + 267q^8\\
    +& 428q^9 + 652q^{10} + \mathcal{O}(q^{11}))\\
    =& j^{\tfrac{1}{6}}_{3^{+}}.
   \end{split}
\end{align}
At central charge $c=6$, both characters are unstable. We here note that the identity character in \ref{c=2_Fricke_3} can be expressed in terms of (semi-)modular forms, which we shall denote by $\mathcal{S}^{0}_{k}(\Gamma)$, of the Hecke group $\Gamma_{0}(3)$ as follows
\begin{align}
    \begin{split}
        \chi_{0}^{(c=2)} = \left(\frac{\Delta_{3}^{0}(2\tau)}{\left(\Delta_{3}^{0}\right)^{\tfrac{1}{2}}(\tau)}\right)^{\tfrac{1}{3}}\left(\frac{1}{\Delta_{3}(\tau)}\right)^{\tfrac{1}{12}},\ \ \ \Delta_{3}^{0}(\tau) = \frac{\eta^{9}(\tau)}{\eta^{3}(3\tau)}\in\mathcal{S}^{0}_{3}(\Gamma_{0}(3)),
    \end{split}
\end{align}
where $\Delta_{3}^{0}(\tau)$ is the $2^{\text{nd}}$ semi-modular form of weight $2$ in $\Gamma_{0}(3)$ \cite{Junichi}.

%\subsubsection{\texorpdfstring{$\ell=1$}{l=1}}
%The $\ell=1$ MLDE takes the following form,
%\begin{align}\label{mlde_F3_p2_l1}
%    \begin{split}
        %&\left[\left(\left(E_4^{(3+)}\right)^2+\mu\Delta_{3^{+},8}\right)\mathcal{D}^2 + \left(\mu_1E_4^{(3+)}E_6^{(3+)}+\mu_2\Delta_{3^{+},10}\right)\mathcal{D}\right.\\
        %&{}\ \ \ \ \ \ \ \ \ \ \ \ \ \ \ \ \ \ \ \ \ \ \ \ \ \ \ \ \ \ \ \ \ \ \ \left.+ \mu_3 (E_4^{(3+)})^3 + \mu_4\frac{(\Delta_{3^{+},8})^2}{E_4^{(3+)}} + \mu_5\Delta_{3^{+},8}E_4^{(3+)}\right]\chi(\tau) = 0.
%    \end{split} 
%\end{align}

\subsubsection{\texorpdfstring{$\ell=2$}{l=2}}
\noindent The (2,2) MLDE takes the following form
\begin{align}
    \left[\mathcal{D}^2 + \mu_1\frac{E_6^{(3+)}}{E_4^{(3+)}}\mathcal{D} + \mu_2E_4^{(3+)} + \mu_3\frac{\Delta_{3^{+},8}}{E_4^{(3+)}}\right]\chi(\tau) = 0, \label{mlde22_3}
\end{align}
where $\chi$ is a weight 0 modular function for $\Gamma^{+}_0(3)$. Substituting for the Fourier expansions of the relevant terms in \ref{mlde22_3}, we obtain the following recursion relation
\begin{align}
    &\sum\limits_{k=0}^i\left[E_{4,k}^{(3+)}(\alpha^j+i-k)(\alpha^j+i-k-1) + E_{4,k}^{(3+)}(\alpha^j+i-k)-\frac{E_{2,k}^{(3+)}E_{4,k}^{(3+)}}{3}(\alpha^j+i-k)\right. \nonumber\\ 
    &\left.\mu_1 E_{6,k}^{(3+)} (\alpha^j+i-k) + \mu_2(E_4^{(3+)})^2_{,k} + \mu_3\Delta_{3^{+},8,k}\right]a^j_{i-k} = 0, \label{recursion_l2F3_2char}
\end{align}
where $a_i$s are Fourier coefficients of the character $q$-series expansion. Setting $i=0$ and $k=0$ in \ref{recursion_l2F3_2char}, we obtain the following indicial equation
\begin{align}
    \alpha^2 + \left(\mu_1-\frac{1}{3}\right)\alpha + \mu_2 = 0. \label{indicial_l2F3_2char}
\end{align}
Setting $\alpha=\alpha_0$ in \ref{indicial_l2F3_2char} and observing Riemann-Roch we get,
\begin{align}
    &\mu_1 = \frac{2}{3}, \label{param2_l2F3}\\
    &\mu_2 = -\left(\alpha_0^2-\frac{\alpha_0}{3}\right).
    \label{param3_l2F3}
\end{align}
Tthe value of $\mu_3$ can now be determined by transforming the $\tau$-MLDE in \ref{mlde22_3} into the corresponding $j_{3^{+}}$-MLDE (about $\tau=\rho_3$),
\begin{align}
    \left[(j_{3^{+}})^2\partial_{j_{3^{+}}}^2 + \frac{j_{3^{+}}}{6}\partial_{j_{3^{+}}} +  \frac{(j_{3^{+}})^2}{2(j_{3^{+}}-108)}\partial_{j_{3^{+}}} + \frac{\mu_2 j_{3^{+}}}{j_{3^{+}}-108} + \frac{\mu_3}{j_{3^{+}}-108}\right]\chi = 0, \label{j3+mlde_p2_l2}
\end{align}
where we have set $\mu_1=\frac{2}{3}$ from the $\ell=2$ indicial equation (see \ref{indicial_l2F3_2char}). Now, note that about $\tau=\rho_3$, we have, $\chi_0\sim j_{3^{+}}^x$ and $j_{3^{+}}\to 0$, with $x=\frac{2}{3}$ for $\ell=2$ (see single-character solutions). Hence, substituting $\chi=j_{3^{+}}^x$ about $j_{3^{+}}=0$, the $j_{3^{+}}$-indicial equation (which is basically, coefficient of $j_{3^{+}}^x$ $=\left.0\right|_{j_{3^{+}}=0}$) reads,
\begin{align}
    x^2-\frac{5}{6}x-\frac{\mu_3}{108} = 0, \label{jindip2l2F3}
\end{align}
which yields $\mu_2 = -12$ for $x=\frac{2}{3}$. Similar analysis gives, $(x,\mu_2)=(\frac{1}{3},-18)$ as solution to \ref{jindip2l2F3}. For $\mu_3=-12$, we get from the $i=1$ equation: $m_1=9-N$ where $N=-12\alpha_0$. This gives the allowed range of central charge as, $0<c<10$ with $c\in 2\mathbb{Z}$. Scanning this space of $c$ we get the following admissible solutions.
\begin{itemize}
    \item $(c,h)=\left(4,0\right)$: 
    \begin{align}\label{18b_character_2}
       \begin{split}
            \chi_0 =& q^{-\tfrac{1}{6}}(1 + 7q + 8q^2 + 22q^3 + 42q^4 + 63q^5 + 106q^6 + 190q^7\\
            &{}\ \ \ \ \ \ \ \ + 267q^8 + 428q^9 + 652q^{10} + \mathcal{O}(q^{11}))\\
            =& j_{3^{+}}^{\tfrac{1}{6}},\\
        \chi_{\text{non-id}} =& \chi_0
       \end{split}
    \end{align}
    \item $(c,h)=\left(6,\frac{1}{6}\right)$: 
    \begin{align}
        \begin{split}
        \chi_0 &= q^{-\tfrac{1}{4}}(1 + 6q + 8q^3 + 18q^4 + 17q^6 + 54q^7 + \mathcal{O}(q^{9})),\\
        \chi_{\frac{1}{6}} &= q^{-\tfrac{1}{12}}(1 + 8q + 17q^2 + 46q^3 + 98q^4 + 198q^5 + 371q^6 + 692q^7 + 1205q^8 + \mathcal{O}(q^{9})).
        \end{split}
    \end{align}
\end{itemize}
\noindent For $\mu_3=-18$, we get from the $i=1$ equation: $m_1=\frac{N^2-17N+108}{8-N}$ where $N=-12\alpha_0$. This gives the allowed range of central charge as, $0<c<8$ with $c\in 2\mathbb{Z}$. Scanning this space of $c$ we get the following admissible solutions.
\begin{itemize}
    \item $(c,h)=\left(4,0\right)$: 
    \begin{align}
        \begin{split}
          \chi_0 =& q^{-\tfrac{1}{6}}(1 + 13q + 50q^2 + 76q^3 + 222q^4 + 405q^5 + 664q^6 + 1222q^7\\
          &{}\ \ \ \ \ \ \ \ + 2121q^8 + 3146q^9 + + \mathcal{O}(q^{10})),\\
        \chi_{\text{non-id}} =& \chi_0  
        \end{split}
    \end{align}
    \item $(c,h)=\left(8,\frac{1}{3}\right)$: 
    \begin{align}\label{9a_character_1}
        \begin{split}
        \chi_0 =& q^{-\tfrac{1}{3}}(1 + 14q + 65q^2 + 156q^3 + 456q^4 + 1066q^5 + 2250q^6 + 4720q^7\\
        &{}\ \ \ \ \ \ \ \ \ + 9426q^8 + 17590q^9 + \mathcal{O}(q^{10})),\\
        =& j_{3^{+}}^{\tfrac{1}{3}},\\
        \chi_{\frac{1}{3}} =& \text{unstable}.
        \end{split}
    \end{align}
    \item $(c,h)=\left(10,\frac{1}{2}\right)$: 
    \begin{align}
        &\chi_0 = j_{3^{+}}^{\tfrac{1}{3}}\otimes \chi_0^{{}^{(3)}\mathcal{W}_2} \nonumber\\
        &\chi_{\frac{1}{2}} = j_{3^{+}}^{\tfrac{1}{3}}\otimes \chi_{\frac{1}{2}}^{{}^{(3)}\mathcal{W}_2}
    \end{align}
    \item $(c,h)=\left(12,\frac{2}{3}\right)$: 
    \begin{align}\label{6b_character_1}
       \begin{split}
        \chi_0 =& q^{-\tfrac{1}{2}}(1 + 21q + 171q^2 + 745q^3 + 2418q^4 + 7587q^5 + 20510q^6 + 51351q^7 + \mathcal{O}(q^{8}))\\
        =& j_{3^{+}}^{\tfrac{1}{2}},\\
        \chi_{\frac{2}{3}} =& \text{unstable}
       \end{split}
    \end{align}
\end{itemize}
We note that the identity character in \ref{18b_character_2}, and also \ref{18b_character_1}, turns out to be the McKay-Thompson series of the $18b$ conjugacy class of $\mathbb{M}$ (OEIS sequence A058537 \cite{A058537}), the identity character in \ref{9a_character_1} is the single character solution we found in \ref{9a_character_OG} which is nothing but the McKay-Thompson series of the $9a$ conjugacy class of $\mathbb{M}$ (OEIS sequence A058092 \cite{A058092}), and the identity character in \ref{6b_character_1} turns out to be the McKay-Thompson series of the $6b$ conjugacy class of $\mathbb{M}$ (OEIS sequence A007261 \cite{A007261}).
\subsection{Three-character solutions}
The $(3,0)$ MLDE reads,
\begin{align}
    \left[\mathcal{D}^3 + \mu_1 E_4^{(3^{+})}\mathcal{D} + \mu_2 E_6^{(3^{+})}\right]\chi(\tau) = 0. \label{30charMLDE_F3}
\end{align}
\noindent The recursion relation corresponding to the above $(3,0)$ MLDE reads
\begin{align}
    &\left[(\alpha^j+i)(\alpha^j+i-1)(\alpha^j+i-2) + 3(\alpha^j+i)(\alpha^j+i-1) + (\alpha^j+i)\right]a^j_i \nonumber\\ 
    &+ \sum\limits_{k=0}^i\left[-E_{2,k}^{(3+)}a^j_{i-k}(\alpha^j+i-k)(\alpha^j+i-k-1) - E_{2,k}^{3+}a^j_{i-k}(\alpha^j+i-k)\right. \nonumber\\
    &\left.+ \frac{1}{18}E_{4,k}^{(3+)}a^j_{i-k}(\alpha^j+i-k) + \frac{1}{6}(E_{2}^{(3+)})^2_{,k}a^j_{i-k}(\alpha^j+i-k) + \mu_1E_{4,k}^{(3+)}a^j_{i-k}(\alpha^j+i-k) + \mu_2E_{6,k}^{(3+)}a^j_{i-k}\right] = 0. \label{recursion3_3}
\end{align}
From which the indicial equation reads,
\begin{align}
    \alpha^3 - \alpha^2 + \left(\mu_1+\frac{2}{9}\right)\alpha + \mu_2 = 0. \label{indicial3_3}
\end{align}
Also,
\begin{align}
&4 m_1 m_2 + 108 m_1 \alpha_0 - 18 m_1^2 \alpha_0 + 180 m_2 \alpha_0 + 6 m_1 m_2 \alpha_0 + 3348 \alpha_0^2 - 594 m_1 \alpha_0^2 \nonumber\\
&- 126 m_1^2 \alpha_0^2 + 360 m_2 \alpha_0^2 - 14472 \alpha_0^3 - 2268 m_1 \alpha_0^3 - 108 m_1^2 \alpha_0^3 + 216 m_2 \alpha_0^3 + 10368 \alpha_0^4 = 0. \label{al1}
\end{align}
The Diophantine equation reads, (which is obtained by multiplying $16^3 \times 2$ with \ref{al1} and then identifying $N=-96\alpha_0$),
\begin{align}
&32768 m_1 m_2 - 9216 m_1 N + 1536 m_1^2 N - 15360 m_2 N \nonumber\\
&- 512 m_1 m_2 N + 2976 N^2 - 528 m_1 N^2 - 112 m_1^2 N^2 + 320 m_2 N^2 + 134 N^3 \nonumber\\
&+ 21 m_1 N^3 + m_1^2 N^3 - 2 m_2 N^3 + N^4 = 0. \label{Diop1_3_3}
\end{align}
By integer root theorem $N=-96\alpha_0$ above should be an integer and hence $N=4c\in\mathbb{Z}$. Next, we have that the discriminant of the cubic indicial equation should be a perfect square, say $k^2$ with $k\in\mathbb{Z}$. Thus we have,
\begin{align}
    &262144 m_1^2 - 163840 m_1 N - 16384 m_1^2 N + 21504 N^2 \nonumber\\
    &+ 8192 m_1 N^2 + 256 m_1^2 N^2 - 448 N^3 - 96 m_1 N^3 - 7 N^4 = k^2. \label{Diop2_3_3}
\end{align}
The three roots of the indicial equation read,
\begin{align}
    &\alpha_0 = -\frac{N}{96}, \label{root1_30_F3} \\
    &\alpha_1 = \frac{1536 m_1 - 672 N + 16 m_1 N - 7 N^2 - 3k}{192(16 m_1 - 7 N)}, \label{roo2_30_F3} \\
    &\alpha_2 = \frac{1536 m_1 - 672 N + 16 m_1 N - 7 N^2 + 3k}{192(16 m_1 - 7 N)}. \label{roo3_30_F3}
\end{align}
Now solving \ref{Diop1_3_3} and \ref{Diop2_3_3}, for the range $1\leq N\leq 96$, yields,
\begin{itemize}
    \item $(c,h_1,h_2)=\left(1,\frac{7}{8},\frac{1}{4}\right)$: 
    \begin{align}
        \chi_0 &= q^{-\tfrac{1}{24}}(1 + q^2 + q^3 + q^4 + q^5 + 2q^6 + q^7 + 2q^8 + 3q^9 + 3q^{10} + \mathcal{O}(q^{11})), \nonumber\\
        \chi_{\frac{7}{8}} &= \text{unstable}, \nonumber\\
        \chi_{\frac{1}{4}} &= q^{\tfrac{5}{24}}(1 + q + q^3 + q^4 + q^5 + 2q^6 + 2q^7 + 2q^8 + 3q^9 + 3q^{10} + \mathcal{O}(q^{11})).
    \end{align}
The $q$-series expansions of the identity and the second non-trivial primary of this theory can be expressed in terms of ratios of Ramanujan theta series following the sequences in \cite{A097242} (OEIS sequence A097242) and \cite{A328796} (OEIS sequence A328796) respectively. 
    \item $(c,h_1,h_2)=\left(2,\frac{3}{4},\frac{1}{2}\right)$: 
    \begin{align}
        \chi_0 &= q^{-\tfrac{1}{12}}(1 + 2q + 2q^2 + 4q^3 + 5q^4 + 6q^5 + 11q^6 + 14q^7 + 17q^8 + 24q^9 + 31q^{10} + \mathcal{O}(q^{11})), \nonumber\\
        \chi_{\frac{3}{4}} &= \text{unstable}, \nonumber\\
        \chi_{\frac{1}{2}} &= q^{\tfrac{5}{12}}(1 + q^2 + 2q^3 + 2q^4 + 2q^5 + 5q^6 + 4q^7 + 7q^8 + 10q^9 + 11q^{10} + \mathcal{O}(q^{11})).
    \end{align}
This is a $(2,0)$ solution found in \ref{c=2_Fricke_3} that makes a reappearance here.
    \item $(c,h_1,h_2)=\left(4,1,\frac{1}{2}\right)$:
    \begin{align}
        \chi_0 &= q^{-\tfrac{1}{6}}(1 + m_1q + m_2q^2 + \mathcal{O}(q^{3})), \nonumber\\
        \chi_{1} &= q^{\tfrac{5}{6}}(1 + 2q^2 + 4q^3 + 5q^4 + 8q^5 + 18q^6 + 20q^7\nonumber\\
        &{}\ \ \ \ \ \ \ \ \ \  + 36q^8 + 56q^9 + 76q^{10} + \mathcal{O}(q^{11})), \nonumber\\
        \chi_{\frac{1}{2}} &= q^{\tfrac{1}{3}}(1 + 2q + 3q^2 + 8q^3 + 13q^4 + 20q^5 + 37q^6 + 56q^7\nonumber\\
        &{}\ \ \ \ \ \ \ \ \ \ + 83q^8 + 134q^9 + 196q^{10} + \mathcal{O}(q^{11})),
    \end{align}
    where for the above identity character we have, $m_1\in\mathbb{N}\cup \{0\}$, $m_2=8$ and $k=1792-256 m_1$ for $m_1< 7$ and $k=-1792 + 256 m_1$ for $m_1\geq 8$.
    \item $(c,h_1,h_2)=\left(6,\frac{5}{4},\frac{1}{2}\right)$:
    \begin{align}
        \chi_0 &= q^{-\tfrac{1}{4}}(1 + 9q + 24q^2 + 56q^3 + 135q^4 + 264q^5 + 497q^6 + 945q^7\nonumber\\
        &{}\ \ \ \ \ \ \ \ \ \ \  \ + 1656q^8 + 2830q^9 + 4815q^{10} + \mathcal{O}(q^{11})), \nonumber\\
        \chi_{\frac{5}{4}} &= \text{unstable}, \nonumber\\
        \chi_{\frac{1}{2}} &= q^{\tfrac{1}{4}}(1 + 7q + 9q^2 + 31q^3 + 66q^4 + 117q^5 + 227q^6 + 436q^7 \nonumber\\
        &{}\ \ \ \ \ \ \ \ \ \ + 702q^8 + 1241q^9 + 2072q^{10} + \mathcal{O}(q^{11})).
    \end{align}
This is a new two-character solution that we did not find in our analysis with $(2,\ell)$ MLDEs.
    \item $(c,h_1,h_2)=\left(7,\frac{9}{8},\frac{3}{4}\right)$:
    \begin{align}
        \chi_0 &= q^{-\tfrac{7}{24}}(1 + 7q + 35q^2 + 70q^3 + 189q^4 + 420q^5 + 833q^6 + 1631q^7\nonumber\\
        &{}\ \ \ \ \ \ \ \ \ \ \ \ \ + 3143q^8 + 5530q^9 + 9863q^{10} + \mathcal{O}(q^{11})), \nonumber\\
        \chi_{\frac{9}{8}} &= \text{unstable}, \nonumber\\
        \chi_{\frac{3}{4}} &= q^{\tfrac{11}{20}}(7 + 22q + 56q^2 + 161q^3 + 343q^4 + 686q^5 + 1421q^6 + 2653q^7\nonumber\\
        &{}\ \ \ \ \ \ \ \ \ \ \ + 4782q^8 + 8638q^9 + 14847q^{10} + \mathcal{O}(q^{11})).
    \end{align}
This is a two-character solution (not yet a $(1,\ell)$ solution).
    \item $(c,h_1,h_2)=\left(7,\frac{7}{4},\frac{1}{8}\right)$:
    \begin{align}
        \chi_0 &= \text{unstable}, \nonumber\\
        \chi_{\frac{7}{4}} &= \text{unstable}, \nonumber\\
        \chi_{\frac{1}{8}} &= q^{-\tfrac{1}{6}}(1 + 7q + 8q^2 + 22q^3 + 42q^4 + 63q^5 + 106q^6 + 190q^7 \nonumber\\
        &{}\ \ \ \ \ \ \ \ \ \ \ \ + 267q^8 + 428q^9 + 652q^{10} + \mathcal{O}(q^{11})).
    \end{align}
The stable non-trivial primary with $h = \tfrac{1}{8}$ is a $c = 4$ single-character theory found in \ref{18b_character_1} that also reappears as a two-character theory with an unstable non-trivial primary in \ref{18b_character_2}. 
    \item $(c,h_1,h_2)=\left(8,1,1\right)$: \begin{align}
        \chi_0 &= q^{-\tfrac{1}{3}}(1 + m_1q + m_2q^2 + \mathcal{O}(q^{3})), \nonumber\\, \nonumber\\
        \chi_{1} &= q^{-\tfrac{2}{3}}(1 + 2q + 8q^2 + 22q^3 + 47q^4 + 102q^5 + 224q^6 + 422q^7 \nonumber\\
        &{}\ \ \ \ \ \ \ \ \ \ \ \ + 815q^8 + 1516q^9 + 2688q^{10} + \mathcal{O}(q^{11})), \nonumber\\
        \chi_{1} &= q^{-\tfrac{2}{3}}(1 + 2q + 8q^2 + 22q^3 + 47q^4 + 102q^5 + 224q^6 + 422q^7 \nonumber\\
        &{}\ \ \ \ \ \ \ \ \ \ \ \ + 815q^8 + 1516q^9 + 2688q^{10} + \mathcal{O}(q^{11})).
    \end{align}
    where for the above identity character we have, $m_1\in\mathbb{N}\cup \{0\}$, $m_2=37+2 m_1$ and $k=0$. Also, note that in the above case the two non-trivial primaries are equal. This is exactly similar to how $D_{4,1}$ (and its GHM dual (\cite{Gaberdiel:2016zke}) with central charge 20) appear as three-character solutions to the $(3,0)$ MLDE in the $\text{SL}(2,\mathbb{Z})$ case (see \cite{Das:2021uvd}).
    \item $(c,h_1,h_2)=\left(16,0,3\right)$: 
    \begin{align}
        \chi_0 &= \text{Log unstable}, \\
        \chi_{0} &= \text{Log unstable}, \nonumber\\
        \chi_{3} &= q^{\tfrac{7}{3}}(1 + 7q + 35q^2 + 140q^3 + 490q^4 + 1547q^5 \nonumber\\ 
        &{}\ \ \ \ \ \ \ \ \ \ \ \ + 4480q^6 + 12192q^7 + 31402q^8\nonumber\\
        &{}\ \ \ \ \ \ \ \ \ \ \ \ + 77119q^9 + 182119q^{10} + \mathcal{O}(q^{11})).
    \end{align}
    \item $(c,h_1,h_2)=\left(20,0,\frac{7}{2}\right)$:
    \begin{align}
        \chi_0 &= q^{-\tfrac{5}{6}}(1 + 47q + 440q^2 + 5782q^3 + 83517q^4 + 239124q^5 + 1070594q^6 + 4192784q^7\nonumber\\
        &{}\ \ \ \ \ \ \ \ \ \ \ \ + 14782019q^8 + 47772551q^9 + 144263738q^{10} + \mathcal{O}(q^{11})), \nonumber\\
        \chi_{0} &=q^{-\tfrac{5}{6}}(1 + 47q + 440q^2 + 5782q^3 + 83517q^4 + 239124q^5 + 1070594q^6 + 4192784q^7\nonumber\\
        &{}\ \ \ \ \ \ \ \ \ \ \ \ + 14782019q^8 + 47772551q^9 + 144263738q^{10} + \mathcal{O}(q^{11})), \nonumber\\
        \chi_{\frac{7}{2}} &= \text{unstable}.
    \end{align}
This is a two-character solution (not yet a $(2,\ell)$ solution).
    \item $(c,h_1,h_2)=\left(20,\frac{1}{2},3\right)$: 
    \begin{align}
        \chi_0 &= \text{Log unstable}, \\
        \chi_{\frac{1}{2}} &= q^{-\tfrac{1}{3}}(1 + 14q + 65q^2 + 156q^3 + 456q^4 + 1066q^5 \nonumber\\ 
        &{}\ \ \ \ \ \ \ \ \ \ \ \ + 2250q^6 + 4720q^7 + 9426q^8\nonumber\\
        &{}\ \ \ \ \ \ \ \ \ \ \ \ + 17590q^9 + 32801q^{10} + \mathcal{O}(q^{11})), \nonumber\\
        \chi_{3} &= \text{unstable}.
    \end{align}
    \item $(c,h_1,h_2)=\left(24,\frac{6}{7},\frac{22}{7}\right)$:
    \begin{align}
        \chi_0 &= q^{-1}(1 + 238q + 783q^2 + 8672q^3 + 65367q^4 + 371520q^5 + 1741655q^6 + 7161696q^7\nonumber\\
        &{}\ \ \ \ \ \ \ \ \ \ \ + 26567946q^8 + 90521472q^9 + 288078201q^{10} + \mathcal{O}(q^{11})), \nonumber\\
        \chi_{\frac{6}{7}} &= \text{unstable}, \nonumber\\
        \chi_{\frac{22}{7}} &= \text{unstable}.
    \end{align}
This is a one-character solution (not yet a $(1,\ell)$ solution).
    \item $(c,h_1,h_2)=\left(24,0,4\right)$: 
    \begin{align}
        \chi_0 &= \text{Log unstable}, \\
        \chi_{0} &= \text{Log unstable}, \nonumber\\
        \chi_{4} &= q^{3}(3185 + 34398q + 231231q^2 + 1201824q^3 + 5321511q^4 + 20914816q^5 \nonumber\\ 
        &{}\ \ \ \ \ \ \ \ \ \ \ \ \ \ + 74740679q^6 + 247766688q^7 + 770976635q^8\nonumber\\
        &{}\ \ \ \ \ \ \ \ \ \ \ \ \ \ + 2272634846q^9 + 6394256424q^{10} + \mathcal{O}(q^{11})).
    \end{align}
\end{itemize}
\section{Lattice relations}\label{sec:Lattice}
\noindent The $\Theta$-function of a $d$-dimensional lattice $\Lambda$ is a weight $\tfrac{d}{2}$ modular form defined as follows
\begin{align}
    \Theta_{\Lambda}(\tau) = \sum\limits_{x\in\Lambda}N(m)q^{m}.
\end{align}
Here, the sum runs over all the vectors $x$ in the lattice $\Lambda$ whose length is $m=x\cdot x$, $N(m)$ denotes the number of vectors of norm $m$, and $q \equiv e^{2\pi i\tau}$. The lattice partition function $\mathcal{Z}$ is defined as follows
\begin{align}\label{partition_theta_lattice}
    \mathcal{Z}(\tau) \equiv \frac{\Theta_{\Lambda}(\tau)}{\eta^{d}(\tau)}.
\end{align}
It is always possible to define a holomorphic modular-covariant CFT with central charge $c$ with a partition function given by \ref{partition_theta_lattice} corresponding to an even self-dual lattice of integer dimension $d$. The packing radius $\rho$, the kissing number $\mathscr{K}$ can be read off from the $q$-series expansion of the $\Theta$-function of a lattice $\Lambda$ as follows
\begin{align}
    \Theta_{\Lambda}(\tau) = 1 + \mathscr{K}q^{4\rho^{2}} + \ldots
\end{align}
Let $\Theta_{2^{+}}(\tau) = 1 + \sum_{m=1}^{\infty}N_{2^{+}}(m)q^{m}$ denote the lattice corresponding to the single character solution of central charge $c = 24$ found in $\Gamma_{0}^{+}(2)$ with partition function $\mathcal{Z}(\tau) = j_{2^{+}}(\tau)$. This $\Theta$-series was found to be related to the $\Theta$-series of odd Leech lattice $O_{24}$ as follows \cite{Umasankar:2022kzs}
\begin{align}\label{odd_Leech_Fricke_2}
    \Theta_{O_{24}}(\tau) = 1 + \sum\limits_{n=1}^{\infty}m^{(23A)}_{n}q^{n} - \Theta_{2^{+}}\left(\frac{\tau}{2}\right).
\end{align}
where $m^{(23A)}_{i}$ indicated the coefficients obtained by linear combinations of character degrees of the McKay-Thompson series of class $23A$ for $\mathbb{M}$. From \ref{j_104_c=24}, we see that the partition function, $\mathcal{Z}(\tau) =j_{2^{+}}(\tau) + \mathcal{N}$, represents a large set of admissible solutions that includes the single character solution with $\mathcal{N} = 0$ at $\ell = 4$, the non-trivial bilinear pairs in \ref{bilin_non_0}, \ref{bilin_non_1}, and \ref{bilin_non_2}  with $\mathcal{N} = -32, -64, -80$ respectively, and possibly more which we haven't found yet in the current analysis. With this partition function, we obtain the following $q$-series expansion of the $\Theta$-function
\begin{align}\label{theta_Gamma_0_2+}
    \eta^{24}(\tau)\left(j_{2^{+}}(\tau)  - 80 + \Tilde{a}\right) = 1 + \Tilde{a}q + (4048 - 24\Tilde{a})q^{2} + (-4096 + 252\Tilde{a})q^{3} + \ldots,
\end{align}
where $\Tilde{a} = \mathcal{N}a$. Now, upon setting $\Tilde{a} = 0$, we find the kissing number and lattice radius to be $\mathscr{K} = 4048$ and $\rho = \tfrac{1}{\sqrt{2}}$ respectively. Thus, a wide set of $c = 24$ single-character solutions and concocted bilinear pairs all possess the same $\Theta$-series that is related to the odd Leech lattice in $d = 24$ via the relation shown in \ref{odd_Leech_Fricke_2}.\\

\noindent 
The space of modular forms of weight $4$ is one-dimensional and it turns out that the modular form is related to a lattice theta function as follows
\begin{align}
    \begin{split}
    \omega^{(2^{+})}_{4}(\tau) =& E_{4}^{(2^{+})}(\tau)\\
    =& \tfrac{1}{4}\left(\theta_{3}^{4}(\tau) + \theta_{4}^{4}(\tau)\right)^{2}\\
    =& \Theta_{\mathbf{D}_{4}\oplus\mathbf{D}_{4}}(\tau).
    \end{split}
\end{align}
The Fricke weight $4$ Eisenstein series of level $2$ is the theta function of two copies of the $\mathbf{D}_{4}$ lattice (OEIS sequence A008658 \cite{A008658}). Such lattice relations are not observed for $E_{k}^{(2^{+}}(\tau)$ with $k\geq 4$. Moving over to $\Gamma_{0}^{+}(3)$, we the following connections between the modular forms lattice theta functions
\begin{align}
    \begin{split}
        \omega^{(3^{+})}_{4}(\tau) =&E_{4}^{(3^{+})}(\tau)= \Theta_{4\cdot\mathbf{H}}(\tau),\\
        E_{2,3^{'}}(\tau) =& \Theta_{2\cdot\mathbf{H}}(\tau)
    \end{split}
\end{align}
The Fricke weight $4$ Eisenstein series of level $3$ is the theta function of six copies of the Hexagonal lattice $\mathbf{H}$ (OEIS sequence A008655 \cite{A008655}) and the modular form $E_{2,3^{'}}(\tau)\in\mathcal{M}_{3}(\Gamma_{0}(3))$ is the theta function of two copies of $\mathbf{H}$ (OEIS sequence A008653 \cite{A008653}. Such lattice relations are not observed for $E_{k}^{(p^{+}}(\tau)$ with $k\geq 4$ and $p = 3,5,7$. However, in $\Gamma_{0}^{+}(7)$, we find that the modular form $E_{2,7^{'}}(\tau)$ (see appendix \ref{appendix: Mod_5_and_7} for definition) is the theta series of the square of Kleinian lattice $\mathbb{Z}\left[\tfrac{-1+\sqrt{-7}}{2}\right]$ (OEIS sequence A028594 \cite{A028594}). Similar relations can be found for other modular forms of Hecke and Fricke groups. 
\section{Discussion}\label{sec:Discussion}
\subsection{Fricke Monsters lurk in the Borcherds product}
\noindent From MLDE analysis for groups $\Gamma_{0}^{+}(2)$ and $\Gamma_{0}^{+}(3)$, we found that the identity characters with unstable or trivial non-trivial primaries possess coefficients equal to a McKay-Thompson series of a certain conjugacy class of $\mathbb{M}$. There exists a Borcherds product interpretation for this observation as pointed out in \cite{Duncan:2022afh} (see table $10$). We summarize this observation here. Consider the following infinite product expansion of the $\text{SL}(2,\mathbb{Z})$ Haputmodul,
\begin{align}\label{j-product}
    j(\tau) - j(z) = q_{\tau}^{-1}\prod\limits_{\substack{m>0\\ n\in\mathbb{Z}}}\left(1 - q_{\tau}^{m}q_{z}^{n}\right)^{c(m,n)},
\end{align}
where $q_{a} = e^{2\pi i a}$, and $c(r)$ is the r$^{\text{th}}$ Fourier coefficient of $J(\tau) = j(\tau) - 744$. When $m=1, n=1$, we find the RHS to read $q_{\tau}^{-1} - q_{z}^{-1}$ which indeed matches the antisymmetric LHS although it may not look it at first sight. This $j$-product formula also called the Koike-Norton-Zagier infinite product identity, was independently discovered by Borcherds, Koike, Norton, and Zagier. This is a generalization of the Weyl denominator formula for the finite-dimensional Lie algebra which reads
\begin{align}\label{Weyl_denominator_formula}
    \sum\limits_{w\in \mathcal{W}}\varepsilon(w)w(P) = q_{\rho}^{-1}\prod\limits_{\alpha>0}\left(1 - q_{\alpha}\right)^{M(\alpha)},
\end{align}
where $\mathcal{W}$ denotes the Weyl group, $\varepsilon(w) = (-1)^\ell(w)$ is the sign function with $\ell(w)$ denoting the length of the Weyl group element, $w(P) = q_{\rho}^{-1}\sum_{k}\varepsilon(k)q_{k}$ where the index $k$ runs over all the sums of pairwise orthogonal simple roots, $\rho$ is the Weyl vector of a group $G$, $\alpha$ are the roots of the Lie algebra, and $M(\alpha)$, which is always unity, is the multiplicity. Comparing \ref{Weyl_denominator_formula} and \ref{j-product}, we see that the infinite sums over all positive vectors correspond to the infinite sum over all positive roots, the $q_{\tau}^{-1}$ is matched by the $q_{\rho}^{-1}$, and expanding the difference of the Hauptmodules, we observe that we obtain an alternating sum that is matched by the alternating sum in the Weyl formula. We direct the reader to section $4$ of \cite{Carnahan2012GeneralizedMI} for a more detailed explanation of the same. The generalized Borcherds product is obtained by taking a product sum of the $j$-product over binary quadratic forms $Q = [a,b,c] = ax^{2} + bxy + cy^{2}$, where $a,b,c\in\mathbb{Z}$, belonging to the orbit space $\mathcal{Q}_{D}^{(N)}/\Gamma_{0}^{+}(N)$ (we note that a similar construction is also applicable to Hecke groups), where $D = b^{2} - 4ac$ is the discriminant of $Q$ and $\mathcal{Q}_{D}$ denotes the space of binary quadratic forms with $D<0$ and $a>0$. For a unique choice of the discriminant, this takes the following form
\begin{align}\label{Borcherds_result}
    \prod\limits_{Q\in\mathcal{Q}_{D_{0}}^{(N)}/\Gamma_{0}^{+}(N)}\left(j_{N^{+}}(\tau) - j_{N^{+}}(\alpha_{Q})\right)^{\frac{1}{\vert s_{Q}\vert}} = \left(j_{N^{+}}(\tau) + \mathcal{N}\right)^{x},
\end{align}
where the root $\alpha_{Q}$ is the complex multiplication point of $Q$, $s_{Q}$ is the stabilizer in $\Gamma_{0}^{+}(N)/\mathbb{Z}_{2}$, and $\mathcal{N}, x\in\mathbb{Z}$. The values of $j(\alpha_{Q})$ solely depends on the $\Gamma_{0}^{+}(N)$-equivalence class of $Q$. See \cite{Kim2002BorcherdsPA} for a detailed construction of the Borcherds product for Fricke groups. For the case of $N = 2$, we notice that the set $(\mathcal{N},x) = \left\{\left(0,\tfrac{1}{4}\right), \left(0,\tfrac{1}{2}\right), (104,1)\right\}$ corresponding to solutions with central charges $c = 6$ (\ref{8C_character_1}, \ref{8C_character_2}, \ref{8C_character_3}), $c = 12$ (\ref{4B_character_1}, \ref{4B_character_2}), and to the single character solution with central charge $c = 24$ (\ref{j_104_c=24}) respectively. Next, when $N = 3$, we find that the set $(\mathcal{N},x) = \left\{\left(0,\tfrac{1}{6}\right), \left(0,\tfrac{1}{3}\right), \left(0,\tfrac{1}{2}\right)\right\}$ that correspond to solutions with central charges $c = 4$ (\ref{18b_character_1}, \ref{18b_character_2}), $c = 8$ (\ref{9a_character_1}), and $c = 12$ (\ref{6b_character_1}).\\

\noindent From \cite{Duncan:2022afh}, we can also predict the admissible solutions with unstable non-trivial primaries for higher-level Fricke groups. This, along with the results thus far for Fricke levels $p = 2,3$ are summarized in table \ref{tab: Summary}. Of the $14$ prime divisor levels, only $10$ turn out to possess admissible identity solutions with an unstable non-trivial primary. At this point, we do not have enough information to state if these solutions correspond to new characters not found by the single character analysis, for this we would need to resort to a complete analysis.
\begin{table}[htb!]
\begin{tabular}{||c|c|c|c||}
\hline
Level $p$          & Conjugacy class  & $c$ & Hauptmodul construction\\[0.5ex]
\hline\hline
\multirow{2}{*}{2} & 8C  & 6  & $j_{2^{+}}^{\tfrac{1}{4}}$\\[0.5ex]
& 4B  & 12  &  $j_{2^{+}}^{\tfrac{1}{2}}$\\[1ex]
\hline
\multirow{3}{*}{3} & 18b & 4 & $j_{3^{+}}^{\tfrac{1}{6}}$\\[0.5ex]
& 9a & 8 & $j_{3^{+}}^{\tfrac{1}{3}}$\\[0.5ex]
& 6b & 12 & $j_{3^{+}}^{\tfrac{1}{2}}$\\[1ex]
\hline
5 & 10a & 12 & $j_{5^{+}}^{\tfrac{1}{2}}$\\[1ex]
\hline
7 & 21C  & 8  &  $\left(j_{7^{+}} + 13\right)^{\tfrac{1}{3}}$  \\[1ex]
\hline   
13 & 39B  & 8 &  $\left(j_{13^{+}} + 5\right)^{\tfrac{1}{3}}$    \\[1ex]
\hline
\end{tabular}
\caption{This table lists all the admissible solutions to $n = 2$ MLDEs, for Fricke groups of prime divisor levels of $\mathbb{M}$, that possess an unstable non-trivial primary. The Hauptmodules are constructed for higher-level groups by using \ref{Borcherds_result} as a guide and playing around with the exponent to obtain the correct sequence. The central charges are found by reading off the value in the exponent of the identity characters (Hauptmodul constructions here), i.e. $j_{p^{+}}^{x} = q^{-\tfrac{c}{24}}(1 + \ldots)$.}
\label{tab: Summary}
\end{table}

\subsection{General formula for identity character with c-chronology}
\noindent For the case of $\Gamma_{0}^{+}(2)$, we found in \ref{Hecke_1_c=1} and \ref{Hecke_1_c=2} that the identity characters with ascending central charges in two- and three-character theories can be expressed in terms of modular forms of the Hecke group $\Gamma_{0}(2)$. We find the following general formula for all the identities one might expect to find with ascending central charges in higher-character theories
\begin{align}\label{c-chronology_Fricke_2}
\begin{split}
        \chi^{(2^{+})}_{0}(\tau) =&          \left(\mathfrak{p}_{2^{+}}\Delta_{2}^{-\tfrac{1}{24}}\right)^{c},\\
        \mathfrak{p}_{2^{+}}(\tau) \equiv& \left(\frac{\left(j_{2}\Delta_{2}^{\infty}\right)(3\tau)}{\left(j_{2}\Delta_{2}^{\infty}\right)^{\tfrac{1}{3}}(\tau)}\right)^{\tfrac{1}{8}}.
\end{split}
\end{align}
We can write down a similar expression for $c$-chronology in $\Gamma_{0}^{+}(3)$ as follows
\begin{align}\label{c-chronology_Fricke_3}
\begin{split}
        \chi^{(3^{+})}_{0}(\tau) =&          \left(\mathfrak{p}_{3^{+}}\Delta_{3}^{-\tfrac{1}{6}}\right)^{c},\\
        \mathfrak{p}_{3^{+}}(\tau) \equiv& \left(\frac{\Delta_{3}^{0}(2\tau)}{\left(\Delta_{3}^{0}\right)^{\tfrac{1}{2}}(\tau)}\right)^{\tfrac{1}{3}}.
\end{split}
\end{align}
Analysis of higher-character theories of $\Gamma_{0}^{+}(2)$ and $\Gamma_{0}^{+}(3)$ would reveal more such solutions. The reason why only the identity corresponding to a theory with a low central charge possesses this nice closed-form expressed would be interesting to study further. On the other hand, MLDE analysis of $\Gamma_{0}(2)$ and $\Gamma_{0}(3)$ could reveal some of these solutions since the Fricke groups are supergroups of Hecke groups.

\subsection{From Fricke solutions to Schellekens' list}
\noindent Schellekens classified all $c = 24$ single-character theories and found that there is a total of $71$ CFTs, all of which possess a partition function of the form \cite{Schellekens:1992db}
\begin{align}
    \begin{split}
        &Z(\tau) = J(\tau) + \mathcal{N},\ \ \ \ \ \mathcal{N} =12m,\\ 
        &m\in\left\{0, 2, 3, 4, 5, 6, 7, 8, 9, 10, 12, 13, 14, 16, 18, 20, 22, 24, 25, 26, 28, 30, 32, 34, 38, 46, 52, 62, 94\right\}
    \end{split}
\end{align}
For $\Gamma_{0}^{+}(2)$, the partition function corresponding to the general solution to $c = 24$ single-character theories we have established in \ref{j_104_c=24} is $Z_{2^{+}}(\tau) = j_{2^{+}}(\tau) + \mathcal{N}$, where $\mathcal{N}\geq -104$. The only solutions we have come across are \ref{bilin_non_0}, \ref{bilin_non_1}, \ref{bilin_non_2}, \ref{the_one_with_N_as15987}, and the $\ell = 4$ solution with $\mathcal{N} = 0$ found in \cite{Umasankar:2022kzs}. This gives us the list $\mathcal{N}\in\{-80, -64, -32, 0, 15987\}$. We understand the relationship between Klein's function, i.e. the Hauptmodul of the modular group, and the Hauptmodul of the Fricke group of level $2$ which reads 
\begin{align}\label{j_to_Fricke_2}
    j(\tau) = \frac{1}{2}\left[\left(j_{2^{+}} - 81\right)\sqrt{j_{2^{+}}(256 - j_{2^{+}})} + \left(j^{2}_{2^{+}} - 207j_{2^{+}} + 3456\right)\right](\tau).
\end{align}
This helps us map $\mathcal{Z}_{2^{+}}(\tau)$ to $Z(\tau)$ for $\mathcal{N}_{2^{+}} = 0$ but does not act as a general map. It would be interesting to complete the list and find a general map between the single-character partition functions. Additionally, since we have the following relations
\begin{align}\label{j-j_Hecke_Fricke}
    \begin{split}
    j_{2^{+}}(\tau) =& \left(j_{2} + 4096j_{2}^{-1} + 128\right)(\tau),\\
    j(\tau) =& \left(\frac{256 + j_{2}}{j_{2}^{2}}\right)(\tau),
    \end{split}
\end{align}
we can utilize these relationships to better understand the maps among the partition functions post an exhaustive analysis of the Hecke group of level $2$.

\subsection{Hecke images}
\noindent We obtain the following Hecke relations between some of our single-character solutions using the Hecke operators constructed in \cite{Harvey:2018rdc}. For single-character theories, the conductor $N$ is computed by observing the smallest positive integer $N$ such that $\left(e^{-\frac{i\pi c}{12}}\right)^N = 1$. Now, if,
\begin{align}
    \chi = \sum\limits_{n\in\mathbb{Z}} a(n) q^{\frac{n}{N}}, \label{nonH}
\end{align}
Then, acting the above with a Hecke operator $T_p$ where $\text{gcd}(p,N)=1$, we get,
\begin{align}
    T_p(\chi) = \sum\limits_{n\in\mathbb{Z}} a^{(p)}(n) q^{\frac{n}{N}}, \label{afH}
\end{align}
where,
\begin{align}
    a^{(p)}(n) = 
    \begin{cases}
        &pa(pn) \, \, \, \, \, \text{if} \, \, p \, \ndiv \, n \\
        &pa(pn) + a(\frac{n}{p}) \, \, \, \, \, \text{if} \, \, p \, | \, n
    \end{cases}
\end{align}
We report a few Hecke images of single-character solutions we found in Fricke groups of levels $2$ and $3$ here.\\
$\mathbf{\Gamma^{+}_0(2)}$:\\
\noindent The conductor is $N=4$ for $j_{2^{+}}^{\frac{1}{4}}$.
\begin{align}
    &T_3\left(j_{2^{+}}^{\frac{1}{4}}\right) = j_{2^{+}}^{\frac{3}{4}}, \label{hF21} \\
    &T_5\left(j_{2^{+}}^{\frac{1}{4}}\right) = j_{2^{+}}^{\frac{1}{4}}\left(j_{2^{+}}-130\right), \label{hF22} \\
    &T_7\left(j_{2^{+}}^{\frac{1}{4}}\right) = j_{2^{+}}^{\frac{3}{4}}\left(j_{2^{+}}-182\right). \label{hF23}
\end{align}
$\mathbf{\Gamma^{+}_0(3)}$:\\
\noindent The conductor is $N=3$ for $j_{3+}^{\frac{1}{3}}$.
\begin{align}
    &T_2\left(j_{3+}^{\frac{1}{3}}\right) = j_{3+}^{\frac{2}{3}}, \label{hF31} \\
    &T_5\left(j_{3+}^{\frac{1}{3}}\right) = j_{3+}^{\frac{2}{3}}\left(j_{3+}-70\right), \label{hF32} \\
    &T_7\left(j_{3+}^{\frac{1}{3}}\right) = j_{3+}^{\frac{1}{3}}\left(j_{3+}^2 -98j_{3+}+917\right). \label{hF33}
\end{align}

%\subsection{\texorpdfstring{$\text{SL}(2,\mathbb{Z})$}{SL(2,Z)} vs. \texorpdfstring{$\Gamma_{0}^{+}(p)$}{Γ0(p)+}}

%\subsection{Monodromy of solutions}

\section{Future directions}\label{sec:Future_work}
\noindent There are various interesting future directions we reserve for upcoming work(s) which we present briefly below while also summarizing the results of this paper.\\
\\
\textbf{\textit{Are there new theories out in the wild?}}\\
We have obtained many single-, two-, and three-character solutions corresponding to the Fricke groups $\Gamma^{+}_0(2)$ and $\Gamma^{+}_0(3)$. Many of the single-character solutions have nice interpretations with regard to the conjugacy classes of $\mathbb{M}$. We noticed the reappearance of certain single-character solutions aided by unstable non-trivial primaries as solutions to $(2,\ell)$ and $(3,0)$ MLDEs which possess a neat description in terms of the Borcherds product. However, these solutions themselves do not describe any CFTs ``yet". It would be nice to see if these solutions, by themselves, could be related to some known theories which have some restricted conformal symmetry such that their corresponding partition functions are only modular (up to a phase if required) with respect to a congruence subgroup rather than the full modular group. For instance, we notice that many of our two-character solutions form nice bilinear pairs with many other two-character solutions to give single-character solutions. This phenomenon was very useful in identifying coset pairs in the $\text{SL}(2,\mathbb{Z})$ case (see \cite{Gaberdiel:2016zke}). This perhaps hints at the fact that there could be some nice coset relations even for these solutions belonging to the congruence subgroups. Since the Fricke groups are an extension of the Hecke group by involution, it would be interesting to see if the admissible two- \& three-character solution of $\Gamma_{0}(p)$ and $\Gamma_{0}^{+}(p)$ can be related. This would also help us understand better the role of the involution at the level of character solutions, and possibly RCFTs. In the CFT literature, it is known that there exist some theories belonging to the Hecke subgroup at level 2, for example, fermionic RCFTs (\cite{Bae:2021mej}), SCFTs and parafermion theories \cite{Anderson:1987ge}. So, it would be natural to expect that there might be theories out in the wild for Fricke groups too.\\
\\
\textbf{\textit{Looking at the bulk and beyond for $\mathbf{\Gamma_{0}^{+}(2)}$ and $\mathbf{\Gamma_{0}^{+}(3)}$.}}\\
For two-character and three-character analysis, we have analyzed the MLDEs and their solutions for $\mathbf{\Gamma_{0}^{+}(2)}$ and $\mathbf{\Gamma_{0}^{+}(3)}$ up to the \textit{bulk point}. There exists a complete classification (see \cite{Chandra:2018pjq}) of \textit{admissible} two-character solutions in the $\text{SL}(2,\mathbb{Z})$ case for $\ell\geq 6$ (note, $\ell=6$ is the bulk point for $\text{SL}(2,\mathbb{Z})$). The way this classification works is that any solution above the bulk point can be written down in terms of the solutions below the bulk point. Performing an MLDE analysis beyond the bulk point for these groups might enable us to get a complete classification of admissible solutions using the solutions obtained above since these solutions are below the bulk point solutions.\\

\noindent
\textbf{\textit{Higher-character MLDEs for $\mathbf{\Gamma_{0}^{+}(2)}$ and $\mathbf{\Gamma_{0}^{+}(3)}$?}}\\
It is natural to extend our analysis to four- and five-character MLDEs and probe the landscape, at least at $\ell = 0$, for admissible solutions (reappearances of other lower-character solutions, minimal models, and of genuine solutions). Additionally, its worth mentioning that in the case of $\Gamma_{0}^{+}(2)$, all the minimal models possess central charges $c \in\tfrac{6}{7}\times\{1, 5, 9, 13\}$, where $\tfrac{6}{7} = \tfrac{4}{7}\times \tfrac{3}{2} = c_{\mathcal{M}(7,2)}\cdot \mu_{0}^{+}$, where $\mathcal{M}(7,2)$ is a three-character minimal model with $(c,h_{1},h_{2}) = \left(\tfrac{4}{7},\tfrac{5}{7},\tfrac{1}{7}\right)$. It would be interesting to explore this relationship further and also to see if such relations show up for minimal models at higher-character theories. The equation would be simple to work out thanks to the ease of computation with modular forms of Fricke levels $2$ and $3$. Additionally, it would be nice to provide a general prescription for modular re-parameterization to $n$-character theories for these levels (which should work out nicely since the re-parameterization setup for these groups mimics the prescription done in the case of $\text{SL}(2,\mathbb{Z})$).\\

\noindent 
\textbf{\textit{What's up with $\mathbf{\Gamma_{0}^{+}(5)}$ and $\mathbf{\Gamma_{0}^{+}(7)}$?}}\\
Although we found that $\Gamma_{0}^{+}(7)$ is quite hard to re-parameterize, we should be able to find three-character solutions to it, and for Fricke level $5$. It would be interesting to see if analysis at $(3,0)$ could reveal some interesting solutions. Firstly, we could check if the identity of a particular three-character solution is expressible in terms of Hecke modular forms of level $5$ and $7$ thereby enabling us to identify the $c$-chronology and devise a general expression akin to \ref{c-chronology_Fricke_2} and \ref{c-chronology_Fricke_3}. Secondly, we can explore if we continue to find minimal models at these levels as we did at levels $p = 2,3$. Lastly, it would be interesting to see if genuine three-character solutions exist at these levels.\\

\noindent
\textbf{\textit{What about the other prime divisors levels, $\mathbf{\Gamma_{0}^{+}(p)}$?}}\\
The most approachable higher Fricke level would be $p = 11$ since the space of modular forms is well studied \cite{Junichi}. Thus, 
An exhaustive analysis at Fricke level $11$ for single-, two-, and three-character solutions should be possible and we can ask the same questions of finding minimal models and the existence of genuine three-character solutions here too. The space of modular forms becomes complicated at levels $p\geq 13$ which unfortunately limits the ease of calculations. Nevertheless, at least an exhaustive single-character and a $(n,\ell) = (2,0)$ analysis for the remaining prime divisor levels could help us establish some common themes among the solutions in $\Gamma_{0}^{+}(p)$. Another very interesting group to consider is the intersection group, $\Gamma_{0}(4)\cap\Gamma(2)$, that is an index $12$ group with the coefficients of its Hauptmodul corresponding to the $8D$ conjugacy class of $\mathbb{M}$ \cite{Magureanu:2022qym, Sebbar2002ModularSF}. The normalizer of this group turns out to be $\Gamma_{0}^{+}(2)$ and studying the solutions should help further classification.\\

\noindent
\textbf{\textit{What about Hecke groups?}}\\
The natural direction to follow post-Fricke would be to investigate the Hecke groups. Taking indications from $c$-chronology, where the identity character could be expressed in terms of Hecke group modular forms, it would be interesting to check if a set of solutions found in our analysis for $\Gamma_{0}^{+}(p)$ reappears in $\Gamma_{0}(p)$. This is natural to expect since Fricke groups are supergroups of Hecke groups. It would be nice to confirm if the ansatz for the quasimodular form made in \ref{quasi_mod_ansatz} holds for all even levels of the Hecke group. It would be nice to check if there exist certain single character solutions found with $(2,\ell)$ that possess unstable non-trivial primaries. If this is the case, then there should also exist a way to read off the (unstable) theories from a Borcherds product with the Hecke Hauptmodul. Lastly, it would be especially interesting to see if we can relate the admissible solutions to existing minimal models, coset theories, or other unitary (s)CFTs.\\

\noindent 
\textbf{Using isomorphism to find relations to $\text{SL}(2,\mathbb{Z})$ theories}\\
\noindent Niwa showed in \cite{Niwa1977OnST} that there exists an isomorphism between the space of holomorphic cusp forms of half-integral weight on $\Gamma_{0}(4N)$, $\mathcal{S}_{k + 1/2}\left(\Gamma_{0}(4N)\right)$, and the space of weight $2k$ cusp forms on $\Gamma_{0}(2N)$, $\mathcal{S}_{2k}\left(\Gamma_{0}(2N)\right)$. Kohnen then defined the subspace $\mathcal{S}_{k + 1/2}^{+}\left(\Gamma_{0}(4)\right)\subset \mathcal{S}_{k + 1/2}\left(\Gamma_{0}(4)\right)$, called the plus space and showed that the newform subspace in this space, $\mathcal{S}_{k + 1/2}^{+,\text{new}}\left(\Gamma_{0}(4)\right)\subset\mathcal{S}_{k + 1/2}^{+}\left(\Gamma_{0}(4)\right)$, is isomorphic to $\mathcal{S}_{2k}^{\text{new}}\left(\Gamma_{0}(N)\right)$ \cite{Kohnen1980ModularFO, Kohnen1982NewformsOH}. For $N = 1$, we find the isomorphism, $\mathcal{M}_{k + 1/2}^{+,\text{new}}(\Gamma_{0}(4))\cong \mathcal{M}_{2k}(\text{SL}(2,\mathbb{Z}))$, which preserves the space of cusp forms. Also, for the case $N=1$, the plus space is an eigenspace of Niwa's Hecke operator, and the space corresponding to the negative eigenvalue is called the minus space, which we shall denote by $\mathcal{M}^{-}_{k + 1/2}\left(\Gamma_{0}(4)\right)$. Hence, we have the following decomposition \cite{Baruch2016NewformsOH}
\begin{align}
    \mathcal{S}_{k + \tfrac{1}{2}}\left(\Gamma_{0}(4)\right) = \mathcal{S}^{+}_{k +  \tfrac{1}{2}}\left(\Gamma_{0}(4)\right)\oplus\Tilde{\mathcal{S}}^{+}_{k +  \tfrac{1}{2}}\left(\Gamma_{0}(4)\right)\oplus \mathcal{S}^{-}_{k +  \tfrac{1}{2}}\left(\Gamma_{0}(4)\right),
\end{align}
where $\Tilde{\mathcal{S}}^{+}_{k +  1/2}\left(\Gamma_{0}(4)\right)$ is the conjugate of the Kohnen plus space. Baruch et al. showed in \cite{Baruch2016NewformsOH} that the minus space at level $4$ possesses the isomorphism, $\mathcal{M}^{-}_{k + 1/2}\left(\Gamma_{0}(4)\right) \cong \mathcal{M}^{\text{new}}_{2k}\left(\Gamma_{0}(2)\right)$. With these isomorphisms, it would be interesting to study character relations between the modular group and the newform space of $\Gamma_{0}(2)$ to the plus and minus space respectively.\\

\noindent 
\textbf{Exploring non-arithmetic Hecke groups.}\\
The group $G(\lambda) = \langle T, S_{\lambda}\rangle$, with $\lambda>0$ and $T = \left(\begin{smallmatrix} 1 & 1\\ 0 & 1 \end{smallmatrix}\right)$ and $S_{\lambda} = \left(\begin{smallmatrix}0 & -1\\ \lambda & 0 \end{smallmatrix}\right)$ is a subgroup of $\text{SL}(2,\mathbb{R})$. This is called a Hecke group where $\lambda = 4\cos^{2}\left(\tfrac{\pi}{p}\right)$. There are exactly four groups that are commensurable with the modular group with $\lambda = 1,2,\sqrt{2},\sqrt{3}$. The groups $G(1) = \Gamma(1)$ and $G(2) = \Gamma_{\theta}$ are called arithmetic Hecke groups and are the familiar groups we have dealt with. The groups $G(\sqrt{2})$ and $G(\sqrt{3})$, called non-arithmetic Hecke groups have a unique multiplier system associated with the vector space, and their basis is comprised of automorphic forms. It should be interesting to study the MLDEs in these groups for many reasons. Firstly, for the sake of classification, and secondly, to investigate the connections to Fricke characters since we have the relation $ M G(\sqrt{N})M^{-1} = \Gamma_{0}^{+}(N),\ M = \left(\begin{smallmatrix}1 & 0\\ 0 & \sqrt{N}
\end{smallmatrix}\right)$ for $N \in\{2, 3, 4\}$.\\

\noindent
\textbf{\textit{Traversing Triangle groups}}\\
Continuing the peripatetic nature of our MLDE analysis, we consider triangle groups that are discrete subgroups of $\text{PSL}(2,\mathbb{R})$ denoted by the triple $\Gamma_{t} = (m_{1},m_{2},m_{3})$, where $m_{i}\in\mathbb{Z}^{+}$, and possess a robust modular structure. It turns out that there exists a deeper connection- the Hecke groups are isomorphic to triangle groups of type $(2, p,\infty)$, where the triple corresponds to a triangle on $\mathbb{H}^{2}$ with angles $\tfrac{\pi}{2}$, $\tfrac{\pi}{p}$, and $\tfrac{\pi}{\infty}$. To begin with, we can consider the nine triangle groups commensurable with the modular group, tabulated in \cite{Doran2013AutomorphicFF}. The conjugation of these groups, $g\Gamma_{t}g^{-1}$ for a specific $g$, turns out to yield known Hecke and Fricke groups. For example, when $\Gamma_{t} = (2,4,\infty)$, we find $g\Gamma_{t}g^{-1} = \Gamma^{+}_{0}(2)$ for $g = \left(\begin{smallmatrix}2 & 0\\0& 1\end{smallmatrix}\right)$. What's more, we also find that the conjugation of groups $(4,4,\infty)$ and $(6,6,\infty)$ yield $\Gamma^{+}_{0}(2)^{2}$ and $\Gamma^{+}_{0}(2)^{2}$, the Fricke groups generated by the squares of all elements $\gamma$  in $\gamma\in\Gamma_{0}^{+}(N)$. It would be interesting to study the connections between the admissible solutions in these subgroups and their Fricke parents.\\

\noindent 
\textbf{\textit{Getting lucky with congruence subgroups.}}\\
Why leave out congruence subgroups $\Gamma_{1}(N)$, the conjugate groups $\Gamma^{0}(N)$ and $\Gamma^{1}(N)$, and the groups $\Gamma_{0}^{0}(N)$? We could fully comprehend how admissible solutions are categorized in these congruence subgroups with the aid of MLDE analysis, but more importantly, this could provide us with new insights into the relationships of character solutions among the subgroups and between the subgroup and modular group characters. For example, the $S$-transform of the $\Gamma_{0}(7)$ Hauptmodul, $j_{7}(\tau)$, is found to be $j_{\Gamma_{0}(7)}(S(\tau)) = 49\left(\tfrac{\eta(\tau)}{\eta\left(\tfrac{\tau}{7}\right)}\right)^{4} = 49j_{\Gamma^{0}(7)}^{-1}(\tau)$. Suppose $j_{\Gamma_{0}(7)}(\tau)$ and $j_{\Gamma^{0}(7)}(\tau)$ are admissible single-character solutions, then we see that analyzing just the Hecke solutions can help us find the related conjugate solutions by mere taking the $S$-transform. Also, Hecke images of $\Gamma_{0}(N)$ characters can be related to those of $\Gamma^{0}(N)$ characters. There also exists other Atkin-Lehner groups $\Gamma_{0}^{*}(N)$ studied in \cite{Junichi} that are desirable candidates to pick for MLDE analysis.\\

\noindent 
\textbf{Fun with the modular normal subgroups.}\\
There exists a unique index-$2$ group  $\Gamma^{2}$ and a unique index-$3$ group $\Gamma^{3}$ that are the normal subgroups of $\Gamma(1)$ generated by squares of elements and cubes of elements of the modular group respectively. The fairly straightforward setup of modular forms and Serre derivatives for these groups and the possibility of finding RCFTs (due to their immediate connection to $\Gamma(1)$) make them attractive candidates for an MLDE study.\\

\noindent 
\textbf{\textit{A modular tower of theories.}}\\
To better comprehend the categorization of CFTs, a more systematic investigation of the two- and three-character MLDE for the Fricke groups and related modular towers, $\Gamma_{0}^{+}(p^{n+2})$ would be interesting. The natural choice of candidates for first consideration would be the towers $\Gamma_{0}^{+}(2)\to \Gamma_{0}^{+}(4)\to \Gamma_{0}^{+}(8)\to \ldots$ (and the corresponding Hecke tower), that follows from the modular tower $X_{0}(2^{n})$, and 
$\Gamma_{0}^{+}(3)\to \Gamma_{0}^{+}(6)\to \Gamma_{0}^{+}(12)\to \ldots$ (and the corresponding Hecke tower), that follows from the modular tower $X^{+}_{0}(3\cdot 2^{n})$. In \cite{Umasankar:2022kzs}, an admissible single-character solution to the first levels of the modular tower $\Gamma_{0}^{+}(7^{n})$ was found in terms of the parameterization in the affine equation of the elliptic modular curve $X_{0}(49)$. Since $\Gamma_{0}^{+}(25)$ is also a Conway-Norton ghost similar to $\Gamma_{0}^{+}(49)$, it would be interesting to examine what occurs for level $5$ Fricke groups.\\

\noindent
\textbf{\textit{Finding new lattices.}}\\
It was reported in \cite{McKay2000FuchsianGA} that the Schwarz derivative of the Hauptmodules of genus-zero Hecke groups and their conjugates with no elliptic elements can be expressed in terms of Eisenstein series or certain lattice $A_{4}$ and $D_{4}$-lattice theta functions.
The Schwarzian of the Fricke level $2$ Hauptmodule is found to possess the following $q$-series expansion
\begin{align}
       \begin{split}
        -\{j_{2^{+}}(\tau),\tau\} =& -\left(\frac{1}{\left(j_{2^{+}}'\right)^{2}}\left(2j'_{2^{+}}j'''_{2^{+}} - 3\left(j''_{2^{+}}\right)^{2}\right) \right)(\tau)\\
        =& 1 +44040 q^2 + 3792000 q^3 + 536995320 q^4 + 57168341760 q^5 + \mathcal{O}(q^{6}),
       \end{split}
\end{align}
where all the coefficients were found to possess positive signs. A first step to try reconstruction of this series would be to use the Jacobi theta-series combination $\tfrac{1}{256}\left(\theta_{3}(q^{4}) + \theta_{4}(q^{4})\right)^{8}$ which has a null-valued coefficient at order $\mathcal{O}(q)$. This ansatz, however, does not turn out to yield the required series. Thus, it could be that the series we are after is some combination of the level $2$ Eisenstein series or it could be constructed out of other integer lattices. Finding such Schwarzian-lattice relations for all Fricke levels would form an interesting study. Additionally, one can also construct $\Theta$-series for the modular invariant partition functions \ref{partintn1}, \ref{partintn3}, and \ref{partintn5} and explore if it can be expressed as a combination of a McKay-Thompson series and a lattice $\Theta$-series akin to what was done with $\Theta_{2^{+}}(\tau)$.
\\
\\
\textbf{\textit{Exhaustive Hecke Relations.}}\\
We observed a few Hecke relations (\cite{Harvey:2018rdc}) among some of our single-character solutions. The analysis done here in terms of Hecke images is far from complete. One natural extension to the analysis provided here is to compute Hecke images for two-character and three-characters solutions obtained above. Since taking Hecke images of higher characters involves the modular $S$ and $T$ matrices, we can in principle ask the opposite question for our case considered above. Say, we construct Hecke images of our two-character and three-character solutions. We would end up with some characters from this procedure. Now, using the knowledge of how Hecke operators act on characters in the $\text{SL}(2,\mathbb{Z})$ case, perhaps we can determine the \textit{restricted} modular $S$ and $T$ matrices which would govern the transformations of these congruence subgroups characters. We aim to return to this kind of Hecke image analysis for other Fricke and Hecke groups in the future.\\
\\
\textbf{\textit{MTCs for Fricke and Hecke groups.}}\\
It is known that every RCFT, whose partition function by definition is modular (up to a phase) with respect to the full modular group, belongs to some Modular Tensor Category (MTC) \cite{rowell2009classification}. Say by following the procedure outlined in the above point, one can find the \textit{restricted} modular $S$ and $T$ matrices for a given character-like solution. Then, this begs the question of which MTC this modular data would belong to. If such a thing would exist then we could find, in principle, MTCs to which our two- and three-character solutions belong. This idea is also motivated by the fact that we observe nice bilinear relations among many of our character-like solutions. In the $\text{SL}(2,\mathbb{Z})$ case, whenever two RCFTs are related by a bilinear relation, it meant that if one belonged to an MTC then the other, in the pair, belonged to the conjugate MTC. Such things might also exist in the case of the Fricke and Hecke groups.\\

\noindent
\textbf{\textit{Kaneko-Zagier Equations.}}\\
It was noted in \cite{Chandra:2018pjq} that the following Kaneko-Zagier equation,
\begin{align}
    \left(\mathcal{D}^2_{(k)}-\frac{k(k+2)}{144}E_4(\tau)\right)f_{(k)}(\tau) = 0, \label{KZSL2z}
\end{align}
can be recast into the usual $(2,0)$ MMS equation,
\begin{align}
    \left(\mathcal{D}^2_{(0)}-\frac{k(k+2)}{144}E_4(\tau)\right)f_{(0)}(\tau) = 0, \label{MMSSL2z}
\end{align}
by doing the following substitution,
\begin{align}
    f_{(k)}(\tau) = \eta(\tau)^{2k}f_{(0)}(\tau), \qquad c=2k. \label{subs1}
\end{align}
We figured out that a similar \textit{Kaneko-Zagier type} equation can be obtained in the $\Gamma^{+}_0(2)$ case. We start with the following \textit{Kaneko-Zagier type} equation,
\begin{align}
    \left(\mathcal{D}^2_{(k)}-\frac{k(k+2)}{64}E^{(2+)}_4(\tau)\right)f_{(k)}(\tau) = 0. \label{KZSG2}
\end{align}
Now we observe that $\theta_q (\eta(\tau)\eta(2\tau)) = \tfrac{1}{8}E_2^{(2+)}(\tau)(\eta(\tau)\eta(2\tau))$ which follows from $\theta_q \Delta_2 = E_2^{(2+)}\Delta_2$ (see \ref{general_cusp_derivative}). So, consider the substitution,
\begin{align}
    f_{(k)} = (\eta(\tau)\eta(2\tau))^k f_{(0)}, \qquad c=3k. \label{subs2}
\end{align}
This then leads to (from \ref{KZSG2}),
\begin{align}
    \left(\mathcal{D}^2_{(0)}-\frac{k(k+2)}{64}E_4(\tau)\right)f_{(0)}(\tau) = 0, \label{G2z}
\end{align}
In \cite{Chandra:2018pjq}, an equation of the type \ref{KZSL2z} was studied to classify quasi-characters. Motivated by this, we are interested to study the equation of the form \ref{KZSG2} to classify quasi-characters that appear as solutions in the $\Gamma^{+}_0(2)$ case. Furthermore, from a mathematical standpoint, equations like \ref{KZSL2z} find applications in the study of Supersingular j-invariants, hypergeometric series, and Atkin's orthogonal polynomials (see \cite{KZ}). It would be interesting to investigate a similar equation (obtained above in \ref{KZSG2}) for the $\Gamma^{+}_0(2)$ case with regards to applications similar to those done in \cite{KZ}. One can in principle also set up \textit{Kaneko-Zagier type} equations for other Fricke groups.\\

\acknowledgments
\noindent We thank Sunil Mukhi for the helpful discussions. Additionally, AD would like to thank Chethan N. Gowdigere, Jagannath Santara and Jishu Das for insightful discussions pertaining to MLDEs. AD would also like to thank Nabil Iqbal for useful discussions on CFTs. AD would also like to express his gratitude to Sigma Samhita for help in \LaTeX-formatting. NU would like to thank Mikhail Isachenkov for multiple discussions, and useful suggestions, and Miranda Cheng and Erik Verlinde for early discussions.

\appendix
\section{Modular forms for levels 5 \& 7}\label{appendix: Mod_5_and_7}
$\mathbf{\Gamma_{0}^{+}(5)}:$\\

\noindent The Fricke group of level $5$ is generated by $\Gamma_{0}^{+}(5) = \langle\left(\begin{smallmatrix}1 & 1\\ 0 & 1\end{smallmatrix}\right), W_{5}, \left(\begin{smallmatrix} & 1\\ 5 & 2\end{smallmatrix}\right)\rangle$. The elliptic points in the fundamental domain are at $\tfrac{i}{\sqrt{5}}$ (also known as the Fricke involution point), $\rho_{5,1} = -\tfrac{1}{2}+\tfrac{i}{2\sqrt{5}}$, and $\rho_{5,2} = \tfrac{-2 + i}{5}$ and this group has a cusp at $\tau = i\infty$. A non-zero $f\in\mathcal{M}_{k}^{!}(\Gamma_{0}^{+}(5))$ satisfies the following valence formula
\begin{align}\label{valence_Fricke_5}
    \nu_{\infty}(f) + \frac{1}{2}\nu_{\tfrac{i}{\sqrt{5}}}(f) + \frac{1}{2}\nu_{\rho_{5,1}}(f) + \frac{1}{2}\nu_{\rho_{5,2}}(f) + \sum\limits_{\substack{p\in\Gamma_{0}^{+}(5)\backslash\mathbb{H}^{2}\\ p\neq \tfrac{i}{\sqrt{5}},\rho_{5,1}, \rho_{5,2}}}\nu_{p}(f) =   \frac{k}{4}.
\end{align}
The Hauptmodul and  modular forms of $\Gamma_{0}^{+}(5)$ read
\begin{align}
    \begin{split}
        j_{5^{+}}(\tau) =& \left(\frac{\left(E_{2,5^{'}}\right)^{2}}{\Delta_{5}}\right)(\tau) = \left(\frac{E_{4}^{(5^{+})}}{\Delta_{5}}\right)(\tau) - \frac{36}{13} = \left(\frac{\eta(\tau)}{\eta(5\tau)}\right)^{6} + 5^{3}\left(\frac{\eta(5\tau)}{\eta(\tau)}\right)^{6} + 22,\\
        \Delta_{5}(\tau) =& \left(\eta(\tau)\eta(5\tau)\right)^{4}\in\mathcal{S}_{4}(\Gamma_{0}^{+}(5)),\\
        E_{2,5^{'}}(\tau) =& \frac{5E_{2,5^{'}}(5\tau) - E_{2}(\tau)}{4}\in\mathcal{M}_{2}(\Gamma_{0}(5)),\\
        E_{k}^{(5^{+})}(\tau) =& \frac{5^{\tfrac{k}{2}}E_{k}(5\tau) + E_{k}(\tau)}{5^{\tfrac{k}{2}} + 1}\in\mathcal{M}_{k}(\Gamma_{0}^{+}(5)),\ \text{for}\ k\geq 4\ \&\ k\in2\mathbb{Z}.
    \end{split}
\end{align}

$\mathbf{\Gamma_{0}^{+}(7)}:$\\

\noindent The Fricke group of level $7$ is generated by $\Gamma_{0}^{+}(7) = \langle\left(\begin{smallmatrix}1 & 1\\ 0 & 1\end{smallmatrix}\right), W_{7}, \left(\begin{smallmatrix}3 & 1\\ 7 & 2\end{smallmatrix}\right)\rangle$. The elliptic points in the fundamental domain are at $\tfrac{i}{\sqrt{7}}$ (also known as the Fricke involution point), $\rho_{7,1} = -\tfrac{1}{2}+\tfrac{i\sqrt{7}}{10}$, and $\rho_{7,2} = \tfrac{-5 + i\sqrt{3}}{14}$ and this group has a cusp at $\tau = i\infty$. A non-zero $f\in\mathcal{M}_{k}^{!}(\Gamma_{0}^{+}(7))$ satisfies the following valence formula
\begin{equation}\label{valence_formula_Fricke}
    \nu_{\infty}(f) + \frac{1}{2}\nu_{\rho_{F}}(f) + \frac{1}{2}\nu_{k_{2}}(f) + \frac{1}{3}\nu_{\rho_{2}}(f) + \sum\limits_{\substack{p\in\ X_{0}^{+}(7)\\ p\neq \tfrac{i}{\sqrt{7}},\rho_{7,1},\rho_{7,2}}}\nu_{p}(f) = \frac{k}{3}.
\end{equation}
The Hauptmodul and some semi-modular and cusp forms in the Hecke group $\Gamma_{0}(7)$ are defined as follows \cite{Umasankar:2022kzs}
\begin{align}\label{Hecke_def}
    \begin{split}
    j_{7}(\tau) =& \left(\frac{\eta(\tau)}{\eta(7\tau)}\right)^{4},\\
    \theta_{7}(\tau) =& \sqrt{E_{2,7^{'}}(\tau)} = \left(\frac{7E_{2}(7\tau) - E_{2}(\tau)}{6}\right)^{\frac{1}{2}},\\
    \mathbf{k}(\tau) =& \frac{\eta^{7}(\tau)}{\eta(7\tau)},\ \ \ \mathbf{t}(\tau) = \frac{1}{j_{7}(\tau)},\\
    \Delta_{7}(\tau) =& \left(\mathbf{k}\mathbf{t}\right)(\tau)\in\mathcal{S}_{3}(\Gamma_{0}(7))
    \end{split}
\end{align}
We now express the Hauptmodul and some semi-modular and cusp forms in $\Gamma_{0}^{+}(7)$ in terms of the definitions in \ref{Hecke_def}
\begin{align}\label{level_7+}
    \begin{split}
        j_{7^{+}}(\tau) =& \left(\frac{E_{1,7^{'}}^{3}}{\Delta_{7}}\right)(\tau) = \left(\frac{E_{4}^{(7^{+})}}{\Delta_{7^{+},4}}\right) - \frac{16}{5} =\left(j_{7} + \frac{49}{j_{7}} + 13\right)(\tau),\\
        E_{1,7^{'}}(\tau) =& \sqrt{E_{2,7^{'}}(\tau)} = \theta_{7}(\tau),\\
        \Delta_{7^{+},4}(\tau) =& \left(\theta_{7}\mathbf{t}\mathbf{k}\right)(\tau) = \left(\sqrt{E_{2,7^{'}}}\Delta_{7}\right)(\tau)\in\mathcal{S}_{4}(\Gamma_{0}^{+}(7)),\\
        \Delta_{7^{+},10}(\tau) =& \frac{559}{690}\left(\frac{41065}{137592}\left(E^{(7^{+})}_{4}(\tau)E^{(7^{+})}_{6}(\tau) - E^{(7^{+})}_{10}(\tau)\right) - E^{(7^{+})}_{6}(\tau)\Delta_{7^{+},4}(\tau)\right)\\
        =& \left(\Delta_{7^{+},4}^{2}\frac{E_{4,7^{'}}}{E_{2,7^{'}}}\right)(\tau)\in\mathcal{S}_{10}(\Gamma_{0}^{+}(7)),\\
        \Delta_{7^{+},6}(\tau) =& \left(\frac{\Delta_{7^{+},10}}{\Delta_{7^{+},4}}\right)(\tau) = \left(\Delta_{7^{+},4}\frac{E_{4,7^{'}}}{E_{2,7^{'}}}\right)(\tau)\in\mathcal{S}_{6}(\Gamma_{0}^{+}(7)),
    \end{split}
\end{align}
where 
\begin{align}
    \begin{split}
        E_{k,7^{'}}(\tau) \equiv& \frac{7^{\tfrac{k}{2}}E_{2}(7\tau) - E_{k}(\tau)}{7^{\tfrac{k}{2}} - 1},\ \text{for}\ k\in2\mathbb{Z},\\
        E_{k}^{(7^{+})}(\tau) \equiv& \frac{7^{\tfrac{k}{2}}E_{k}(7\tau) + E_{k}(\tau)}{7^{\tfrac{k}{2}} + 1},\ \text{for}\ k\geq 2\ \&\ k\in2\mathbb{Z}.
    \end{split}
\end{align} 
\section{Modular re-parameterization for \texorpdfstring{$\mathbf{\Gamma_{0}^{+}(5)}$}{Γ0(5)+}}\label{appendix:B}
\noindent From \ref{derivative_1}, we find 
\begin{align}
    E_{4}^{(5^{+})}(\tau) = \left(1 - \frac{36}{13}\frac{1}{j_{5^{+}}(\tau)}\right)E_{2,5^{'}}^{2}(\tau).
\end{align}
We now make the following definitions
\begin{align}\label{A5+_def}
    A_{5^{+}}(\tau) \equiv \left(\frac{E_{6}^{(5^{+})}}{E_{2,5^{'}}^{2}}\right)(\tau),\ \ \ P_{5^{+}}(\tau)\equiv \frac{36}{13}\frac{1}{j_{5^{+}}(\tau)},
\end{align}
where we see that the definition of $A_{5^{+}}$ naturally follows from observing the form of the derivative of the Hauptmodul. The derivative of $P_{5^{+}}$ is found to be
\begin{align}
   q\frac{d}{dq}P_{5^{+}}(\tau) = \left(P_{5^{+}}A_{5^{+}}\right)(\tau). 
\end{align}
With this we find the following expressions for $\theta_{q}$ and $\mathcal{D}^{2}$,
\begin{align}
    \begin{split}
        \theta_{q} =& A_{5^{+}}\theta_{P_{5^{+}}},\\
        \mathcal{D}^{2} =& \left(\theta_{q} - \frac{1}{2}E_{2}^{(5^{+})}\right)\theta_{q}\\
        =& A_{5^{+}}^{2}\theta_{P_{5^{+}}}^{2} + \left[E_{2,5^{'}}^{2}\left(\frac{1250}{81}P_{5^{+}}^{2} - \frac{116}{9}(1 - P_{5^{+}})P_{5^{+}} - \frac{3}{2}(1 - P_{5^{+}})^{2}\right) + \frac{A_{5^{+}}^{2}}{1 - P_{5^{+}}}\right]\theta_{P_{5^{+}}}.
    \end{split}
\end{align}
The modular forms $E_{6}^{(5^{+})}(\tau)$
and $E_{2,5^{'}}^{2}(\tau)$ can be expressed in terms of $K_{5^{+}}\equiv \tfrac{22 + 10\sqrt{5}}{j_{5^{+}}}$, but the relationship is more complicated unlike what we observed previously at levels $2$ and $3$. These modular forms are related to the inverse Hauptmodul $K_{5^{+}}$ via Heun functions as shown below
\begin{align}
        E_{6}^{(5^{+})}(\tau) = \sqrt{1 - \frac{143}{9}P_{5^{+}} - \frac{169}{81}P^{2}_{5^{+}}}\ H\ell_{5}^{6}(\tau),\ \ \ 
        E_{2,5^{'}}^{2}(\tau) = H\ell_{5}^{4}(\tau).
\end{align}
Substituting this in the definition \ref{A5+_def}, we set
\begin{align}
    A_{5^{+}}(\tau) \equiv& \frac{1}{9}\sqrt{81 - 1287P_{5^{+}} - 169P_{5^{+}}^{2}}\ H\ell_{5}^{2}(\tau)\\
    =& f(P_{5^{+}})H\ell_{5}^{2}(\tau).
\end{align}
With $\omega_{4}(\tau) = \mu_{1}E_{2,5^{'}}^{2}(\tau) + \mu_{2}\Delta_{5}(\tau)$, the re-parameterized MLDE reads
\begin{align}
    \begin{split}
    \left[\theta_{P_{5^{+}}}^{2} + \left( \frac{4345P_{5^{+}} - 1602P_{5^{+}} - 243}{162f(P_{5^{+}})} + \frac{1}{1 - P_{5^{+}}}\right) \theta_{P_{5^{+}}}+ \frac{36\mu_{1} -13\mu_{2}P_{5^{+}}}{36f(P_{5^{+}})}\right]f(\tau) = 0.
    \end{split}
\end{align}
\section{Useful relations in level 7 groups}\label{appendix:modular_forms}
\noindent Some useful expressions of the $q$-derivatives of forms in $\Gamma_{0}(7)$ and $\Gamma_{0}^{+}(7)$ are listed below
\begin{align}\label{j'_Hecke_7}
    \begin{split}
        &q\frac{d}{dq}j_{7}(\tau) =-\left(j_{7}\theta_{7}^{2}\right)(\tau),\\
        &q\frac{d}{dq}\mathbf{t}(\tau) = \left(\mathbf{t}\theta_{7}^{2}\right)(\tau),\\
        &q\frac{d}{dq}\mathbf{k}(\tau)= \frac{7}{24}\mathbf{k}(\tau)\left(E_{2}(\tau) - E_{2}(7\tau)\right)\\
        &{}\ \ \ \ \ \ \ \ \ \ \ = \mathbf{k}(\tau)\left[E_{2}^{(7^{+})} - \left(\frac{E_{4,7^{'}}}{E_{2,7^{'}}}\right) - E_{2,7^{'}}\frac{13\mathbf{t} + 98\mathbf{t}^{2}}{1 + 13\mathbf{t} + 49\mathbf{t}^{2}}\right](\tau),\\
        &q\frac{d}{dq}\Delta_{7}(\tau) =\left(\Delta_{7}E_{2}^{(7^{+})}\right)(\tau)\\
        &q\frac{d}{dq}E_{2,7^{'}}(\tau) = \frac{2}{3}\left(E_{2,7^{'}}E_{2}^{(7^{+})} - E_{4,7^{'}}\right)(\tau),\\
        &q\frac{d}{dq}E_{4,7^{'}}(\tau) = \frac{19}{8}\left(\frac{7^{3}E_{2}(7\tau)E_{4}(7\tau) - E_{2}(\tau)E_{4}(\tau)}{342} - E_{6,7^{'}}(\tau)\right),\\
        &q\frac{d}{dq}\Delta_{7^{+},4}(\tau) = \frac{4}{3}\Delta_{7^{+},4}(\tau)\left(E_{2}^{(7^{+})} - \frac{1}{4}\frac{E_{4,7^{'}}}{E_{2,7^{'}}}\right)(\tau),\\
        &q\frac{d}{dq}\Delta_{7^{+},10}(\tau) = \Delta_{10,7^{'}}(\tau)\left[2\left(E_{2}^{(7^{+})} - \frac{E_{4,7^{'}}}{E_{2,7^{'}}}\right)(\tau) + q\frac{d}{dq}\log{E_{4,7'}}(\tau)\right]\\
        &{}\ \ \ \ \ \ \ \ \ \ \ \ \ \ \ \ \ \ = \Delta_{10,7^{'}}(\tau)\left[2\left(E_{2}^{(7^{+})} - \frac{E_{4,7^{'}}}{E_{2,7^{'}}}\right)(\tau) + \frac{57}{8E_{4,7^{'}}(\tau)}q\frac{d}{dq}\left(7^{3}E_{4}(7\tau) - E_{4}(\tau)\right)\right],
    \end{split}
\end{align}
where we used $q\tfrac{d}{dq}E_{4}(\tau) = \tfrac{1}{3}\left(E_{2}E_{4} - E_{6}\right)(\tau)$ to obtain the last line. 

\section{Heun function relations in \texorpdfstring{$\mathbf{\Gamma_{0}^{+}(7)}$}{Γ0(7)+} re-parameterization}\label{appendix:level_7+_simplifications}
\noindent We suspect that it should be possible to simplify the covariant derivative $\mathcal{D}^{2}$ by expressing modular forms $E_{k,7^{'}}(\tau)$ in terms of those belonging to $\Gamma_{0}(7)$ since we know that the $\text{SL}(2,\mathbb{Z})$ Eisenstein series can be expressed in terms of $\mathbf{k}(\tau)$ and $\mathbf{t}(\tau)$ as follows \cite{Liu}
\begin{align}\label{forE10}
    \begin{split}
        E_{4}(\tau)=& \mathbf{k}^{\tfrac{4}{3}}\left(1 + 245\mathbf{t} + 2401\mathbf{t}^{2}\right)\left(1 + 13\mathbf{t} + 49\mathbf{t}^{2}\right)^{\tfrac{1}{3}}(\tau),\\
        E_{4}(7\tau) =& \mathbf{k}^{\tfrac{4}{3}}\left(1 + 5\mathbf{t} + \mathbf{t}^{2}\right)\left(1 + 13\mathbf{t} + 49\mathbf{t}^{2}\right)^{\tfrac{1}{3}}(\tau),\\
        E_{6}(\tau) =& \mathbf{k}^{2}\left(1 - 7(5 + 2\sqrt{7}\mathbf{t} - 7^{3}(21 + 8\sqrt{7}))\mathbf{t}^{2}\right)\left(1 - 7(5 - 2\sqrt{7}\mathbf{t} - 7^{3}(21 - 8\sqrt{7}))\mathbf{t}^{2}\right)(\tau),\\
        E_{6}(7\tau) =& \mathbf{k}^{2}\left(1 + (7 + 2\sqrt{7})\mathbf{t} + (21 + 8\sqrt{7})\mathbf{t}^{2}\right)\left(1 + (7 - 2\sqrt{7})\mathbf{t} + (21 - 8\sqrt{7})\mathbf{t}^{2}\right)(\tau).
    \end{split}
\end{align}
From this, we see that the modular forms $E_{2,7^{'}}^{(7^{+})}(\tau)$, $E_{4}^{(7^{+})}(\tau)$, $E_{4,7^{'}}(\tau)$, and it's logarithmic derivative can be rewritten using \ref{forE10} as follows
\begin{align}\label{k,t-relations}
       \begin{split}
            E_{2,7^{'}}(\tau) =& \mathbf{k}^{\tfrac{2}{3}}\left(1 + 13\mathbf{t} + 49\mathbf{t}^{2}\right)^{\tfrac{2}{3}}(\tau),\\
            E_{4}^{(7^{+})}(\tau) =& \frac{1}{5}\mathbf{k}^{\tfrac{4}{3}}\left(5 + 49\mathbf{t} + 245\mathbf{t}^{2}\right)\left(1 + 13\mathbf{t} + 49\mathbf{t}^{2}\right)^{\tfrac{1}{3}}(\tau),\\
            E_{4,7^{'}}(\tau) =& \mathbf{k}^{\tfrac{4}{3}}\left(1 - 49\mathbf{t}^{2}\right)\left(1 + 13\mathbf{t} + 49\mathbf{t}^{2}\right)^{\tfrac{1}{3}}(\tau)\\
            q\frac{d}{dq}\log{E_{4,7^{'}}}(\tau) =& \frac{4}{3}\left[E_{2}^{(7^{+})} - \left(\frac{E_{4,7^{'}}}{E_{2,7^{'}}}\right) - \frac{13\mathbf{t} + 98\mathbf{t}^{2}}{1 + 13\mathbf{t} + 49\mathbf{t}^{2}}E_{2,7^{'}}\right](\tau)\\ &- \left[\frac{\mathbf{t}\left(19208\mathbf{t}^{3} + 4459\mathbf{t}^{2} + 196\mathbf{t} - 13\right)}{(1- 49\mathbf{t}^{2})(1 + 13\mathbf{t}+ 49\mathbf{t}^{2})}\frac{E_{2,7^{'}}}{3}\right](\tau)
       \end{split}
\end{align}
Using these expressions, the covariant derivative $\mathcal{D}^{2}$ now reads
\begin{align}
    \mathcal{D}^{2} = A_{7^{+}}^{2}\theta_{K_{7^{+}}}^{2} + A_{7^{+}}\left(\frac{4}{3}E_{2}^{(7^{+})} - \frac{13 + 196\mathbf{t} + 637\mathbf{t}^{2}}{(1-49\mathbf{t}^{2})(1 + 13\mathbf{t} + 49\mathbf{t}^{2})}E_{2,7^{'}}\right)\theta_{K_{7^{+}}}.
\end{align}
\noindent Now, since $\text{dim}\ \mathcal{M}_{4}(\Gamma_{0}^{+}(7)) = 2$, we make the choice $\omega_{4}(\tau) = \mu_{1}E_{2,7^{'}}^{2} + \mu_{2}\Delta_{7}E_{1,7^{'}}$. From \cite{Sakai2014TheAO}, we have the following expressions for the Eisenstein series and the cusp form in terms of Heun's function,
\begin{align}
    \begin{split}
        \Delta_{7}(\tau) =& \frac{1}{j_{7^{+}}}H\ell_{7}^{3}(\tau),\\    
        E_{2,7^{'}}^{2}(\tau) =& H\ell_{7}^{4}(\tau),\\
        E_{6}^{(7^{+})}(\tau) =& \sqrt{\left(1 + \frac{1}{j_{7^{+}}}\right)\left(1 - \frac{27}{j_{7^{+}}}\right)}\ H\ell_{7}^{6}(\tau),\\
        H\ell_{7}(\tau) \equiv& H\ell\left(-27, -2;\frac{1}{3},\frac{2}{3},1,\frac{1}{2};K_{7^{+}}(\tau)\right).
    \end{split}
\end{align}
We now want to find similar Heun function relations for other modular forms. Let us begin with the cusp form $\Delta_{7^{+},4}(\tau)$. From its definition in \ref{level_7+}, we have
\begin{align}
    \begin{split}
        \Delta_{7^{+},4}(\tau) =& \left(\sqrt{E_{2,7^{'}}}\Delta_{7}\right)(\tau) = \frac{1}{j_{7^{+}}}H\ell_{7}^{5}(\tau).
    \end{split}
\end{align}
Next, consider the Eisenstein series $E_{4}^{(7^{+})}(\tau)$ which is found to possess the following Heun function relation
\begin{align}
    E_{4}^{(7^{+})}(\tau) = \left(E_{2,7^{'}}^{2} - \frac{16}{5}\Delta_{7^{+},4}\right)(\tau) = \left(H\ell_{7}^{4} - \frac{16}{5j_{7^{+}}}H\ell_{7}^{5}\right)(\tau).
\end{align}
Consider now the cusp form $\Delta_{7^{+},10}(\tau)$. From its definition in \ref{level_7+}, we see that the Eisenstein series $E_{10}^{(7^{+})}(\tau)$ is the only modular form we do not possess a Heun function relation of. Consider now the basis decomposition for the space of modular forms of weight $10$ reads
\begin{align}
    \mathcal{M}_{10}(\Gamma_{0}^{+}(7)) = \mathbb{C}E_{4}^{(7^{+})}E_{6}^{(7^{+})}\oplus\mathbb{C}E_{4}^{(7^{+})}\Delta_{7^{+},6}\oplus\mathbb{C}\Delta_{7^{+},10}.
\end{align}
Considering an ansatz with three unique coefficients and comparing its $q$-series expansion with that of $E_{10}^{(7^{+})}(\tau)$, we can find an explicit expression for the latter to be
\begin{align}
    \begin{split}
        E_{10}^{(7^{+})}(\tau) =& \left(E_{4}^{(7^{+})}E_{6}^{(7^{+})}- \frac{137592}{41065}\frac{\Delta_{7^{+},4}}{E_{2,7^{'}}}E_{4}^{(7^{+})}E_{4,7^{'}} - \frac{63504}{4775}\frac{\Delta_{7^{+},4}^{2}}{E_{2,7^{'}}}E_{4,7^{'}}\right)(\tau)\\
        =& \left(E_{4}^{(7^{+})}E_{6}^{(7^{+})}- \frac{137592}{41065}\frac{\Delta_{7^{+},4}}{E_{2,7^{'}}}\left(E_{4}^{(7^{+})}\right)^{2}\left(\frac{5(1-49\mathbf{t}^{2})}{(5 + 49\mathbf{t} + 245\mathbf{t}^{2})}\right)\right.\\
        &{}\ \ \ \ \ \ \ \ \ \ \ \ \ \ \ \ \ \ \ \ \ \ \ \ \left.- \frac{63504}{4775}\frac{\Delta_{7^{+},4}^{2}}{E_{2,7^{'}}}E_{4}^{7^{+}}\left(\frac{5(1-49\mathbf{t}^{2})}{(5 + 49\mathbf{t} + 245\mathbf{t}^{2})}\right)\right)(\tau),
    \end{split}
\end{align}
where we have used expressions in \ref{k,t-relations} to express $E_{4,7^{'}}(\tau)$ in terms of $E_{4}^{(7^{+})}(\tau)$. Using this in the definition of $\Delta_{7^{+},10}(\tau)$, we find
\begin{align}
    \Delta_{7^{+},10}(\tau) =& \left[\frac{559}{690}\left(\frac{5(j_{7} - 49)}{245 + 49j_{7} + 5j_{7}^{2}} - \sqrt{\left(1 - \frac{1}{j_{7^{+}}}\right)\left(1 - \frac{27}{j_{7^{+}}}\right)}\right)\frac{1}{j_{7^{+}}}H\ell_{7}^{11}\right.\\
    &\ \ \ \ \ \ \ \ \ \left.- \frac{5(j_{7} - 49)}{245 + 49j_{7} + 5j_{7}^{2}}\left(\frac{688}{345j_{7^{+}}^{3}}H\ell_{7}^{13} + \frac{3397}{1725j_{7^{+}}^{2}}H\ell_{7}^{12}\right)\right](\tau),
\end{align}
where we have expressed all $\mathbf{t}(\tau)$ in terms of $j_{7}(\tau)$, the Hauptmodul of $\Gamma_{0}(7)$. Using these relations, it should be possible to express the re-parameterized MLDE in terms of the Hauptmodul $j_{7^{+}}(\tau)$. 
From this, we find the following useful relations
\begin{align}
    \begin{split}
        \left(\frac{\Delta_{7}^{4}}{E_{2,7^{'}}^{6}}\right)(\tau) =& \mathfrak{a} = \frac{1}{j_{7^{+}}(\tau)},\\
        \left(\frac{\left(E_{6}^{(7^{+})}\right)^{2}}{\Delta_{7}^{2}}\right)(\tau) =& \mathfrak{b} = \left(j_{7^{+}}^{2} - 26j_{7^{+}} - 27\right)(\tau),\\
        \left(\frac{\left(E_{6}^{(7^{+})}\right)^{2}}{E^{6}_{2,7^{'}}}\right)(\tau) =& \mathfrak{c}  = \left(1 - \frac{27}{j_{7^{+}}^{2}} - \frac{26}{j_{7^{+}}}\right)(\tau).
    \end{split}
\end{align}

\section{Group data}
\begin{table}[htb!]
    \centering
    \begin{tabular}{||c|c|c|c|c|c|c|c|c|c|c|c|c|c|c||}
    \hline
     $p$ & 2 & 3 & 5 & 7 & 11 & 13 & 17 & 19 & 23 & 29 & 31 & 41 & 59 & 71 \\ [0.5ex]
     \hline\hline
     $\overline{\mu}_{0}^{+}$ & $\tfrac{3}{2}$ & 2 & 3 & 4 & 6 & 7 & 9 & 10 & 12 & 15 & 16 & 21 & 30 & 36\\ [1ex]
     \hline
    \end{tabular}
    \caption{Index data for Fricke groups of prime divisor levels of $\mathbb{M}$ with $\overline{\mu}_{0}^{+} = \left[\text{PSL}(2,\mathbb{Z}),\overline{\Gamma}_{0}^{+}(p)\right]$}
    \label{tab:Fricke_index_data}
\end{table}

\begin{table}[htb!]
    \centering
    \begin{tabular} {||c|c|c|c|c|c|c|c|c|c|c|c|c|c|c|c|c|c|c|c|c||}
    \hline
     $p$ & 2 & 3 & 4 & 5 & 6 & 7 & 8 & 9 & 10 & 11 & 12 & 13 & 17 & 19 & 23 & 29 & 31 & 41 & 59 & 71\\ [0.5ex]
     \hline\hline
     $\overline{\mu}_{0}$ & 3 & 4 & 6 & 6 & 12 & 8 & 12 & 12 & 18 & 12 & 24 & 14 & 18 & 20 & 24 & 30 & 32 & 42 & 60 & 72\\ [1ex]
     \hline
    \end{tabular}
    \caption{Index data for Fricke groups of prime divisor levels of $\mathbb{M}$ with $\overline{\mu}_{0} = \left[\text{PSL}(2,\mathbb{Z}),\overline{\Gamma}_{0}(p)\right]$}
    \label{tab:Hecke_index_data}
\end{table}
\section{\texorpdfstring{$(2,0)\ \mathbf{\Gamma_{0}^{+}(2)}$}{(2,0)} 
 MLDE specifics}\label{appendix:detailed_derivation}
\noindent Substituting mode expansions in the MLDE, we obtain
\begin{align}\label{MLDE_mode_Gamma_0_2+}
    q^{\alpha}\sum\limits_{n=0}^{\infty}f_{n}q^{n}\left[(n+\alpha)^{2} - (n + \alpha)\sum\limits_{m=0}^{\infty}\frac{1}{4}E^{(2^{+})}_{2,m}q^{m} + \mu\sum\limits_{m=0}^{\infty} E_{4,m}^{(2^{+})}q^{m}\right] = 0.
\end{align}
When $n=0,m=0$, with $E^{(2^{+})}_{2,0} = E^{(2^{+})}_{4,0} = 1$, we obtain the following indicial equation
\begin{align}
    \alpha^{2} -\frac{1}{4} \alpha +  \mu = 0.
\end{align}
The parameter $\mu$ can now be fixed using this equation. Let us denote the roots of this equation, in increasing order, as $\alpha_{0}$ and $\alpha_{1}$. Solving this indicial equation gives us
\begin{align}\label{roots_n=2_l=0_Gamma_0_2+}
    \begin{split}
        \alpha_{0} =& \frac{1}{8}\left(1 - \sqrt{1 - 64\mu}\right) \equiv \frac{1}{8}(1 - x),\\
        \alpha_{1} =& \frac{1}{12}\left(1 + \sqrt{1 - 64\mu}\right) \equiv \frac{1}{8}(1 + x),
    \end{split}
\end{align}
where we have set $x = \sqrt{1 - 64\mu}$. The smaller solution $\alpha_{0} = \tfrac{1}{8}(1-x)$ corresponds to the identity character which behaves as $f(\tau) \sim q^{\tfrac{1-x}{8}}\left(1 + \mathcal{O}(q)\right)$. We know that the identity character $\chi_{0}$, associated with a primary of weight $h_{0} = 0$ behaves as $\chi_{0}\sim q^{-\tfrac{c}{24}}\left(1 + \mathcal{O}(q)\right)$. Comparing the two behaviours, we obtain the following expression for the central charge
\begin{align}\label{central_charge_x}
    c = 3(x-1).
\end{align}
To find the conformal dimension $h$, we compare the behaviours with the larger solution for $\alpha$, i.e. $f(\tau)\sim q^{\tfrac{1 + x}{8}}\left(1  +\mathcal{O}(q)\right)$ and $\chi\sim q^{-\tfrac{c}{24} + h}\left(1 + \mathcal{O}(q)\right)$. This gives us
\begin{align}
    h = \frac{x}{4}.
\end{align}
Next, to obtain a recurrence relation among the coefficients $f_{n}$, we use the Cauchy product of two infinite series which reads
\begin{align}
    \left(\sum\limits_{i=0}^{\infty}\alpha_{i}\right)\cdot\left(\sum\limits_{j=0}^{\infty}\beta_{j}\right) = \sum\limits_{k=0}^{\infty}\gamma_{k},\ \gamma_{k} = \sum\limits_{p=0}^{k}\alpha_{p}\beta_{k-p}.
\end{align}
This gives us
\begin{align}
    \begin{split}
         \left(\sum\limits_{m=0}^{\infty}E^{(2^{+})}_{4,m}(\tau)q^{m}\right)\left(\sum\limits_{n=0}^{\infty}f_{n}q^{n}\right) =& \sum\limits_{k=0}^{\infty}\left(\sum\limits_{p=0}^{k}E^{(2^{+})}_{4,m}f_{k-p}\right)q^{k},\\
        \left(\sum\limits_{m=0}^{n}E^{(2^{+})}_{2,m}q^{m}\right)\left(\sum\limits_{n=0}^{\infty}f_{n}q^{n}(n+\alpha)\right) =& \sum\limits_{k=0}^{\infty}\left(\sum\limits_{p=0}^{k}E^{(2^{+})}_{2,p}(k+\alpha - p)f_{k-p}\right)q^{k}.
    \end{split}
\end{align}
Substituting this back into the MLDE, we get
\begin{align}\label{recursion_l=0_Gamma_0_2+}
    f_{n} = \left((n + \alpha)^{2} - \frac{1}{4}(n + \alpha) + \mu\right)^{-1}\sum\limits_{m=1}^{n}\left(\frac{(n + \alpha - m)}{4}E^{(2^{+})}_{2,m} - \mu E^{(2^{+})}_{4,m}\right)f_{n-m}.
\end{align}
When $n = 1$, we solve for the ratio $\tfrac{f_{1}}{f_{0}}$ in terms of $\alpha$ with coefficients $E^{(2^{+})}_{2,1} = -8$ and $E_{4,1}^{(2^{+})} = 48$ to obtain 
\begin{align}
    m^{(i)}_{1} = \frac{f^{(i)}_{1}}{f^{(i)}_{0}} = \frac{8\alpha_{i}(-7 + 24\alpha_{i})}{3 + 8\alpha_{i}},
\end{align}
for $ i = 0,1$ corresponding to ratios of $m^{(0)}_{1}$ and $m^{(1)}_{1}$ taken with respect to values $\alpha_{0}$ and $\alpha_{1}$ respectively. Restricting to $i=0$ and assuming that $\alpha_{i} = -\tfrac{c}{24}$, we find
\begin{align}
    m_{1}^{(0)} = \frac{c(7+c)}{9-c}.
\end{align}
Dropping the script associated with the index $i$, we see that for $m_{1}\geq 0$, we require $c<9$. Rewriting this, we have
\begin{align}
    c^{2} + c(m_{1} + 7) = 9m_{1}.
\end{align}
This tells us that $c$ is an integer. Hence, for two character theories with $\ell = 0$, we have $c\in\mathbb{Z}$ and the bound $c< 9$. Solving the quadratic equation, we notice that for $c$ to be rational, the following determinant has to be rational
\begin{align}
    \sqrt{49 + 50m_{1} + m_{1}^{2}}.
\end{align}
Now, since we require $m_{1}\in\mathbb{Z}$, we must demand that the square root is an integer and this gives us
\begin{align}\label{recast_prior}
    \left(m_{1} + 25\right)^{2} - 576 = k^{2},
\end{align}
where $k^{2}\in\mathbb{Z}$. We define $p$ such that we shift $k$ by an integer amount that absorbs the first two terms on the left-hand side
\begin{align}
    p = 25 + m_{1} - k,
\end{align}
and we recast \ref{recast_prior} as follows
\begin{align}
    m_{1} + 25 = \frac{576 + p^{2}}{2p} = \frac{288}{p} + \frac{p}{2}.
\end{align}
This tells us that $p$ must be even and that it must divide $288$. Restricting $k$ to be positive, we see that we have $k\geq m_{1}$ which implies that $p<25$. We conclude that all possible values of $m_{1}$ are found by those values of $p$ below $26$ that divide $288$ and are even. The list of these values is $p = \{2,4,6,8,12,16,18,24\}$. We note that at $p = 24$, we obtain $(m_{1}, c, h) = (-1, -3,0)$ which we ignore since we only want unitary theories. Table \ref{tab:theory_Fricke_2+} contains CFT data.
\begin{table}[htb!]
    \centering
    \begin{tabular}{||c|c|c|c|c||}
    \hline
    $p$ & $\mu$ & $m_{1}$ & $c$ & $h$\\ [0.5ex]
    \hline\hline
    2  & $-\tfrac{7}{36}$ & 120 & 8 & $\tfrac{11}{12}$\\[0.5ex]
    4 & $-\tfrac{91}{576}$ & 49 & 7 & $\tfrac{5}{6}$\\[0.5ex]
    6 & $-\tfrac{1}{8}$ & 26 & 6 & $\tfrac{3}{4}$\\[0.5ex]
    8 &$-\tfrac{55}{576}$ & 15 & 5 & $\tfrac{2}{3}$\\[0.5ex]
    12 & $-\tfrac{3}{64}$ & 5 & 3 & $\tfrac{1}{2}$\\[0.5ex]
    16 & $-\tfrac{7}{576}$ & 1 & 1 & $\tfrac{1}{3}$\\[0.5ex]
    18 & $0$ & 0 & 0 & $\tfrac{1}{4}$\\[1ex]
    \hline
    \end{tabular}
    \caption{$c$ and $h$ data corresponding to the Fricke group $\Gamma_{0}^{+}(2)$ for the choice $\phi_{0} = \mu E^{(2^{+})}_{4}(\tau)$ with $\ell = 0$.}
    \label{tab:theory_Fricke_2+}
\end{table}
\noindent
The ratio $m_{1}$ being non-negative is not sufficient proof to convince ourselves that this data might indeed be related to a CFT. Hence, we compute $m_{2}$ using recursion relation \ref{recursion_l=0_Gamma_0_2+}. If this turns out to be negative or fractional, then we rule out the theory as a viable candidate. When $n = 2$, we get
\begin{align}
     m_{2}^{(i)} \equiv \frac{f_{2}^{(i)}}{f_{0}^{(i)}} = \frac{4\alpha_{i}(312\alpha_{i} -83)}{8\alpha_{i} + 7} + \frac{4m_{1}^{(i)}(-1 + \alpha_{i}(-7 + 24\alpha_{i}))}{8\alpha_{i} + 7}.
\end{align}
The values of $m_{2}$ for $i = 0$ are tabulated in table
\ref{tab:theory_Fricke_2+_m2}.
\begin{table}[htb!]
    \centering
    \begin{tabular}{||c|c|c||}
    \hline
    $p$ & $c$ & $m_{2}$\\ [0.5ex]
    \hline\hline
    2  & 8 & $\tfrac{6508}{13}$\\[0.5ex]
    4 & 7 & 173\\[0.5ex]
    6 & 6 & 79\\[0.5ex]
    8 & 5 & 40\\[0.5ex]
    12 & 3 & 11\\[0.5ex]
    16 & 1 & 2\\[0.5ex]
    18 & 0 & 0\\[1ex]
    \hline
    \end{tabular}
    \caption{Values of $m_{2}$ for corresponding to the Fricke group $\Gamma_{0}^{+}(2)$ for the choice $\phi_{0} = \mu E^{(2^{+})}_{4}(\tau)$ with $\ell = 0$.}
    \label{tab:theory_Fricke_2+_m2}
\end{table}
\noindent
We discard the case $p = 2$ since the coefficient turns out to be fractional. Furthermore, we also discard the case $p = 18$ since further computation yields all coefficients $m_{i}$ to be null valued which makes this a trivial solution. We have now restricted to cases $p = \{4,6,8,12,16\}$. Next, we compute $m_{3}$ using recursion relation \ref{recursion_l=0_Gamma_0_2+} and this turns out to be
\begin{align}
     m_{3}^{(i)} \equiv \frac{f_{3}^{(i)}}{f_{0}^{(i)}} = \frac{32\alpha_{i}(168\alpha_{i} -43)}{24\alpha_{i} + 33} + \frac{8m_{1}^{(i)}(-5 + \alpha_{i}(-83 + 312\alpha_{i}))}{24\alpha_{i} + 33} + \frac{8m_{2}^{(i)}(-2 + \alpha_{i}(-7 + 24\alpha_{i}))}{24\alpha_{i} + 33}.
\end{align}
\noindent
The integral values of $m_{3}$ for $i = 0$ are shown in table \ref{tab:theory_Fricke_2+_m3}.
\begin{table}[htb!]
    \centering
    \begin{tabular}{||c|c|c||}
    \hline
    $p$ & $c$ & $m_{2}$\\ [0.5ex]
    \hline\hline
    6 & 6 & 326\\[0.5ex]
    8 & 5 & 135\\[0.5ex]
    12 & 3 & 20\\ [0.5ex]
    16 & 1 & 1\\[0.5ex]
    \hline
    \end{tabular}
    \caption{Values of $m_{3}$ for corresponding to the Fricke group $\Gamma_{0}^{+}(2)$ for the choice $\phi_{0} = \mu E^{(2^{+})}_{4}(\tau)$ with $\ell = 0$.}
    \label{tab:theory_Fricke_2+_m3}
\end{table}
\noindent
In a similar fashion, computing coefficients for $i = 0,1$, we obtain the coefficients of the two characters.

\bibliographystyle{utphys}
\bibliography{MLDE}

\providecommand{\href}[2]{#2}\begingroup\raggedright\begin{thebibliography}{10}

\bibitem{Umasankar:2022kzs}
N.~B. Umasankar, ``{Modular Linear Differential Equations for Hecke and Fricke Groups},'' \href{http://arxiv.org/abs/2210.07186}{{\ttfamily arXiv:2210.07186 [hep-th]}}.

\bibitem{Anderson:1987ge}
G.~Anderson and G.~W. Moore, ``{Rationality in Conformal Field Theory},''
\href{http://dx.doi.org/10.1007/BF01223375}{{\em Commun. Math. Phys.} {\bfseries 117} (1988) 441}.
%%CITATION = CMPHA,117,441;%%.

\bibitem{Moore:1988qv}
G.~W. Moore and N.~Seiberg, ``{Classical and Quantum Conformal Field Theory},'' \href{http://dx.doi.org/10.1007/BF01238857}{{\em Commun. Math. Phys.} {\bfseries 123} (1989) 177}.

\bibitem{Mathur:1988na}
S.~D. Mathur, S.~Mukhi, and A.~Sen, ``{On the Classification of Rational Conformal Field Theories},''
\href{http://dx.doi.org/10.1016/0370-2693(88)91765-0}{{\em Phys. Lett.} {\bfseries B213} (1988) 303}.
%%CITATION = PHLTA,B213,303;%%.

\bibitem{Schellekens:1996tg}
A.~N. Schellekens, ``{Introduction to conformal field theory},'' {\em Fortsch. Phys.} {\bfseries 44} (1996) 605--705.

\bibitem{Hampapura:2015cea}
H.~R. Hampapura and S.~Mukhi, ``{On 2d Conformal Field Theories with Two Characters},'' \href{http://dx.doi.org/10.1007/JHEP01(2016)005}{{\em JHEP} {\bfseries 01} (2016) 005},
\href{http://arxiv.org/abs/1510.04478}{{\ttfamily arXiv:1510.04478 [hep-th]}}.
%%CITATION = ARXIV:1510.04478;%%.

\bibitem{Das:2022uoe}
A.~Das, C.~N. Gowdigere, and S.~Mukhi, ``{Meromorphic cosets and the classification of three-character CFT},'' \href{http://dx.doi.org/10.1007/JHEP03(2023)023}{{\em JHEP} {\bfseries 03} (2023) 023}, \href{http://arxiv.org/abs/2212.03136}{{\ttfamily arXiv:2212.03136 [hep-th]}}.

\bibitem{Bantay:2005vk}
P.~Bantay and T.~Gannon, ``{Conformal characters and the modular representation},'' \href{http://dx.doi.org/10.1088/1126-6708/2006/02/005}{{\em JHEP} {\bfseries 02} (2006) 005},
\href{http://arxiv.org/abs/hep-th/0512011}{{\ttfamily arXiv:hep-th/0512011 [hep-th]}}.
%%CITATION = HEP-TH/0512011;%%.

\bibitem{Mason:2007}
G.~Mason, ``{Vector-valued Modular Forms and Linear Differential Operators},'' \href{http://dx.doi.org/10.1142/S1793042107000973}{{\em International Journal of Number Theory} {\bfseries 03} (2007) 377--390}.

\bibitem{Gannon:2013jua}
T.~Gannon, ``{The theory of vector-modular forms for the modular group},'' \href{http://dx.doi.org/10.1007/978-3-662-43831-2_9}{{\em Contrib. Math. Comput. Sci.} {\bfseries 8} (2014) 247--286},
\href{http://arxiv.org/abs/1310.4458}{{\ttfamily arXiv:1310.4458 [math.NT]}}.
%%CITATION = ARXIV:1310.4458;%%.

\bibitem{Mathur:1988rx}
S.~D. Mathur, S.~Mukhi, and A.~Sen, ``{Differential Equations for Correlators and Characters in Arbitrary Rational Conformal Field Theories},''
\href{http://dx.doi.org/10.1016/0550-3213(89)90022-9}{{\em Nucl. Phys.} {\bfseries B312} (1989) 15}.
%%CITATION = NUPHA,B312,15;%%.

\bibitem{Mathur:1988gt}
S.~D. Mathur, S.~Mukhi, and A.~Sen, ``{Reconstruction of Conformal Field Theories From Modular Geometry on the Torus},''
\href{http://dx.doi.org/10.1016/0550-3213(89)90615-9}{{\em Nucl. Phys.} {\bfseries B318} (1989) 483}.
%%CITATION = NUPHA,B318,483;%%.

\bibitem{Naculich:1988xv}
S.~G. Naculich, ``{Differential Equations for Rational Conformal Characters},''
\href{http://dx.doi.org/10.1016/0550-3213(89)90150-8}{{\em Nucl. Phys.} {\bfseries B323} (1989) 423}.
%%CITATION = NUPHA,B323,423;%%.

\bibitem{Kiritsis:1988kq}
E.~B. Kiritsis, ``{Fuchsian Differential Equations for Characters on the Torus: A Classification},''
\href{http://dx.doi.org/10.1016/0550-3213(89)90475-6}{{\em Nucl. Phys.} {\bfseries B324} (1989) 475}.
%%CITATION = NUPHA,B324,475;%%.

\bibitem{Mason:2008}
G.~Mason, ``{2-Dimensional vector-valued modular forms},'' \href{http://dx.doi.org/10.1007/s11139-007-9054-4}{{\em The Ramanujan Journal} {\bfseries 17} no.~3, (Dec, 2008) 405--427}. \url{https://doi.org/10.1007/s11139-007-9054-4}.

\bibitem{Bantay:2007zz}
P.~Bantay and T.~Gannon, ``{Vector-valued modular functions for the modular group and the hypergeometric equation},''
\href{http://dx.doi.org/10.4310/CNTP.2007.v1.n4.a2}{{\em Commun. Num. Theor. Phys.} {\bfseries 1} (2007) 651--680}.
%%CITATION = 00649,1,651;%%.

\bibitem{Tuite:2008pt}
M.~P. Tuite, ``{Exceptional Vertex Operator Algebras and the Virasoro Algebra},'' {\em Contemp. Math.} {\bfseries 497} (2009) 213--225,
\href{http://arxiv.org/abs/0811.4523}{{\ttfamily arXiv:0811.4523 [math.QA]}}.
%%CITATION = ARXIV:0811.4523;%%.

\bibitem{Bantay:2010uy}
P.~Bantay, ``{Modular differential equations for characters of RCFT},'' \href{http://dx.doi.org/10.1007/JHEP06(2010)021}{{\em JHEP} {\bfseries 06} (2010) 021},
\href{http://arxiv.org/abs/1004.2579}{{\ttfamily arXiv:1004.2579 [hep-th]}}.
%%CITATION = ARXIV:1004.2579;%%.

\bibitem{Marks:2011}
C.~Marks, ``{Irreducible vector-valued modular forms of dimension less than six},'' {\em Illinois J. Math.} {\bfseries 55} no.~4, (2011) 1267--1297. \url{https://projecteuclid.org:443/euclid.ijm/1373636684}.

\bibitem{Kawasetsu:2014}
K.~Kawasetsu, ``{The Intermediate Vertex Subalgebras of the Lattice Vertex Operator Algebras},'' \href{http://dx.doi.org/10.1007/s11005-013-0658-x}{{\em Letters in Mathematical Physics} {\bfseries 104} no.~2, (Feb, 2014) 157--178}. \url{https://doi.org/10.1007/s11005-013-0658-x}.

\bibitem{Franc:2016}
C.~Franc and G.~Mason, ``{Hypergeometric Series, Modular Linear Differential Equations, and Vector-valued Modular Forms},'' {\em Ramanujan Journal} {\bfseries 41} (2016) 233.

\bibitem{Gaberdiel:2016zke}
M.~R. Gaberdiel, H.~R. Hampapura, and S.~Mukhi, ``{Cosets of Meromorphic CFTs and Modular Differential Equations},'' {\em JHEP} {\bfseries 04} (2016) 156,
\href{http://arxiv.org/abs/1602.01022}{{\ttfamily arXiv:1602.01022 [hep-th]}}.
%%CITATION = ARXIV:1602.01022;%%.

\bibitem{Hampapura:2016mmz}
H.~R. Hampapura and S.~Mukhi, ``{Two-dimensional RCFT's without Kac-Moody symmetry},'' \href{http://dx.doi.org/10.1007/JHEP07(2016)138}{{\em JHEP} {\bfseries 07} (2016) 138},
\href{http://arxiv.org/abs/1605.03314}{{\ttfamily arXiv:1605.03314 [hep-th]}}.
%%CITATION = ARXIV:1605.03314;%%.

\bibitem{Arike:2016ana}
Y.~Arike, M.~Kaneko, K.~Nagatomo, and Y.~Sakai, ``{Affine Vertex Operator Algebras and Modular Linear Differential Equations},''
\href{http://dx.doi.org/10.1007/s11005-016-0837-7}{{\em Lett. Math. Phys.} {\bfseries 106} no.~5, (2016) 693--718}.
%%CITATION = LMPHD,106,693;%%.

\bibitem{Tener:2016lcn}
J.~E. Tener and Z.~Wang, ``{On classification of extremal non-holomorphic conformal field theories},'' \href{http://dx.doi.org/10.1088/1751-8121/aa59cd}{{\em J. Phys.} {\bfseries A50} no.~11, (2017) 115204},
\href{http://arxiv.org/abs/1611.04071}{{\ttfamily arXiv:1611.04071 [math-ph]}}.
%%CITATION = ARXIV:1611.04071;%%.

\bibitem{Mason:2018}
G.~{Mason}, K.~{Nagatomo}, and Y.~{Sakai}, ``{Vertex Operator Algebras with Two Simple Modules - the Mathur-Mukhi-Sen Theorem Revisited},'' \href{http://arxiv.org/abs/1803.11281}{{\ttfamily arXiv:1803.11281 [math.QA]}}.

\bibitem{Harvey:2018rdc}
J.~A. Harvey and Y.~Wu, ``{Hecke Relations in Rational Conformal Field Theory},'' \href{http://dx.doi.org/10.1007/JHEP09(2018)032}{{\em JHEP} {\bfseries 09} (2018) 032},
\href{http://arxiv.org/abs/1804.06860}{{\ttfamily arXiv:1804.06860 [hep-th]}}.
%%CITATION = ARXIV:1804.06860;%%.

\bibitem{Chandra:2018pjq}
A.~R. Chandra and S.~Mukhi, ``{Towards a Classification of Two-Character Rational Conformal Field Theories},'' \href{http://dx.doi.org/10.1007/JHEP04(2019)153}{{\em JHEP} {\bfseries 04} (2019) 153},
\href{http://arxiv.org/abs/1810.09472}{{\ttfamily arXiv:1810.09472 [hep-th]}}.
%%CITATION = ARXIV:1810.09472;%%.

\bibitem{Chandra:2018ezv}
A.~R. Chandra and S.~Mukhi, ``{Curiosities above c = 24},'' \href{http://dx.doi.org/10.21468/SciPostPhys.6.5.053}{{\em SciPost Phys.} {\bfseries 6} no.~5, (2019) 053},
\href{http://arxiv.org/abs/1812.05109}{{\ttfamily arXiv:1812.05109 [hep-th]}}.
%%CITATION = ARXIV:1812.05109;%%.

\bibitem{Bae:2018qfh}
J.-B. Bae, K.~Lee, and S.~Lee, ``{Monster Anatomy},'' \href{http://dx.doi.org/10.1007/JHEP07(2019)026}{{\em JHEP} {\bfseries 07} (2019) 026}, \href{http://arxiv.org/abs/1811.12263}{{\ttfamily arXiv:1811.12263 [hep-th]}}.

\bibitem{Bae:2018qym}
J.-B. Bae, S.~Lee, and J.~Song, ``{Modular Constraints on Superconformal Field Theories},'' \href{http://dx.doi.org/10.1007/JHEP01(2019)209}{{\em JHEP} {\bfseries 01} (2019) 209}, \href{http://arxiv.org/abs/1811.00976}{{\ttfamily arXiv:1811.00976 [hep-th]}}.

\bibitem{franc2020classification}
C.~Franc and G.~Mason, ``Classification of some vertex operator algebras of rank 3,'' {\em Algebra \& Number Theory} {\bfseries 14} no.~6, (2020) 1613--1667.

\bibitem{Bae:2020xzl}
J.-B. Bae, Z.~Duan, K.~Lee, S.~Lee, and M.~Sarkis, ``{Fermionic rational conformal field theories and modular linear differential equations},'' \href{http://dx.doi.org/10.1093/ptep/ptab033}{{\em PTEP} {\bfseries 2021} no.~8, (2021) 08B104}, \href{http://arxiv.org/abs/2010.12392}{{\ttfamily arXiv:2010.12392 [hep-th]}}.

\bibitem{Mukhi:2020gnj}
S.~Mukhi, R.~Poddar, and P.~Singh, ``{Rational CFT with three characters: the quasi-character approach},'' \href{http://dx.doi.org/10.1007/JHEP05(2020)003}{{\em JHEP} {\bfseries 05} (2020) 003}, \href{http://arxiv.org/abs/2002.01949}{{\ttfamily arXiv:2002.01949 [hep-th]}}.

\bibitem{Kaidi:2020ecu}
J.~Kaidi and E.~Perlmutter, ``{Discreteness and integrality in Conformal Field Theory},'' \href{http://dx.doi.org/10.1007/JHEP02(2021)064}{{\em JHEP} {\bfseries 02} (2021) 064}, \href{http://arxiv.org/abs/2008.02190}{{\ttfamily arXiv:2008.02190 [hep-th]}}.

\bibitem{Das:2020wsi}
A.~Das, C.~N. Gowdigere, and J.~Santara, ``{Wronskian Indices and Rational Conformal Field Theories},'' \href{http://dx.doi.org/10.1007/JHEP04(2021)294}{{\em JHEP} {\bfseries 04} (2021) 294}, \href{http://arxiv.org/abs/2012.14939}{{\ttfamily arXiv:2012.14939 [hep-th]}}.

\bibitem{Das:2021uvd}
A.~Das, C.~N. Gowdigere, and J.~Santara, ``{Classifying three-character RCFTs with Wronskian index equalling 0 or 2},'' \href{http://dx.doi.org/10.1007/JHEP11(2021)195}{{\em JHEP} {\bfseries 11} (2021) 195}, \href{http://arxiv.org/abs/2108.01060}{{\ttfamily arXiv:2108.01060 [hep-th]}}.

\bibitem{Mukhi:2022bte}
S.~Mukhi and B.~C. Rayhaun, ``{Classification of Unitary RCFTs with Two Primaries and Central Charge Less Than 25},'' \href{http://arxiv.org/abs/2208.05486}{{\ttfamily arXiv:2208.05486 [hep-th]}}.

\bibitem{Das:2022slz}
A.~Das, C.~N. Gowdigere, and S.~Mukhi, ``{New meromorphic CFTs from cosets},'' \href{http://dx.doi.org/10.1007/JHEP07(2022)152}{{\em JHEP} {\bfseries 07} (2022) 152}, \href{http://arxiv.org/abs/2207.04061}{{\ttfamily arXiv:2207.04061 [hep-th]}}.

\bibitem{Kaidi:2021ent}
J.~Kaidi, Y.-H. Lin, and J.~Parra-Martinez, ``{Holomorphic modular bootstrap revisited},'' \href{http://dx.doi.org/10.1007/JHEP12(2021)151}{{\em JHEP} {\bfseries 12} (2021) 151}, \href{http://arxiv.org/abs/2107.13557}{{\ttfamily arXiv:2107.13557 [hep-th]}}.

\bibitem{Bae:2021mej}
J.-B. Bae, Z.~Duan, K.~Lee, S.~Lee, and M.~Sarkis, ``{Bootstrapping fermionic rational CFTs with three characters},'' \href{http://dx.doi.org/10.1007/JHEP01(2022)089}{{\em JHEP} {\bfseries 01} (2022) 089}, \href{http://arxiv.org/abs/2108.01647}{{\ttfamily arXiv:2108.01647 [hep-th]}}.

\bibitem{Duan:2022ltz}
Z.~Duan, K.~Lee, and K.~Sun, ``{Hecke relations, cosets and the classification of 2d RCFTs},'' \href{http://dx.doi.org/10.1007/JHEP09(2022)202}{{\em JHEP} {\bfseries 09} (2022) 202}, \href{http://arxiv.org/abs/2206.07478}{{\ttfamily arXiv:2206.07478 [hep-th]}}.

\bibitem{Rayhaun:2023pgc}
B.~C. Rayhaun, ``{Bosonic Rational Conformal Field Theories in Small Genera, Chiral Fermionization, and Symmetry/Subalgebra Duality},'' \href{http://arxiv.org/abs/2303.16921}{{\ttfamily arXiv:2303.16921 [hep-th]}}.

\bibitem{Gowdigere:2023xnm}
C.~N. Gowdigere, S.~Kala, and J.~Santara, ``{Classifying three-character RCFTs with Wronskian index equalling 3 or 4},'' \href{http://arxiv.org/abs/2308.01149}{{\ttfamily arXiv:2308.01149 [hep-th]}}.

\bibitem{Pan:2023jjw}
Y.~Pan and Y.~Wang, ``{Flavored modular differential equations},'' \href{http://arxiv.org/abs/2306.10569}{{\ttfamily arXiv:2306.10569 [hep-th]}}.

\bibitem{Schellekens:1992db}
A.~N. Schellekens, ``{Meromorphic c = 24 conformal field theories},'' \href{http://dx.doi.org/10.1007/BF02099044}{{\em Commun. Math. Phys.} {\bfseries 153} (1993) 159--186},
\href{http://arxiv.org/abs/hep-th/9205072}{{\ttfamily arXiv:hep-th/9205072 [hep-th]}}.
%%CITATION = HEP-TH/9205072;%%.

\bibitem{kainberger}
P.~Kainberger, ``Drawing fundamental domains for congruence subgroups using mathematica,''.

\bibitem{Lin2021QuasimodularFA}
C.~Lin and Y.~Yang, ``Quasimodular forms and modular differential equations which are not apparent at cusps: I,''
\newblock 2021.

\bibitem{Junichi}
J.~Shigezumi, ``{On the zeros of certain modular functions for the normalizers of congruence subgroups of low levels I},'' \href{http://arxiv.org/abs/0802.1307}{{\ttfamily arXiv:0802.1307 [math.NT]}}.

\bibitem{DiFrancesco:1997nk}
P.~Di~Francesco, P.~Mathieu, and D.~Senechal, \href{http://dx.doi.org/10.1007/978-1-4612-2256-9}{{\em {Conformal Field Theory}}}.
\newblock Graduate Texts in Contemporary Physics. Springer-Verlag, New York, 1997.
\newblock
\url{http://www-spires.fnal.gov/spires/find/books/www?cl=QC174.52.C66D5::1997}.
\newblock
%%CITATION = INSPIRE-454643;%%.

\bibitem{rowell2009classification}
E.~Rowell, R.~Stong, and Z.~Wang, ``On classification of modular tensor categories,'' {\em Communications in Mathematical Physics} {\bfseries 292} no.~2, (2009) 343--389, \href{http://arxiv.org/abs/0712.1377}{{\ttfamily arXiv:0712.1377}}.

\bibitem{Rose2014IntroductionTM}
S.~C.~F. Rose, ``Introduction to modular forms,'' {\em arXiv: Number Theory} (2014) 423--444.

\bibitem{A004011}
{OEIS Foundation Inc. (2022)}, ``Entry {A}004011 in the {O}n-line {E}ncyclopedia of {I}nteger {S}equences,.'' \url{https://oeis.org/A004011}.

\bibitem{Zudilin2003TheHE}
W.~Zudilin, ``The hypergeometric equation and ramanujan functions,'' {\em The Ramanujan Journal} {\bfseries 7} (2003) 435--447.

\bibitem{Sakai2014TheAO}
Y.~Sakai, ``The atkin orthogonal polynomials for the fricke groups of levels 5 and 7,'' {\em International Journal of Number Theory} {\bfseries 10} (2014) 2243--2255.

\bibitem{Das:2023qns}
A.~Das, C.~N. Gowdigere, S.~Mukhi, and J.~Santara, ``{Modular Differential Equations with Movable Poles and Admissible RCFT Characters},'' \href{http://arxiv.org/abs/2308.00069}{{\ttfamily arXiv:2308.00069 [hep-th]}}.

\bibitem{A052241}
{OEIS Foundation Inc. (2022)}, ``Entry {A}052241 in the {O}n-line {E}ncyclopedia of {I}nteger {S}equences,.'' \url{https://oeis.org/A052241}.

\bibitem{A070047}
{OEIS Foundation Inc. (2022)}, ``Entry {A}070047 in the {O}n-line {E}ncyclopedia of {I}nteger {S}equences,.'' \url{https://oeis.org/A070047}.

\bibitem{A007247}
{OEIS Foundation Inc. (2022)}, ``Entry {A}007247 in the {O}n-line {E}ncyclopedia of {I}nteger {S}equences,.'' \url{https://oeis.org/A007247}.

\bibitem{A058092}
{OEIS Foundation Inc. (2022)}, ``Entry {A}058092 in the {O}n-line {E}ncyclopedia of {I}nteger {S}equences,.'' \url{https://oeis.org/A058092}.

\bibitem{A058537}
{OEIS Foundation Inc. (2022)}, ``Entry {A}058537 in the {O}n-line {E}ncyclopedia of {I}nteger {S}equences,.'' \url{https://oeis.org/A058537}.

\bibitem{A007261}
{OEIS Foundation Inc. (2022)}, ``Entry {A}007261 in the {O}n-line {E}ncyclopedia of {I}nteger {S}equences,.'' \url{https://oeis.org/A007261}.

\bibitem{A097242}
{OEIS Foundation Inc. (2022)}, ``Entry {A}097242 in the {O}n-line {E}ncyclopedia of {I}nteger {S}equences,.'' \url{https://oeis.org/A097242}.

\bibitem{A328796}
{OEIS Foundation Inc. (2022)}, ``Entry {A}328796 in the {O}n-line {E}ncyclopedia of {I}nteger {S}equences,.'' \url{https://oeis.org/A328796}.

\bibitem{A008658}
{OEIS Foundation Inc. (2022)}, ``Entry {A}008658 in the {O}n-line {E}ncyclopedia of {I}nteger {S}equences,.'' \url{https://oeis.org/A008658}.

\bibitem{A008655}
{OEIS Foundation Inc. (2022)}, ``Entry {A}008655 in the {O}n-line {E}ncyclopedia of {I}nteger {S}equences,.'' \url{https://oeis.org/A008655}.

\bibitem{A008653}
{OEIS Foundation Inc. (2022)}, ``Entry {A}008653 in the {O}n-line {E}ncyclopedia of {I}nteger {S}equences,.'' \url{https://oeis.org/A008653}.

\bibitem{A028594}
{OEIS Foundation Inc. (2022)}, ``Entry {A}028594 in the {O}n-line {E}ncyclopedia of {I}nteger {S}equences,.'' \url{https://oeis.org/A028594}.

\bibitem{Duncan:2022afh}
J.~F.~R. Duncan, J.~A. Harvey, and B.~C. Rayhaun, ``{Two New Avatars of Moonshine for the Thompson Group},'' \href{http://arxiv.org/abs/2202.08277}{{\ttfamily arXiv:2202.08277 [math.RT]}}.

\bibitem{Carnahan2012GeneralizedMI}
S.~Carnahan, ``Generalized moonshine, ii: Borcherds products,'' {\em Duke Mathematical Journal} {\bfseries 161} (2012) 893--950.

\bibitem{Kim2002BorcherdsPA}
C.~H. Kim, ``Borcherds products associated with certain thompson series,'' {\em Compositio Mathematica} {\bfseries 140} (2002) 541 -- 551.

\bibitem{Magureanu:2022qym}
H.~Magureanu, ``{Seiberg-Witten geometry, modular rational elliptic surfaces and BPS quivers},'' \href{http://dx.doi.org/10.1007/JHEP05(2022)163}{{\em JHEP} {\bfseries 05} (2022) 163}, \href{http://arxiv.org/abs/2203.03755}{{\ttfamily arXiv:2203.03755 [hep-th]}}.

\bibitem{Sebbar2002ModularSF}
A.~Sebbar, ``Modular subgroups, forms, curves and surfaces,'' {\em Canadian Mathematical Bulletin} {\bfseries 45} (2002) 294 -- 308.

\bibitem{Niwa1977OnST}
S.~Niwa, ``On shimura's trace formula,'' {\em Nagoya Mathematical Journal} {\bfseries 66} (1977) 183 -- 202.

\bibitem{Kohnen1980ModularFO}
W.~Kohnen, ``Modular forms of half-integral weight on $\gamma_{0}$(4),'' {\em Mathematische Annalen} {\bfseries 248} (1980) 249--266.

\bibitem{Kohnen1982NewformsOH}
W.~Kohnen, ``Newforms of half-integral weight.,'' {\em Journal f{\"u}r die reine und angewandte Mathematik (Crelles Journal)} {\bfseries 1982} (1982) 32 -- 72.

\bibitem{Baruch2016NewformsOH}
E.~M. Baruch and S.~Purkait, ``Newforms of half-integral weight: The minus space counterpart,'' {\em Canadian Journal of Mathematics} {\bfseries 72} (2016) 326 -- 372.

\bibitem{Doran2013AutomorphicFF}
C.~F. Doran, T.~Gannon, H.~Movasati, and K.~M. Shokri, ``Automorphic forms for triangle groups,'' {\em arXiv: Number Theory} (2013) .

\bibitem{McKay2000FuchsianGA}
J.~McKay and A.~Sebbar, ``Fuchsian groups, automorphic functions and schwarzians,'' {\em Mathematische Annalen} {\bfseries 318} (2000) 255--275.

\bibitem{KZ}
M.~Kaneko and D.~Zagier, ``{Supersingular j-invariants, hypergeometric series, and Atkin's orthogonal polynomials},'' {\em AMS/IP Studies in Advanced Mathematics} {\bfseries 7} (1998) 97--126.

\bibitem{Liu}
Z.-G. Liu, ``Some eisenstein series identities related to modular equations of the seventh order,'' \href{http://dx.doi.org/10.2140/pjm.2003.209.103}{{\em Pacific Journal of Mathematics - PAC J MATH} {\bfseries 209} (03, 2003) 103--130}.

\end{thebibliography}\endgroup

\end{document}